\documentclass[a4paper,12pt,notitlepage]{article}
\usepackage{amsfonts}

\usepackage{amsmath}
\usepackage{graphicx,lscape}
\usepackage{epstopdf}

\usepackage{rotating}
\usepackage[english]{babel}
\usepackage[latin1]{inputenc}
\usepackage{latexsym}
\usepackage{amssymb}
\usepackage{amsthm}
\theoremstyle{plain}

\newtheorem{theorem}{Theorem}

\newtheorem*{algorithm*}{Algorithm}

\newtheorem{assumption}{Assumption}

\newtheorem{corollary}{Corollary}

\evensidemargin\oddsidemargin
\input{tcilatex}

\begin{document}

\title{
Recovering Latent Variables by Matching\thanks{%
We thank Tincho Almuzara and Beatriz Zamorra for excellent research assistance. We thank Colin Mallows, Alfred Galichon, Jiaying Gu, Kei Hirano, Pierre Jacob, Roger Koenker, Thibaut Lamadon, Guillaume Pouliot, Azeem Shaikh, Tim Vogelsang, Daniel Wilhelm, and audiences at various places for comments. Arellano acknowledges research funding from the Ministerio de Econom\'{\i}a y Competitividad, Grant ECO2016-79848-P. Bonhomme acknowledges support from the NSF, Grant SES-1658920.}}
\author{Manuel Arellano\thanks{%
CEMFI, Madrid.} \and St\'ephane
Bonhomme\thanks{%
University of Chicago.}}
\date{December 2019}
\maketitle

\begin{abstract}
\noindent    

We propose an optimal-transport-based matching method to nonparametrically estimate linear models with independent latent variables. The method consists in generating pseudo-observations from the latent variables, so that the Euclidean distance between the model's predictions and their matched counterparts in the data is minimized. We show that our nonparametric estimator is consistent, and we document that it performs well in simulated data. We apply this method to study the cyclicality of permanent and transitory income shocks in the Panel Study of Income Dynamics. We find that the dispersion of income shocks is approximately acyclical, whereas the skewness of permanent shocks is procyclical. By comparison, we find that the dispersion and skewness of shocks to hourly wages vary little with the business cycle. 
\bigskip


\bigskip


\end{abstract}

\baselineskip21pt

\bigskip

\bigskip

\setcounter{page}{0}\thispagestyle{empty}

\newpage

\section{Introduction\label{Intro_sec}}

In this paper we propose a method to nonparametrically estimate a class of models with latent variables. We focus on linear factor models whose latent factors are mutually independent. These models have a wide array of economic applications, including measurement error models, fixed-effects models, and error components models. We briefly review the existing literature in Section \ref{Models_sec}. In many empirical settings, such as in our application to the study of the cyclical behavior of income shocks, it is appealing not to restrict the functional form of the distributions of latent variables and adopt a nonparametric approach.

Nonparametric estimation based on empirical characteristic functions has been extensively studied in the literature (e.g., Carroll and Hall, 1988; Stefanski and Carroll, 1990). However, while such Fourier-based methods apply to general mutivariate linear factor models with independent components (Horowitz and Markatou, 1996; Li and Vuong, 1998; Bonhomme and Robin, 2010), they tend to be sensitive to the choice of regularization parameters, and they do not guarantee that the estimated densities be non-negative and integrate to one. Recently, Efron (2016) motivated his ``parametric g-modeling'' approach by the difficulties of nonparametric estimation in this context; see also Efron and Hastie (2016, Chapter 21) and Koenker and Gu (2019).  

In this paper we propose a novel nonparametric estimator, and provide evidence that it performs well even in relatively small samples. Our approach differs from the literature in two main aspects. First, we generate a sample of \emph{pseudo-observations} that may be interpreted as the order statistics of the latent variables. Moments, densities, or other functionals can then be estimated based on them. In particular, densities will be non-negative and integrate to one by construction. Means or other features of the distribution of the latent variables conditional on the data, such as optimal predictors, can also be directly estimated. 

The second main feature of our approach is that it is based on \emph{matching}. Specifically, we generate pseudo-observations from the latent variables so that the Euclidean distance between the model's predictions and their matched counterparts in the data is minimized. The model predictions are computed as independent combinations of the pseudo latent observations. This ``observation matching'' estimation approach can be interpreted as a nonparametric counterpart to (simulated) method-of-moments estimators, which are commonly used in parametric econometric models. Our nonparametric approach, which amounts to minimizing a quadratic \emph{Wasserstein} distance between empirical distribution functions, exploits linearity and independence to provide a computationally convenient estimator.

\begin{figure}[tbp]
	\caption{Illustration of the estimation algorithm}
	\label{Fig_illus}
	\begin{center}
		\begin{tabular}{ccc}
			\multicolumn{3}{c}{Data and matched predicted observations}\\
			1st iteration &2nd iteration & 5th iteration\\
			\includegraphics[width=50mm, height=40mm]{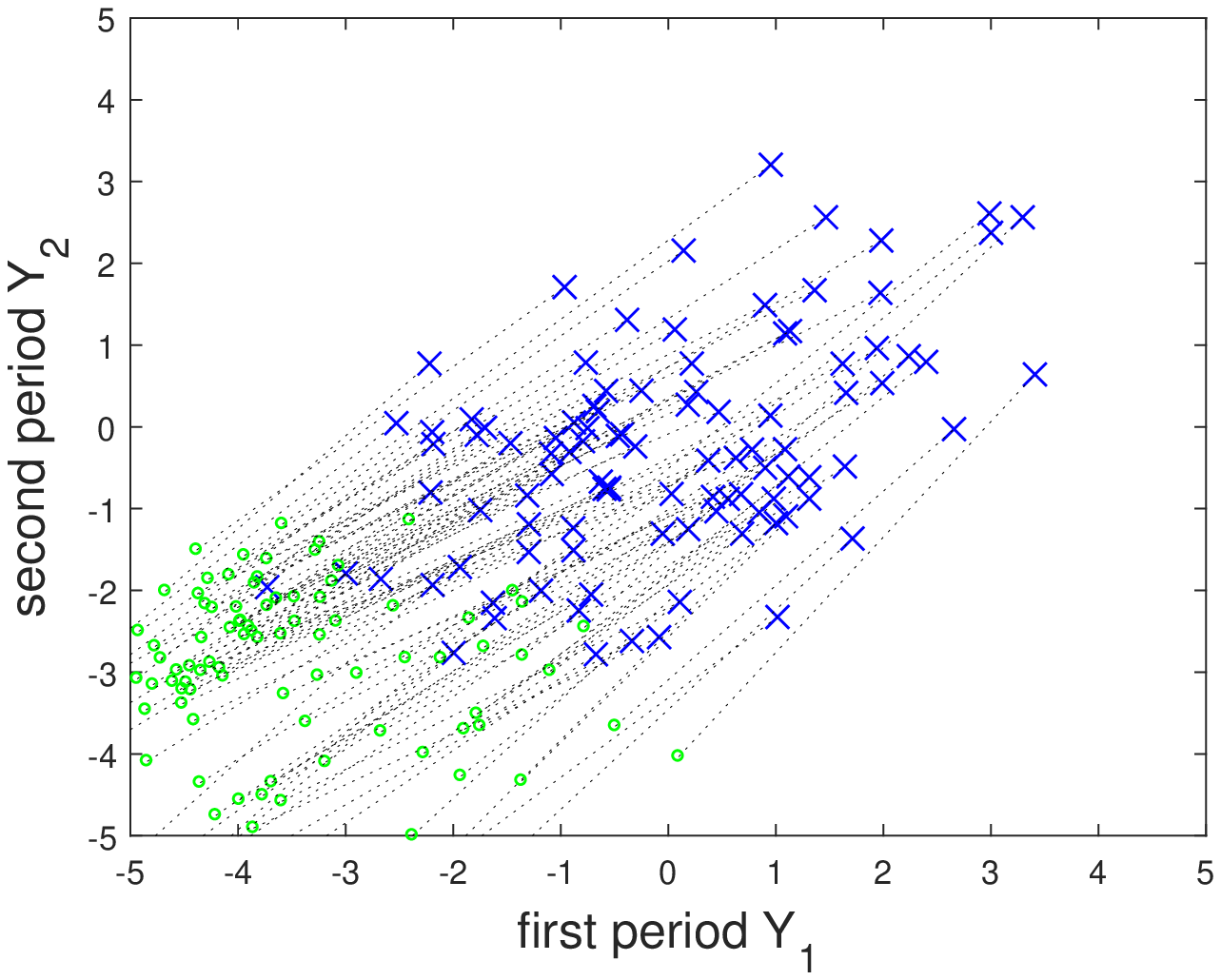} & %
			\includegraphics[width=50mm, height=40mm]{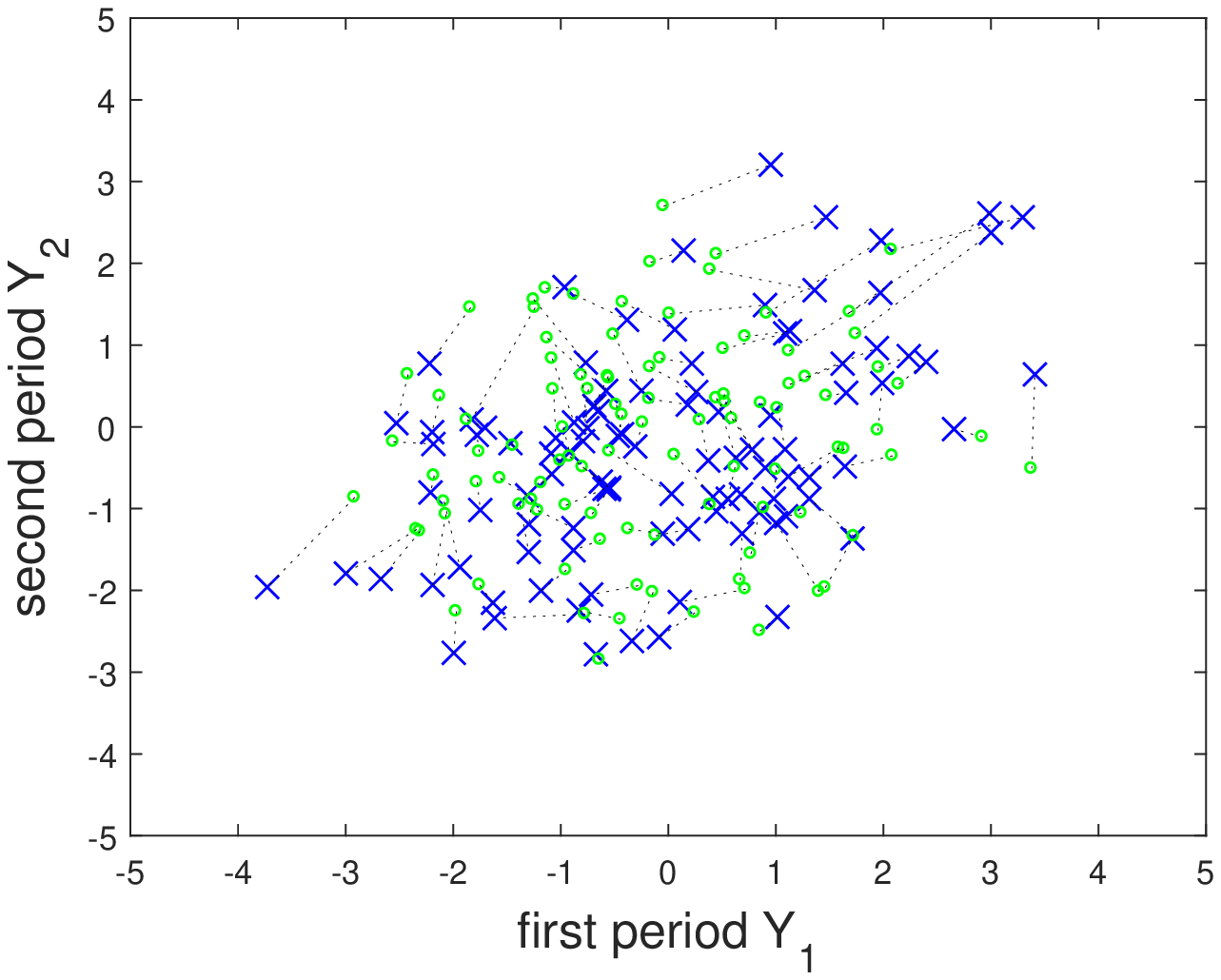} & 
			\includegraphics[width=50mm, height=40mm]{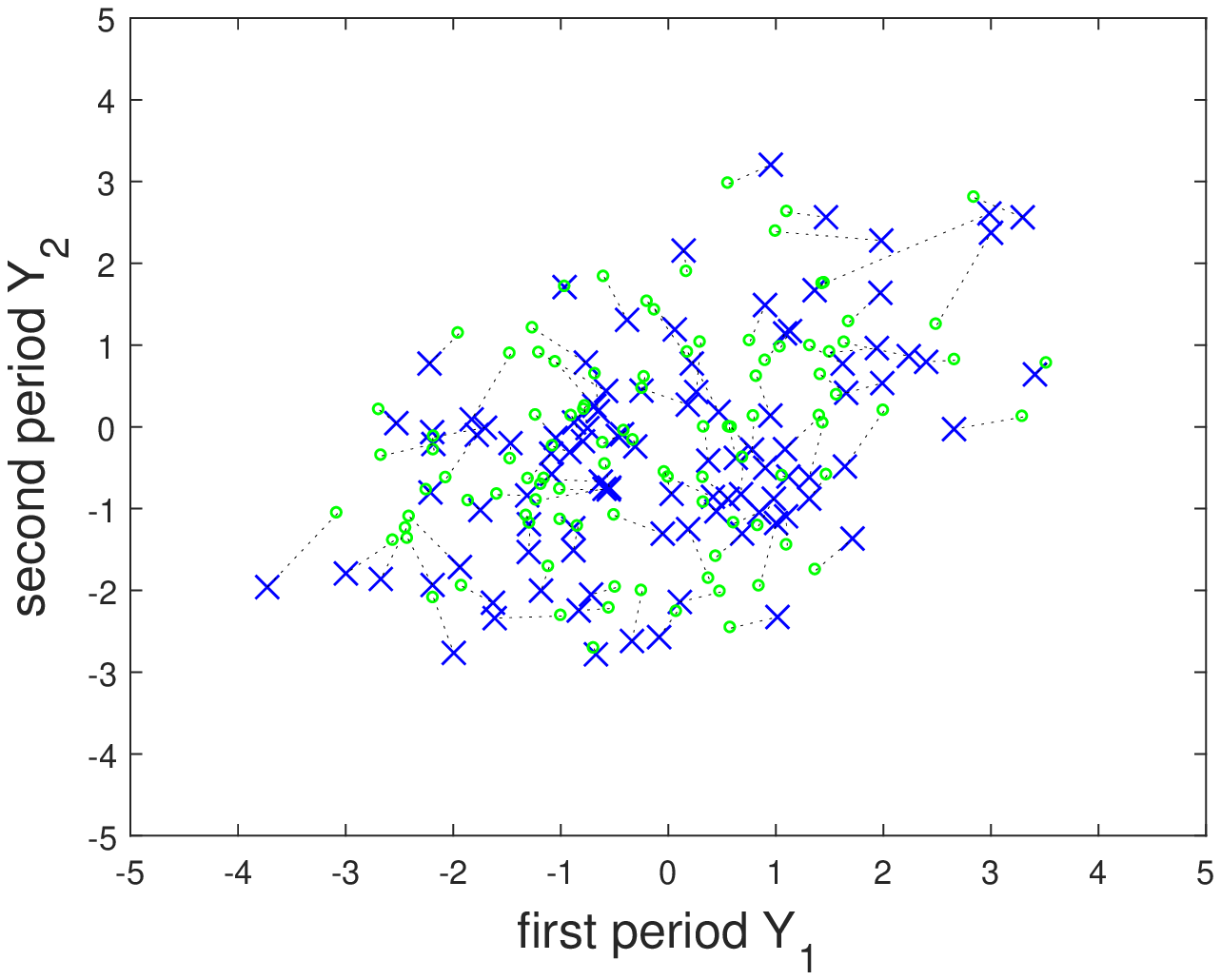}\\
			&  &\\
			\multicolumn{3}{c}{True versus estimated latent values}\\
			1st iteration &2nd iteration & 5th iteration\\
			\includegraphics[width=50mm, height=40mm]{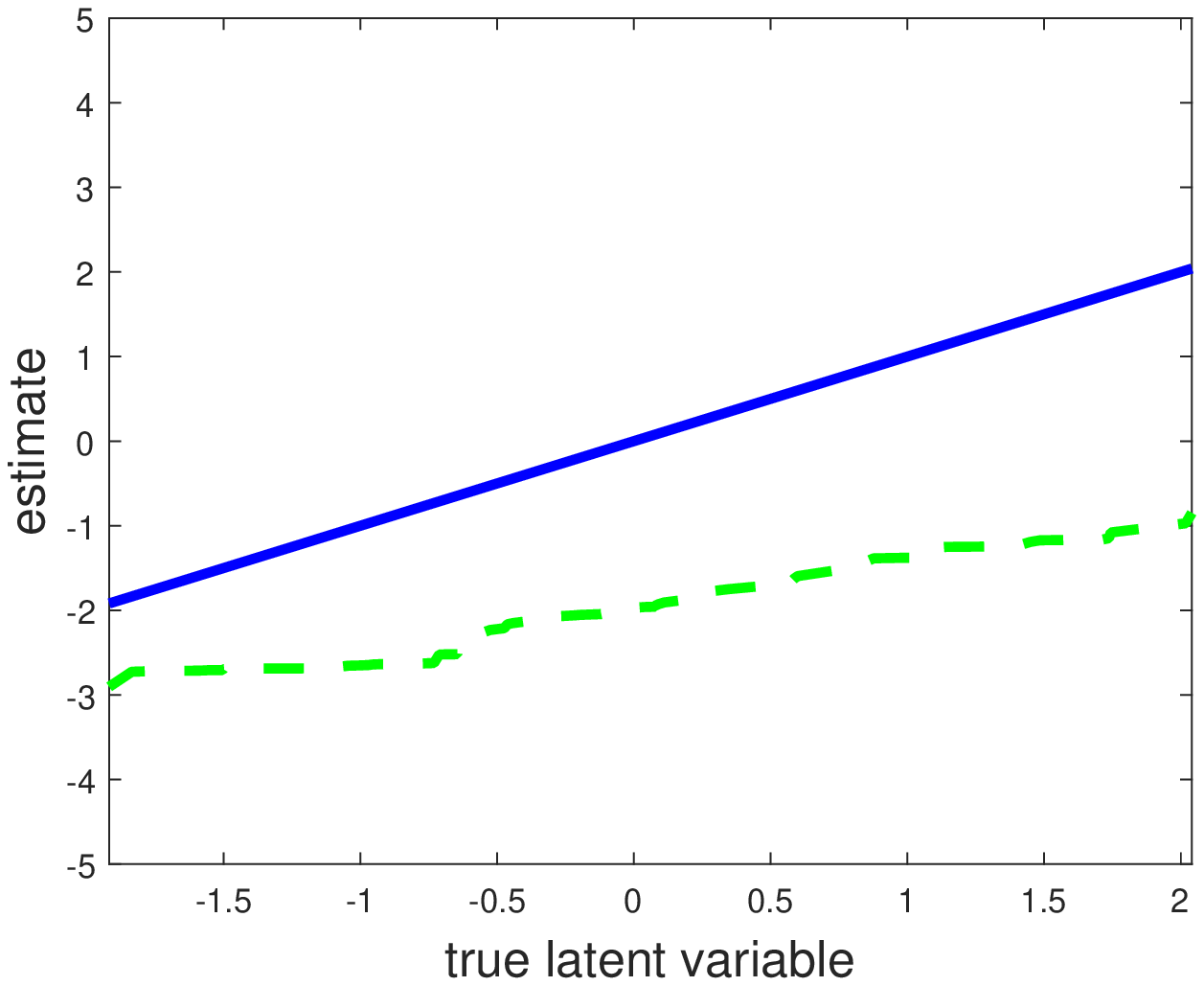} & %
			\includegraphics[width=50mm, height=40mm]{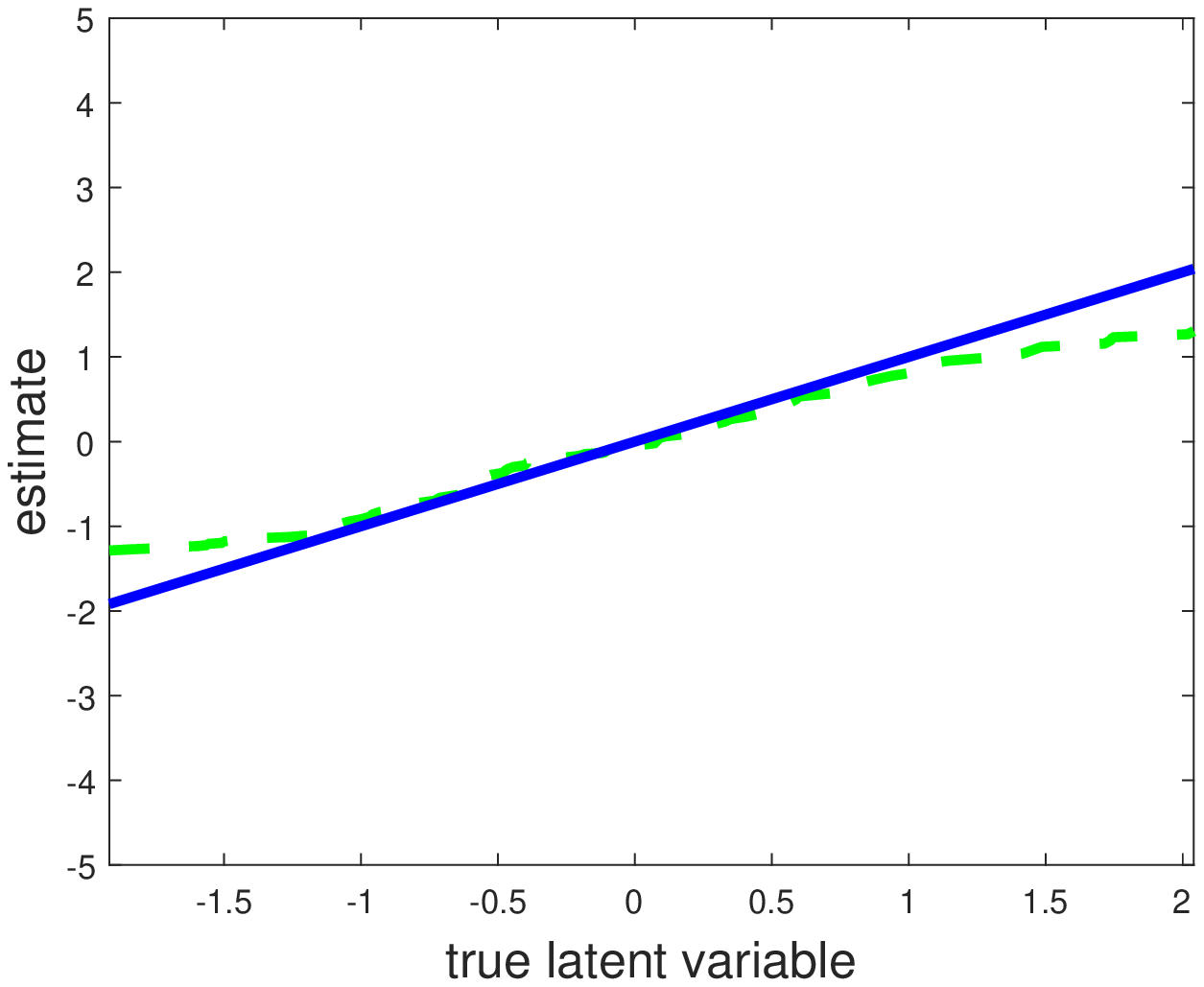} & 
			\includegraphics[width=50mm, height=40mm]{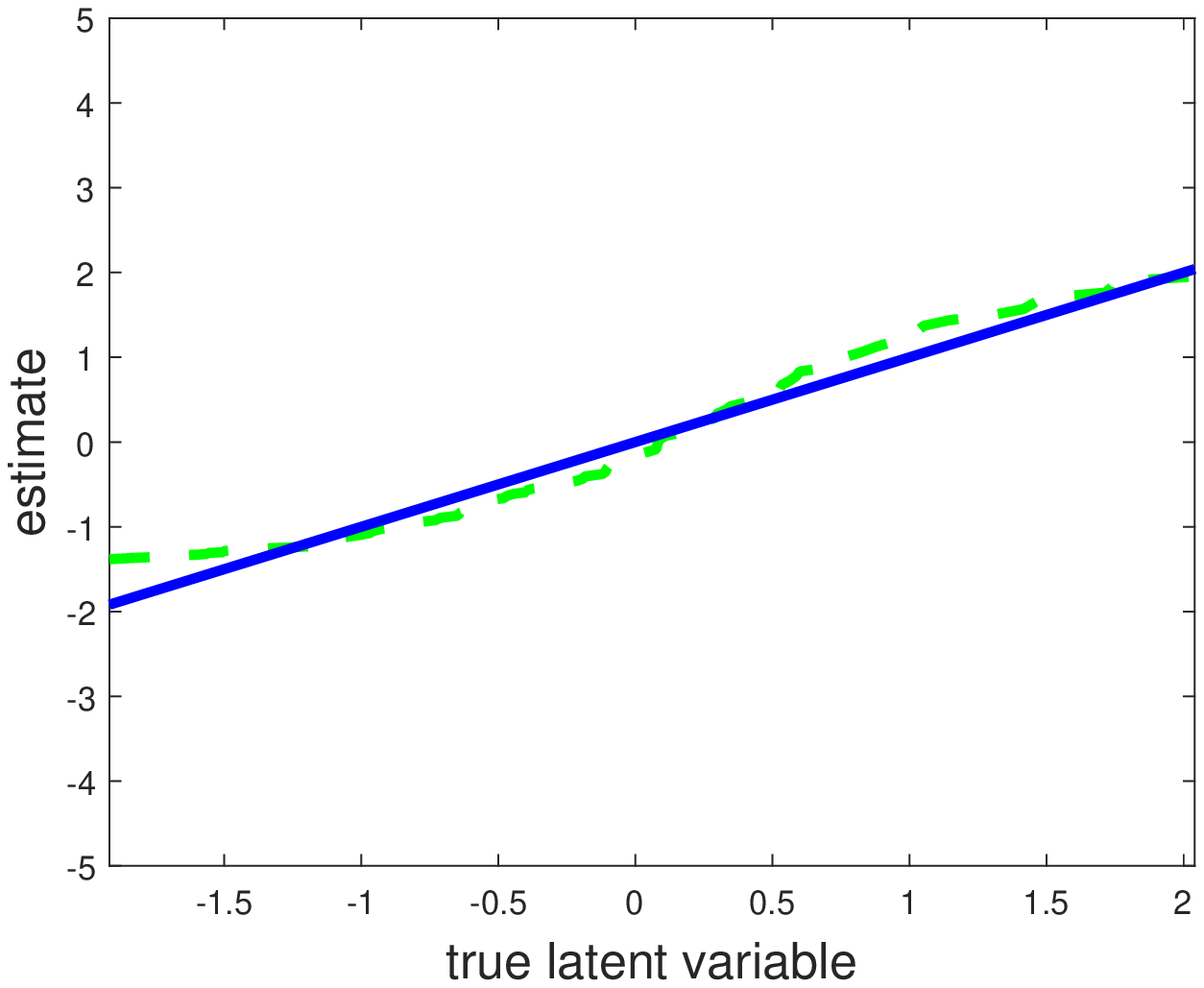}
			\\\end{tabular}
	\end{center}
	\par
	\textit{{\footnotesize Notes: The graphs correspond to one simulation from a fixed-effects model with two observation periods $Y_1=X_1+X_2$, $Y_2=X_1+X_3$, with $X_1,X_2,X_3$ mutually independent (Kotlarski, 1967). In the data generating process the $X$'s are standardized Beta(2,2), and there are $N=100$ observations. The top panel shows the observations $Y_1,Y_2$ (crosses) and the predicted observations $Y_1^{pred},Y_2^{pred}$ (circles), with a link between them when they are matched to each other. The bottom panel shows the estimates of $X_1$ values sorted in ascending order on the y-axis against the population values on the x-axis (dashed), and the 45 degree line (solid). See Section \ref{Estim_sec} for details about the algorithm. 		}}
	\end{figure}

As an illustration, in Figure \ref{Fig_illus} we show the results of several iterations of our algorithm, in a fixed-effects model with two observation periods and 100 individuals. We start the algorithm from parameter values that are far from the true ones (in the left column). As shown on the top panel, the outcome observations in the data (in crosses) are first matched to model-based predictions (in circles). Pseudo-observations of the latent variables are then updated based on the matched outcome values. The objective function we aim to minimize is the sum of squares of the segments shown on the top panel. The bottom panel shows the estimates of the latent individual-specific effect sorted in ascending order (on the y-axis), against the true values (on the x-axis). We see that, within a few iterations, the model's predictions and the empirical observations tend to agree with each other (in the top panel), and that the distribution of the pseudo latent observations gets close to the population distribution (in the bottom panel).

Our approach builds on and generalizes an important idea due to Colin Mallows (2007), who proposed a ``deconvolution by simulation'' method based on iterating between sorts of the data and random permutations of pseudo-observations of a latent variable. Mallows (2007) focused on the classical deconvolution model with scalar outcome and known error distribution. Our main goal in this paper is to extend Mallows' insight by proposing a framework to analyze estimators based on matching predicted values from the model to data observations.

In particular, as an extension of Mallows' (2007) original idea, we show how our method can handle multivariate outcomes, hence extending the scope of application to fixed-effects models and multi-factor models. While a number of estimation methods are available for scalar nonparametric deconvolution with known error distribution, the multivariate case -- which is of interest in many economic applications -- remains challenging. Our estimator exploits that the multi-factor models we consider are linear in the independent latent variables, even though they imply nonlinear restrictions on density functions.

A key step in our analysis is to relate the estimation problem to optimal transport theory. Optimal transport is the subject of active research in mathematics, see for example Villani (2003, 2008). Economic applications of optimal transport are many fold, as documented in Galichon (2016). In our context, optimal transport provides a natural way to estimate models with multivariate outcomes via ``generalized sorting'' algorithms (i.e., matching algorithms) based on linear programming. 

To establish the consistency of our estimator we use that, in large samples, our estimator minimizes the Wasserstein distance between the population distribution of the data and the one implied by the model. This problem has a unique solution under suitable conditions on the characteristic functions of the factors (Sz\'ekely and Rao, 2000). Consistency then follows from verifying the conditions for the consistency of sieve extremum estimators (e.g., Chen, 2007) in this setting. When analyzing the multivariate case, our arguments rely on properties of Wasserstein distances established in the optimal transport literature.

We illustrate the performance of our estimator on simulated data. Under various specifications of a nonparametric fixed-effects model, we find that our estimator recovers accurately the true underlying quantile functions and densities, even for samples with only 100 individual observations. This finite-sample performance is remarkable in a fully nonparametric model with multiple latent variables. In addition, we find that our estimator outperforms characteristic-function based estimators, particularly due to improved estimation of the tails of the distributions. In contrast with Fourier methods, our estimator imposes that quantile functions be monotone, and that densities be non-negative. In the related problem of nonparametric instrumental variables estimation, Chetverikov and Wilhelm (2017) show that imposing monotonicity in estimation can help alleviating ill-posedness issues. We conjecture that this feature contributes to explain the finite-sample performance of our estimator.

We then apply our method to study the cyclicality of permanent and transitory income shocks in the US. Answering this question is important, since a well-calibrated cyclical income process is a key input to many models of business cycle dynamics. Storesletten \textit{et al.} (2004) estimate on the Panel Study of Income Dynamics (PSID) that the dispersion of persistent shocks is countercyclical. However, using a nonparametric descriptive analysis, Guvenen \textit{et al.} (2014) find using administrative data that the dispersion of log-income growth is acyclical, whereas skewness is procyclical, and Busch \textit{et al.} (2018) find similar results using the PSID. 

We revisit this debate by working with a permanent-transitory model of log-income dynamics, and estimating the annual densities of permanent and transitory shocks nonparametrically. Using the PSID, we estimate that income shocks are not normally distributed, confirming previous evidence using other nonparametric methods. Our main finding is that the dispersion of income shocks is approximately acyclical, whereas the skewness of permanent shocks is procyclical. By comparison, our nonparametric estimates suggest that the dispersion and skewness of shocks to hourly wages vary little with the business cycle.

Our matching-based, minimum Wasserstein distance estimator is related to recent work in machine learning and statistics on the estimation of parametric generative models (see Bernton \textit{et al.}, 2017; Genevay \textit{et al.}, 2017; Bousquet \textit{et al.}, 2017). In contrast with this emerging literature, the models we consider here are nonparametric. In an early theoretical contribution, Bassetti \textit{et al.} (2006) study consistency in minimum Wasserstein distance estimation. Recently, Rigollet and Weed (2019) develop a minimum Wasserstein deconvolution approach for uncoupled isotonic regression, and Rigollet and Weed (2018) relate maximum-likelihood scalar deconvolution under Gaussian noise to entropic regularized optimal transport. Lastly, our general estimation strategy is also related to Galichon and Henry's (2011) analysis of partially identified models.

As we show at the end of the paper, our matching approach can be generalized to nonparametric estimation of other latent variables models. We briefly describe such generalizations to cross-sectional random coefficients models with exogenous covariates (Beran and Hall, 1992; Ichimura and Thompson, 1998), panel data random coefficients models (Arellano and Bonhomme, 2012), nonparametric deconvolution under heteroskedasticity (Delaigle and Meister, 2008), and nonparametric finite mixture models (Hall and Zhou, 2003).

The outline of the paper is as follows. In Section \ref{Models_sec} we describe linear independent factor models, and we briefly review applications and existing estimation approaches. In Section \ref{Estim_sec} we introduce our matching estimator. In Sections \ref{Comput_sec} and \ref{Asympt_sec} we study computation and consistency, respectively. In Sections \ref{Simu_sec} and \ref{Appli_sec} we present the simulation exercises and empirical application. In Section \ref{Extens_sec} we outline several extensions. Lastly, we conclude in Section \ref{Conclu_sec}. Proofs and additional material are collected in the appendix.

\section{Independent factor models\label{Models_sec}}

We focus on linear independent factor models of the form $Y=AX$, where $Y=(Y_1,...,Y_T)'$, $X=(X_{1},...,X_{K})'$, $A$ is a known or consistently estimable $T\times K$ matrix, and the components $X_1,...,X_K$ are mutually independent. In this section we review several examples of models and applications that have such a structure. We focus on the case $K>T$, so the system is singular and the realizations of the latent variables are not identifiable, although under suitable conditions their distributions will be. 

\paragraph{Nonparametric deconvolution.}

When $T=1$, $Y=X_1+X_2$, and $X_2$ has a known or consistently estimable distribution, one obtains the scalar nonparametric deconvolution model. This model has been extensively studied in statistics and econometrics. Nonparametric deconvolution is often used to deal with the presence of measurement error. In such settings, $Y$ is an error-ridden variable, $X_1$ is the true value of the variable, and $X_2$ is an independent, classical measurement error (e.g., Carroll \textit{et al.}, 2006; Chen \textit{et al.}, 2011; Schennach, 2013a). Other economic applications of nonparametric deconvolution are the estimation of the heterogeneous effects of an exogenous binary treatment under the assumption that the potential outcome in the absence of treatment is independent of the gains from treatment (Heckman \textit{et al.}, 1997; Wu and Perloff, 2006), and the estimation of the distribution of time-invariant random coefficients of binary treatments in panel data models (Arellano and Bonhomme, 2012). 

The statistical literature on nonparametric deconvolution provides conditions under which the distribution of $X_1$ is nonparametrically identified, alongside numerous estimation approaches such as kernel deconvolution estimators (Carroll and Hall, 1988; Delaigle and Gijbels, 2002; Fan, 1991), wavelet methods (Pensky and Vidakovic, 1999; Fan and Koo, 2002), regularization techniques (Carrasco and Florens, 2011), and nonparametric maximum likelihood methods (Kiefer and Wolfowitz, 1956; Gu and Koenker, 2017).

\paragraph{Nonparametric distribution of fixed effects.}

A leading example of a linear independent factor model is the fixed-effects model:
\begin{equation}\label{eq_Kotlarski}
Y_t=\underset{\equiv X_1}{\underbrace{\alpha}}+\underset{\equiv X_{t+1}}{\underbrace{\varepsilon_t}},\quad t=1,...,T,
\end{equation}  
where $Y_1,...,Y_T$ are observed outcomes and $\alpha,\varepsilon_1,...,\varepsilon_T$ are latent and mutually independent. Working with $T=2$, Kotlarski (1967) provided simple conditions under which the density functions of the latent factors are nonparametrically identified in model (\ref{eq_Kotlarski}).

This fixed-effects structure arises frequently in economic applications. As an example, $\alpha$ can be a latent skill of an individual, measured with error (as in Cunha \textit{et al.}, 2010). In other applications, researchers may be interested in estimating the distribution of worker, teacher, firm, school, hospital, neighborhood, or bank-specific fixed-effects, for example. Compared to commonly used Gaussian specifications (e.g., Kane and Staiger, 2008; Angrist \textit{et al.}, 2017; Chetty and Hendren, 2018), a nonparametric estimator of the distribution of $\alpha$ in (\ref{eq_Kotlarski}) will be robust to functional form assumptions under the maintained assumption of mutual independence. Non-Gaussianity, such as skewness or fat tail behavior, is relevant in many empirical settings. The fixed-effects model (\ref{eq_Kotlarski}) and its generalizations are sometimes estimated using flexible parametric specifications such as finite Gaussian mixtures (e.g., Carneiro \textit{et al.}, 2003). Alternatively, nonparametric estimators based on empirical characteristic functions can be constructed, by mimicking and extending Kotlarski's proof (Li and Vuong, 1998; Li, 2002; Horowitz and Markatou; 1996).

\paragraph{Error components: generalized nonparametric deconvolution.}

A prominent error component model is the permanent-transitory model for the dynamics of log-income: $Y_t=\eta_t+\varepsilon_t$, where $\eta_t=\eta_{t-1}+v_t$ is a random walk with independent innovations, and all $\varepsilon_t$'s and $v_t$'s are independent over time and independent of each other and of the initial $\eta_0$ (e.g., Hall and Mishkin, 1982; Blundell \textit{et al.}, 2008). This model is a special case of a linear independent factor model $Y=AX$, where $Y=(Y_1,...,Y_T)'$ are observed outcomes, $X=(X_1,...,X_K)'$ are mutually independent latent factors, and $A$ is a known $T\times K$ matrix. Identification of such generalized deconvolution models is established in Sz\'ekely and Rao (2000). Bonhomme and Robin (2010) propose nonparametric characteristic-function based estimators of factor densities, and apply them to study income dynamics; see also Botosaru and Sasaki (2015).\footnote{Quantile-based estimation in linear and nonlinear factor models is introduced in Arellano and Bonhomme (2016), and applied in Arellano \textit{et al.} (2017) to document income dynamics in the PSID.} In such settings, a nonparametric approach is able to capture the skewness and kurtosis of income shocks. In addition, an important application of error components models is to relax independence in fixed-effects models such as (\ref{eq_Kotlarski}). This can be done provided $T$ is large enough.\footnote{Modeling $\varepsilon_t$ in (\ref{eq_Kotlarski}) as a finite-order moving average or autoregressive process with independent innovations preserves the linear independent factor structure of the model (Arellano and Bonhomme, 2012; see also Hu \textit{et al.}, 2019). Ben Moshe (2017) shows how to allow for arbitrary subsets of dependent factors, and proposes characteristic-function based estimators. In addition, in model (\ref{eq_Kotlarski})  Schennach (2013b) points out that full independence between the factors is not necessary, and that sub-independence suffices to establish identification.} Such specifications can be estimated using the methods we introduce in this paper.

\section{Latent variable estimation by matching\label{Estim_sec}}

In this section, to introduce the main ideas we start by describing our estimator in the scalar nonparametric deconvolution model. We then show how the same approach can be used to estimate linear multi-factor models with independent factors. 

\subsection{Nonparametric deconvolution}

Let $Y=X_1+X_2$ be a scalar outcome, where $X_1$ and $X_2$ are independent, $X_1$ is unobserved to the econometrician, and its distribution is unspecified. We assume that $Y$, $X_1$ and $X_2$ are continuously distributed, and postpone more specific assumptions until Section \ref{Asympt_sec}. Let $F_Z$ denote the cumulative distribution function (cdf) of any random variable $Z$. We assume that two random samples, $Y_1,...,Y_N$ and $X_{12},...,X_{N2}$, drawn from $F_Y$ and $F_{X_2}$, respectively, are available.\footnote{The sample sizes being the same for $Y$ and $X_2$ is not essential and can easily be relaxed. In a setting where the cdf $F_{X_2}$ is known, one can draw a sample from it, or alternatively work with an integral counterpart to our estimator.} 

Our goal is to estimate a sample of \emph{pseudo-observations} $\widehat{X}_{11},...,\widehat{X}_{N1}$, whose empirical cdf is asymptotically distributed as $F_{X_1}$ as $N$ tends to infinity. To do so, we minimize a distance between the sample of observed $Y$'s and a sample of $Y$'s predicted by the model. We rely on the quadratic \emph{Wasserstein distance} (see, e.g., Chapter 7 in Villani, 2003), which is the minimum Euclidean distance between observed $Y$'s and predicted $Y$'s with respect to all possible reorderings of the observations. 

Formally, assume without loss of generality that $Y_i\leq Y_{i+1}$ and $X_{i2}\leq X_{i+1,2}$ for all $i$. Let $\Pi_N$ denote the set of permutations $\pi$ of $\{1,...,N\}$. Moreover, let $\overline{C}_N>0$ and $\underline{C}_N>0$ be two constants, and let ${\cal{X}}_{N}$ be the set of parameter vectors $X_1=(X_{11},...,X_{N1})\in\mathbb{R}^N$ such that $|X_{i1}|\leq \overline{C}_N$ and $\underline{C}_N\leq (N+1)(X_{i+1,1}-X_{i1})\leq \overline{C}_N$ for all $i$. The constants $\underline{C}_N$ and $\overline{C}_N$ play a role in our consistency argument below, and we will study how their choice affects our estimator in simulations. We propose to compute:
\begin{align}\label{dec_est}
\widehat{X}_1=\underset{X_1\in{\cal{X}}_N}{\limfunc{argmin}}\,\left\{\underset{\pi\in \Pi_N}{\limfunc{min}}\, \sum_{i=1}^N\left(Y_{\pi(i)}-X_{\sigma(i),1}-X_{i,2}\right)^2\right\},
\end{align}
where $\sigma$ is a random permutation in $\Pi_N$ (i.e., a uniform draw on $\Pi_N$), independent of $Y_1,...,Y_N,X_{12},...,X_{N2}$.

To interpret the objective function on the right-hand side of (\ref{dec_est}), note that, for any random permutation $\sigma$, $Z_i\equiv X_{\sigma(i),1}+X_{i,2}$, $i=1,...,N$, are $N$ draws from the model. Predicted values from the model could be generated in other ways. For example, one could instead compute $X_{i1}+\widetilde{X}_{i2}$, where $\widetilde{X}_{i2}$ are i.i.d. draws from the empirical distribution of $X_{i2}$. Alternatively, one could generate $R>1$ predictions per observation $i$, although here we take $R=1$ to minimize computation cost.\footnote{Specifically, one could compute $X_{\sigma(i,r),1}+X_{i2}$, with $\sigma(\cdot,1),...\sigma(\cdot,R)$ being $R$ independent permutations. In that case, $\pi$ would be a generalized permutation (or ``pure matching''), mapping $\{1,...,N\}^R$ to $\{1,...,N\}$.}

A simple way to reduce the dependence of the estimator on the random $\sigma$ draw is to compute $\widehat{X}^{(m)}_{i1}$, for $i=1,...,N$ and $m=1,...,M$, where $\sigma^{(1)},...,\sigma^{(M)}$ are independent random permutations drawn from $\Pi_N$, and to report the averages: $\widehat{X}_{i1}=\frac{1}{M}\sum_{m=1}^M\widehat{X}^{(m)}_{i1}$, for $i=1,...,N$. For fixed $M$, such averages will be consistent as $N$ tends to infinity under similar conditions as our baseline estimator.

The estimator $\widehat{X}_1$ in (\ref{dec_est}) minimizes the Wasserstein distance between the empirical distributions of the model predictions $Z_i=X_{\sigma(i),1}+X_{i2}$ and the outcome observations $Y_i$. The Wasserstein distance is defined as:
\begin{equation}W_2(\widehat{F}_Y,\widehat{F}_Z)=\left\{\underset{\pi\in \Pi_N}{\limfunc{min}}\,\sum_{i=1}^N\left(Y_{\pi(i)}-Z_i\right)^2\right\}^{\frac{1}{2}}.\label{eq_wass_emp}\end{equation}
Since $Y_i$ and $Z_i$ are scalar, the Hardy-Littlewood-P\'olya rearrangement inequality implies that the solution to (\ref{eq_wass_emp}) is to sort $Y_i$'s and $Z_i$'s in the same order. That is, letting $\widehat{\pi}$ denote the minimum argument in (\ref{eq_wass_emp}), $\widehat{\pi}(i)=\limfunc{Rank}(Z_i)\equiv N\widehat{F}_Z(Z_i)$ is the \emph{rank} of $Z_i$.

\subsection{Nonparametric factor models}

We now apply the same idea to a general linear independent multi-factor model $Y=AX$, where $A$ is a $T\times K$ matrix with generic element $a_{tk}$, and $X=(X_1,...,X_K)'$ with $X_1,...,X_K$ mutually independent. For simplicity we assume that $X$ and $Y$ have zero mean.\footnote{It is common in applications to assume that some of the $X_k$'s have zero mean while leaving the remaining means unrestricted. For example, in the fixed-effects model, assuming that $\mathbb{E}(X_1)=0$ suffices for identification. Our algorithm can easily be adapted to such cases.} We seek to compute pseudo-observations $\widehat{X}_{11},...,\widehat{X}_{N1}$, ..., $\widehat{X}_{1K},...,\widehat{X}_{NK}$, which minimize the Wasserstein distance between the sample of observed $Y$'s, which here are $T\times 1$ vectors, and the sample of $Y$'s predicted by the factor model.

As before, let $\overline{C}_N>0$ and $\underline{C}_N>0$ be two constants, and let ${\cal{X}}_N$ be the set of $(X_1,...,X_N)\in\mathbb{R}^{NK}$ such that $|X_{i,k}|\leq \overline{C}_N$ and $\underline{C}_N\leq (N+1)(X_{i+1,k}-X_{ik})\leq \overline{C}_N$ for all $i$ and $k$, and $\sum_{i=1}^N X_{ik}=0$ for all $k$. We define:
\begin{align}\label{dec_est_gen}
&\widehat{X}=\underset{X\in{\cal{X}}_N}{\limfunc{argmin}}\, \left\{\underset{\pi\in \Pi_N}{\limfunc{min}}\, \sum_{i=1}^N\sum_{t=1}^T\left(Y_{\pi(i),t}-\sum_{k=1}^Ka_{tk}X_{\sigma_k(i),k}\right)^2\right\},
\end{align}
where $\sigma_1,...,\sigma_K$ are independent random permutations in $\Pi_N$, independent of $Y_{11},...,Y_{NT}$.

As in the scalar case, $Z_{it}\equiv\sum_{k=1}^Ka_{tk}X_{\sigma_k(i),k}$, $i=1,...,N$, $t=1,...,T$, are $NT$ predicted values from the factor model. Hence, as before, the vector $\widehat{X}$ minimizes the Wasserstein distance between the empirical distributions of the data $(Y_{i1},...,Y_{iT})$ and of the model predictions $(Z_{i1},...,Z_{iT})$. A difference with the scalar deconvolution model is that, when $Y_i$ are multivariate, the minimization with respect to $\pi$ inside the brackets in (\ref{dec_est_gen}) does not have an explicit form in general. However, from optimal transport theory it is well-known that the solution can be obtained as the solution to a linear program. We will exploit this feature in our estimation algorithm.

\paragraph{Densities and expectations.}

In Section \ref{Asympt_sec} we will provide conditions under which $\widehat{X}_{ik}$, $i=1,...,N$, consistently estimate the quantile function of $X_k$. More precisely, we will show that $\max_{i=1,...,N}\, |\widehat{X}_{ik}-F_{X_k}^{-1}(\frac{i}{N+1})|$ tends to zero in probability asymptotically. This provides uniformly consistent estimators of the quantile functions of the latent variables, which can in turn be used for density estimation under a slight modification of the parameter space ${\cal{X}}_N$. Indeed, let us restrict the parameter space to elements $X=(X_1,...,X_N)$ in ${\cal{X}}_N$ which satisfy the following additional restrictions on second-order differences: $(N+1)^2\left|X_{i+2,k}-2X_{i+1,k}+X_{i,k}\right|\leq \overline{C}_N$, for all $i$ and $k$. Let us then define, for a bandwidth parameter $b>0$ and a kernel function $\kappa\geq 0$ that integrates to one:
\begin{equation}\label{f_hat_estim}
\widehat{f}_{X_k}(x)=\frac{1}{Nb}\sum_{i=1}^N\kappa\left(\frac{\widehat{X}_{ik}-x}{b}\right),\quad x\in\mathbb{R}.
\end{equation}
We will show that $\widehat{f}_{X_k}$ is uniformly consistent for the density of $X_k$, under standard conditions on the kernel $\kappa$ and bandwidth $b$. 

In addition, our estimator delivers simple consistent estimators of unconditional and conditional expectations, as we show in Appendix \ref{App_Extens}. As an example of practical interest, in the fixed-effects model (\ref{eq_Kotlarski}) the best predictor of $X_1$ under squared loss can be estimated as:
\begin{equation}
\label{eq_cond_exp}\widehat{\mathbb{E}}(X_1\,|\, Y=Y_i)=\sum_{i=1}^N\widehat{\omega}_{i}\widehat{X}_{i1},
\end{equation}
where the weights $\widehat{\omega}_{i}$ are given by: 
$$\widehat{\omega}_{i}=\frac{\prod_{t=1}^T\widehat{f}_{X_{t+1}}(Y_{it}-\widehat{X}_{i1})}{\sum_{j=1}^N\prod_{t=1}^T\widehat{f}_{X_{t+1}}(Y_{jt}-\widehat{X}_{j1})},\quad i=1,...,N.$$

\section{Computation\label{Comput_sec}}

The optimization problems in (\ref{dec_est}) and (\ref{dec_est_gen}) are mixed integer quadratic programs. Although the literature on mixed integer programming has recently made substantial progress (e.g., Bliek \textit{et al.}, 2014), exact algorithms are currently limited in the dimensions they can allow for. Here we describe a simple, practical method to minimize (\ref{dec_est}) and (\ref{dec_est_gen}).


\subsection{Algorithm}

The algorithm we propose is based on the observation that, for given $X_{11},...,X_{NK}$ values, (\ref{dec_est_gen}) is a \emph{linear assignment} (or \emph{discrete optimal transport}) problem, hence it can be solved by any linear programming routine. In turn, given $\pi$, (\ref{dec_est_gen}) is a monotone least squares problem. Our estimation algorithm is as follows. Here we focus on the general form (\ref{dec_est_gen}), since the estimator for the scalar deconvolution model (\ref{dec_est}) is a special case of it.

\begin{algorithm*}$\quad$
	
		\begin{itemize}
		\item Start with initial values $\widehat{X}_1^{(1)},...,\widehat{X}_N^{(1)}$ in $\mathbb{R}^K$. Iterate the following two steps on $s=1,2,...$ until convergence.
		\item (Matching step) Given $\widehat{X}_1^{(s)},...,\widehat{X}_N^{(s)}$, compute:\footnote{Notice that, since $\pi$ is a permutation, $\sum_{i=1}^N\sum_{t=1}^TY_{\pi(i),t}^2=\sum_{i=1}^N\sum_{t=1}^TY_{it}^2$ does not depend on $\pi$.}
		\begin{align}\label{eq_step1}
		\widehat{\pi}^{(s+1)}&=\underset{\pi\in \Pi_N}{\limfunc{argmin}}\, \sum_{i=1}^N\sum_{t=1}^T\left(Y_{\pi(i),t}-\sum_{k=1}^Ka_{tk}\widehat{X}^{(s)}_{\sigma_k(i),k}\right)^2\notag\\&=\underset{\pi\in {\Pi}_{N}}{\limfunc{argmax}}\,\,\, \sum_{i=1}^N\sum_{t=1}^T \left(\sum_{k=1}^K a_{tk} \widehat{X}^{(s)}_{\sigma_k(i),k}\right)Y_{\pi(i),t}.
		\end{align}
		\item (Update step) Compute:
		\begin{equation}\label{eq_step2}
		\widehat{X}^{(s+1)}=\underset{X\in {\cal{X}}_N}{\limfunc{argmin}}\,\,\, \sum_{i=1}^N\sum_{t=1}^T\left(Y_{\widehat{\pi}^{(s+1)}(i),t}-\sum_{k=1}^Ka_{tk}X_{\sigma_k(i),k}\right)^2.		\end{equation}
	\end{itemize}
\end{algorithm*} 
 
Both steps in the algorithm are straightforward to implement. The matching step (\ref{eq_step1}) can be computed by a linear programming routine, due to the fact that the linear programming relaxation of a discrete optimal transport problem has integer-valued solutions.\footnote{See for example Chapter 3 in Galichon (2016) on discrete Monge-Kantorovitch problems, and Conforti \textit{et al.} (2014) on integer programming problems and perfect formulations.} Formally, $\widehat{\pi}^{(s+1)}$ in (\ref{eq_step1}) is a solution to the following \emph{linear program}:
\begin{equation*}
\underset{P\in {\cal{P}}_{N}}{\limfunc{max}}\,\,\, \sum_{i=1}^N \sum_{t=1}^T\left(\sum_{k=1}^K a_{tk} \widehat{X}^{(s)}_{\sigma_k(i),k}\right)\left(\sum_{j=1}^{N}P_{ij}Y_{j t}\right),
\end{equation*}
where ${\cal{P}}_{N}$ denotes the set of $N\times N$ matrices with non-negative elements, whose rows and columns all sum to one. In the scalar nonparametric deconvolution case (\ref{dec_est}), this gives $\widehat{\pi}^{(s+1)}(i)=\widehat{\limfunc{Rank}}\left(\widehat{X}^{(s)}_{\sigma(i),1}+X_{i2}\right)$ for all $i$. 

In fact, it is possible to write $\widehat{X}=(\widehat{X}_{1},...,\widehat{X}_{N})$ in (\ref{dec_est_gen}) as the solution to a \emph{quadratic program}:
\begin{align*}
&(\widehat{X},\widehat{P})=\underset{X\in {\cal{X}}_N,\, P\in \cal{P}_{N}}{\limfunc{argmin}} \sum_{i=1}^N\sum_{t=1}^T\Bigg\{ \left(\sum_{k=1}^K a_{tk} X_{\sigma_k(i),k}\right)^2
-2 \left(\sum_{k=1}^K a_{tk} X_{\sigma_k(i),k}\right)\left(\sum_{j=1}^{N}P_{ij}Y_{j t}\right)\Bigg\},
\end{align*}
which is not convex in general. Our estimation algorithm is a method to solve this non-convex quadratic program. However, the algorithm is not guaranteed to reach a global minimum in (\ref{dec_est_gen}). Our implementation is based on starting the algorithm from multiple random values. We will assess the impact of starting values on simulated data.

\subsection{Comparison to Mallows (2007)}

Our algorithm may be seen as a generalization of Mallows' (2007) ``deconvolution by simulation'' method. To highlight the connection, consider the scalar nonparametric deconvolution model. The two steps in our algorithm take the following form:
\begin{align*}
&\widehat{\pi}^{(s+1)}(i)=\widehat{\limfunc{Rank}}\left(\widehat{X}^{(s)}_{\sigma(i),1}+X_{i2}\right),\quad  i=1,...,N,\\
&\widehat{X}^{(s+1)}_{1}=\underset{X_1\in {\cal{X}}_N}{\limfunc{argmin}}\,\sum_{i=1}^N\left(Y_{\widehat{\pi}^{(s+1)}(i)}-X_{\sigma(i),1}-X_{i2}\right)^2.
\end{align*}

The Mallows (2007) algorithm is closely related to this algorithm. The main difference is that, instead of minimizing an objective function for fixed values of the random permutation $\sigma$, random permutations are re-drawn in each step of the algorithm. In addition, the ordering of the $X_{i1}$'s is not restricted, and neither are the values and increments of the $X_{i1}$'s. Formally, the sub-steps of the Mallows algorithm are the following:
\begin{itemize}
	\item Draw a random permutation $\sigma^{(s)}\in\Pi_{N}$.
\item Compute $\widehat{\pi}^{(s+1)}(i)=\widehat{\limfunc{Rank}}\left(\widehat{X}^{(s)}_{{\sigma}^{(s)}(i),1}+X_{i2}\right)$, $i=1,...,N$.
\item Compute $\widehat{X}^{(s+1)}_{{\sigma}^{(s)}(i),1}=Y_{\widehat{\pi}^{(s+1)}(i)}-X_{i2}$, $i=1,...,N$.\footnote{Strictly speaking, Mallows (2007) redefines $\widehat{X}^{(s+1)}_{i1}\equiv \widehat{X}^{(s+1)}_{{\sigma}^{(s)}(i),1}$ for all $i=1,...,N$ at the end of step $s$, and then applies the random permutation $\sigma^{(s+1)}$ to the new $\widehat{X}^{(s+1)}$ values. This difference with the algorithm outlined here turns out to be immaterial, since the composition of $\sigma^{(s+1)}$ and $\sigma^{(s)}$ is also a random permutation of $\{1,...,N\}$.}
\end{itemize}

To provide intuition about this algorithm, Mallows (2007) observes that, starting with draws from the true latent $X_1$, one expects the iteration to continue to draw from that distribution. However, starting from different values, the $\widehat{X}_1$ vectors implied by the algorithm will follow a complex, $N$-dimensional Markov Chain. Moreover, the consistency properties of the Mallows estimator are currently unknown. Lastly, note that the methods introduced in this paper naturally deliver counterparts to the Mallows algorithm for other models beyond deconvolution, such as general linear independent factor models.

\section{Consistency analysis\label{Asympt_sec}}

In this section we provide conditions under which the estimators introduced in Section \ref{Estim_sec} are consistent. 

For $k\in\{1,...,K\}$, let us denote the quantile function of $X_k$ as:
$$F_{X_k}^{-1}(\tau)=\limfunc{inf}\, \{x\in\limfunc{Supp}(X_k)\, :\, F_{X_k}(x)\geq \tau\},\text{ for all }\tau\in(0,1).$$
In addition, for any candidate quantile function $H_k$ that maps the unit interval to the real line, let us define the following Sobolev sup-norms: $$\|H_k\|_{\infty} =\sup_{\tau\in(0,1)}\, |H_k(\tau)|, \,\,\,\text{ and }\,\,\, \|H_k\| =\max_{m\in\{0,1\}}\sup_{\tau\in(0,1)}\, |\nabla^{m}H_k(\tau)|,$$ where $\nabla^{m}H_k$ denotes the $m$-th derivative of $H_k$ (when it exists). We will simply denote $\nabla=\nabla^1$ for the first derivative.

To a solution $\widehat{X}_k$ to (\ref{dec_est_gen}),\footnote{It is not necessary for $\widehat{X}_k$ to be an exact minimizer of (\ref{dec_est_gen}). As we show in the proof, it suffices that the value of the objective function at $(\widehat{X}_1,...,\widehat{X}_K)$ be in an $\epsilon_N$-neighborhood of the global minimum, for $\epsilon_N$ tending to zero as $N$ tends to infinity.} we will associate an interpolating quantile function $\widehat{H}_k$ such that $\widehat{H}_k\left(\frac{i}{N+1}\right)=\widehat{X}_{ik}$ for all $i$. We will then show that $  \|\widehat{H}_k-F_{X_k}^{-1}\|_{\infty}=o_p(1)$. This result will be obtained as an application of the consistency theorem for sieve extremum estimators in Chen (2007).


We make the following assumptions.

\begin{assumption}\label{ass_consis_gen}$\quad$
	
	$(i)$ (Continuity and support) $Y$ and $X$ have compact supports in $\mathbb{R}^T$ and $\mathbb{R}^K$, respectively, and admit absolutely continuous densities $f_Y,f_X$ that are bounded away from zero and infinity. Moreover, $f_Y$ is differentiable.  
	

	$(ii)$ (Identification) The characteristic function of $X_k$ does not vanish on the real line for any $k$, and the vectors $\limfunc{vec}A_kA_k'$, $k=1,...,K$, are linearly independent.

	$(iii)$ (Penalization) $\overline{C}_N$ is increasing and $\underline{C}_N$ is decreasing with $\limfunc{lim}_{N\rightarrow +\infty}\, \overline{C}_N =\overline{C}$ and
	$\limfunc{lim}_{N\rightarrow +\infty}\, \underline{C}_N =\underline{C}$, where $\overline{C}$ and $\underline{C}<\overline{C}$ are such that, for all $k$, $\|F_{X_k}^{-1}\|\leq \overline{C}$ and $\nabla F_{X_k}^{-1}(\tau)\geq \underline{C}$ for all $\tau\in(0,1)$.
	
	$(iv)$ (Sampling) $(Y_{i1},...,Y_{iT})$, $i=1,...,N$, are i.i.d.

\end{assumption}

Though convenient for the derivations, the compact supports assumption in part $(i)$ is strong. This could be relaxed by working with weighted norms, at the cost of achieving a weaker consistency result. The simulation experiments we report below suggest that the estimator continues to perform well when supports are unbounded. Part $(ii)$ is a sufficient condition for the distributions of latent variables $X_k$ to be nonparametrically identified (Sz\'ekely and Rao, 2000). The constants $\underline{C}_N$ and $\overline{C}_N$ appearing in part $(iii)$ ensure that the $\widehat{X}_{ik}$ values are bounded and of bounded variation. Lastly, the independent random permutations $\sigma_1,...,\sigma_K$ in (\ref{dec_est_gen}) depend on $N$, although we have omitted this dependence for conciseness.


Consistency is established in the following theorem. Proofs are in Appendix \ref{App_proofs}.

\begin{theorem}\label{theo_consis_gen}
	Consider the independent factor model $Y=AX$. Let Assumption \ref{ass_consis_gen} hold. Then, as $N$ tends to infinity:
	$$\underset{i\in\{1,...,N\}}{\limfunc{max}}\, \left|\widehat{X}_{ik}-F_{X_{k}}^{-1}\left(\frac{i}{N+1}\right)\right|=o_p(1),\quad \text{ for all }k=1,...,K.$$
	
\end{theorem}

While Theorem \ref{theo_consis_gen} does not formally cover the scalar deconvolution model, the same proof arguments can be used to show the following result, under similar assumptions to those of Theorem \ref{theo_consis_gen}.

\begin{corollary}\label{theo_consis}
 Consider the scalar deconvolution model $Y=X_1+X_2$, where one observes two samples $Y_1,...,Y_N$ and $X_{21},...,X_{2N}$ from $Y$ and $X_2$, respectively. Let Assumption \ref{ass_consis} in Appendix \ref{App_proofs} hold. Then, as $N$ tends to infinity:
	$$\underset{i\in\{1,...,N\}}{\limfunc{max}}\, \left|\widehat{X}_{i1}-F_{X_1}^{-1}\left(\frac{i}{N+1}\right)\right|=o_p(1).$$
\end{corollary}

An important step in the proof of Theorem \ref{theo_consis_gen} is to define the population counterpart to the estimation problem (\ref{dec_est_gen}). Let $\mu_Y$ denote the population measure of $Y$. Moreover, for any candidate quantile functions $H=(H_1,...,H_K)$, let $\mu_{AH}$ denote the population measure of the random vector $Z\equiv \sum_{k=1}^KA_kH_k(V_k)$, where $V_1,...,V_K$ are independent standard uniform random variables on the unit interval. Finally, let ${\cal{M}}(\mu_Y,\,\mu_{AH})$ denote the set  all possible joint distributions, or \emph{couplings}, of the random vectors $Y$ and $\sum_{k=1}^KA_k H_k\left(V_k\right)$, with marginals $\mu_Y$ and $\mu_{AH}$. The population objective function is then:
$$Q(H)\equiv\underset{\pi\in {\cal{M}}(\mu_Y,\,\mu_{AH})}{\limfunc{inf}}\,\mathbb{E}_{\pi}\left[\sum_{t=1}^T\left
(Y_t-\sum_{k=1}^Ka_{tk} H_k\left(V_k\right)\right)^2\right],$$
which is the quadratic Wasserstein distance between the population distribution of the data and the one implied by the model. Under part $(ii)$ in Assumption \ref{ass_consis_gen} that ensures identification, $Q(H)$ is minimized at the true quantile functions $H_k=F_{X_k}^{-1}$. 

In the scalar deconvolution model, the population objective takes the explicit form:
$$Q(H_1)\equiv\mathbb{E}\left[\left
(F_Y^{-1}\left(\int_0^1 F_{X_2}\left(H_1(V_1)+F_{X_2}^{-1}(V_2)-H_1(\tau)\right)d\tau\right)-H_1(V_1)-F_{X_2}^{-1}(V_2)\right)^2\right],$$
where the expectation is taken with respect to independent standard uniform random variables $V_1$ and $V_2$. Note that the integral in this expression is simply the population rank of $H_1(V_1)+F_{X_2}^{-1}(V_2)$. When the characteristic function of $X_2$ has no real zeros, $Q(H_1)$ is minimized at $H_1=F_{X_1}^{-1}$.

\paragraph{Densities and expectations.}

Under slightly stronger assumptions, Theorem \ref{theo_consis_gen} can be modified to obtain consistent estimators of both $F_{X_k}^{-1}$ and its derivative, which can then be used for density estimation. To see this, let us denote as ${\cal{X}}^{(2)}_N$ the set of $X$ in ${\cal{X}}_N$ which satisfy the restrictions on second-order differences: $(N+1)^2\left|X_{i+2,k}-2X_{i+1,k}+X_{ik}\right|\leq \overline{C}_N$, for all $i$ and $k$, and replace the minimization in (\ref{dec_est_gen}) by a minimization with respect to $X\in {\cal{X}}^{(2)}_N$. Imposing in Assumption \ref{ass_consis_gen} that the densities of $X_k$ have bounded second-order derivatives, and modifying the proof of Theorem \ref{theo_consis_gen} accordingly, we obtain that:
\begin{equation}\underset{i\in\{1,...,N\}}{\limfunc{max}}\, \left|(N+1)(\widehat{X}_{i+1,k}-\widehat{X}_{ik})-\nabla \left(F_{X_k}^{-1}\right)\left(\frac{i}{N+1}\right)\right|=o_p(1), \text{ for all }k=1,...,K.\label{eq_der}\end{equation}


We then have the following result.
\begin{corollary}\label{coro_density_consis}
	Let $b$ in (\ref{f_hat_estim}) be such that $b\rightarrow 0$ and $Nb\rightarrow +\infty$ as $N$ tends to infinity. Let $\kappa$ be a Lipschitz kernel that integrates to one and has finite first moments. Then, provided Theorem \ref{theo_consis_gen} and equation (\ref{eq_der}) hold, we have:
	\begin{equation}\underset{x\in\mathbb{R}}{\limfunc{sup}}\, \left|\widehat{f}_{X_k}(x)-f_{X_k}(x)\right|=o_p(1),\quad  \text{ for all }k=1,...,K.\label{f_hat_consis}\end{equation}
	
\end{corollary} 

Lastly, given Corollary \ref{coro_density_consis} it can readily be checked that conditional expectations estimators, such as (\ref{eq_cond_exp}) and those in Appendix \ref{App_Extens}, are consistent in sup-norm for their population counterparts. 

\paragraph{Remark: convergence rates and inference.}

It follows from existing convergence rates in nonparametric deconvolution models (e.g., Fan, 1991; Hall and Lahiri, 2008) that neither $\widehat{X}_{ik}$ (as an estimator of the quantile function of $X_k$) nor its functionals will converge at the root-$N$ rate in general. Bertail \textit{et al.} (1999) propose an inference method under the condition that the estimator is $N^{\beta}$-consistent with a continuous asymptotic distribution, for some $\beta>0$. Their rate-adaptive method is attractive in our setting, although polynomial convergence rates may rule out cases of severe ill-posedness. Completing the characterization of the asymptotic behavior of our estimator is an important task for future work.

\section{Performance on simulated data\label{Simu_sec}}

In this section we illustrate the finite-sample performance of our estimator on data simulated from a nonparametric fixed-effects model. In Appendix \ref{App_Additional_Simu} we report additional results for a scalar nonparametric deconvolution model.

\begin{figure}[tbp]
	\caption{Monte Carlo results for $X_1$ in  the fixed-effects model, $N=100$, $T=2$}
	\label{Fig_MC_Kot}
	\begin{center}
		\begin{tabular}{cccc}
			\multicolumn{2}{c}{Quantile functions}&  \multicolumn{2}{c}{Densities}\\
			Strong constraint & Weak constraint & Strong constraint & Weak constraint \\
			\\ 
			\multicolumn{4}{c}{$(X_1,X_2,X_3) \sim $ Beta(2,2)}\\
			\includegraphics[width=40mm, height=30mm]{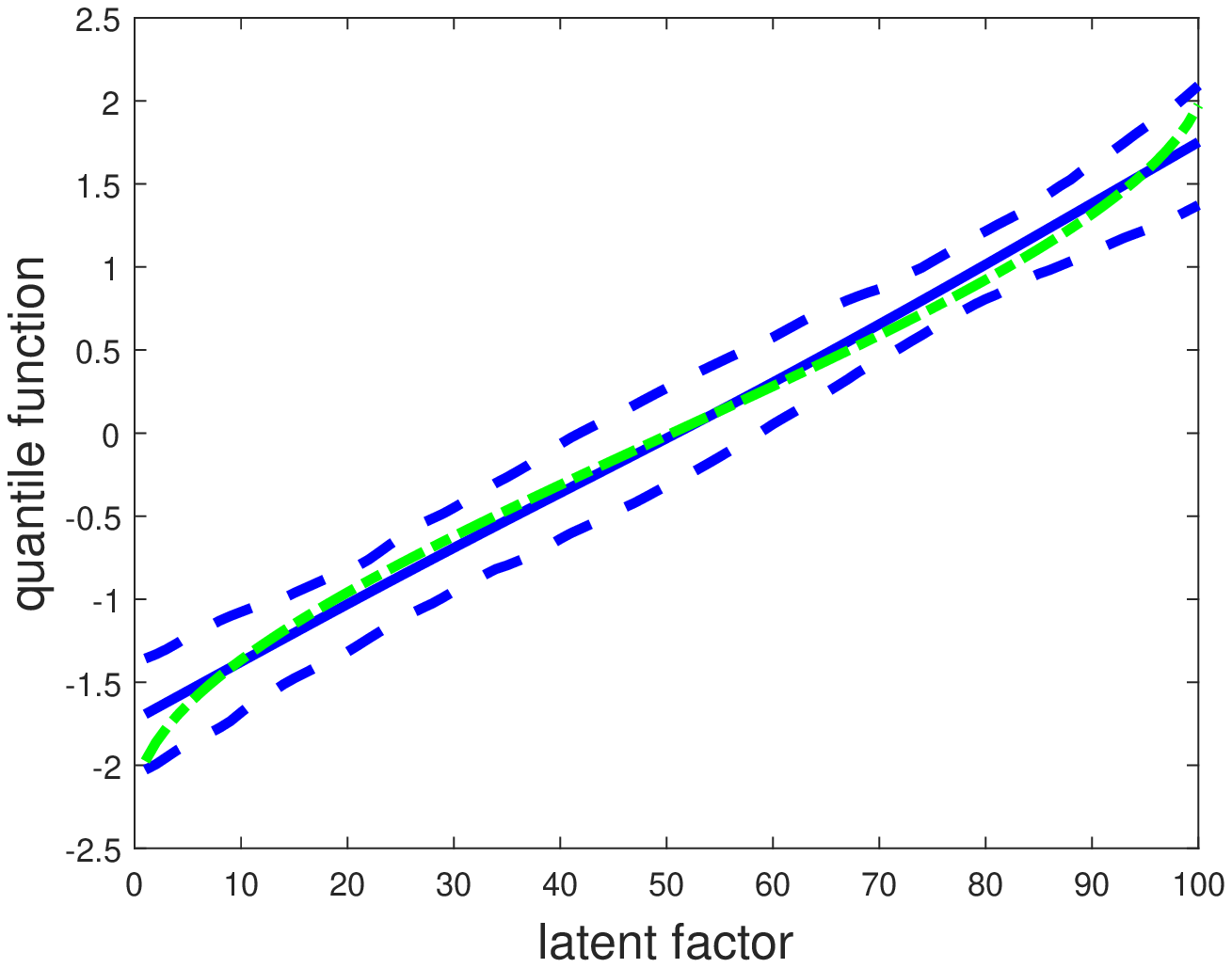} & %
			\includegraphics[width=40mm, height=30mm]{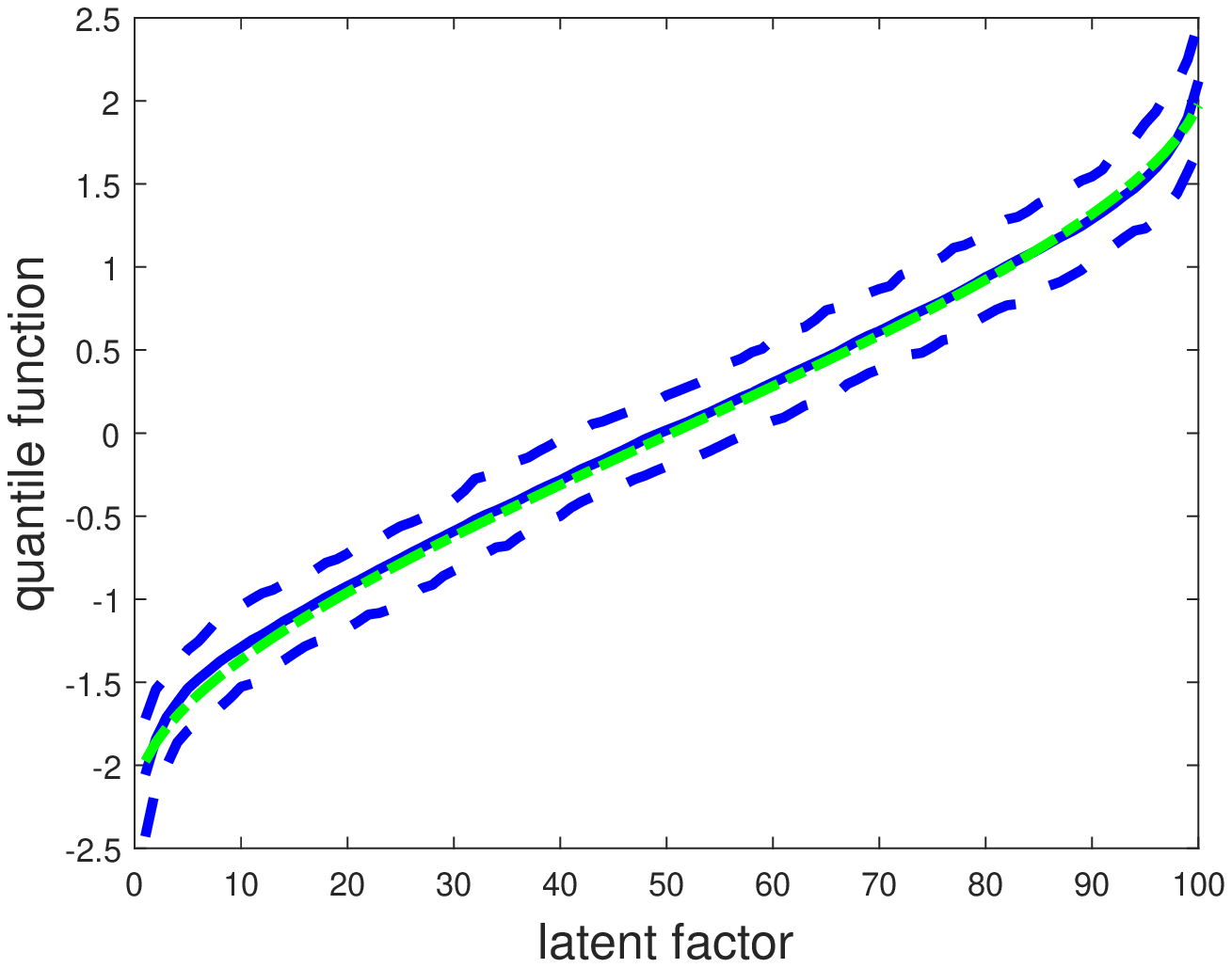} & 
			\includegraphics[width=40mm, height=30mm]{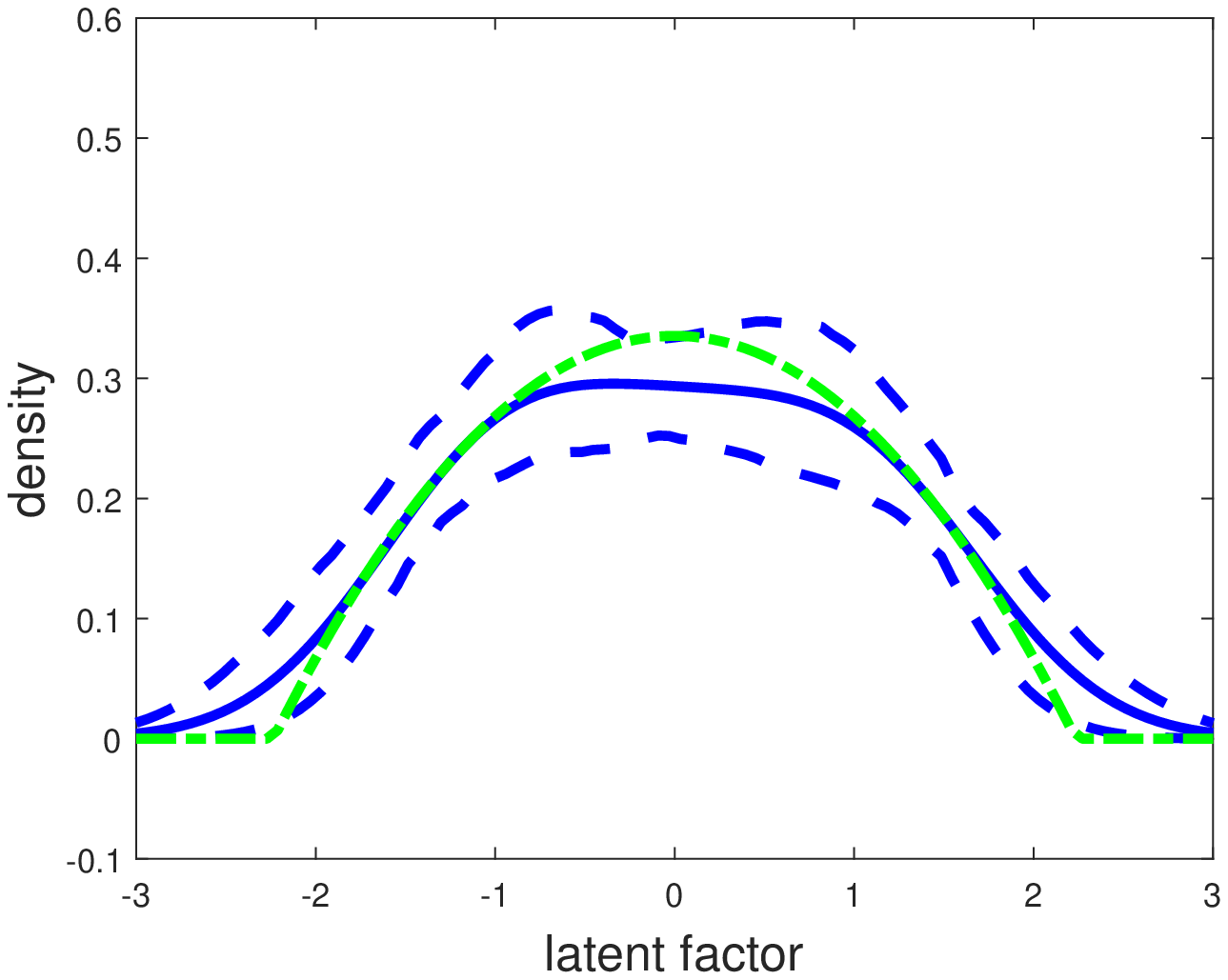} & %
			\includegraphics[width=40mm, height=30mm]{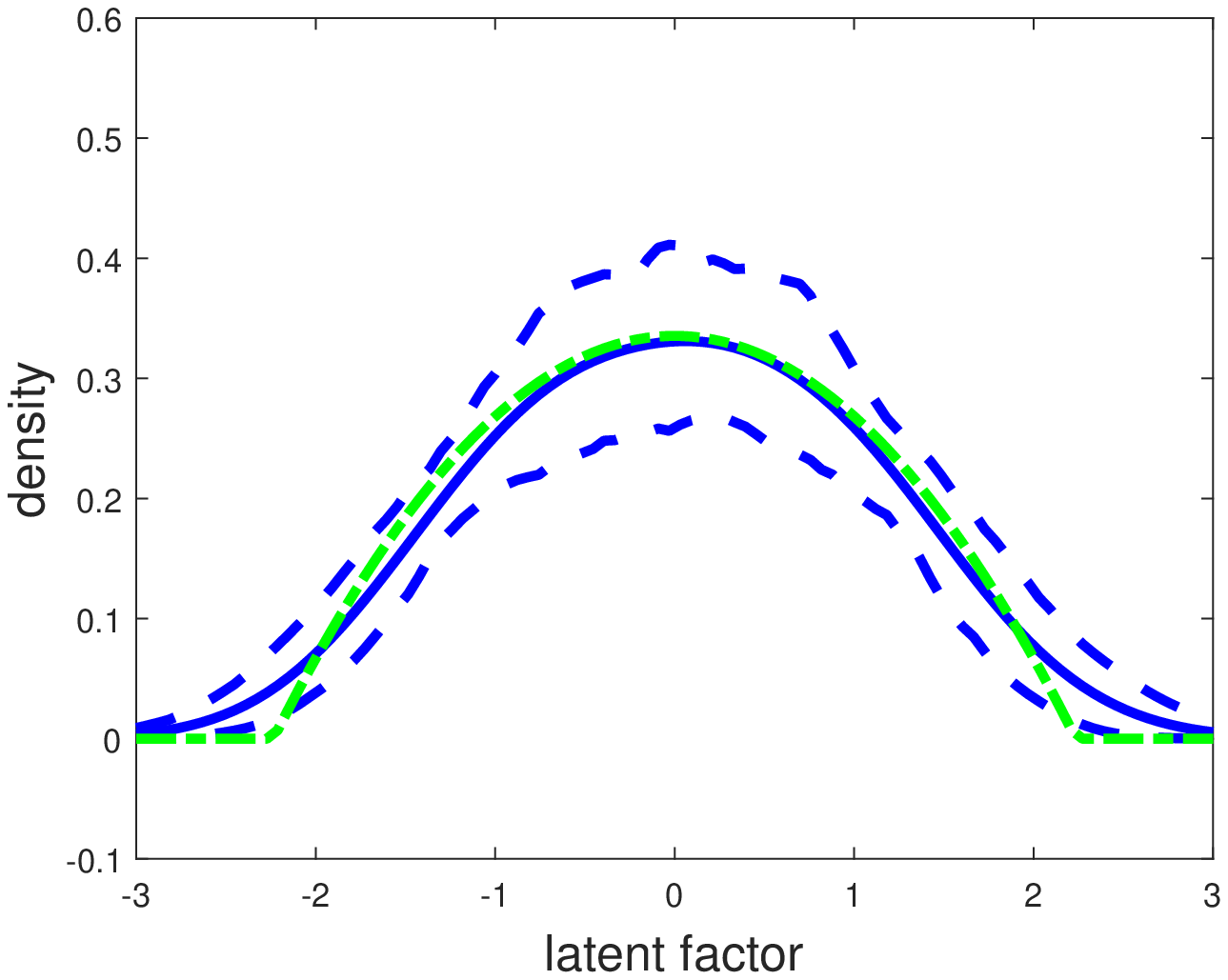}\\
			&  &\\
			\multicolumn{4}{c}{$(X_1,X_2,X_3) \sim $ Beta(5,2)}\\
			\includegraphics[width=40mm, height=30mm]{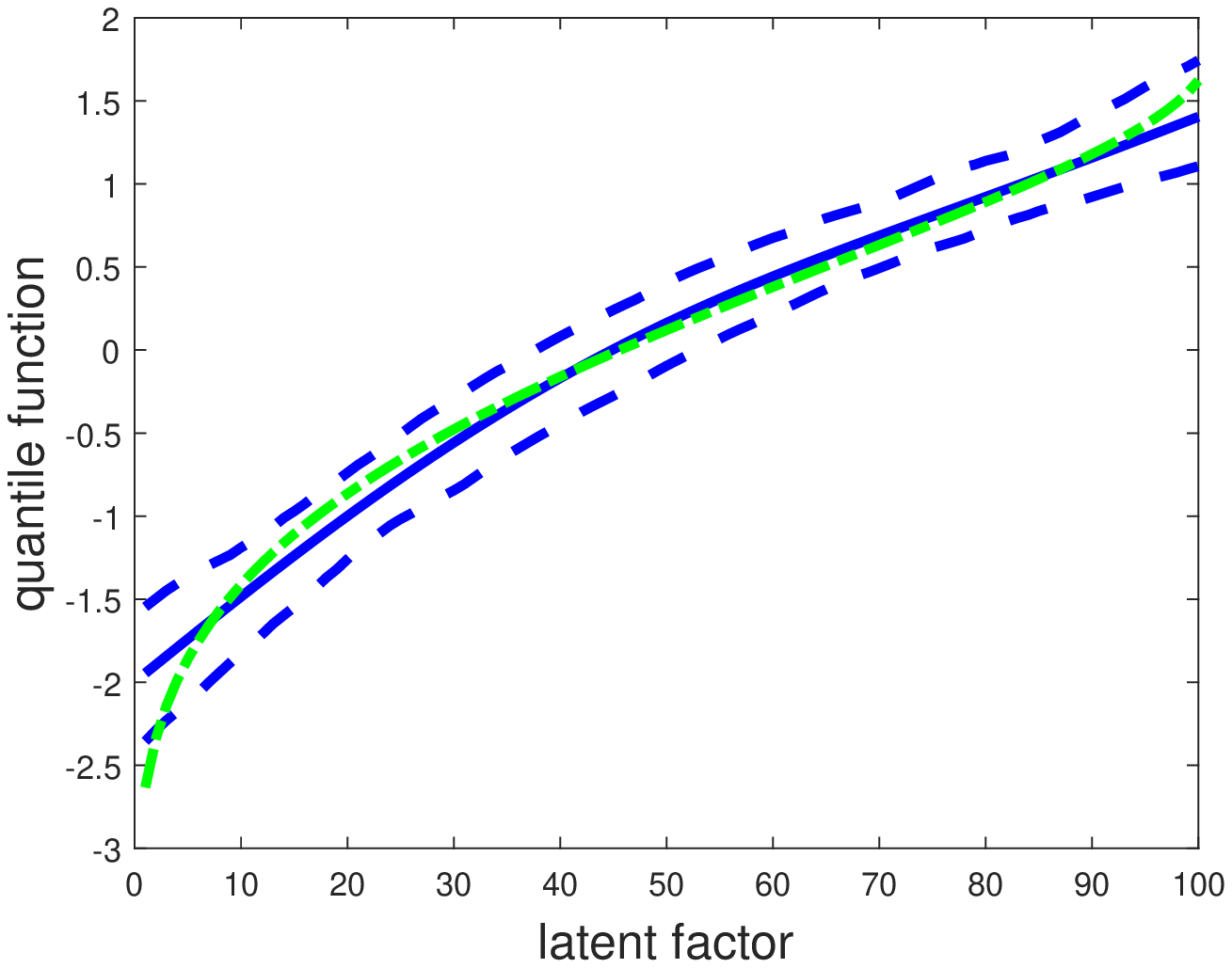} & %
			\includegraphics[width=40mm, height=30mm]{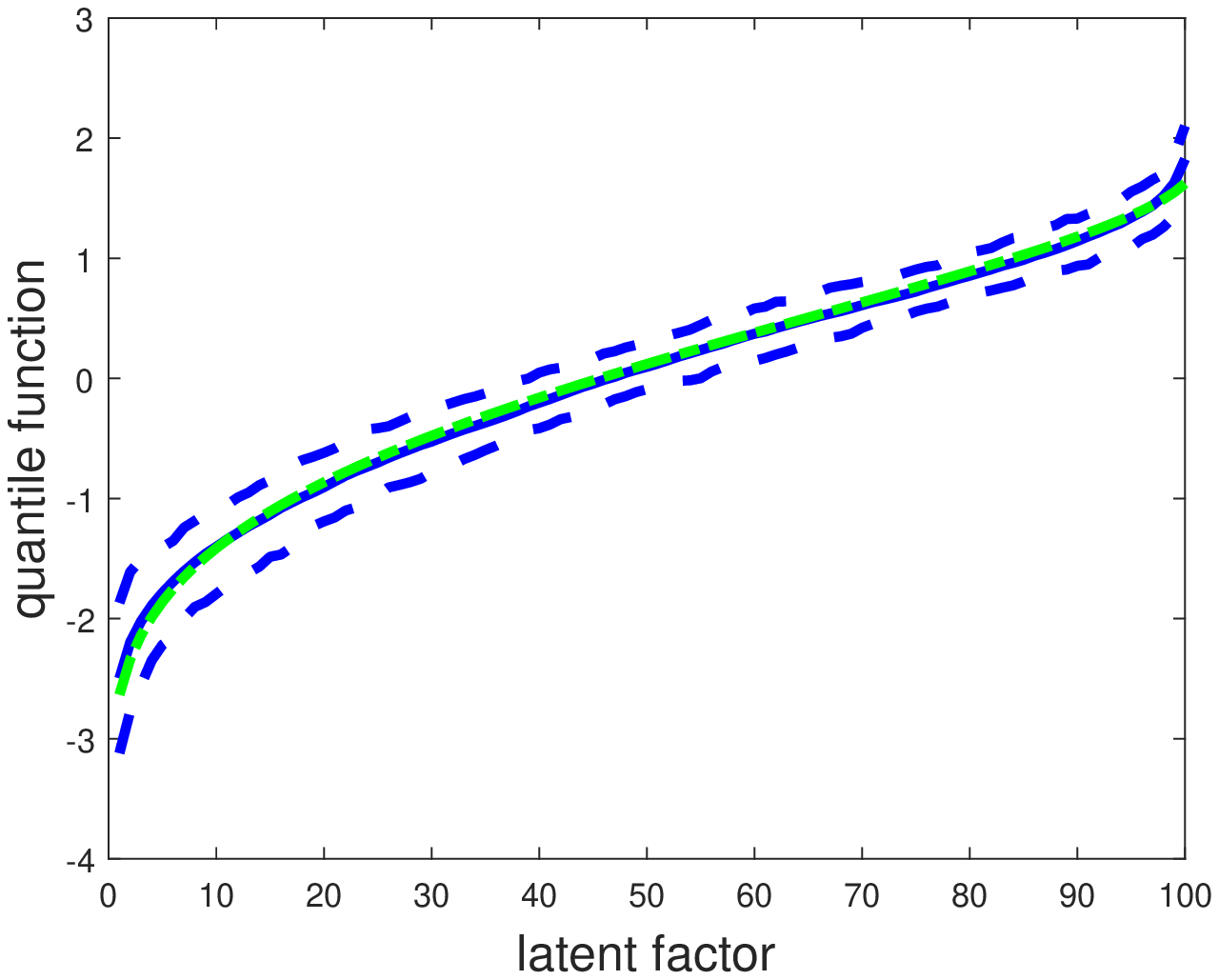} & 
			\includegraphics[width=40mm, height=30mm]{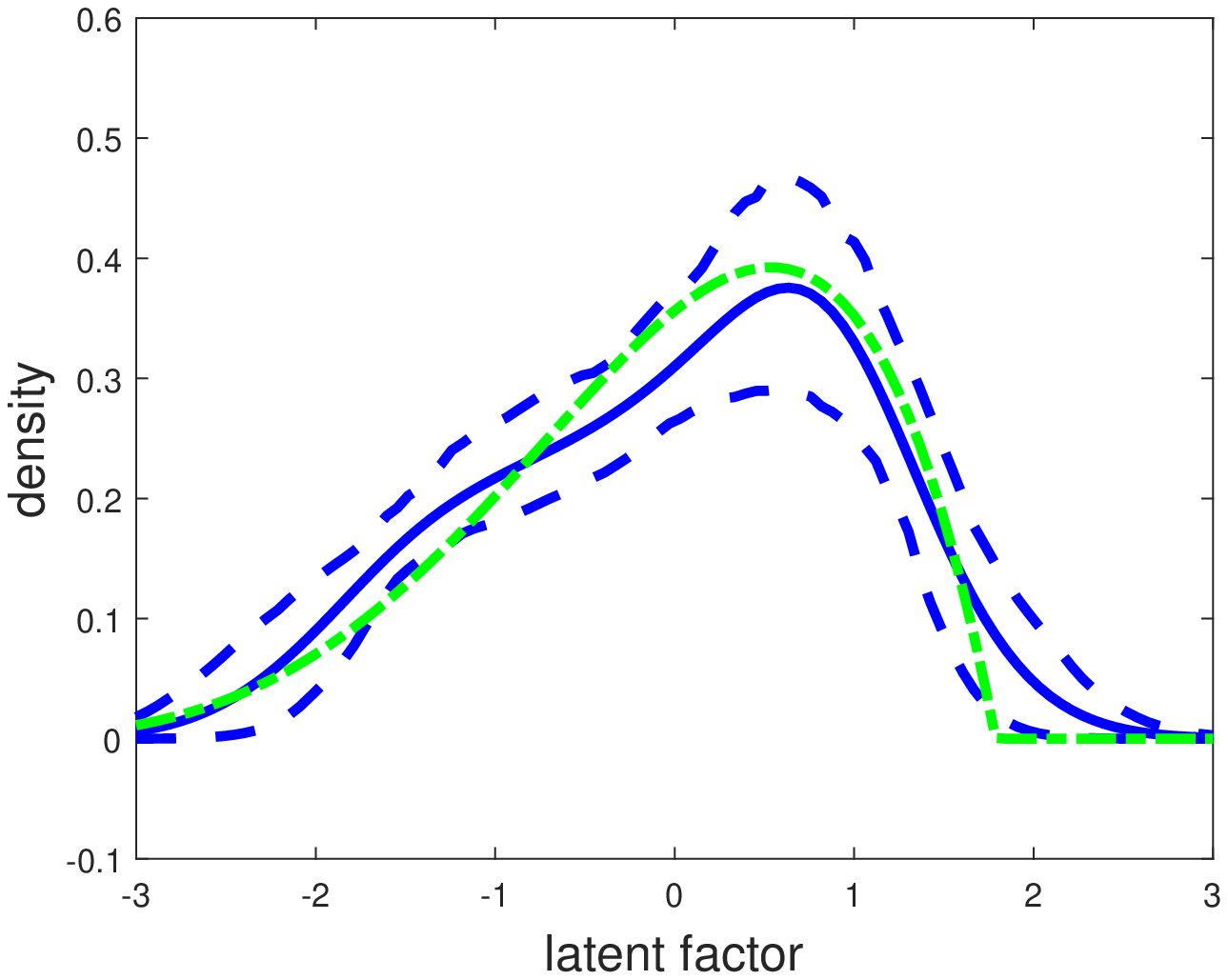} & %
			\includegraphics[width=40mm, height=30mm]{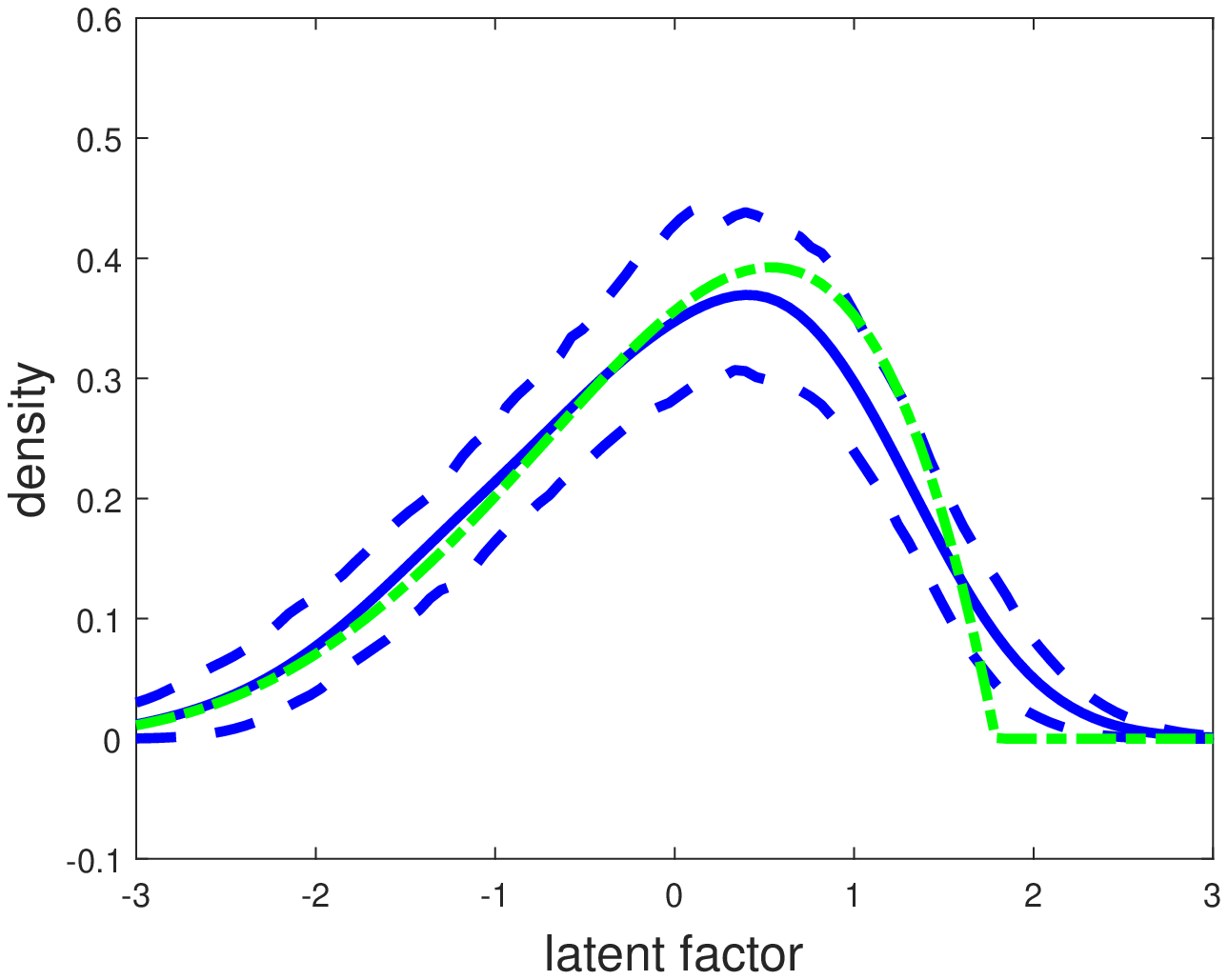}\\
			&  &\\
			\multicolumn{4}{c}{$(X_1,X_2,X_3) \sim {\cal{N}}(0,1)$}\\
			\includegraphics[width=40mm, height=30mm]{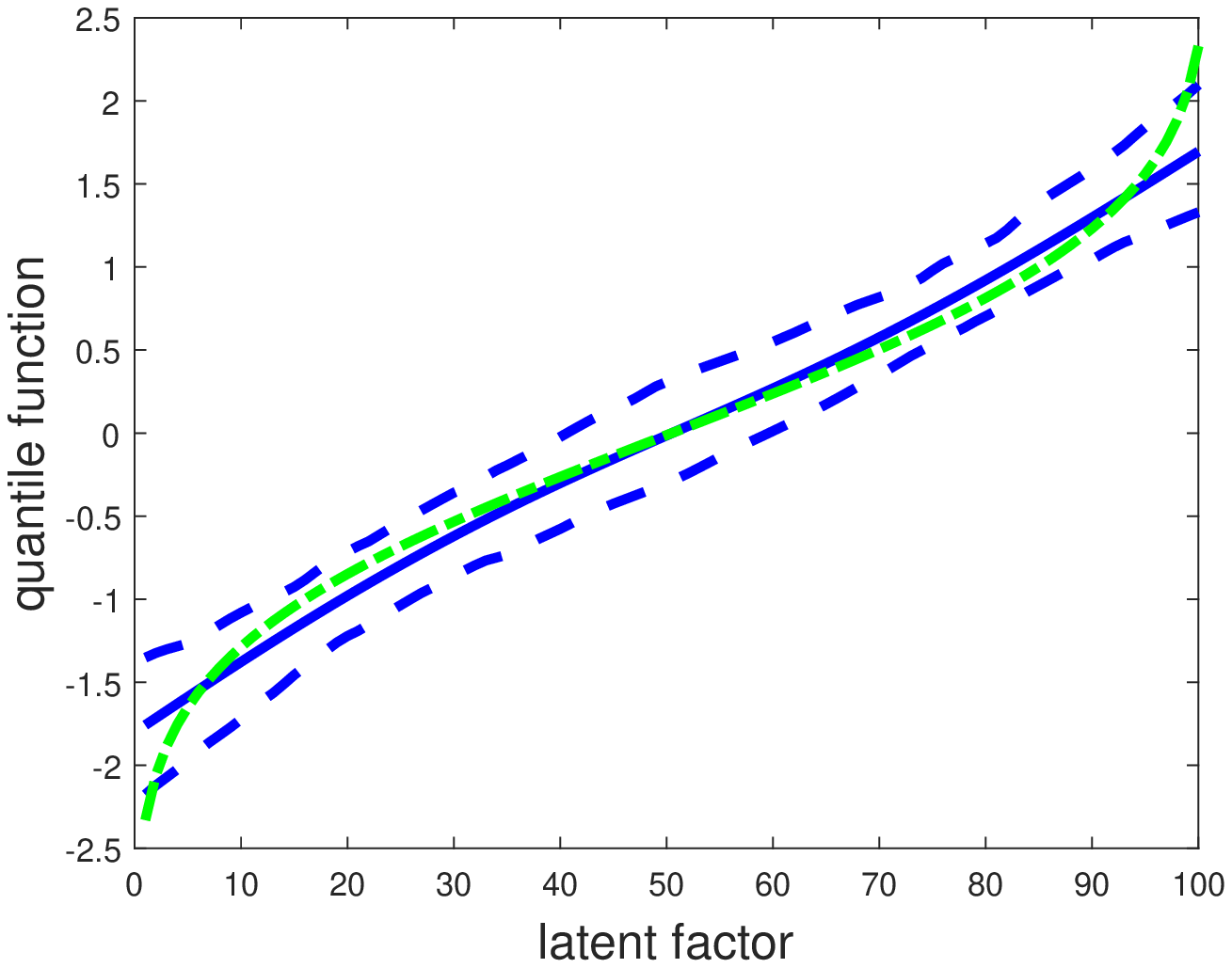} & %
			\includegraphics[width=40mm, height=30mm]{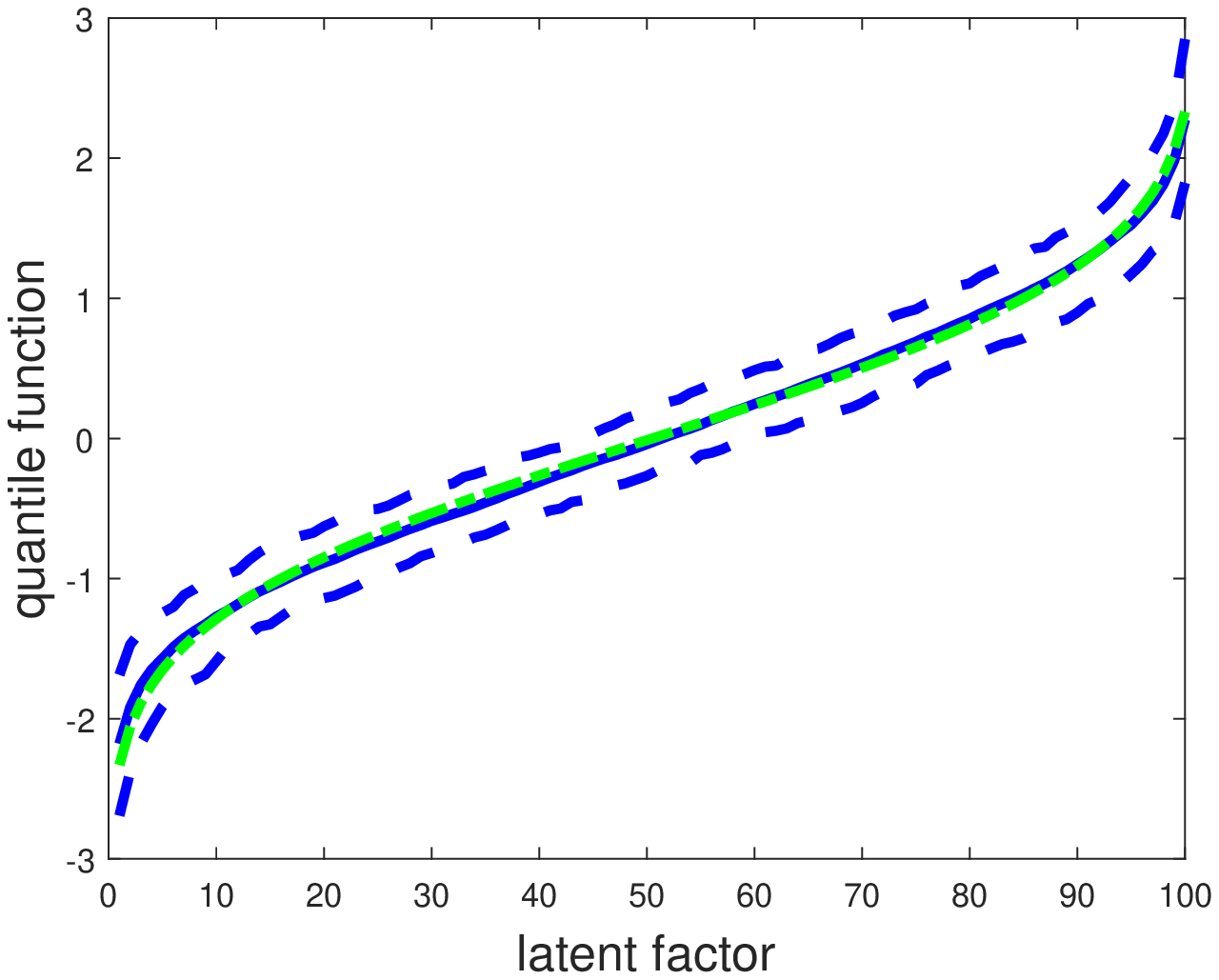} & 
			\includegraphics[width=40mm, height=30mm]{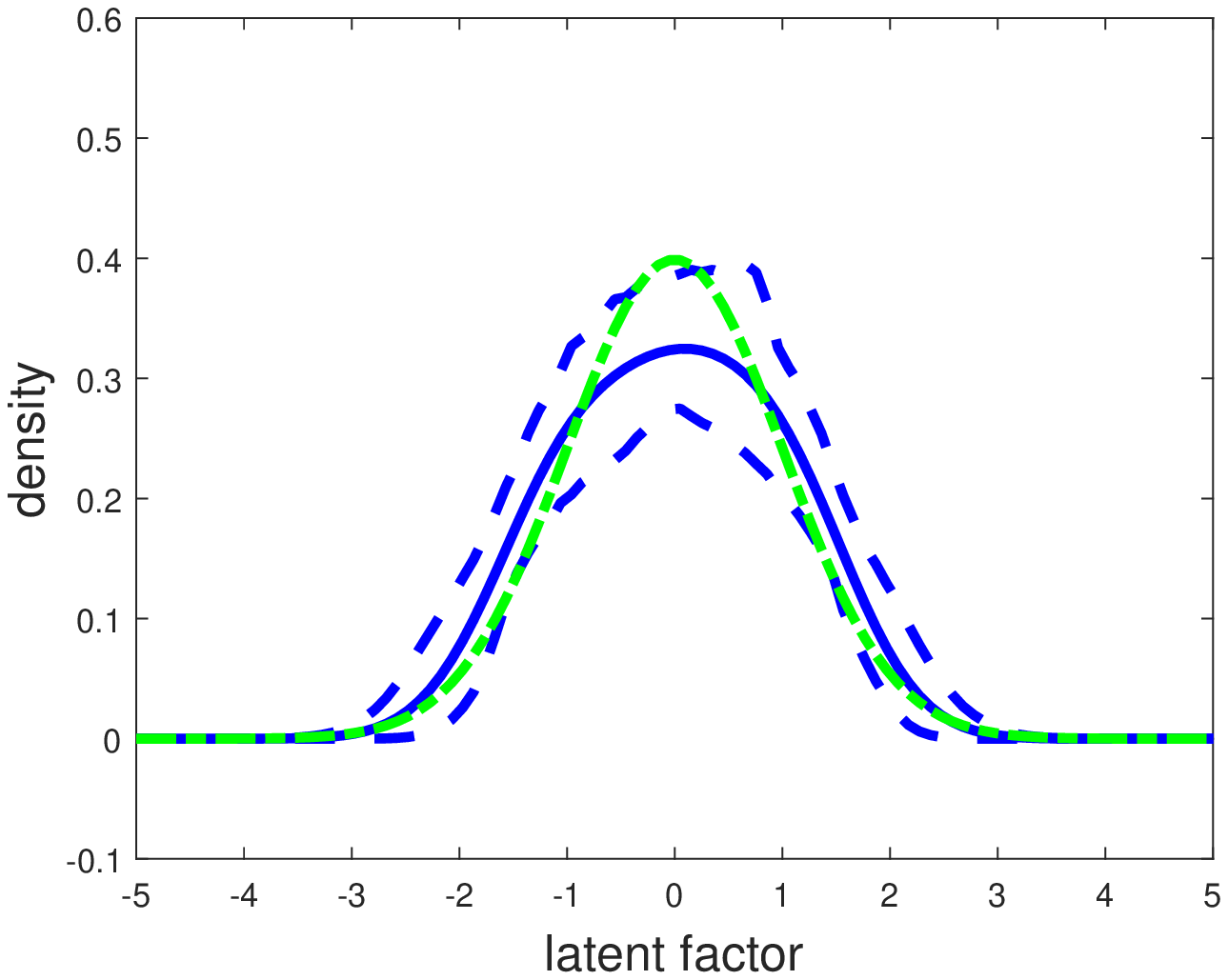} & %
			\includegraphics[width=40mm, height=30mm]{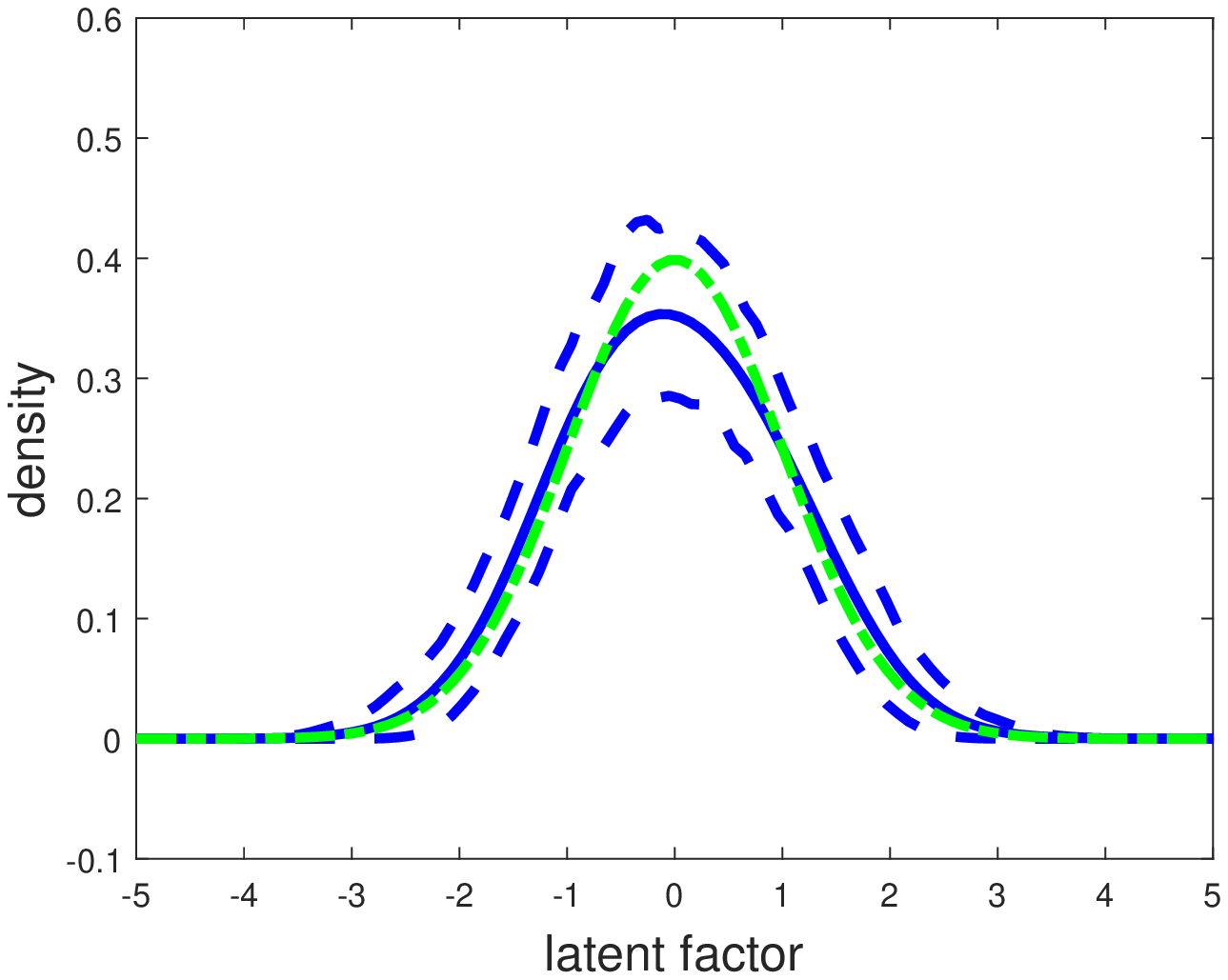}\\
			&  &\\
			\multicolumn{4}{c}{$(X_1,X_2,X_3) \sim  \limfunc{exp}{\cal{N}}(0,1)$}\\
			\includegraphics[width=40mm, height=30mm]{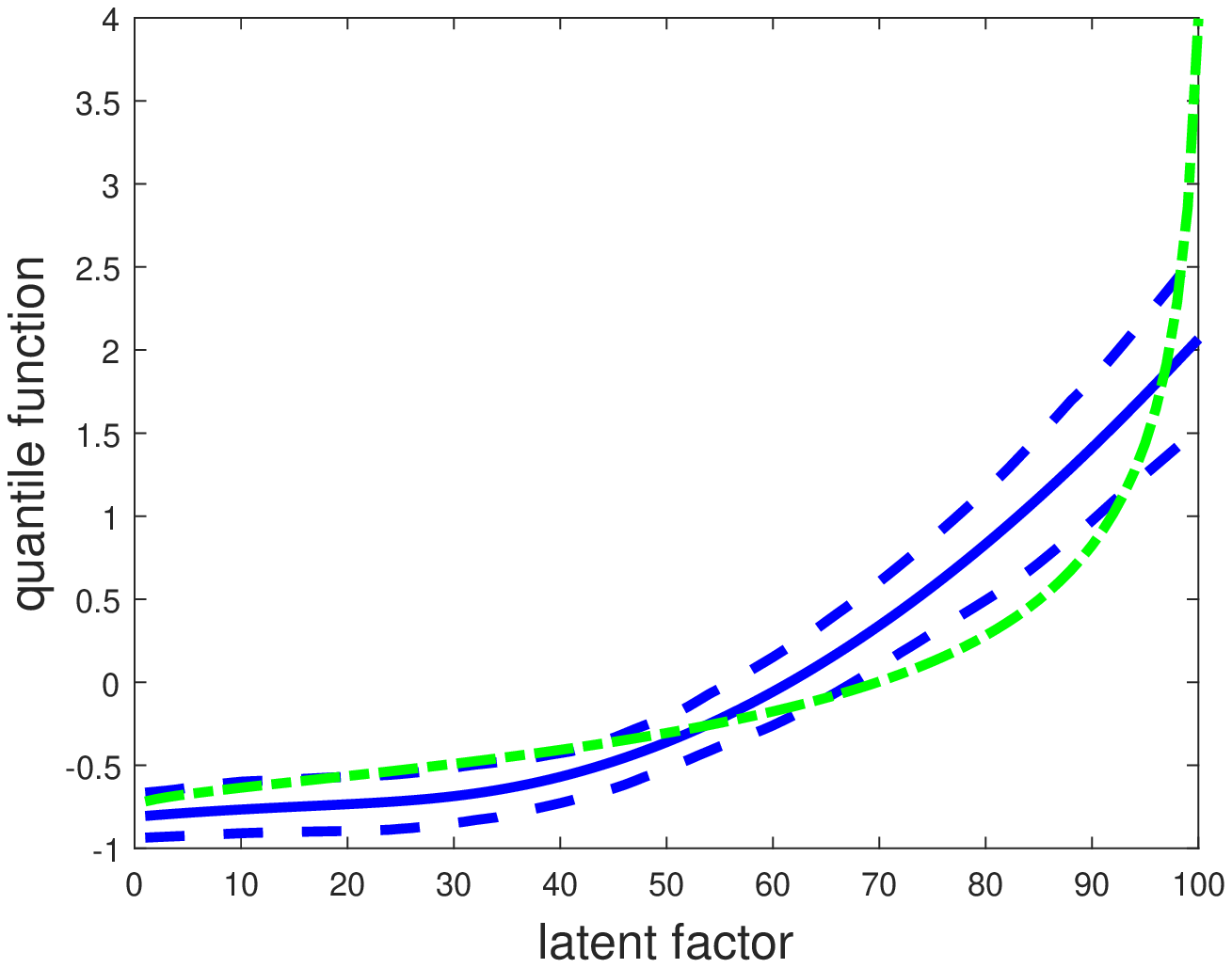} & %
			\includegraphics[width=40mm, height=30mm]{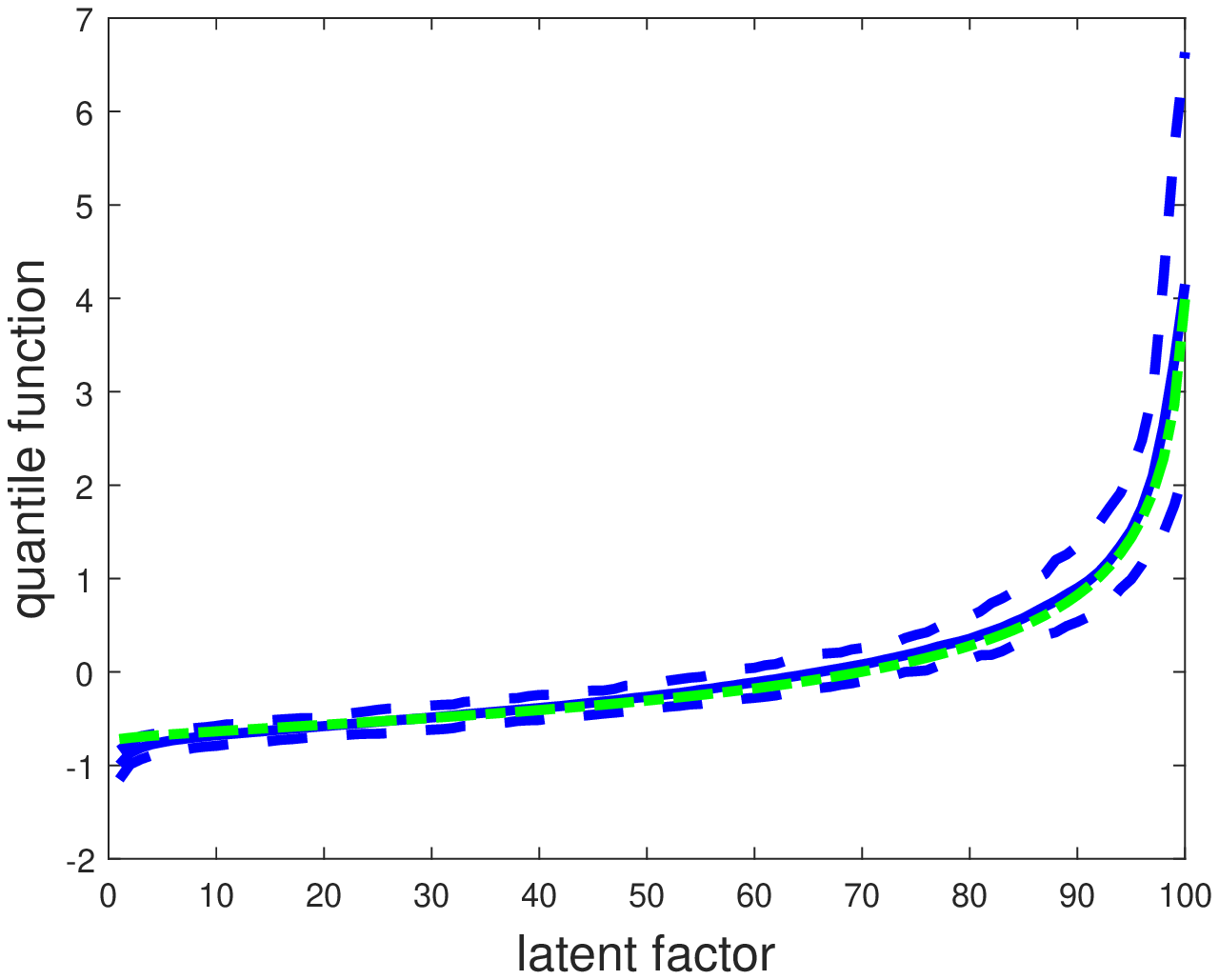} & 
			\includegraphics[width=40mm, height=30mm]{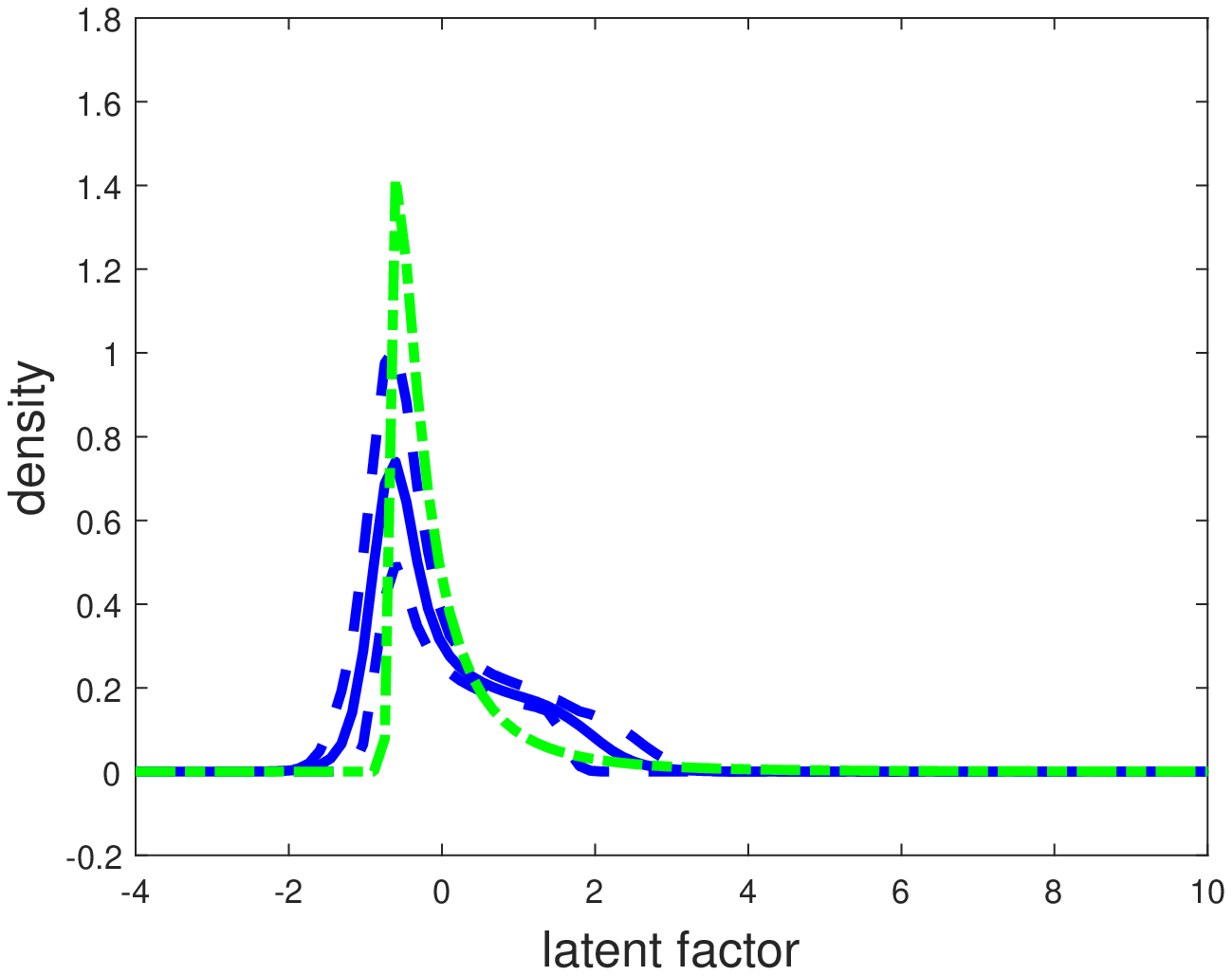} & %
			\includegraphics[width=40mm, height=30mm]{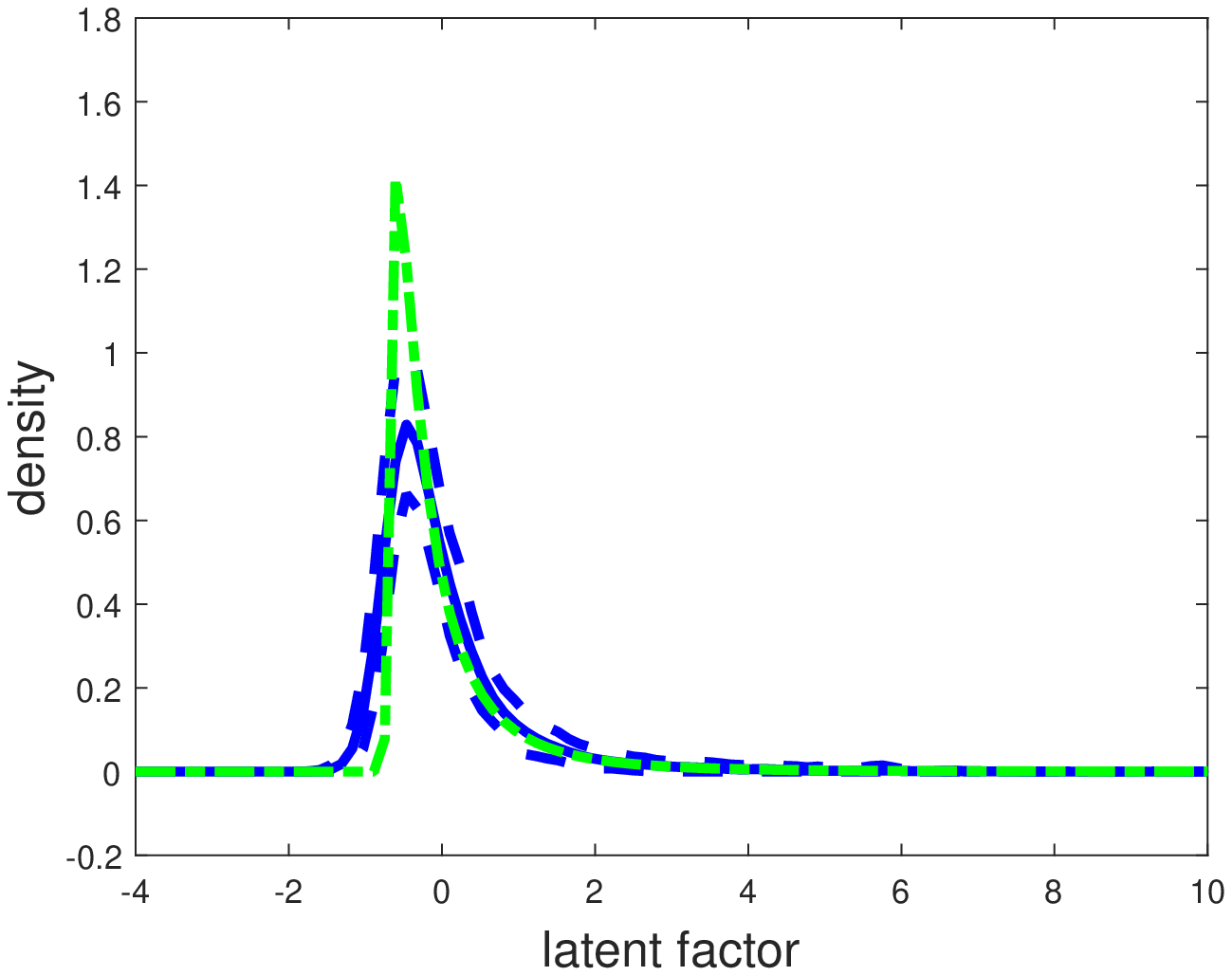}\\
			&  &\\
		\end{tabular}%
	\end{center}
	\par
	\textit{{\footnotesize Notes: Simulated data from the fixed-effects model, results for the first factor $X_1$. The mean across simulations is shown in solid, 10 and 90 percent pointwise quantiles are shown in dashed, and the true quantile function or density is shown in dashed-dotted. $100$ simulations. $10$ random starting values. $M=10$ averages over $\sigma$ draws.}}
\end{figure}

We focus on the model $Y_1=X_1+X_2$, $Y_2=X_1+X_3$, where $X_1,X_2,X_3$ are independent of each other and have identical distributions. We consider four specifications for the distribution of $X_k$ for all $k$: Beta$(2,2)$, Beta$(5,2)$, normal, and log-normal, all standardized so that $X_k$ has mean zero and variance one. To restrict the maximum values of $\widehat{X}_{ik}$, its increments, and its second-order differences, we consider two choices for the penalization constants: $(\underline{C}_N,\overline{C}_N)=(.1,10)$ (``strong constraint''), and $(\underline{C}_N,\overline{C}_N)=(0,10000)$ (``weak constraint''). To minimize the objective function in (\ref{dec_est_gen}) we start with $10$ randomly generated starting values, drawn from widely dispersed mixtures of five Gaussian distributions, and keep the solution corresponding to the minimum value of the objective. Lastly, we draw $M=10$ independent random permutations in $\Pi_N$, and average the resulting $M$ sets of estimates $\widehat{X}^{(m)}_{i1}$, for $i=1,...,N$.

In Appendix \ref{App_Additional_Simu} we study the sensitivity of the estimates to the penalization constants, the starting values, and the number $M$ of $\sigma$ draws, in a nonparametric deconvolution model. We find that the estimator is quite robust to these choices. In particular, we document that taking conservative choices for $\underline{C}_N$ and $\overline{C}_N$ (such as in the ``weak constraint'' case) results in a well-behaved estimator, suggesting that our matching procedure induces an implicit regularization, even in the absence of additional constraints on parameters. At the same time, we find that such a conservative choice may not be optimal in terms of mean squared errors of quantile estimates. The optimal choice of penalization constants is an interesting question for future work.\footnote{A simple recommendation for practice could be based on a truncated normal distribution. Let $\widehat{\sigma}_k$ denote a consistent estimate of the standard deviation of $X_k$, e.g. obtained by covariance-based minimum distance, and let $c>0$ be a tuning parameter. Possible penalization constants are: $2.3c\widehat{\sigma}_k$ (upper bound on quantile values), $2.5c^{-1}\widehat{\sigma}_k$ and $37c\widehat{\sigma}_k$ (lower and upper bounds for first derivatives), and $3275c\widehat{\sigma}_k$ (upper bound on second derivatives). When $c=1$, these constants are binding when $X_k$ follows a normal truncated at the 99th percentiles. A default choice could be $c=2$.}

In the first two columns in Figure \ref{Fig_MC_Kot} we show the estimates of the quantile functions $\widehat{X}_{i1}=\widehat{F}_{X_1}^{-1}\left(\frac{i}{N+1}\right)$, for the four specifications and both penalization parameters. The results for the other two factors are similar and omitted for brevity. The solid and dashed lines correspond to the mean and 10 and 90 percentiles across 100 simulations, respectively, while the dashed-dotted line corresponds to the true quantile function. The sample size is $N=100$. Even for such a small sample size, our nonparametric estimator performs well, especially under a weaker constraint on the parameters (second column). In the last two columns of Figure \ref{Fig_MC_Kot} we show density estimates for the same specifications. We take a Gaussian kernel and set the bandwidth based on Silverman's rule. Although there are some biases in the strong constraint case, our nonparametric estimator reproduces the shape of the unknown densities well.

\begin{table}[tbp]\caption{Monte Carlo simulation, mean integrated squared and absolute errors of density estimators in the fixed-effects model, results for $X_1$\label{Tab_MSE_FE}}
	\begin{center}
		\begin{tabular}{r||cccccc}
			&MISE & MIAE & MISE & MIAE & MISE & MIAE\\\hline
			&\multicolumn{6}{c}{$(X_1,X_2,X_3) \sim $ Beta(2,2)}\\\hline \hline
			&\multicolumn{2}{c}{Strong constraint} & \multicolumn{2}{c}{Weak constraint} & \multicolumn{2}{c}{Fourier}\\
			&  0.0036  & 0.0654 &  0.0035 & 0.0631 & 0.0123 &0.2274
			\\\hline\hline
			&\multicolumn{6}{c}{$(X_1,X_2,X_3) \sim $ Beta(5,2)}\\\hline \hline
			&\multicolumn{2}{c}{Strong constraint} & \multicolumn{2}{c}{Weak constraint} & \multicolumn{2}{c}{Fourier}\\
			&    0.0050 &  0.0750&  0.0042 & 0.0677 & 0.0249 &0.2979
			\\\hline\hline
			&\multicolumn{6}{c}{$(X_1,X_2,X_3) \sim  {\cal{N}}(0,1)$}\\\hline \hline
			&\multicolumn{2}{c}{Strong constraint} & \multicolumn{2}{c}{Weak constraint} & \multicolumn{2}{c}{Fourier}\\
			&    0.0056&  0.0796&  0.0040& 0.0674 & 0.0122 &0.2372
			\\\hline\hline
			&\multicolumn{6}{c}{$(X_1,X_2,X_3) \sim  \exp[{\cal{N}}(0,1)]$}\\\hline \hline
			&\multicolumn{2}{c}{Strong constraint} & \multicolumn{2}{c}{Weak constraint} & \multicolumn{2}{c}{Fourier}\\
			&    0.1003&  0.2415&  0.0536& 0.1492 & 0.3344  &0.8613
			\\\hline\hline
		\end{tabular}
	\end{center}
	{\footnotesize \textit{Notes: Mean integrated squared and absolute errors across $100$ simulations from the fixed-effects model. $N=100$, $T=2$. ``Fourier'' is the characteristic-function based estimator of Bonhomme and Robin (2010). Results for the first factor $X_1$.}}
\end{table}

In Table \ref{Tab_MSE_FE} we report the mean integrated squared and absolute errors (MISE and MIAE, respectively) of our density estimators, for the four distributional specifications and $N=100$. We see that the estimator performs better under the weak constraint. Moreover, interestingly, as shown by the last two columns of Table \ref{Tab_MSE_FE} our estimator outperforms characteristic-function based density estimators. Here the ``Fourier'' results are based on the estimator of Bonhomme and Robin (2010), and we use their recommended choice to set the regularization parameter in each replication. Inspection of the estimates suggests that the differences are mainly driven by estimates of the tails of the densities. Characteristic-function based estimators do not guarantee that densities be non-negative, and we find that their values tend to oscillate in the left and right tails. From results in Chetverikov and Wilhelm (2017), we conjecture that finite-sample performance benefits from the fact that our estimator enforces monotonicity of quantile functions and non-negativity of densities. However, proving this conjecture would require providing convergence rate results in addition to our consistency analysis.

Lastly, in Appendix \ref{App_Additional_Simu} we present numerical calculations of the rate of convergence of our estimator of latent quantiles, in data simulated from a nonparametric scalar deconvolution model. The results suggest the rate ranges between $N^{-\frac{3}{10}}$ and $N^{-\frac{7}{10}}$ in the data generating processes that we study. We also compare the performance of our method to Mallows' (2007) ``deconvolution by simulation'' estimator.

\section{Empirical application: income risk over the business cycle in the PSID \label{Appli_sec}}

In this section we use our nonparametric method to study the cyclical behavior of income risk in the US. In an influential contribution, Storesletten \textit{et al.} (2004) report using the PSID that the dispersion of idiosyncratic income shocks increases substantially in recessions. Guvenen \textit{et al.} (2014) re-examine this finding, using US administrative data and focusing on log-income growth. They find that the dispersion of log-income growth is acyclical, and that its skewness is procyclical. Recently, Busch \textit{et al.} (2018) find similar results using the PSID and data from Sweden and Germany. Nakajima and Smyrnyagin (2019) use an approach similar to the one in Storesletten \textit{et al.} (2004), making use of a larger PSID sample and different measures of income, and find that log-income shocks exhibit countercyclical dispersion and procyclical skewness. This literature is motivated by the key quantitative role of the cyclical behavior of the income process when calibrating models of business cycle dynamics.  

Here we revisit this question, by estimating a nonparametric permanent-transitory model where log-income, net of the effect of some covariates, is the sum of a random walk $\eta_{it}=\eta_{i,t-1}+v_{it}$ and an independent innovation $\varepsilon_{it}$. In first-differences we have, denoting log-income growth as $\Delta Y_{it}=Y_{it}-Y_{i,t-1}$:
\begin{equation} \Delta Y_{it}=v_{it}+\varepsilon_{it}-\varepsilon_{i,t-1},\quad t=1,...,T.\label{mod_perm_trans}\end{equation}
Model (\ref{mod_perm_trans}) is a linear factor model with $2T-1$ independent factors. Indeed, we have: $$\underset{\equiv Y}{\underbrace{\left(\begin{array}{c}\Delta Y_1\\\Delta Y_2\\\Delta Y_3\\...\\\Delta Y_T\end{array}\right)}}=\underset{\equiv A}{\underbrace{\left(\begin{array}{cccccccc}1 & 0 & ... & 0& 1 & 0 & ... & 0\\ 0 & 1 &  ... & 0 & -1 & 1 & ... & 0\\ 0 & 0 & ... & 0 & 0 & -1 & ...& 0\\... & ... & ... & ... & ...& ...& ...& ...\\0 & 0 & ... & 1 & 0 & 0 & ... & -1\end{array}\right)}}\,\underset{\equiv X}{\underbrace{\left(\begin{array}{c}v_1-\varepsilon_0\\v_{2}\\...\\v_T+\varepsilon_T\\\varepsilon_1\\\varepsilon_2\\...\\\varepsilon_{T-1}\end{array}\right)}}.$$ We leave the distributions of $v_{it}$ and $\varepsilon_{it}$ unrestricted. Our aim is to document the behavior of these distributions over the business cycle.

There are several differences between our model and estimation approach and the ones in Storesletten \textit{et al.} (2004). A substantive difference is that we estimate the densities of the shocks nonparametrically, while they use a parametric model under Gaussian assumptions. This is important, since estimates of non-Gaussian models (e.g., Horowitz and Markatou, 1996; Geweke and Keane, 2000; Bonhomme and Robin, 2010; Arellano \textit{et al.}, 2017) and descriptive evidence (e.g., Guvenen \textit{et al.}, 2014; Guvenen \textit{et al.}, 2016) both suggest that income shocks are strongly non-Gaussian in the US. Another difference is that we rely on first-differences of log-income in estimation, while Storesletten \textit{et al.} (2004) estimate the model in levels. This choice allows them to exploit a long past history of recessions, even before the PSID started to be collected, since past recessions and expansions affect the cross-sectional variance of log-income. On the other hand, income levels may also reflect other differences between cohorts, and our estimation in differences is robust to those. A last, less substantive difference is that we impose that the persistent component follows a unit root, while Storesletten \textit{et al.} (2004) use an autoregressive process whose baseline value for the autoregressive coefficient is 0.96.  


\begin{figure}[h!]
	\caption{Quantile functions and densities of income shocks, averaged over years\label{Fig_PSID_QuantDens}}
	\begin{center}
		\begin{tabular}{cc}
			\multicolumn{2}{c}{Permanent shocks}\\
			Quantile function & Density\\
			\includegraphics[width=60mm, height=40mm]{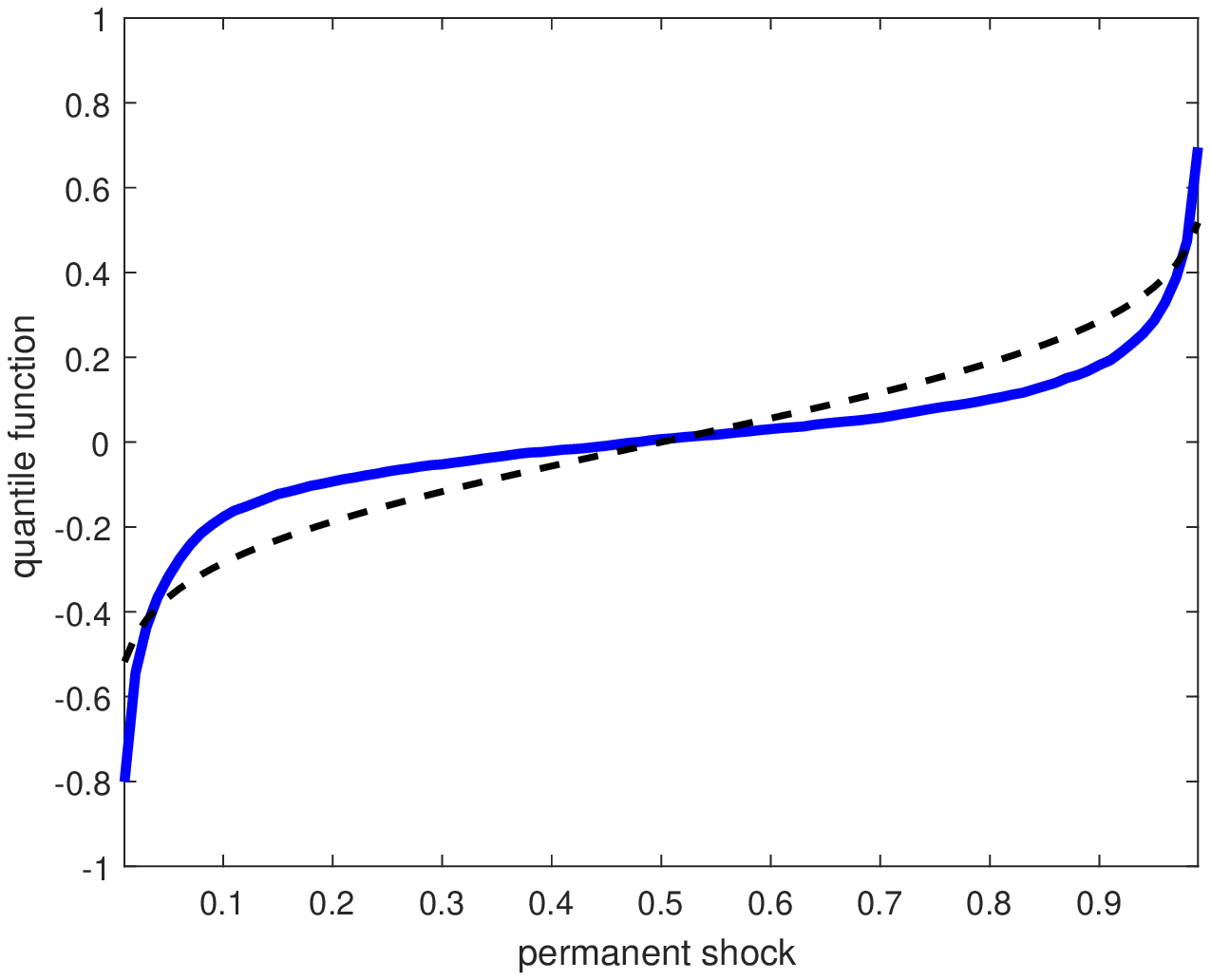} & %
			\includegraphics[width=60mm, height=40mm]{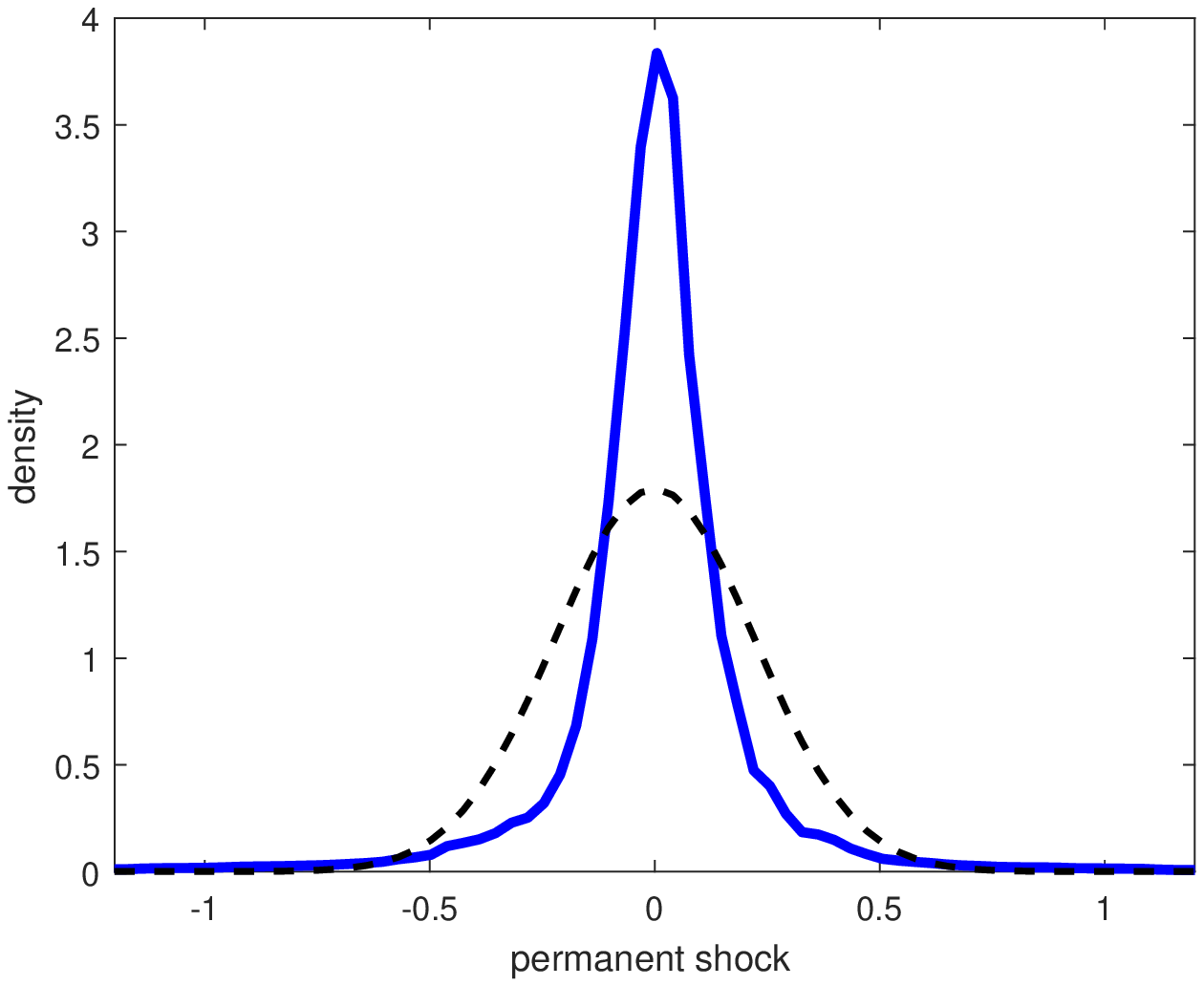} \\
			\multicolumn{2}{c}{Transitory shocks}\\
			Quantile function & Density\\
			\includegraphics[width=60mm, height=40mm]{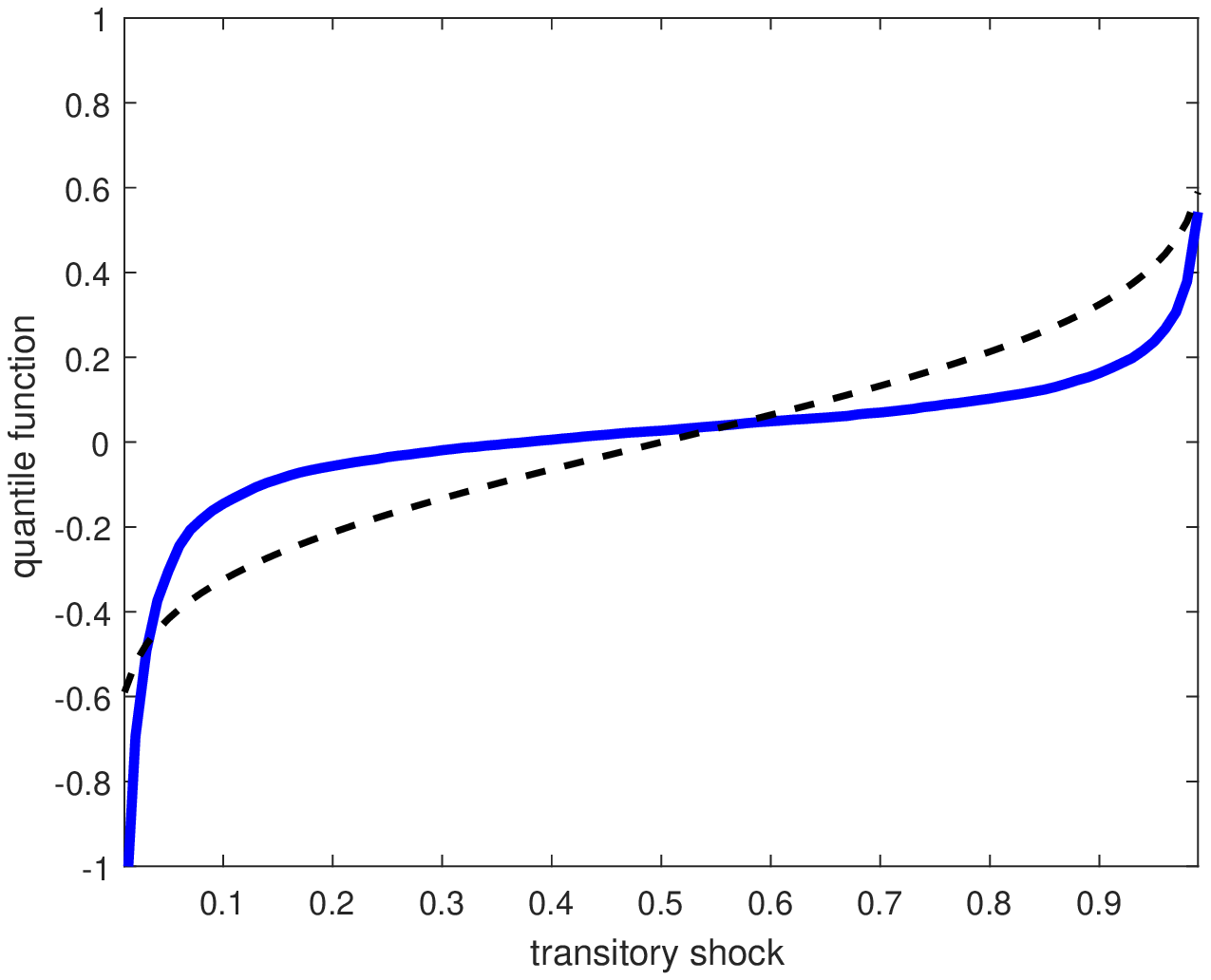} & %
			\includegraphics[width=60mm, height=40mm]{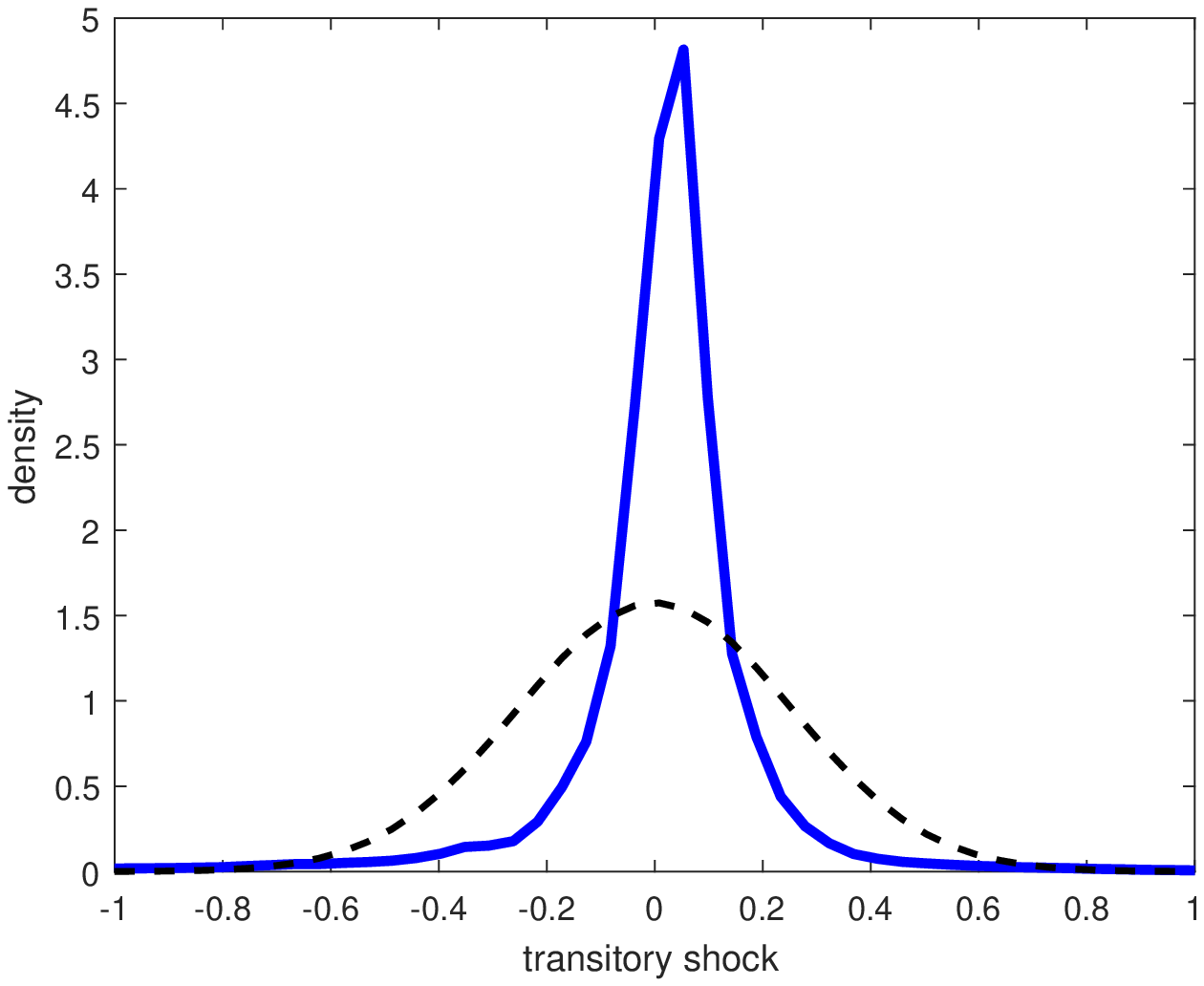} 			
		\end{tabular}%
	\end{center}
	\par
	\textit{{\footnotesize Notes: Sequence of balanced four-year subpanels from the PSID, 1969-1992. Nonparametric estimates of the quantile functions and densities of permanent and transitory income shocks to log-household annual labor income residuals, averaged over years. Normal fits are shown in dashed.}}
\end{figure}

\begin{figure}[h!]
	\caption{Dispersion and skewness of income shocks over the business cycle\label{Fig_PSID_Cycle}}
	\begin{center}
		\begin{tabular}{cc}
			\multicolumn{2}{c}{Permanent shocks}\\
			Dispersion & Skewness\\
			\includegraphics[width=70mm, height=50mm]{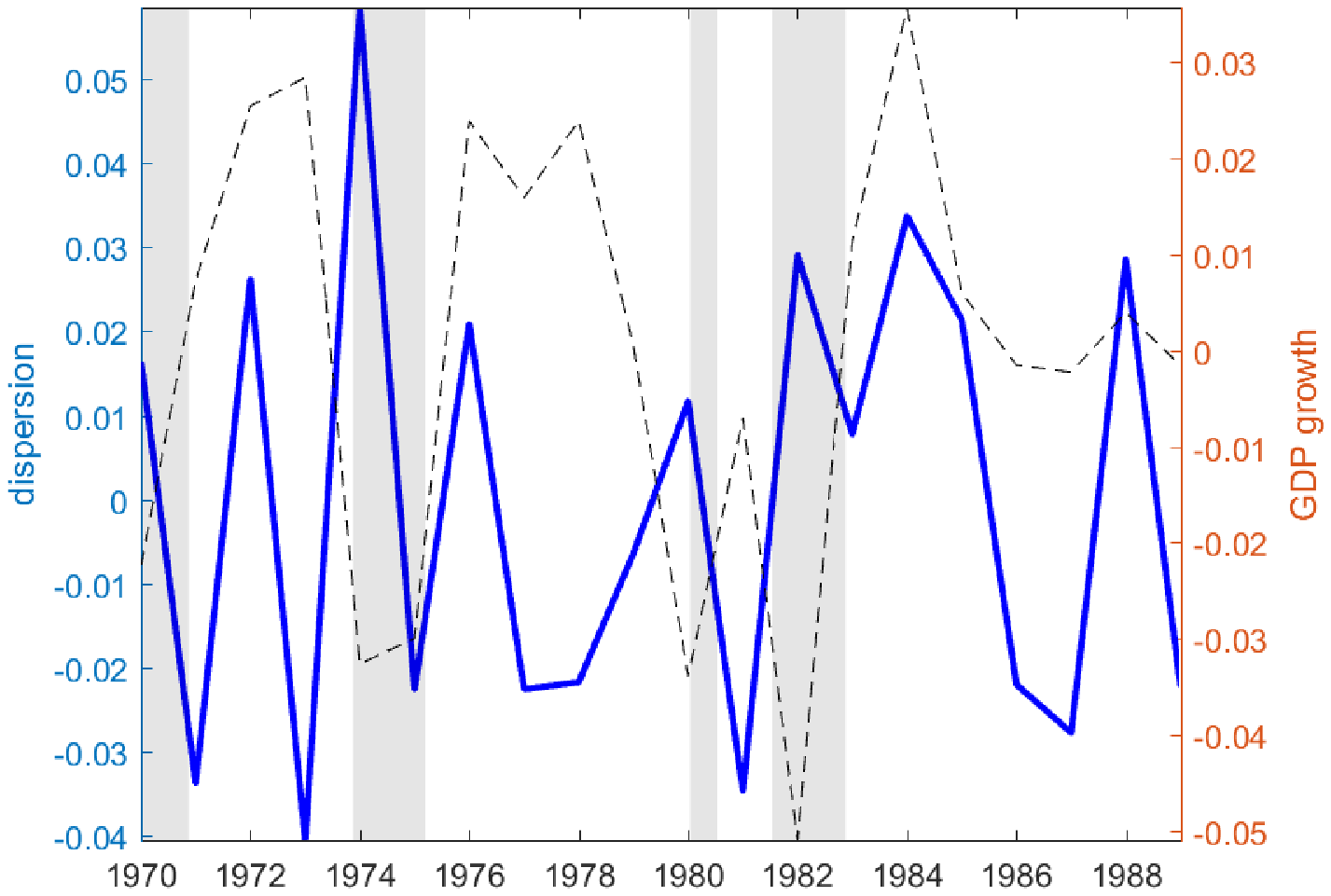} & %
			\includegraphics[width=70mm, height=50mm]{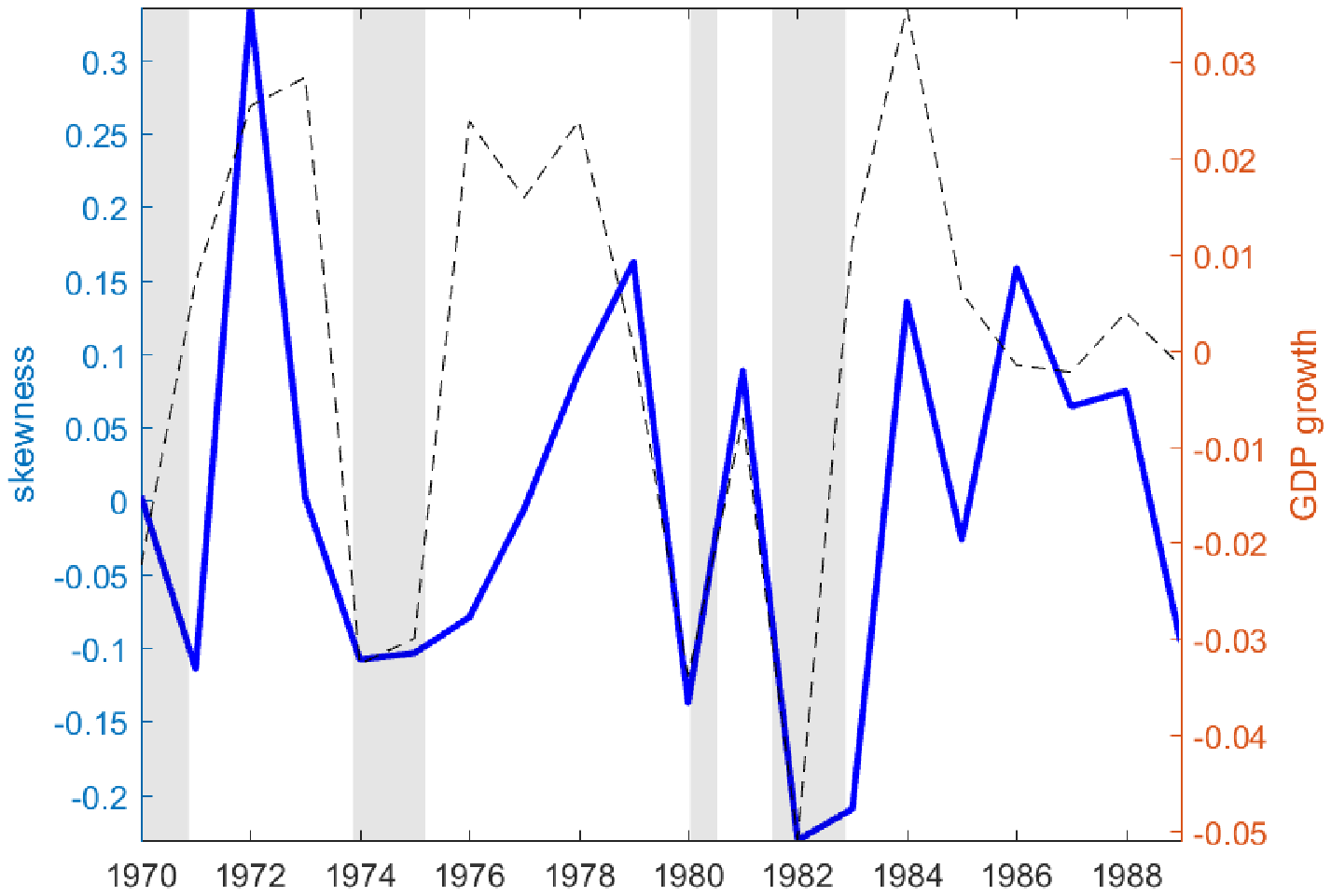} \\
			\multicolumn{2}{c}{Transitory shocks}\\
			Dispersion & Skewness\\
			\includegraphics[width=70mm, height=50mm]{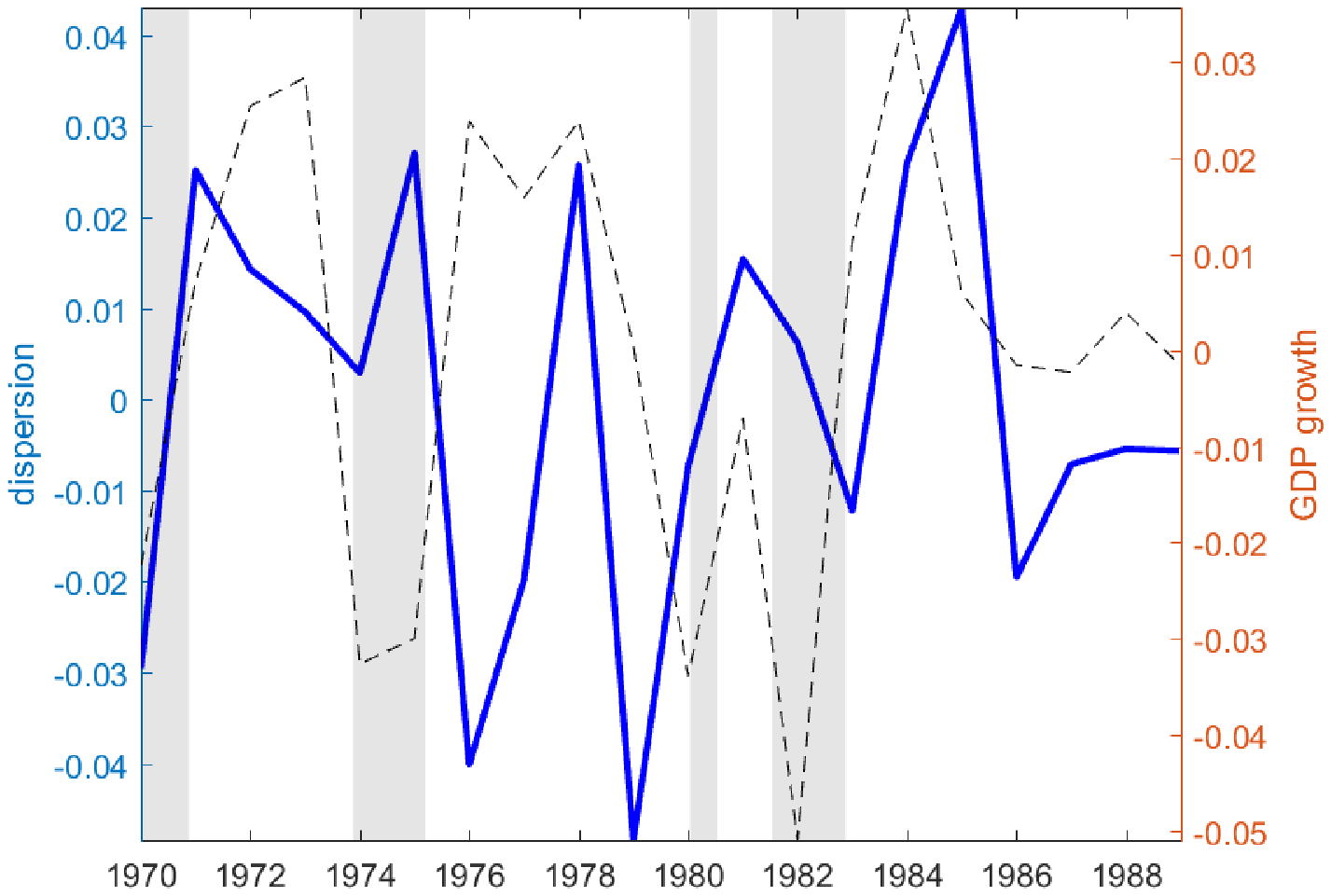} & %
			\includegraphics[width=70mm, height=50mm]{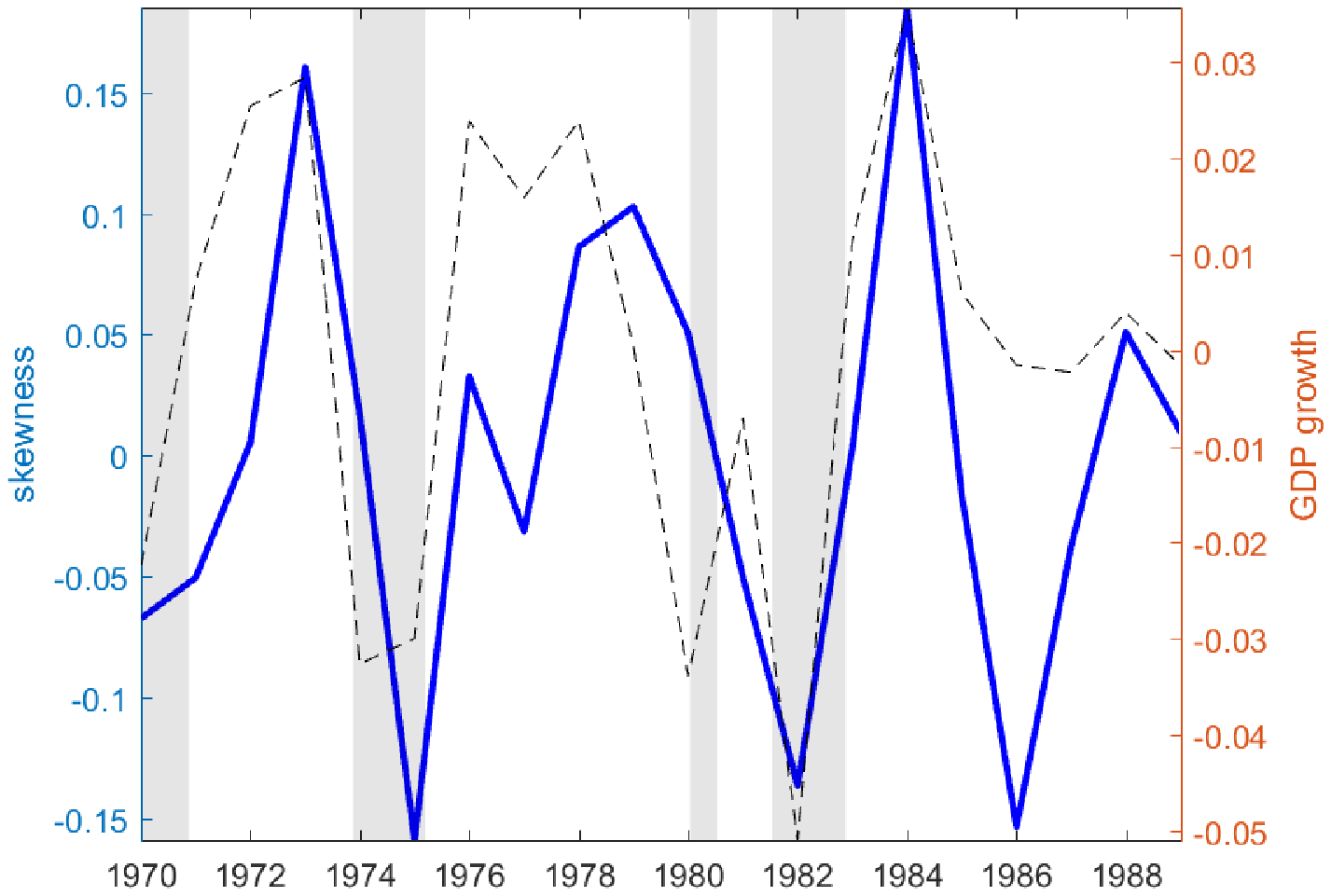} \\
		\end{tabular}%
	\end{center}
	\par
	\textit{{\footnotesize Notes: See notes to Figure \ref{Fig_PSID_QuantDens}. Dispersion ($P_{90}-P_{10}$) and skewness (Bowley-Kelley) are indicated in solid, log-real GDP growth is in dashed.}}
\end{figure}

Studying aggregate dynamics using survey panel data like the PSID is complicated by attrition and confounding age effects. To minimize the impact of these factors, we follow the approach pioneered by Storesletten \textit{et al.} (2004) and construct a sequence of balanced, four-year subpanels. In every subpanel, we require that households have non-missing data on income and demographics and comply with standard selection criteria: the household has positive annual labor income during the four years, the head is between 23 and 60 years old, and is not part of the SEO low-income sample or the immigrant sample. We estimate model (\ref{mod_perm_trans}) on 21 subpanels, whose base years range between 1969 and 1989. Log-household income growth is net of indicators for age (of head), education, gender, race, marital status, state of residence, number of children, and family size. In estimation we set conservative values for the penalization constants (that is, we use the ``weak constraint'' values of the simulation section), we use a single starting value in the algorithm, and we average the results of $M=10$ draws.

Our first finding is that income shocks are strongly non-Gaussian. In Figure \ref{Fig_PSID_QuantDens} we report the estimated quantile functions and densities of permanent shocks $v_{it}$ and transitory shocks $\varepsilon_{it}$, averaged over years (in solid), together with normal fits (in dashed). The excess kurtosis of both shocks is in line with previous evidence reported in the literature (e.g., Geweke and Keane, 2000; Bonhomme and Robin, 2010).

\begin{table}[tbp]\caption{Cyclicality of the distributions of income shocks\label{Tab_PSID_Cycle}}
	\begin{center}
		\begin{tabular}{r||cccc|cccc}
			& \multicolumn{8}{c}{Income}\\
			& \multicolumn{4}{c}{Permanent} & \multicolumn{4}{c}{Transitory}\\
			&Dispersion & Skewness & Upper & Lower &Dispersion & Skewness & Upper & Lower \\\hline
			Coeff. &-0.2528  & 3.0752  & 0.4647   &-0.7175  &  0.0752      & 2.3612      &  0.4133  & -0.3381   
			\\
			St. Er.& 0.3011 &  0.7576	&  0.2023 &  0.2167& 0.2380&0.6239 &   0.1536 &  0.1568
			\\\hline \hline
			& \multicolumn{8}{c}{Wages}\\
			& \multicolumn{4}{c}{Permanent} & \multicolumn{4}{c}{Transitory}\\
			&Dispersion & Skewness & Upper & Lower&Dispersion & Skewness & Upper & Lower  \\\hline
			Coeff.  &   0.1629&   0.5235   &  0.1680&  -0.0051 &0.2374    &  0.7793    & 0.2594   & -0.0220     \\
			St. Er. &    0.3750 & 0.5558 &  0.2627 &  0.1295 &  0.2453 &  0.7093 &  0.2150 &  0.1351 	\\\hline \hline
		\end{tabular}
	\end{center}
	{\footnotesize \textit{Notes: See notes to Figure \ref{Fig_PSID_Cycle}. The coefficients are obtained from a regression of $P_{90}-P_{10}$ dispersion (respectively, Bowley-Kelley skewness, upper tail $P_{90}-P_{50}$, or lower tail $P_{50}-P_{10}$) on log-real GDP growth and a linear time trend. Newey-West standard errors (one lag).}}
\end{table}

\begin{figure}[h!]
	\caption{Quantiles over the business cycle\label{Fig_PSID_Cycle_Quant}}
	\begin{center}
		\begin{tabular}{cc}
			Permanent shocks & Transitory shocks\\
			\includegraphics[width=70mm, height=50mm]{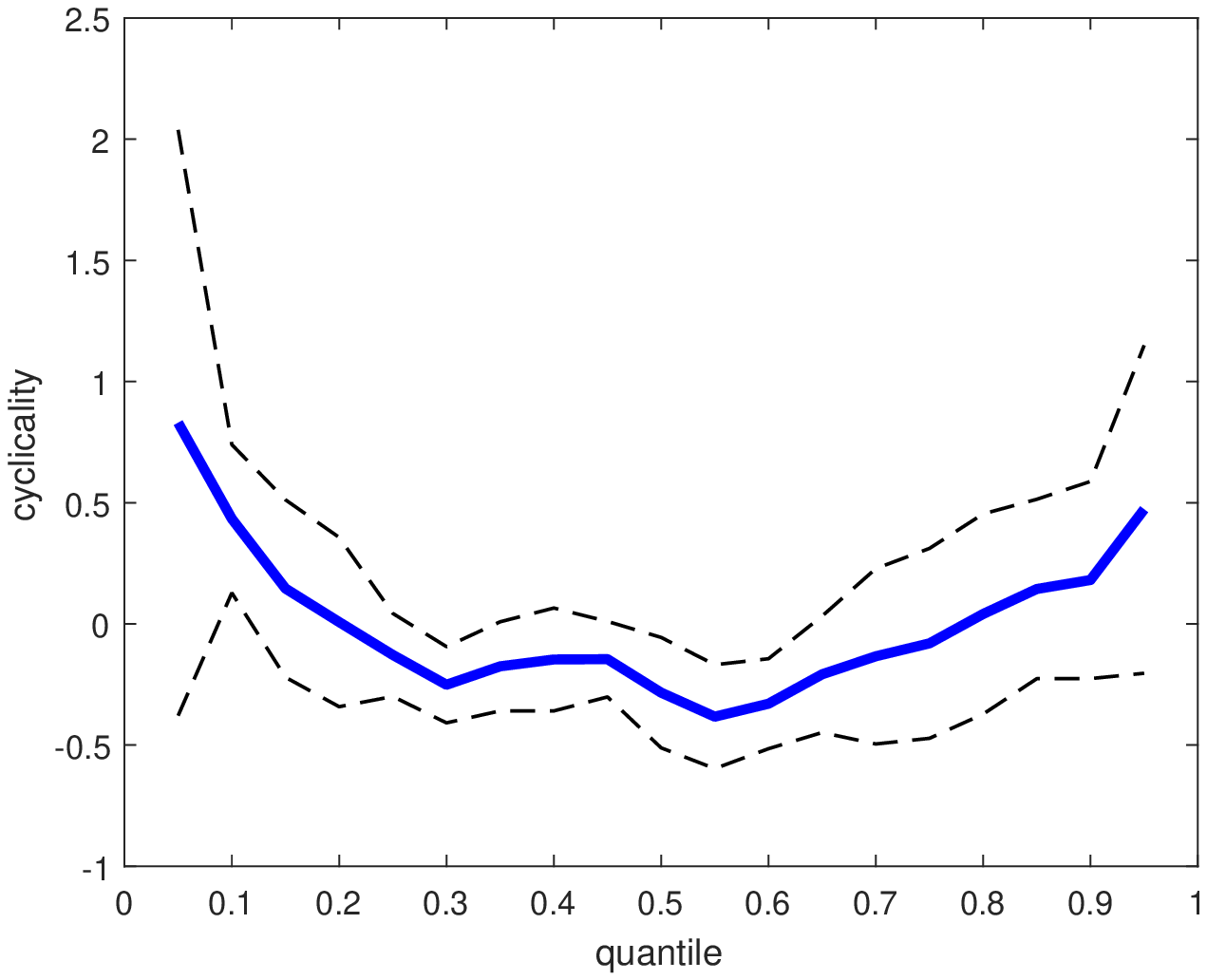} & %
			\includegraphics[width=70mm, height=50mm]{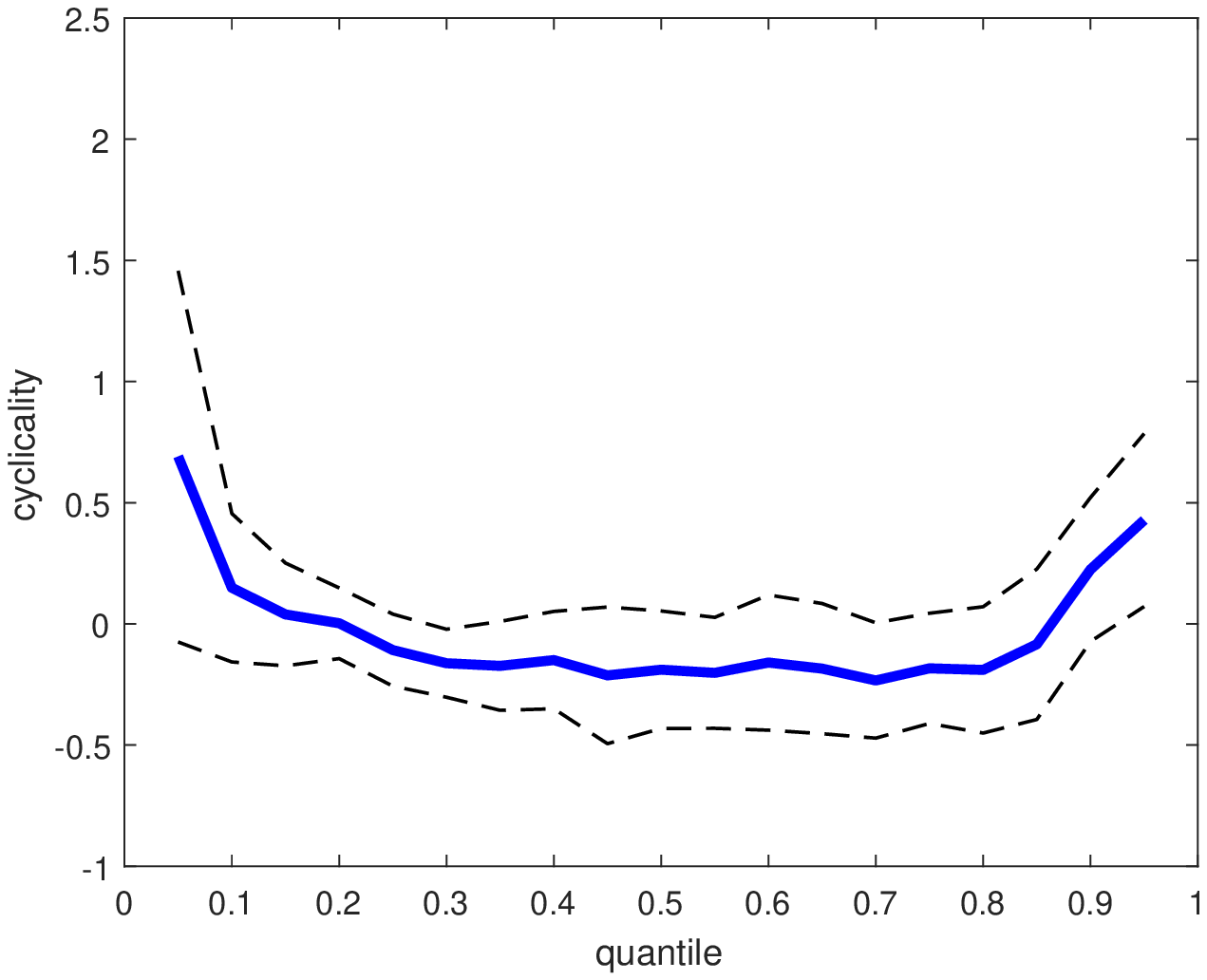} \\
		\end{tabular}%
	\end{center}
	\par
	\textit{{\footnotesize Notes: See notes to Figure \ref{Fig_PSID_Cycle}. On the y-axis we report estimates of the coefficient of log-real GDP growth in the regression of quantiles of permanent or transitory shocks in a regression that includes a time trend. The quantiles are shown on the x-axis. Newey-West 95\% confidence intervals are shown in dashed.}}
\end{figure}

We are interested in how features of these distributions vary with the business cycle. In the left column of Figure \ref{Fig_PSID_Cycle} we plot the 90/10 percentile difference of log-income $P_{90}-P_{10}$ (a common measure of dispersion, in solid) together with log-GDP growth (in dashed), both of them net of a linear time trend. While permanent and transitory shocks tend to move countercyclically in the first part of the period, the relationship tends to become procyclical in the 1980's. As we report in Table \ref{Tab_PSID_Cycle}, the coefficient of log-GDP growth in a regression of the dispersion of permanent income shocks on log-GDP growth and a time trend is -0.25, with a Newey-West standard error of 0.30.\footnote{We compute the Newey-West formula with one lag. Using two or three lags instead has little impact. In the computation we do not account for the fact that the quantiles are estimated, our rationale being that the cross-sectional sizes are large relative to the length of the time series.} Hence, overall we do not find significant evidence that the dispersion of permanent shocks varies systematically with the business cycle. This result based on first-differenced estimation and a nonparametric approach contrasts with the main finding in Storesletten \textit{et al.} (2004). In addition, we neither find that the dispersion of transitory shocks varies with the cycle.

\begin{figure}[h!]
	\caption{Fit to densities and quantile cyclicality of log-income/wage growth\label{Fig_PSID_Fit}}
	\begin{center}
		\begin{tabular}{cc}
			\multicolumn{2}{c}{Income}\\
			Density & Quantile cyclicality\\
			\includegraphics[width=60mm, height=40mm]{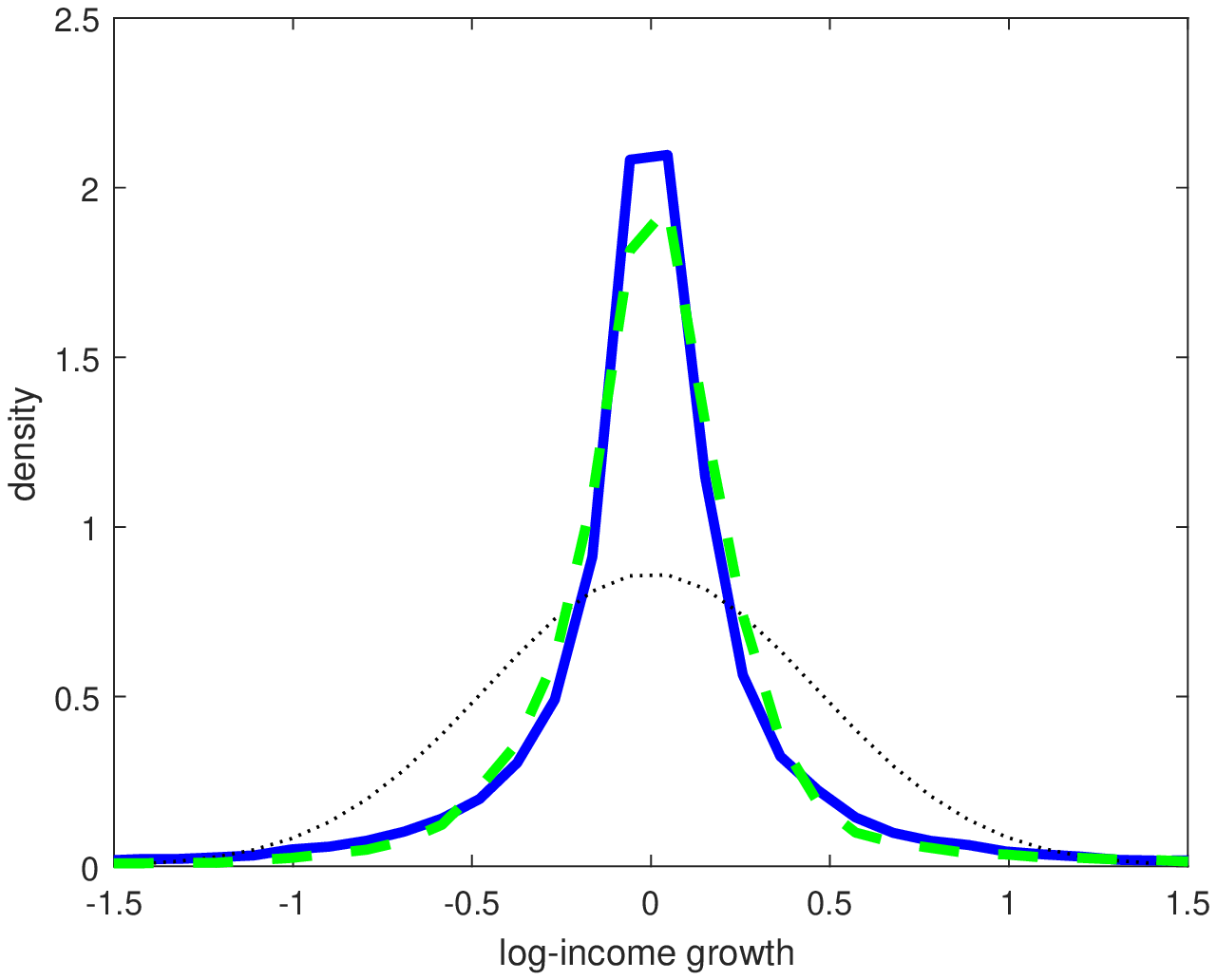} & %
			\includegraphics[width=60mm, height=40mm]{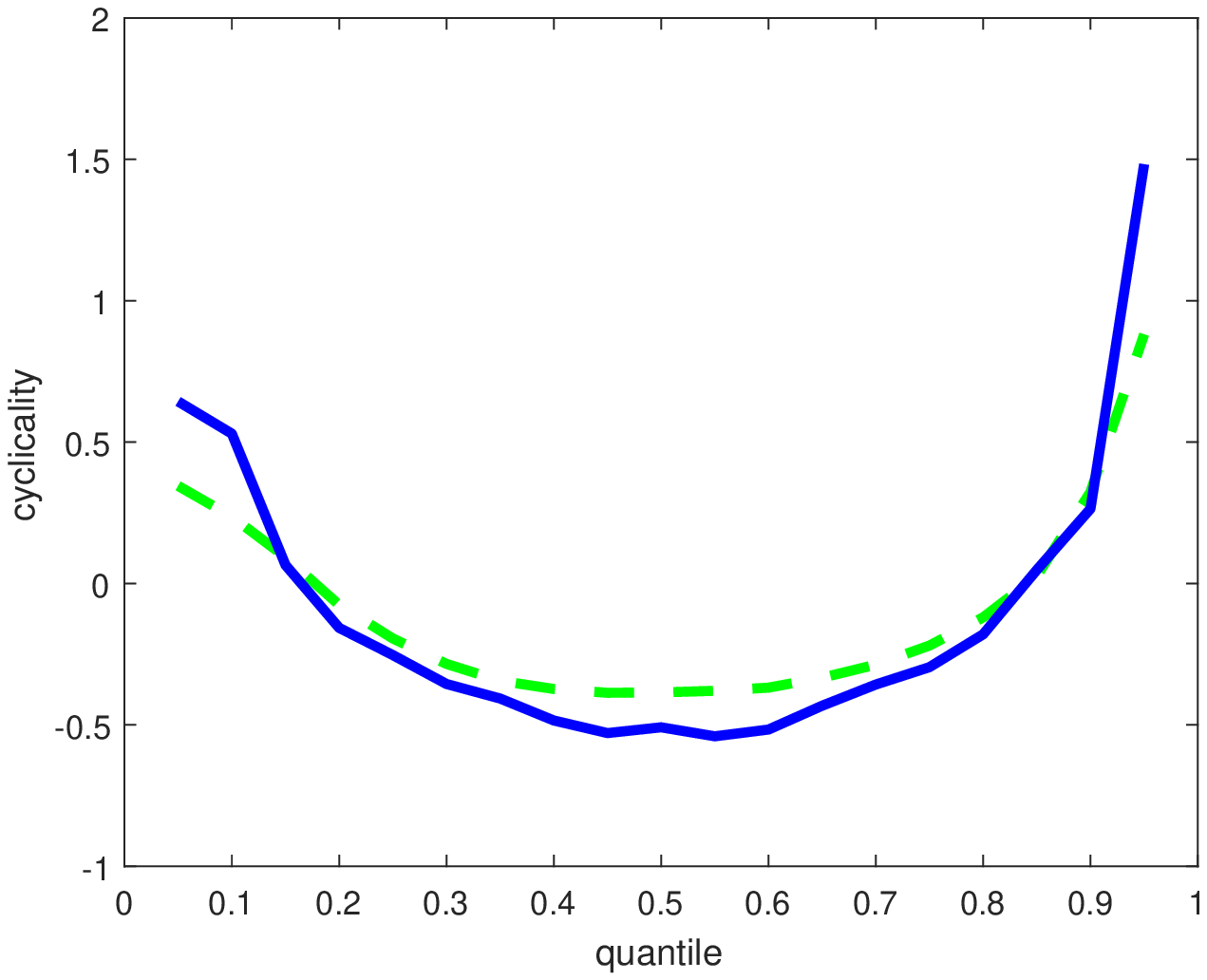} \\
			\multicolumn{2}{c}{Wages}\\
			Density & Quantile cyclicality\\
			\includegraphics[width=60mm, height=40mm]{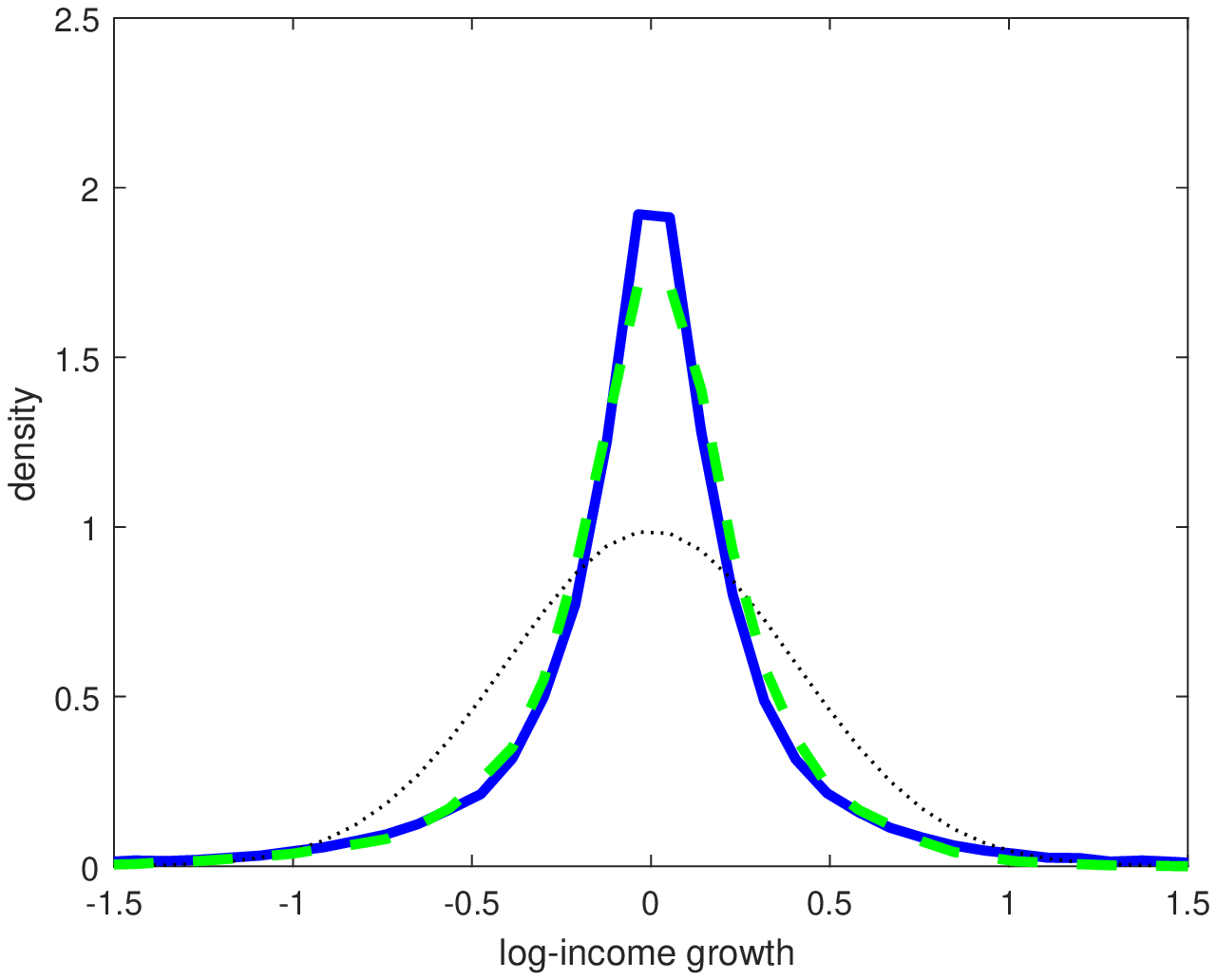} & %
			\includegraphics[width=60mm, height=40mm]{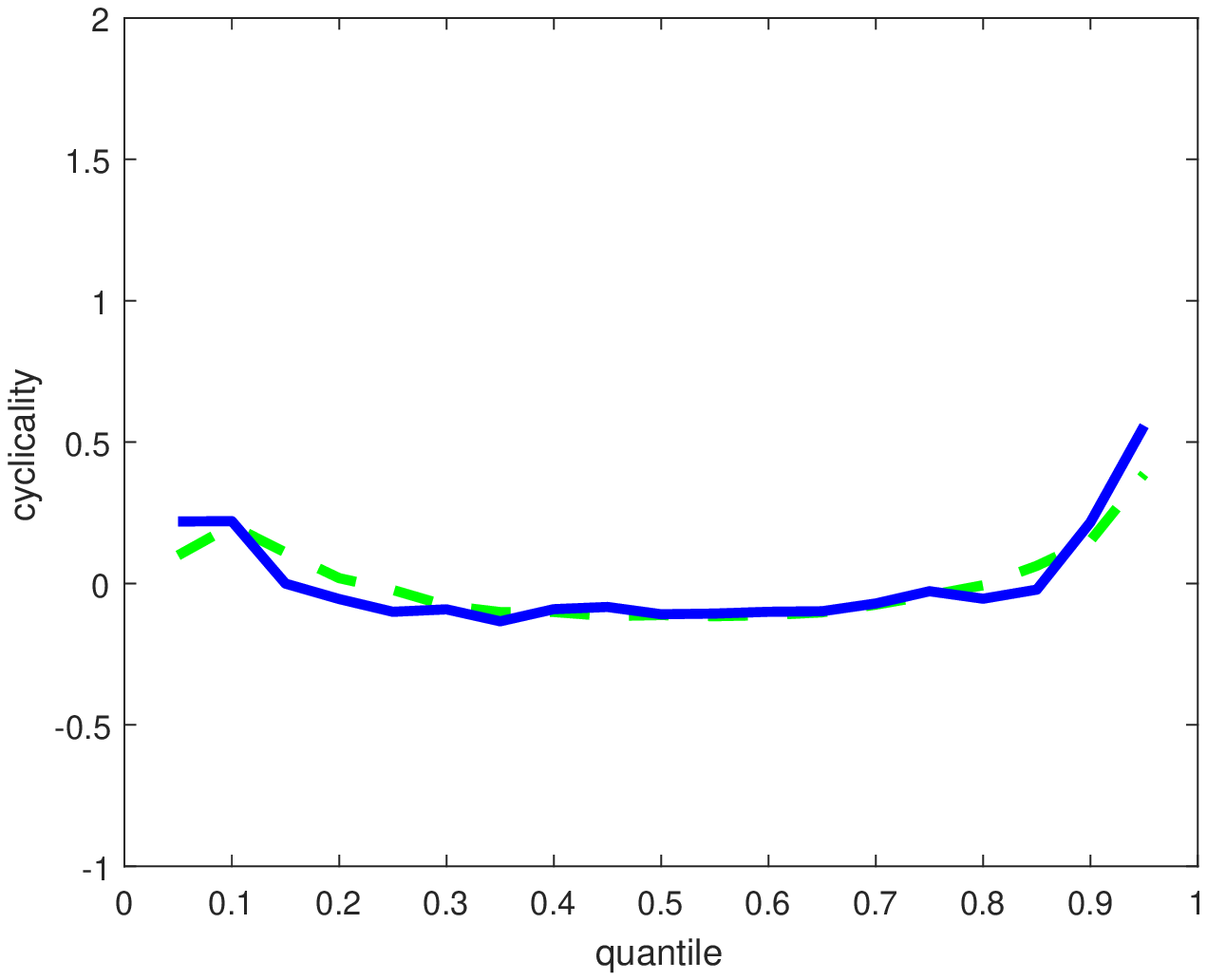} 			
		\end{tabular}%
	\end{center}
	\par
	\textit{{\footnotesize Notes: See notes to Figure \ref{Fig_PSID_Cycle}. In the upper left panel we show the density of log-income growth in the data (in solid), and as predicted by our model (in dashed), with a normal fit (in dotted). In the upper right panel we show a measure of quantile cyclicality similar to the one in Figure \ref{Fig_PSID_Cycle_Quant} for log-income growth, in the data (in solid), and as predicted by the model (in dashed). In the bottom panels we show results for hourly wages.}}
\end{figure}
Next, in the right column of Figure \ref{Fig_PSID_Cycle_Quant} we plot the Bowley-Kelley quantile measure of skewness $[(P_{90}-P_{50})-(P_{50}-P_{10})]/(P_{90}-P_{10})$. The graphs of permanent and transitory income shocks suggest that skewness is procyclical. This is confirmed in Table \ref{Tab_PSID_Cycle}, which shows that the coefficient of log-GDP growth in the skewness regression is 3.07 for permanent shocks, and 2.36 for transitory shocks, significant at the 5\% level in both cases. Our nonparametric estimates of a permanent-transitory model of income dynamics thus suggest that dispersion is approximately acyclical, and skewness is procyclical, in line with the conclusions of the descriptive evidence in Guvenen \textit{et al.} (2014) and Busch \textit{et al.} (2018).

  As graphical way to illustrate the distributional dynamics of income over the business cycle, in Figure \ref{Fig_PSID_Cycle_Quant} we plot the coefficients of log-GDP growth in regressions of the quantiles of permanent or transitory income shocks on log-GDP growth and a time trend. The estimates suggest a U-shape pattern along the distribution, both for permanent and transitory shocks. Expansions are associated with increases at the top and bottom of the distribution, while recessions are associated with the opposite pattern and a relative increase of the middle quantiles. In the upper panel of Figure \ref{Fig_PSID_Fit} we show how the model fits the distributions of log-income growth, suggesting that our model is able to reproduce the density and quantile cyclicality of log-income growth that we observe in the data.

  We performed several exercises to probe the robustness of these findings, using the ``strong constraint'' penalization of Section \ref{Simu_sec},  measuring business cycle conditions using the unemployment rate instead of log-GDP growth, and varying the choice of starting values in the algorithm. While we found the year-to-year variation in Figure \ref{Fig_PSID_Cycle} to depend on the chosen specification, in all our checks we found a lack of systematic cyclical variability of the dispersion of income shocks, and a significant procyclicality of the skewness of permanent shocks. Among the results reported in Table \ref{Tab_PSID_Cycle}, we found the procyclicality of the skewness of transitory shocks to be most sensitive to specification changes.

  \paragraph{Hourly wages.}
  
  We next use the information in the PSID about hours worked to compute similar measures of cyclicality based on hourly wages of household heads. Evidence from Italy and France (Hoffmann and Malacrino, 2019; Pora and Wilner, 2019) suggests that days and hours worked may contribute significantly to the observed cyclical patterns of skewness. For the US, Nakajima and Smyrnyagin (2019) obtain similar conclusions. In contrast, Busch \textit{et al.} (2018) find a moderate role of hours worked in Germany. In the bottom panel of  Table \ref{Tab_PSID_Cycle} we see that the skewnesses of permanent and transitory shocks to hourly wages do not vary significantly with the cycle, and that the point estimates are greatly reduced compared to the case of total income. This suggests that hours worked largely contribute to the distributional income dynamics that we document. In the lower panels in Figure \ref{Fig_PSID_Fit} we show the model fit to log-hourly wage growth. The estimates show that quantiles of log-hourly wage growth vary little with the business cycle in our sample, and that our model is able to reproduce this pattern.

\section{Extensions\label{Extens_sec}}

In this section we briefly outline several extensions of our matching approach to random coefficients models, finite mixture models, and deconvolution models with heteroskedasticity. These extensions show that the idea of matching data observations to model predictions is applicable to a variety of settings with latent variables.

\subsection{Random coefficients}

Consider the linear cross-sectional random coefficients model:
\begin{equation}\label{RC_model}
Y=X_1+\sum_{k=2}^{K}W_k X_k,
\end{equation}
where $(W_2,...,W_K)$ is independent of $(X_1,...,X_K)$, the scalar outcome $Y$ and the covariates $W_2,...,W_K$ are observed, and $(X_1,...,X_K)$ is a latent vector with an unrestricted joint distribution (e.g., Beran and Hall, 1992; Hoderlein \textit{et al.}, 2010). To construct a matching estimator in this case, we augment (\ref{RC_model}) with: $W_k=V_k$, $k=2,...,K$, where the $V_k$'s are \emph{auxiliary latent variables} independent of the $X_k$'s. In this augmented model, the joint distributions of $(X_1,...,X_K)$ and $(V_2,...,V_K)$ can be estimated by minimizing the Euclidean distance between the model's predictions of $Y,W$ observations, and their matched values in the data. A similar approach can be used in binary choice models with random coefficients (Ichimura and Thompson, 1998; Gautier and Kitamura, 2013)  

To see how to adapt this idea to panel data random coefficients models, consider the random trends model:
\begin{equation}
\label{RC_trend}Y_{it}=\alpha_i+\beta_i t+\varepsilon_{it},
\end{equation}
where $(\alpha_i,\beta_i)$, $\varepsilon_{i1}$, ..., $\varepsilon_{iT}$ are mutually independent. Our matching approach applies directly to this case, by minimizing the following objective:
\begin{align}\label{emp_obj}
(\widehat{\alpha},\widehat{\beta},\widehat{\varepsilon}_1,...\widehat{\varepsilon}_T)=\underset{(\alpha,\beta,\varepsilon_1,...,\varepsilon_T)}{\limfunc{argmin}}\, \left\{\underset{\pi\in \Pi_N}{\limfunc{min}}\, \sum_{i=1}^N\sum_{t=1}^T\left(Y_{\pi(i),t}-\alpha_i-\beta_i t-\varepsilon_{\sigma_t(i),t}\right)^2 \right\},
\end{align}
where $\sigma_1,...,\sigma_T$ are independent random permutations of $\{1,...,N\}$. In this case our algorithm consists in alternating optimal transport (matching) steps and least squares (update) steps. Note that in this case the algorithm delivers bivariate pseudo-observations $(\widehat{\alpha}_i,\widehat{\beta}_i)$, and that those can no longer be interpreted as estimates of order statistics; see Chernozhukov \textit{et al.} (2017) for an optimal transport approach to multivariate quantiles.\footnote{When the trend $t$ in (\ref{RC_trend}) is replaced by a strictly exogenous regressor $X_{it}$, we can augment the model with auxiliary latent variables $V_{it}$ using the same strategy as in the cross-sectional case, and minimize the distance between the model's predictions of $Y,X$ and their matched values in the data. Note that in that case $X_{it}$ and $(\alpha_i,\beta_i)$ are allowed to be dependent.}

\subsection{Finite mixtures}

Consider next a finite mixture model with $G$ groups, for a $T$-dimensional outcome $Y$:
\begin{equation}\label{FM_model}
Y_t=\sum_{g=1}^G Z_g X_{gt},\quad t=1,...,T,
\end{equation}
where $Z_1,...,Z_G$ and $X_{11},...,X_{GT}$ are unobserved, $Z_g\in\{0,1\}$ with $\sum_{g=1}^G Z_g=1$, and $(Z_1,...,Z_G)$ and all $X_{11}$, ..., $X_{GT}$ are mutually independent. The nonparametric model (\ref{FM_model}) has been extensively analyzed in the literature (e.g., Hall and Zhou, 2003; Hu, 2008; Allmann \textit{et al.}, 2009; Bonhomme \textit{et al.}, 2016).

To construct a matching estimator in model (\ref{FM_model}) we first note that, by the threshold crossing representation, there exist a parameter vector $\mu=(\mu_1,...,\mu_{G-1})$ and a standard uniform random variable $V$ such that $Z_g=Z_g(V,\mu)$, where $Z_1(V,\mu)=1$ if and only if $V\leq \mu_1$, $Z_g(V,\mu)=1$ if and only if $\mu_{g-1}<V\leq \mu_g$ for $g=2,...,G-1$, and $Z_{G}(V,\mu)=1$ if and only if $\mu_{G-1}<V$. We denote as ${\cal{M}}_{G-1}$ the set of vectors $\mu\in \mathbb{R}^{G-1}$ such that $0\leq \mu_1\leq \mu_2\leq ...\leq \mu_{G-1}\leq 1$. We then define the following estimator:
\begin{align}\label{est_FM_model_simu}
&(\widehat{X},\widehat{\mu})=\underset{X\in {\cal{X}}_N,\, \mu\in {\cal{M}}_{G-1}}{\limfunc{argmin}}\,\,\,\Bigg\{\underset{\pi\in \Pi_N}{\limfunc{min}}\,   \sum_{i=1}^N\sum_{t=1}^T\left(Y_{\pi(i),t}-\sum_{g=1}^G Z_{g}(V_i,\mu)X_{\sigma_{gt}(i),gt}\right)^2\Bigg\},
\end{align}
where $V_1,...,V_N$ are standard uniform draws, and $\sigma_{gt}$ are random permutations in $\Pi_N$ for all $g=1,...,G$, $t=1,...,T$, all independent of each other. 

For given $\mu$, we use an algorithm analogous to the one described in Section \ref{Comput_sec} to compute $\widehat{X}$. The outer minimization with respect to $\mu$ can be performed using simulated annealing or other methods to minimize non-differentiable objective functions. In Appendix \ref{App_Additional} we report simulation results for a nonparametric two-component mixture model. In that case grid search is a viable option. Moreover, a similar approach can be used to estimate finite mixtures of linear independent factor models (also known as ``mixtures of factor analyzers''); see Ghahramani and Hinton (1997) and McLachlan \textit{et al.} (2003).

\subsection{Heteroskedastic deconvolution}

Finally, consider the model
\begin{equation}Y=X_1+SX_2,\label{eq_CH}\end{equation}
where $(X_1,S)$ is independent of $X_2$, and $X_2\sim F$, where $F$ is known and has zero mean. The econometrician observes a sample $Y_1,\widetilde{S}_1,...,Y_N,\widetilde{S}_N$ from $(Y,\widetilde{S})$, where $\widetilde{S}_i$ is a consistent estimator of $S_i$ for all $i$.

To motivate this setup, consider the estimation of neighborhood effects on income in Chetty and Hendren (2018), where $i$ is a commuting zone or county, and $Y_i$ is a neighborhood-specific estimate of the ``causal effect'' of place $i$. Within-$i$, a central limit theorem-type argument suggests that $Y_i$ is approximately normally distributed, with mean $X_{i1}$ and standard deviation $S_i$. Chetty and Hendren report, alongside $Y_i$ estimates, standard deviation estimates $\widetilde{S}_i$. In this example $F$ is the standard normal distribution. See Azevedo \textit{et al.} (2019) for other examples and a parametric estimation approach.

To estimate the distribution of $X_{1}$ by matching, we minimize the following objective:
\begin{align}\label{emp_obj}
(\widehat{X}_1,\widehat{S})=\underset{(X_1,S)\in{\cal{X}}_N\times {\cal{S}}_N}{\limfunc{argmin}}\, \left\{\underset{\pi\in \Pi_N}{\limfunc{min}}\, \sum_{i=1}^N\left(Y_{\pi(i)}-X_{i1}-S_iX_{\sigma(i),2}\right)^2 +\lambda \sum_{i=1}^N\left(\widetilde{S}_{\pi(i)}-S_i\right)^2\right\},
\end{align}
where $\sigma$ is a random permutation of $\{1,...,N\}$, ${\cal{S}}_N$ is the parameter space for $S$, and $\lambda>0$ is a constant. In this case our algorithm again consists in alternating optimal transport steps and least squares steps. As an illustration, in Appendix \ref{App_Additional} we estimate the density of neighborhood effects across US commuting zones using the data from Chetty and Hendren (2018).

\section{Conclusion\label{Conclu_sec}}

In this paper we have proposed an approach to nonparametrically estimate models with latent variables. The method is based on matching predicted values from the model to the empirical observations. We have provided a simple algorithm for computation, and established consistency. We have also documented remarkable performance of our nonparametric estimator in small samples, and we have used it to shed new light on the cyclicality of permanent and transitory shocks to income and wages in the US. Progress on computation might be possible by leveraging recent advances on regularized optimal transport (Cuturi, 2013; Peyré and Cuturi, 2019). Finally, an important question for future work will be to characterize rates of convergence and asymptotically valid confidence sets for our estimator.

	\baselineskip21pt
	\clearpage
	
	\appendix
	
\renewcommand{\theequation}{\thesection \arabic{equation}}

\renewcommand{\thelemma}{\thesection \arabic{lemma}}

\renewcommand{\theproposition}{\thesection \arabic{proposition}}

\renewcommand{\thecorollary}{\thesection \arabic{corollary}}

\renewcommand{\thetheorem}{\thesection \arabic{theorem}}

\renewcommand{\theassumption}{\thesection \arabic{assumption}}

\renewcommand{\thefigure}{\thesection \arabic{figure}}

\renewcommand{\thetable}{\thesection \arabic{table}}

\setcounter{equation}{0}
\setcounter{table}{0}
\setcounter{figure}{0}
\setcounter{assumption}{0}
\setcounter{proposition}{0}
\setcounter{lemma}{0}
\setcounter{corollary}{0}
\setcounter{theorem}{0}
	
	{\small \baselineskip15pt }
	
	\begin{center}
		{\small {\LARGE APPENDIX} }
	\end{center}
	
	\section{Proofs\label{App_proofs}}

\subsection{Proofs of Theorem \ref{theo_consis_gen} and Corollary \ref{theo_consis}}

Before proving Theorem \ref{theo_consis_gen} for multi-factor models, we first prove Corollary \ref{theo_consis} for the scalar deconvolution case where explicit expressions for Wasserstein distances are available.

\subsubsection{Scalar deconvolution: Corollary \ref{theo_consis}} 

We first state the following assumption, where for conciseness we denote $H\equiv H_1$. Part $(ii)$ is an identification condition that is commonly assumed in nonparametric deconvolution.

\begin{assumption}\label{ass_consis}$\quad$
	
	$(i)$ (Continuity and support) $Y$, $X_1$ and $X_2$ have compact supports in $\mathbb{R}$, and admit absolutely continuous densities $f_Y,f_{X_1},f_{X_2}$ that are bounded away from zero and infinity. Moreover, $f_Y$ is differentiable.  
	
%
	$(ii)$ (Identification) The characteristic function of $X_2$ does not vanish on the real line.

	$(iii)$ (Penalization) $\overline{C}_N$ is increasing and $\underline{C}_N$ is decreasing with $\limfunc{lim}_{N\rightarrow +\infty}\, \overline{C}_N =\overline{C}$ and
	$\limfunc{lim}_{N\rightarrow +\infty}\, \underline{C}_N =\underline{C}$, where $\overline{C}$ and $\underline{C}<\overline{C}$ are such that $\|F_{X_1}^{-1}\|\leq \overline{C}$ and $\nabla F_{X_1}^{-1}(\tau)\geq \underline{C}$ for all $\tau\in(0,1)$.
	
	$(iv)$ (Sampling) $Y_1,...,Y_N$ and $X_{12},...,X_{N2}$ are i.i.d.

\end{assumption}


We now prove Corollary \ref{theo_consis}. Define the empirical objective function, for any candidate quantile function $H$, as:
\begin{align*}
&\widehat{Q}(H)=\underset{\pi\in\Pi_N}{\limfunc{min}}\,\frac{1}{N}\sum_{i=1}^N\left
(Y_{\pi(i)}-H\left(\frac{\sigma(i)}{N+1}\right)-X_{i2}\right)^2\\
&\quad \quad \quad =\frac{1}{N}\sum_{i=1}^N\left
(\widehat{F}^{-1}_Y\left(\frac{1}{N}\,\widehat{\limfunc{Rank}}\left(H\left(\frac{\sigma(i)}{N+1}\right)+X_{i2}\right)\right)-H\left(\frac{\sigma(i)}{N+1}\right)-X_{i2}\right)^2,
\end{align*}
where $\widehat{F}^{-1}_Y(\tau)=\limfunc{inf}\, \{y\in\limfunc{Supp}(Y)\, :\, \widehat{F}_Y(y)\geq \tau\}$, and $\widehat{\limfunc{Rank}}(Z_i)=N\widehat{F}_Z(Z_i)$. The second equality follows from Hardy-Littlewood-P\'olya. With some abuse of notation, for all $X\in\mathbb{R}^N$ we will denote $\widehat{Q}(X)=\widehat{Q}(H)$ for any function $H$ such that $H\left(\frac{i}{N+1}\right)=X_i$ for all $i$.

Define the population counterpart to $\widehat{Q}$, for any $H\in{\cal{H}}$, as:
$$Q(H)=\mathbb{E}\left[\left
(F_Y^{-1}\left(\int_0^1 F_{X_2}\left(H(V)+X_2-H(\tau)\right)d\tau\right)-H(V)-X_2\right)^2\right],$$
where the expectation is taken with respect to pairs $(V,X_2)$ of independent random variables, where $V$ is standard uniform and $X_2\sim F_{X_2}$.

\paragraph{Parameter space.}

Let ${\cal{H}}$ be the closure of the set $\{H\in {\cal{C}}^1\, : \, \nabla H\geq \underline{C},\,  \|H\|\leq \overline{C}\}$ under the norm $\|\cdot\|_{\infty}$. ${\cal{H}}$ is compact with respect to $\|\cdot\|_{\infty}$ (Gallant and Nychka, 1987).\footnote{Compactness can be preserved when sup-norms are replaced by weighted Sobolev sup-norms (e.g., using polynomial or exponential weights); see for example Theorem 7 in Freyberger and Masten (2015).}

\paragraph{Sieve construction.}

For any $N$, let us define the \emph{sieve space}:
$${\cal{H}}_N=\left\{H\in{\cal{H}}\,:\, \left|H\left(\frac{i}{N+1}\right)\right|\leq \overline{C}_N,\,\underline{C}_N\leq (N+1)\left(H\left(\frac{i+1}{N+1}\right)-H\left(\frac{i}{N+1}\right)\right)\leq \overline{C}_N\right\}.$$
Let $\widehat{X}\in{\cal{X}}_N$ be such that:
\begin{equation*}
\widehat{Q}(\widehat{X})\leq\underset{X\in{\cal{X}}_N}{\limfunc{min}}\,\widehat{Q}(X)+\epsilon_N.
\end{equation*}
We first note that there exists an $\widehat{H}\in{\cal{H}}_N$ such that $\widehat{H}\left(\frac{i}{N+1}\right)=\widehat{X}_i$ for all $i$.\footnote{Take a smooth interpolating function of the $\widehat{X}_i$'s, arbitrarily close in sup-norm to the piecewise-linear interpolant of the $\widehat{X}_i$'s extended to have slope $(\underline{C}+\overline{C})/2$ on the intervals $[0,1/(N+1)]$ and $[N/(N+1),1]$. This is always possible since $\overline{C}_N<\overline{C}$ and $\underline{C}_N>\underline{C}$.} Hence:
\begin{equation}\widehat{Q}(\widehat{H})=\widehat{Q}(\widehat{X})\leq\underset{X\in{\cal{X}}_N}{\limfunc{min}}\,\widehat{Q}(X)+\epsilon_N\leq \underset{H\in{\cal{H}}_N}{\limfunc{min}}\,\widehat{Q}(H)+\epsilon_N.\label{eq_Hhat}\end{equation}

Let $H_0=F_{X_1}^{-1}$. To show Corollary \ref{theo_consis} it is thus sufficient to show that, when $\widehat{H}$ satisfies (\ref{eq_Hhat}), we have $\|\widehat{H}-H_0\|_{\infty}=o_p(1)$. This will follow from verifying conditions (3.1''), (3.2), (3.4), and (3.5(i)) in Chen (2007). 

\paragraph{$\boldsymbol{\cal{H}}$ is compact under $\boldsymbol{\|\cdot \|_{\infty}}$ and $\boldsymbol{Q(H)}$ is upper semicontinuous on $\boldsymbol{\cal{H}}$.} Compactness holds as indicated above. (3.4) in Chen (2007) follows, since ${\cal{H}}_N$ is a closed subset of ${\cal{H}}$. To show that $Q(H)$ is continuous on ${\cal{H}}$ under $\|\cdot \|_{\infty}$, let $H_1,H_2$ in ${\cal{H}}$. By Assumption \ref{ass_consis} $(i)$, $F_Y^{-1}$  and $F_{X_2}$ are Lipschitz. It follows that, for some constant $\widetilde{C}$, $|Q(H_2)-Q(H_1)|\leq \widetilde{C}\|H_2-H_1\|_{\infty}$. This implies continuity of $Q$. This shows (3.1'') in Chen (2007). 

\paragraph{$\boldsymbol{{\cal{H}}_N\subset {\cal{H}}_{N+1}\subset {\cal{H}}}$ for all $\boldsymbol{N}$, and there exists a sequence $\boldsymbol{H_N\in{\cal{H}}_N}$ such that $\boldsymbol{\|H_N-H_0\|_{\infty}=o_p(1)}$.} Since $\overline{C}>\underline{C}$ there is an $\epsilon>0$ such that $\overline{C}>\underline{C}+\epsilon$. Let $G_0$ be linear with 
slope $\underline{C}+\epsilon$, such that $G_0(1/2)=0$. For an increasing sequence $\lambda_N$ tending to one as $N$ tends to infinity, let $H_N=\lambda_N H_0+(1-\lambda_N)G_0$. Taking $1-\lambda_N\geq \limfunc{max}\left\{\frac{\underline{C}_N-\underline{C}}{\epsilon},\frac{\overline{C}-\overline{C}_N}{\overline{C}-(\underline{C}+\epsilon)}\right\}$, we have $|H_N|\leq \overline{C}_N$ and $ \underline{C}_N\leq \nabla H_N\leq  \overline{C}_N$, hence $H_N\in{\cal{H}}_N$. Moreover: 
$$ \|H_N-H_0\|_{\infty}\leq (1-\lambda_N) \|H_0\|_{\infty}+(1-\lambda_N)\|G_0\|_{\infty}=o(1).$$ 	
This shows (3.2) in Chen (2007).


\paragraph{$\boldsymbol{Q(H)}$ is uniquely minimized at $\boldsymbol{H_0}$ on $\boldsymbol{\cal{H}}$, and $\boldsymbol{Q(H_0)<+\infty}$.} We have $Q(H)\geq Q(H_0)=0$ for all $H\in{\cal{H}}$. Suppose that $Q(H)=0$. Then, $(V,X_2)$-almost surely we have:
$$F_Y^{-1}\left(\int_0^1 F_{X_2}\left(H(V)+X_2-H(\tau)\right)d\tau\right)=H(V)+X_2.$$
Since the left-hand side in this equation is distributed as $F_Y$, it thus follows that, almost surely:
$$F_{H(V)+X_2}\left(H(V)+X_2\right)=F_Y\left(H(V)+X_2\right).$$
It follows that $F_{H(V)+X_2}=F_Y$ almost everywhere on the real line. Since $Y$ and $X_2$ have densities $f_Y$ and $f_{X_2}$, this also implies that $f_Y(y)=\int_0^1 f_{X_2}(y-H(\tau))d\tau$, $y$-almost everywhere. Now, since $H\in{\cal{H}}$, the function $f_{\widetilde{X}}(x)\equiv 1/\nabla H(H^{-1}(x))$ is well-defined, continuous and bounded. We then have by a change of variables, $f_Y(y)=\int_{-\infty}^{+\infty} f_{X_2}(y-x)f_{\widetilde{X}}(x)dx$. Taking Fourier transforms in this equation yields, denoting as $\Psi_Z$ the characteristic function of any random variable $Z$:
$$\Psi_Y(s)=\Psi_{X_1}(s)\Psi_{X_2}(s)=\Psi_{\widetilde{X}}(s)\Psi_{X_2}(s),\text{ for all }s\in\mathbb{R}.$$
As $\Psi_{X_2}$ is non-vanishing we thus have $\Psi_{X_1}=\Psi_{\widetilde{X}}$. It follows that $f_{X_1}=f_{\widetilde{X}}$, hence that $H=H_0$. This shows (3.1''(ii)) in Chen (2007).

\paragraph{$\boldsymbol{\limfunc{plim}_{N\rightarrow +\infty}\, \sup_{H\in{\cal{H}}}\, |\widehat{Q}(H)-Q(H)|=0}$.} First, notice that since ${\cal{H}}$ consists of Lipschitz functions its $\epsilon$-bracketing entropy is finite for any $\epsilon>0$ (e.g., Corollary 2.7.2 in van der Vaart and Wellner, 1996). Hence ${\cal{H}}$ is Glivenko Cantelli for the $\|\cdot\|_{\infty}$ norm. 

Let now:
$$G_H(v,x)\equiv \left
(F_Y^{-1}\left(\int_0^1 F_{X_2}\left(H(v)+x-H(\tau)\right)d\tau\right)-H(v)-x\right)^2.$$
Notice that, for all $H\in {\cal{H}}$ and as $N$ tends to infinity:
\begin{align}\label{eq_argument_permut}
&\frac{1}{N}\sum_{i=1}^NG_H\left(\frac{\sigma(i)}{N+1},X_{i2}\right)=\frac{1}{N}\sum_{i=1}^NG_H\left(\frac{i}{N+1},X_{\sigma^{-1}(i),2}\right)\notag\\&=\int_0^1 \mathbb{E}\left(  G_H\left(\tau,X_2\right)\right)d\tau+o_p(1) =Q(H)+o_p(1).
\end{align}
Moreover, as $H\mapsto G_H$ is Lipschitz on ${\cal{H}}$ (since $f_Y$ is bounded away from zero and $f_{X_2}$ is bounded away from infinity), and as ${\cal{H}}$ is Glivenko Cantelli, the set of functions $\{G_H\,:\, H\in{\cal{H}}\}$ is also Glivenko Cantelli. Hence:
$$\underset{H\in{\cal{H}}}{\limfunc{sup}}\, \left|\frac{1}{N}\sum_{i=1}^NG_H\left(\frac{\sigma(i)}{N+1},X_{i2}\right)-Q(H)\right|=o_p(1).$$   

Next, we are going to show that:
\begin{align}\label{eq_rank}
&\underset{H\in{\cal{H}}}{\limfunc{sup}}\, \left|\frac{1}{N}\sum_{i=1}^N\frac{1}{N}\,\widehat{\limfunc{Rank}}\left(H\left(\frac{\sigma(i)}{N+1}\right)+X_{i2}\right)-\int_0^1 F_{X_2}\left(H\left(\frac{\sigma(i)}{N+1}\right)+X_{i2}-H(\tau)\right)d\tau\right|=o_p(1). 
\end{align}
From (\ref{eq_rank}) and the fact that $F_Y^{-1}$ is Lipschitz we will then have:
\begin{align*}&\underset{H\in{\cal{H}}}{\limfunc{sup}}\, \left|\frac{1}{N}\sum_{i=1}^N\left
(F_Y^{-1}\left(\frac{1}{N}\,\widehat{\limfunc{Rank}}\left(H\left(\frac{\sigma(i)}{N+1}\right)+X_{i2}\right)\right)-H\left(\frac{\sigma(i)}{N+1}\right)-X_{i2}\right)^2-Q(H)\right|=o_p(1).\end{align*}

To show (\ref{eq_rank}) we are going to show that:
\begin{equation}\label{eq_rank2}
\underset{H\in{\cal{H}},\, a\in\mathbb{R}}{\limfunc{sup}}\, \left|\frac{1}{N}\sum_{i=1}^N \boldsymbol{1}\left\{H\left(\frac{\sigma(i)}{N+1}\right)+X_{i2}\leq a \right\}-\int_0^1 F_{X_2}\left(a-H(\tau)\right)d\tau\right|=o_p(1).
\end{equation}
Pointwise convergence in (\ref{eq_rank2}) is readily verified (similarly to (\ref{eq_argument_permut})). Uniform convergence follows provided we can show that ${\cal{G}}=\{g_{H,a}\,:\, H\in{\cal{H}},a\in\mathbb{R}\}$ is Glivenko Cantelli, where $g_{H,a}(v,u)\equiv\boldsymbol{1}\{H(v)+u\leq a\}$. We are going to show this using a bracketing technique from empirical process theory. Fix an $\epsilon>0$. Since ${\cal{H}}$ has finite $\epsilon$-bracketing entropy there exists a set of functions $H_j$, $j=1,...,J$, such that for all $H\in {\cal{H}}$ there is a $j$ such that $H_j(\tau)\leq H(\tau)\leq H_{j+1}(\tau)$ for all $\tau$, and $\|H_j-H_{j-1}\|_{\infty}<\epsilon$ for all $j$. Moreover, there exists a set of scalars $a_k$, $k=1,...,K$, such that the real line is covered by the intervals $[a_k,a_{k+1}]$, and $F_{X_2}(a_{k+1})-F_{X_2}(a_k)<\epsilon$ for all $k$. Since $X_2$ has bounded support we can assume without loss of generality that $a_{k+1}-a_k<\epsilon$. Hence for all $H$ and $a$ there exist $j$ and $k$ such that $\boldsymbol{1}\{H_{j+1}(v)+u\leq a_{k}\}\leq g_{H,a}(v,u)\leq \boldsymbol{1}\{H_{j}(v)+u\leq a_{k+1}\}$ for all $(u,v)$. Since $\int_0^1 F_{X_2}(a_{k+1}-H_j(\tau))d\tau-\int_0^1 F_{X_2}(a_{k}-H_{j+1}(\tau))d\tau< \widetilde{C} \epsilon$, where $\widetilde{C}>0$ is finite as $f_{X_2}$ is bounded away from infinity, ${\cal{G}}$ is Glivenko Cantelli and (\ref{eq_rank2}) has been shown.

Lastly, since $f_Y$ is bounded away from zero and infinity and differentiable, the empirical quantile function of $Y$ is such that (e.g., Corollary 1.4.1 in Cs\"org\"o, 1983):
$$\left\|\widehat{F}_Y^{-1}(\tau)-F_Y^{-1}(\tau) \right\|_{\infty}=o_p(1).$$
Hence:
\begin{align*}&\underset{H\in{\cal{H}}}{\limfunc{sup}}\, \left|\frac{1}{N}\sum_{i=1}^N\left
(\widehat{F}_Y^{-1}\left(\frac{1}{N}\,\widehat{\limfunc{Rank}}\left(H\left(\frac{\sigma(i)}{N+1}\right)+X_{i2}\right)\right)-H\left(\frac{\sigma(i)}{N+1}\right)-X_{i2}\right)^2-Q(H)\right|=o_p(1). 
\end{align*}

This shows (3.5(i)) in Chen (2007), and ends the proof of Corollary \ref{theo_consis}.

\subsubsection{Factor models: Theorem \ref{theo_consis_gen}} 
	
		We now prove Theorem \ref{theo_consis_gen}. For any $H=(H_1,...,H_K)$, let us denote the empirical objective function as:
		\begin{align*}
		&\widehat{Q}(H)=\underset{\pi\in\Pi_N}{\limfunc{min}}\,\frac{1}{N}\sum_{i=1}^N\left\|Y_{\pi(i)}-\sum_{k=1}^KA_k H_k\left(\frac{\sigma_k(i)}{N+1}\right)\right\|^2,
		\end{align*}
		where $Y_i=(Y_{i1},...,Y_{iT})'$ is a $T\times 1$ vector for all $i$, $A=(A_1,...,A_K)$ with $A_k$ a $T\times 1$ vector for all $k$, and $\|\cdot\|$ is the Euclidean norm on $\mathbb{R}^T$. Denote as $\widehat{\mu}_Y$ the empirical measure of $Y_i$, $i=1,...,N$, with population counterpart $\mu_Y$, and as $\widetilde{\mu}_{AH}$ the empirical measure of $\sum_{k=1}^KA_k H_k\left(\frac{\sigma_k(i)}{N+1}\right)$, $i=1,...,N$, with population counterpart $\mu_{AH}$. Then $\widehat{Q}(H)^{\frac{1}{2}}= W_2\left(\widehat{\mu}_Y,\widetilde{\mu}_{AH}\right)$ is the quadratic Wasserstein distance between $\widehat{\mu}_Y$ and $\widetilde{\mu}_{AH}$. See Chapter 7 in Villani (2003) for some properties of Wasserstein distances.
		
		Likewise, let us define the population counterpart to $\widehat{Q}$, for any $H=(H_1,...,H_K)$, as:
		$$Q(H)=\underset{\pi\in {\cal{M}}(\mu_Y,\,\mu_{AH})}{\limfunc{inf}}\,\mathbb{E}_{\pi}\left[\left
		\|Y-\sum_{k=1}^KA_k H_k\left(V_k\right)\right\|^2\right],$$ 
		where the infimum is taken over all possible joint distributions of the random vectors $Y$ and $\sum_{k=1}^KA_k H_k\left(V_k\right)$, with marginals $\mu_Y$ and $\mu_{AH}$. In this case $Q(H)^{\frac{1}{2}}=W_2\left({\mu}_Y,{\mu}_{AH}\right)$ is the Wasserstein distance between the two population marginals.


		
		
		
		The proof follows the steps of the proof of Corollary \ref{theo_consis}. The differences are as follows.
		
		\paragraph{Parameter space.}
		
		Let ${\cal{H}}$ be the closure of the set $\{H\in {\cal{C}}^1\, : \, \nabla H\geq \underline{C},\,  \|H\|\leq \overline{C}\}$ under $\|\cdot\|_{\infty}$. Then, let us define:
		 $${\cal{H}}_K\equiv\left\{(H_1,...,H_K)\, :\, H_k\in{\cal{H}} \text{ and } \sum_{i=1}^NH_k\left(\frac{i}{N+1}\right)=0\,  \text{ for all }k \right\}.$$
		 ${\cal{H}}_K$ is compact with respect to $\|\cdot\|_{\infty}$. The sieve construction is then similar to the scalar case.

		\paragraph{$\boldsymbol{Q(H)}$ is continuous on $\boldsymbol{{\cal{H}}_K}$.} Let $H_1$ and $H_2$ in ${{\cal{H}}_K}$. Since $Y$ has bounded support, and $H_{1k}$ and $H_{2k}$ are bounded for all $k$, we have: 
		$$|Q(H_2)-Q(H_1)|\leq \widetilde{C}\left|Q(H_2)^{\frac{1}{2}}-Q(H_1)^{\frac{1}{2}}\right|=\widetilde{C}\left|W_2\left({\mu}_Y,{\mu}_{AH_2}\right)-W_2\left({\mu}_Y,{\mu}_{AH_1}\right)\right|,$$
		for some constant $\widetilde{C}>0$. Hence, since $W_2$ satisfies the triangle inequality (see Theorem 7.3 in Villani, 2003):
		$$|Q(H_2)-Q(H_1)|\leq\widetilde{C}W_2\left({\mu}_{AH_1},{\mu}_{AH_2}\right).$$
		
		Next, we use that, since supports are bounded, $W_2\left({\mu}_{AH_1},{\mu}_{AH_2}\right)$ is bounded (up to a multiplicative constant) by the Kantorovich-Rubinstein distance: 
		$$W_1({\mu}_{AH_1},{\mu}_{AH_2})=\underset{\pi\in {\cal{M}}({\mu}_{AH_1},\,{\mu}_{AH_2})}{\limfunc{inf}}\,\mathbb{E}_{\pi}\left(\left
		\|\sum_{k=1}^KA_k H_{1k}\left(V_{1k}\right)-\sum_{k=1}^KA_k H_{2k}\left(V_{2k}\right)\right\|\right).$$ 
		
		Now, using the dual representation of the Kantorovich-Rubinstein distance, $W_1$ can be equivalently written as (see Theorem 1.14 in Villani, 2003):
		$$W_1({\mu}_{AH_1},{\mu}_{AH_2})=\underset{\varphi \textit{1-Lipschitz}}{\limfunc{sup}}\,\,\,\mathbb{E}\left(\varphi\left(\sum_{k=1}^KA_k H_{1k}\left(V_{1k}\right)\right)\right)-\mathbb{E}\left(\varphi\left(\sum_{k=1}^KA_k H_{2k}\left(V_{2k}\right)\right)\right),$$ 
		where $\varphi$ are $1$-Lipschitz functions on $\mathbb{R}^T$; that is, such that $|\varphi(y_2)-\varphi(y_1)|\leq \|y_2-y_1\|$ for all $(y_1,y_2)\in\mathbb{R}^{T}\times \mathbb{R}^{T}$.

		Hence:
		\begin{align*}
		W_1({\mu}_{AH_1},{\mu}_{AH_2}) &=\underset{\varphi \textit{1-Lipschitz}}{\limfunc{sup}}\,\int...\int \left[\varphi\left(\sum_{k=1}^KA_k H_{1k}\left(\tau_{k}\right)\right)-\varphi\left(\sum_{k=1}^KA_k H_{2k}\left(\tau_{k}\right)\right)\right]d\tau_1...d\tau_K\\
		& \leq \int...\int \left\|\sum_{k=1}^KA_k H_{1k}\left(\tau_{k}\right)-\sum_{k=1}^KA_k H_{2k}\left(\tau_{k}\right)\right\|d\tau_1...d\tau_K\\
		&\leq  \sum_{k=1}^K\|A_k\| \, \|H_{1k}-H_{2k}\|_{\infty}.
		\end{align*}

		
		This implies that $H\mapsto Q(H)$ is continuous on ${{\cal{H}}_K}$.
		
		
		\paragraph{$\boldsymbol{Q(H)}$ is uniquely minimized at $\boldsymbol{H_0}$ on $\boldsymbol{\cal{H}}_K$.} Let $H$ be such that $Q(H)=0$. Then $W_2\left({\mu}_Y,{\mu}_{AH}\right)=0$. By Theorem 7.3 in Villani (2003) this implies that ${\mu}_Y={\mu}_{AH}$. Hence the cdfs of $Y=\sum_{k=1}^KA_k H_{0k}\left(V_k\right)$ and $\sum_{k=1}^KA_k H_k\left(V_k\right)$ are equal. By Assumption \ref{ass_consis_gen} $(ii)$, it follows from the identification result in Bonhomme and Robin (2010) that $H_k=H_{0k}$ for all $k$.  
		
		\paragraph{$\boldsymbol{\limfunc{plim}_{N\rightarrow +\infty}\, \sup_{H\in{\cal{H}}_K}\, |\widehat{Q}(H)-Q(H)|=0}$.} Using similar arguments to the ones we used to show the continuity of $Q(H)$, we have:
		\begin{align*}
		\sup_{H\in{\cal{H}}_K}\, |\widehat{Q}(H)-Q(H)|&\leq \widetilde{C} \sup_{H\in{\cal{H}}_K}\, |W_2\left(\widehat{\mu}_Y,\widetilde{\mu}_{AH}\right)-W_2\left({\mu}_Y,{\mu}_{AH}\right)|\\
		&\leq \widetilde{C} \sup_{H\in{\cal{H}}_K}\, \left(W_2\left(\mu_Y,\widehat{\mu}_Y\right)+W_2\left({\mu}_{AH},\widetilde{\mu}_{AH}\right)\right),
		\end{align*}
		where we have used the triangle inequality.
		
		Now, there is a positive constant $\widetilde{C}$ (different from the previous one) such that:
		\begin{align*}
		&W_2\left(\mu_Y,\widehat{\mu}_Y\right)\leq \widetilde{C}W_1\left(\mu_Y,\widehat{\mu}_Y\right)= \widetilde{C}\underset{\varphi \textit{1-Lipschitz}}{\limfunc{sup}}\,\,\,\left[\mathbb{E}\left(\varphi\left(Y\right)\right)-\frac{1}{N}\sum_{i=1}^N\varphi\left(Y_i\right)\right]\, =o_p(1),
		\end{align*}
		where the last equality follows from the set of 1-Lipschitz functions $\varphi$ being Glivenko Cantelli.
		
		Next, we have:
		\begin{align*}
		&\sup_{H\in{\cal{H}}_K}\,W_2\left({\mu}_{AH},\widetilde{\mu}_{AH}\right)\leq \widetilde{C}\sup_{H\in{\cal{H}}_K}\,W_1\left({\mu}_{AH},\widetilde{\mu}_{AH}\right)\\
		& = \widetilde{C}\sup_{H\in{\cal{H}}_K}\,\underset{\varphi \textit{1-Lipschitz}}{\limfunc{sup}}\,\,\,\left[\mathbb{E}\left(\varphi\left(\sum_{k=1}^KA_k H_{k}\left(V_{k}\right)\right)\right)-\frac{1}{N}\sum_{i=1}^N\varphi\left(\sum_{k=1}^KA_k H_{k}\left(\frac{\sigma_k(i)}{N+1}\right)\right)\right]\, =o_p(1),
		\end{align*}
		where the last equality follows from the fact that the following set of functions is Glivenko Cantelli:
		$$\left\{\varphi \circ  \left(\sum_{k=1}^KA_k H_{k}\right)\,:\,  \varphi \textit{ is 1-Lipschitz},\, H=(H_1,...,H_K)\in{\cal{H}}_K\right\}.$$
		
		This concludes the proof of Theorem \ref{theo_consis_gen}.

		\subsection{Proof of Corollary \ref{coro_density_consis}}
		
			Let ${\cal{H}}_K^{(2)}$ denote the set of functions $(H_1,...,H_K)\in{\cal{H}}_K$ which additionally satisfy $\|\nabla^{2}H_k\|_{\infty}\leq \overline{C}$ for all $k$. Let $k\in\{1,...,K\}$. Let $\widehat{H}_k\in{\cal{H}}^{(2)}_N$ be such that $\widehat{H}_k\left(\frac{i}{N+1}\right)=\widehat{X}_{ik}$ for all $i$, where ${\cal{H}}^{(2)}_N=\left\{H\in{\cal{H}}^{(2)}_K \,:\, \left\{H_k\left(\frac{i}{N+1}\right)\, :\, i=1,...,N,\, k=1,...,K\right\}\in {\cal{X}}_N^{(2)}\right\}$. We have:
			\begin{align*}
			&\left|\frac{1}{Nb}\sum_{i=1}^N\kappa\left(\frac{\widehat{H}_k\left(\frac{i}{N+1}\right)-x}{b}\right)-\frac{1}{b}\int_0^1\kappa\left(\frac{\widehat{H}_k\left(u\right)-x}{b}\right)du\right|\\
			& =\left|\frac{1}{Nb}\sum_{i=1}^N\int_{\frac{i-1}{N}}^{\frac{i}{N}}\left[\kappa\left(\frac{\widehat{H}_k\left(\frac{i}{N+1}\right)-x}{b}\right)-\kappa\left(\frac{\widehat{H}_k\left(u\right)-x}{b}\right)\right]du\right|\\
			& \leq \frac{C}{Nb^2}\sum_{i=1}^N\int_{\frac{i-1}{N}}^{\frac{i}{N}}\left|\widehat{H}_k\left(\frac{i}{N+1}\right)-\widehat{H}_k\left(u\right)\right| du\\
			& \leq \frac{\widetilde{C}}{Nb^2}\sum_{i=1}^N\int_{\frac{i-1}{N}}^{\frac{i}{N}}\left|\frac{i}{N+1}-u\right| du=O(N^{-2}b^{-2})=o(1),
			\end{align*}
			where $C>0$ and $\widetilde{C}>0$ are constants, and we have used that $\kappa$ is Lipschitz, $\nabla \widehat{H}_k$ is uniformly bounded, and $Nb\rightarrow +\infty$.
			
			Now, using the change of variables $\omega=\frac{\widehat{H}_k\left(u\right)-x}{b}$, we obtain:
				\begin{align*}
				&\frac{1}{b}\int_0^1\kappa\left(\frac{\widehat{H}_k\left(u\right)-x}{b}\right)du=\int_{-\infty}^{+\infty} \kappa(\omega)\frac{1}{\nabla \widehat{H}_k\left(\widehat{H}_k^{-1}(x+b\omega)\right)}d\omega =\frac{1}{\nabla \widehat{H}_k\left(\widehat{H}_k^{-1}(x)\right)}+o(1),
				\end{align*}
			where we have used that $x\mapsto 1/\nabla \widehat{H}_k(\widehat{H}_k^{-1}(x))$ is differentiable with uniformly bounded derivative, $\kappa$ has finite first moments, $b\rightarrow 0$, and $\kappa$ integrates to one.

	Lastly, note that $f_{X_k}(x)=1/\nabla H_{0k}(H_{0k}^{-1}(x))$, where by Theorem \ref{theo_consis_gen} and equation (\ref{eq_der}) we have $\|\widehat{H}_k-H_{0k}\|_{\infty}=o_p(1)$, $\|\widehat{H}_k^{-1}-H_{0k}^{-1}\|_{\infty}=o_p(1)$, and  $\|\nabla\widehat{H}_k-\nabla H_{0k}\|_{\infty}=o_p(1)$.

			This shows Corollary \ref{coro_density_consis}.

		\clearpage
		
		\begin{center}
		{\Large \textbf{SUPPLEMENTARY APPENDIX }}
	 \end{center}
	 \section{Expectations\label{App_Extens}}
	 
	 For any Lipschitz function $h$, the expectation $\mathbb{E}(h(X_k))$ can be consistently estimated as: $$\frac{1}{N}\sum_{i=1}^N h\left(\widehat{X}_{ik}\right).$$ Likewise, for all $t$, the expectation $\mathbb{E}(h(X_k,Y_t))$ is consistently estimated as: $$\frac{1}{N}\sum_{i=1}^N h\left(\widehat{X}_{\sigma_k(i),k},\sum_{\ell=1}^Ka_{t\ell}\widehat{X}_{\sigma_{\ell}(i),\ell}\right),$$ for independent random permutations $\sigma_1,...,\sigma_K$ in $\Pi_N$. 
	 
	 Conditional expectations are of particular interest in prediction problems. Given the $\widehat{X}_{ik}$'s and the $\widehat{f}_{X_k}$'s, a consistent estimator of the conditional expectation $\mathbb{E}\left(X_k\,|\, Y=y\right)$ is readily constructed. To see this, suppose the matrix formed by all the columns of $A$ except the $k$-th one has rank $T$ (which ensures that the conditional density of $Y$ given $X_k$ is not degenerate). Partition $A$ into a $T\times (K-T)$ submatrix $B_k$ and a non-singular $T\times T$ submatrix $C_k$, where the $k$-th column of $A$ is one of the columns of $B_k$. Denote as $X^{B_k}$ (resp., $\widehat{X}_{\sigma(i)}^{B_k}$) and $X^{C_k}$ (resp., $\widehat{X}_{\sigma(i)}^{C_k}$) the subvectors of $X$ (resp., $(\widehat{X}_{\sigma_1(i)},...,\widehat{X}_{\sigma_K(i)})'$) corresponding to $B_k$ and $C_k$. An estimator of $\mathbb{E}\left(X_k\,|\, Y=y\right)$ is then: 
	 \begin{equation}\label{Cond_mean_gen}
	 \widehat{\mathbb{E}}\left(X_k\,|\, Y=y\right)=\frac{\sum_{i=1}^N \widehat{f}_{X^{B_k}}\left(\widehat{X}_{\sigma(i)}^{B_k}\right)\widehat{f}_{X^{C_k}}\left(C_k^{-1}\left[y-B_k \widehat{X}_{\sigma(i)}^{B_k}\right]\right)	
	 	\widehat{X}_{\sigma_k(i),k}}{\sum_{i=1}^N \widehat{f}_{X^{B_k}}\left(\widehat{X}_{\sigma(i)}^{B_k}\right)\widehat{f}_{X^{C_k}}\left(C_k^{-1}\left[y-B_k \widehat{X}_{\sigma(i)}^{B_k}\right]\right)	
	 }.\end{equation}

	 As an example, in the fixed-effects model (\ref{eq_Kotlarski}), a consistent estimator of $\mathbb{E}\left(X_1\,|\, Y=y\right)$ is, for $y=(y_1,...,y_T)$:
	 \begin{equation}\label{Cond_mean_FE}
	 \widehat{\mathbb{E}}\left(X_1\,|\, Y=y\right)=\frac{\sum_{i=1}^N \prod_{t=1}^T\widehat{f}_{X_{t+1}}\left(y_t-\widehat{X}_{\sigma_1(i),1}\right)\widehat{X}_{\sigma_1(i),1}}{\sum_{i=1}^N \prod_{t=1}^T\widehat{f}_{X_{t+1}}\left(y_t-\widehat{X}_{i1}\right)}=\frac{\sum_{i=1}^N \prod_{t=1}^T\widehat{f}_{X_{t+1}}\left(y_t-\widehat{X}_{i1}\right)\widehat{X}_{i1}}{\sum_{i=1}^N \prod_{t=1}^T\widehat{f}_{X_{t+1}}\left(y_t-\widehat{X}_{i1}\right)}.\end{equation}
	 More generally, the densities $\widehat{f}_{X^{B_k}}$ and $\widehat{f}_{X^{C_k}}$ in (\ref{Cond_mean_gen}) are products of marginal densities of individual latent factors.

	 \paragraph{Remark: constrained prediction.}
	 
	 In the present setting, an alternative to the usual prediction problem consists in minimizing expected square loss subject to the constraint that the cross-sectional distribution of the predicted values coincide with the population distribution of the latent variable. The resulting \emph{constrained} optimal predictor can be estimated as: $\widetilde{X}_{ik}=\widehat{X}_{\pi^*(i),k}$, $i=1,...,N$, where the $\widetilde{X}_{i}$'s are equal to the $\widehat{X}_{j}$'s sorted in the same order as the $\widehat{\mathbb{E}}(X_{k}\,|\, Y=Y_i)$'s; that is: $\pi^*={\limfunc{argmin}}_{\pi\in \Pi_{N}}\, \sum_{i=1}^N\left(\widehat{\mathbb{E}}(X_{k}\,|\, Y=Y_i)-\widehat{X}_{\pi(i)}\right)^2$. In a similar spirit, one can construct a matching-based alternative to $\widehat{\mathbb{E}}(X_{k}\,|\, Y=Y_i)$ as: $\frac{1}{M}\sum_{j=1}^N\sum_{m=1}^M\boldsymbol{1}\{\widehat{\pi}^{(m)}(j)=i\} \widehat{X}_{\sigma^{(m)}_k(j),k},$	where $\sigma_k^{(m)}$, $m=1,...,M$, are independent random permutations in $\Pi_N$, and $\widehat{\pi}^{(m)}={\limfunc{argmin}}_{\pi\in \Pi_{N}}\, \sum_{i=1}^N\sum_{t=1}^T\left(Y_{\pi(i),t}-\sum_{k=1}^Ka_{tk}X_{\sigma_k(i),k}\right)^2$. We leave the characterization of the properties of such constrained predictors to future work.

	 \section{Additional simulation results\label{App_Additional_Simu}}
	 
	 In this section of the appendix we show simulation results for two a scalar nonparametric deconvolution model. Consider the model $Y=X_1+X_2$, where $X_1$ and $X_2$ are scalar, independent, and follow identical distributions. As for the fixed-effects model in the main text, we consider four specifications: Beta$(2,2)$, Beta$(5,2)$, normal, and log-normal, and we consider two choices for the penalization constants: $(\underline{C}_N,\overline{C}_N)=(.1,10)$ (``strong constraint''), and $(\underline{C}_N,\overline{C}_N)=(0,10000)$ (``weak constraint''). We use $10$ randomly generated starting values, and average $M=10$ sets of estimates.

\begin{figure}[tbp]
	 	\caption{Monte Carlo results, deconvolution model, $N=100$	\label{Fig_MC_Deconv}}
	 	\begin{center}
	 		\begin{tabular}{cccc}
	 			\multicolumn{2}{c}{Quantile functions}&  \multicolumn{2}{c}{Densities}\\
	 			Strong constraint & Weak constraint & Strong constraint & Weak constraint \\
	 			\\ 
	 			\multicolumn{4}{c}{$(X_1,X_2) \sim $ Beta(2,2)}\\
	 			\includegraphics[width=40mm, height=30mm]{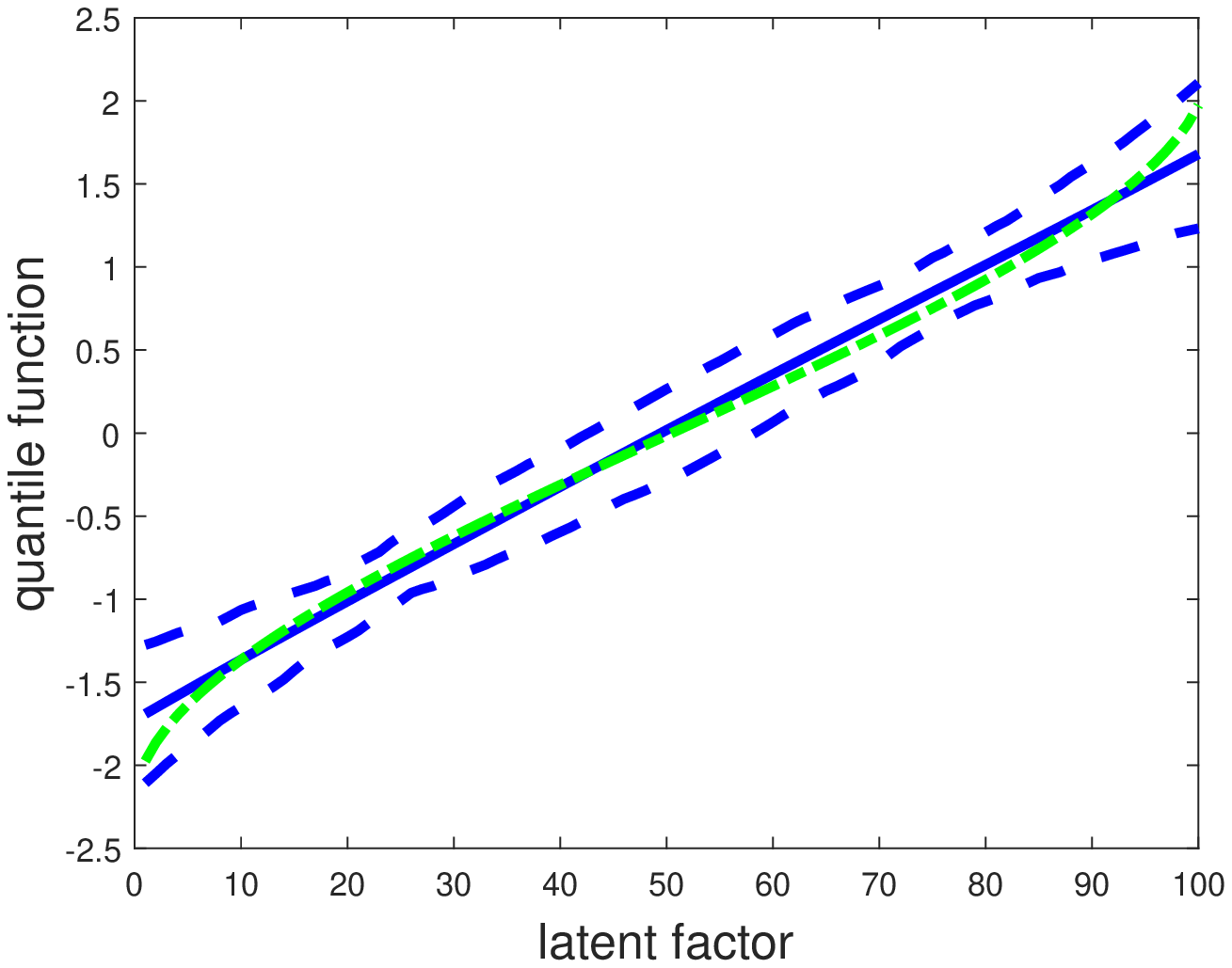} & %
	 			\includegraphics[width=40mm, height=30mm]{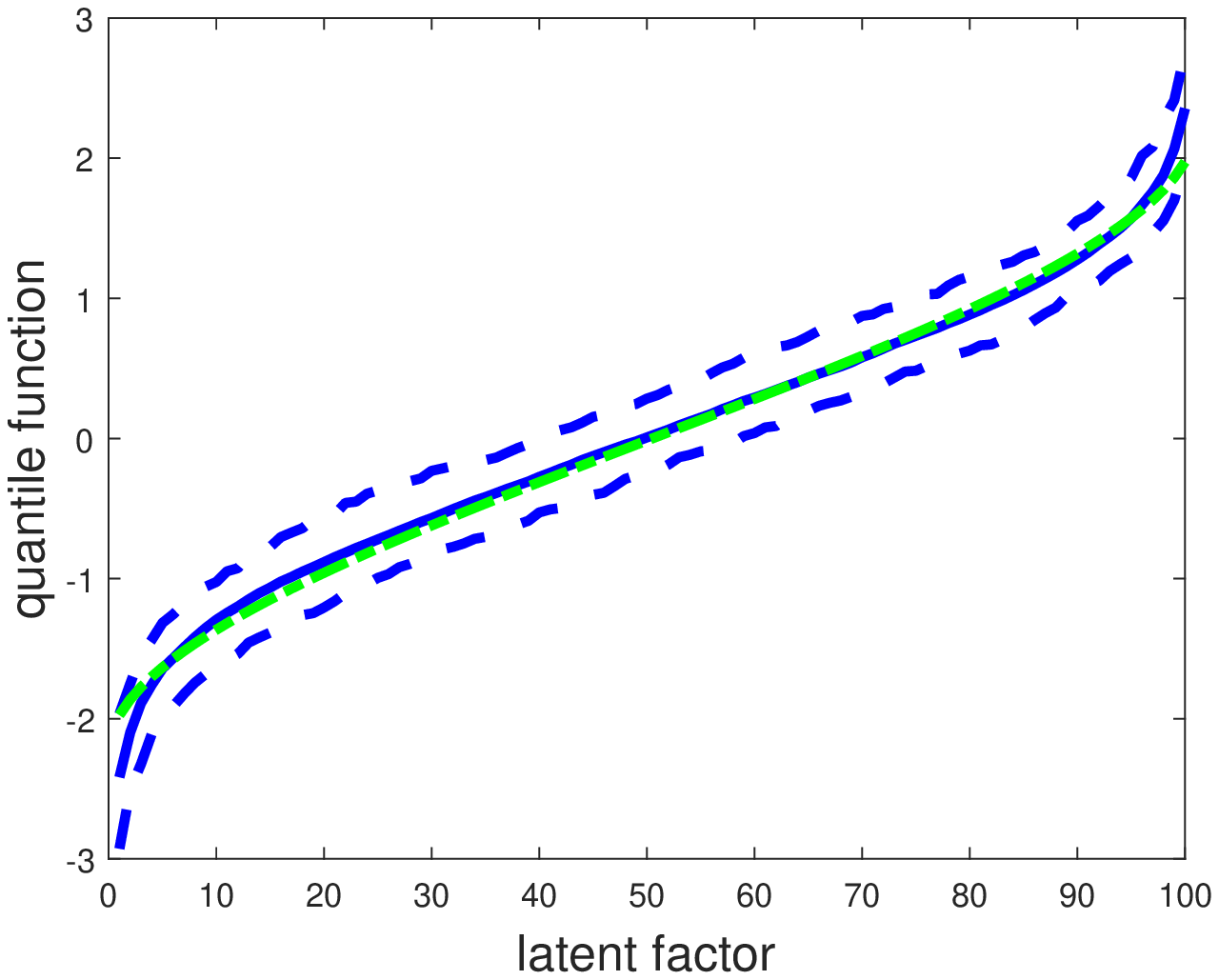} & 
	 			\includegraphics[width=40mm, height=30mm]{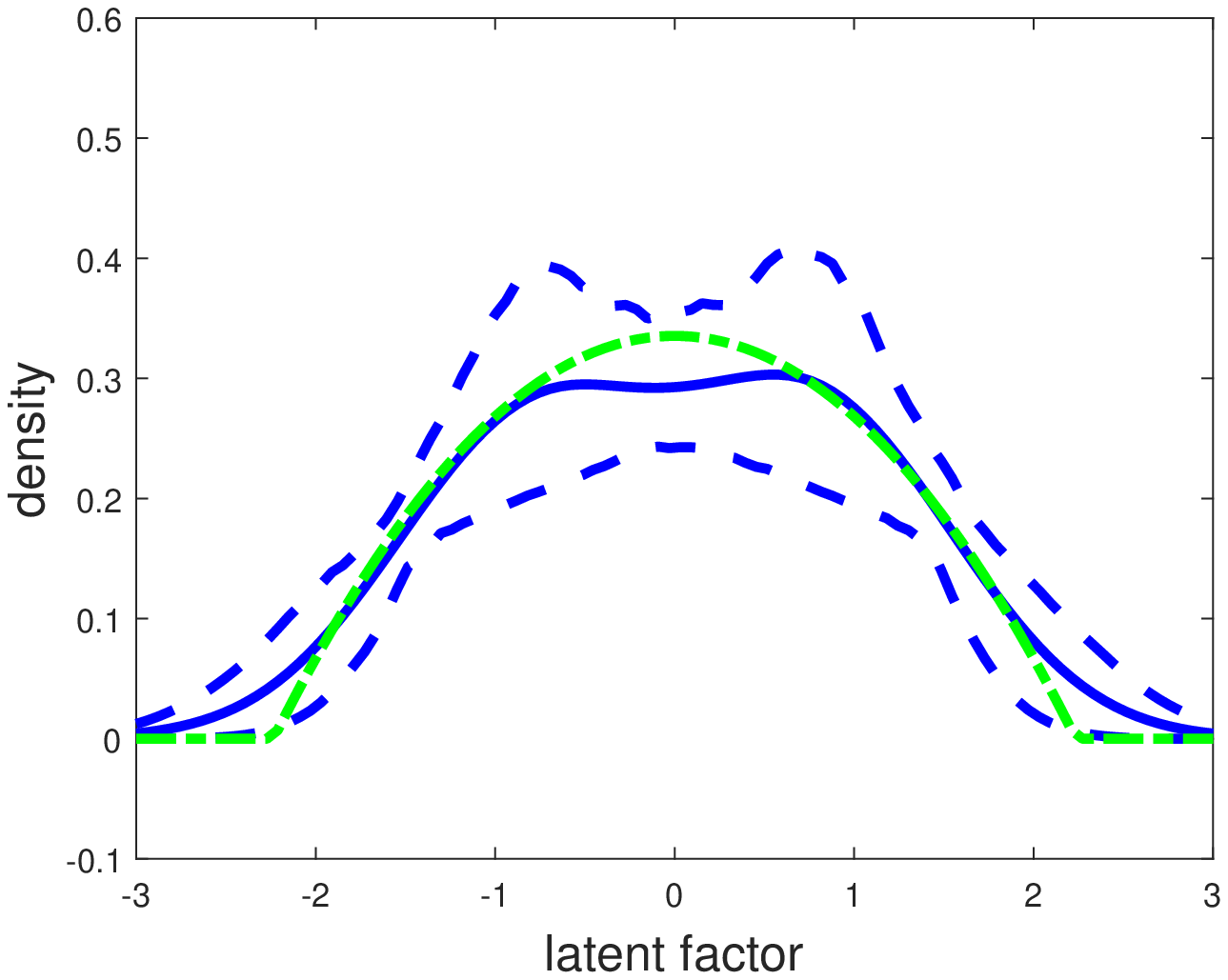} & %
	 			\includegraphics[width=40mm, height=30mm]{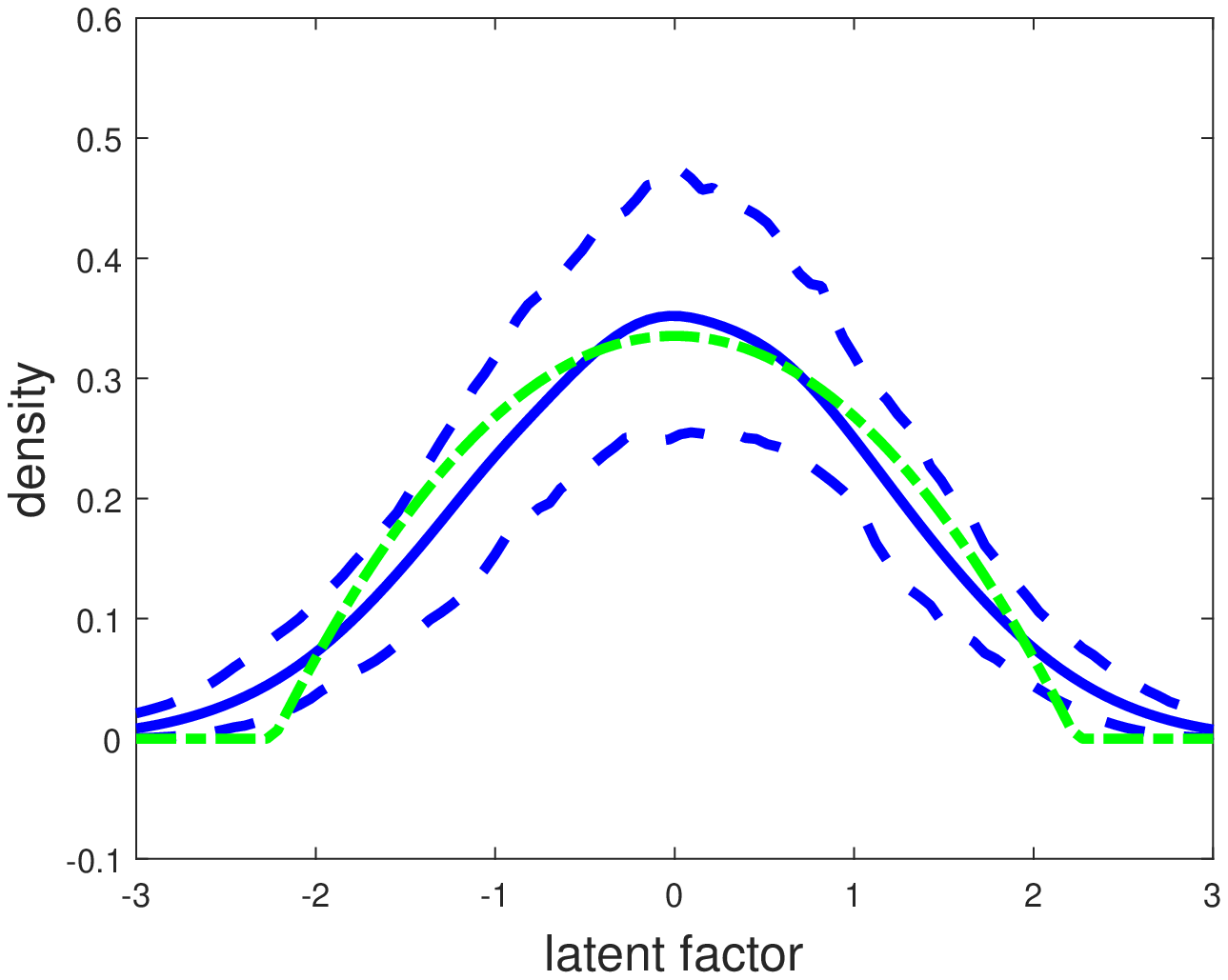}\\
	 			&  &\\
	 			\multicolumn{4}{c}{$(X_1,X_2) \sim $ Beta(5,2)}\\
	 			\includegraphics[width=40mm, height=30mm]{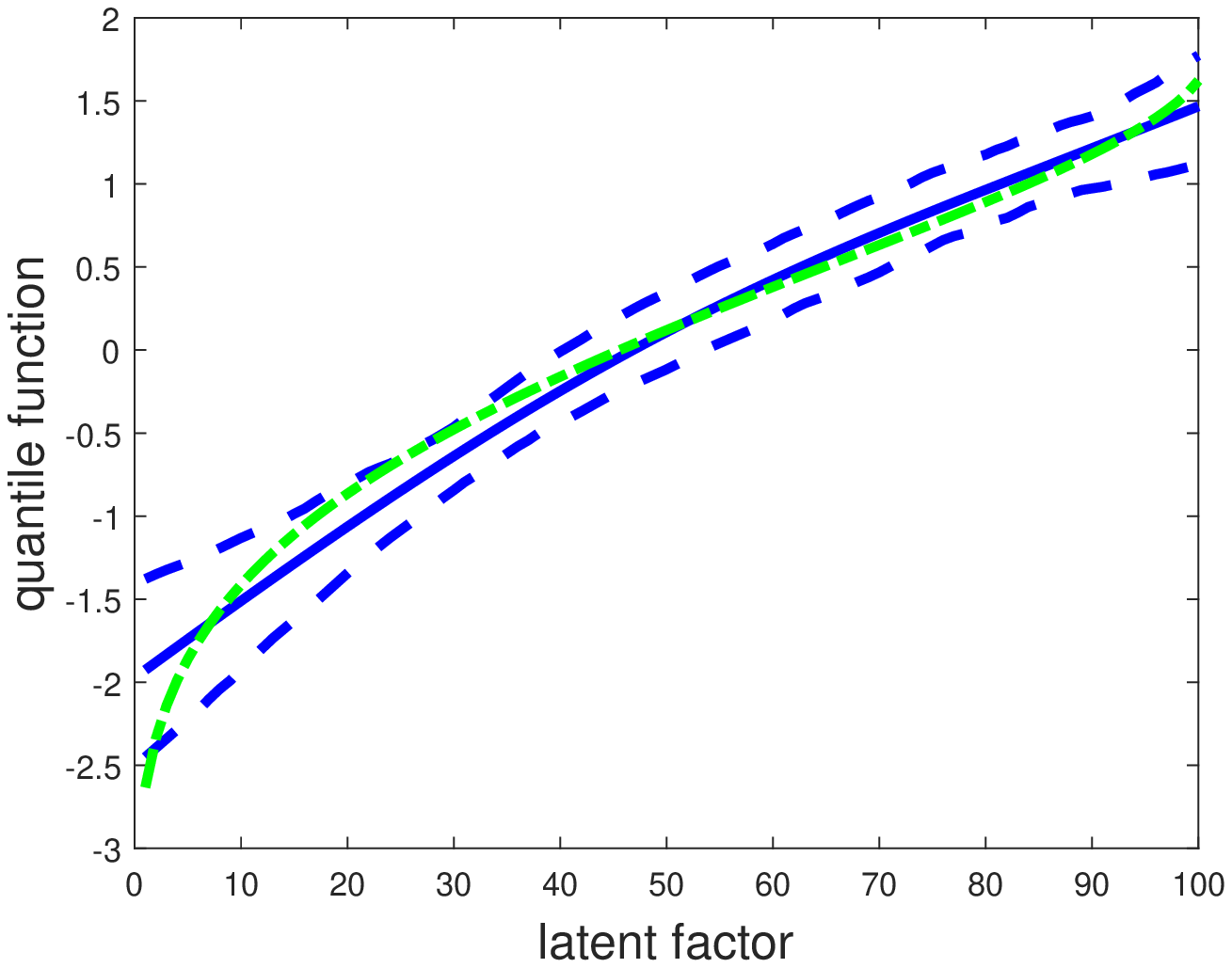} & %
	 			\includegraphics[width=40mm, height=30mm]{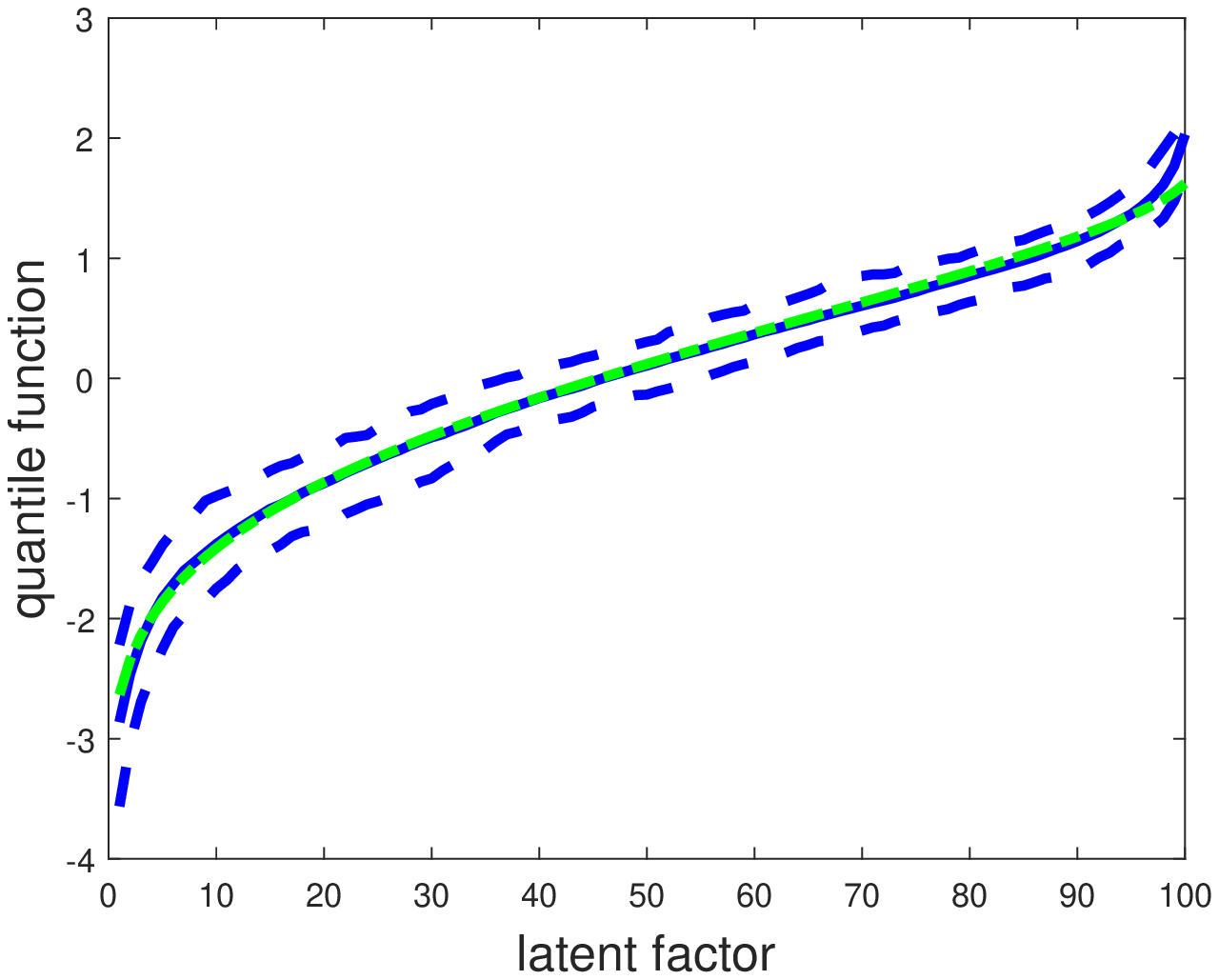} & 
	 			\includegraphics[width=40mm, height=30mm]{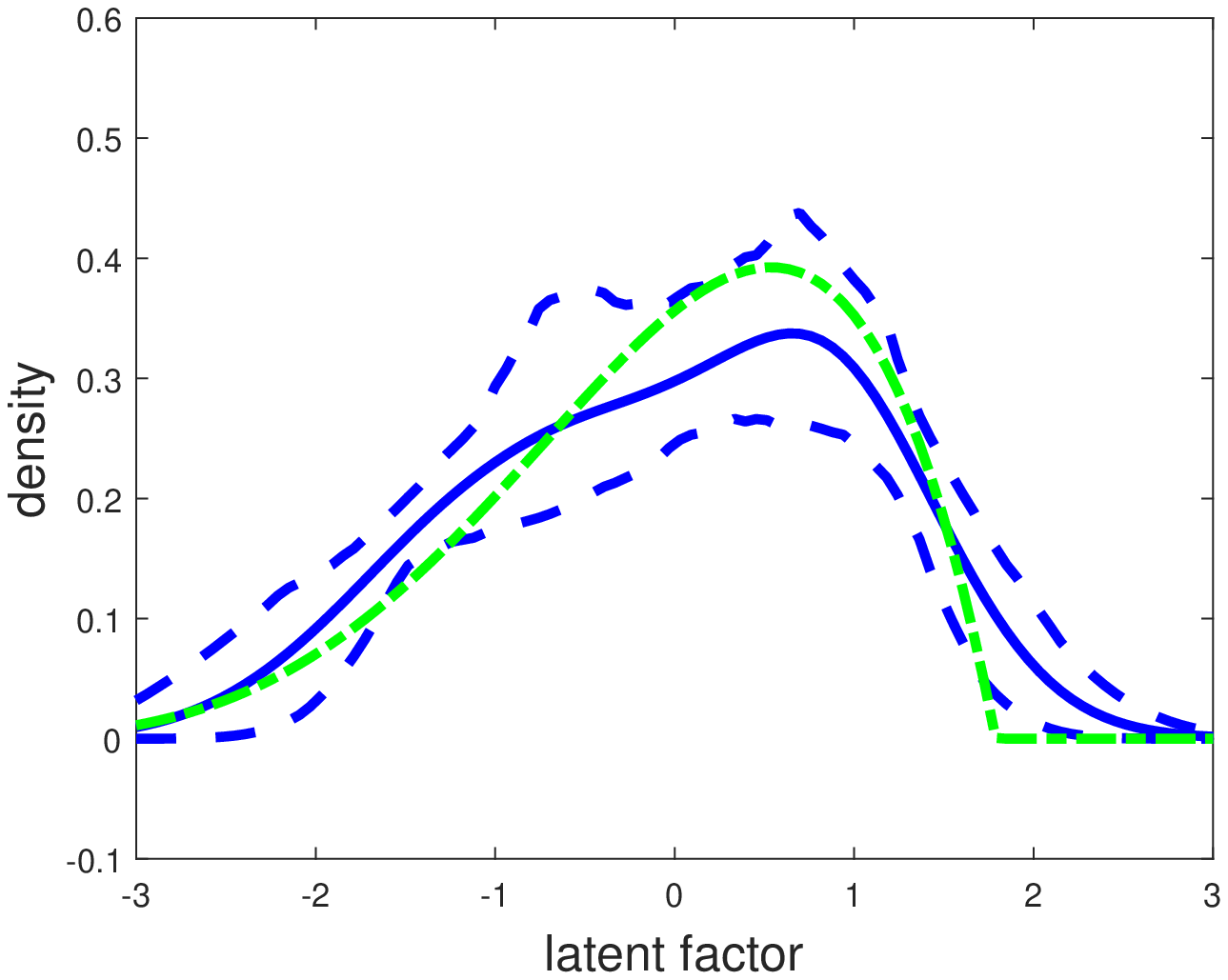} & %
	 			\includegraphics[width=40mm, height=30mm]{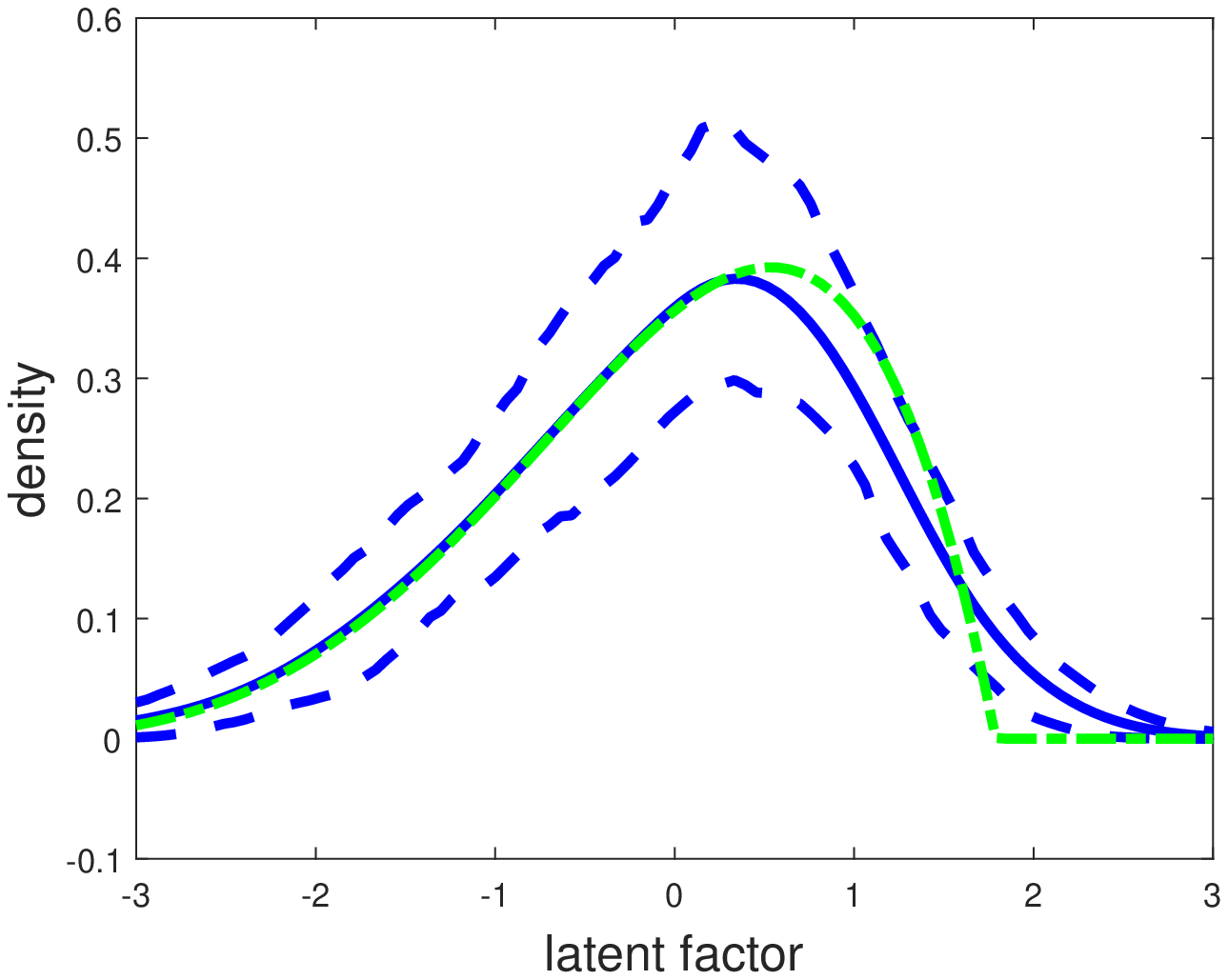}\\
	 			&  &\\
	 			\multicolumn{4}{c}{$(X_1,X_2) \sim {\cal{N}}(0,1)$}\\
	 			\includegraphics[width=40mm, height=30mm]{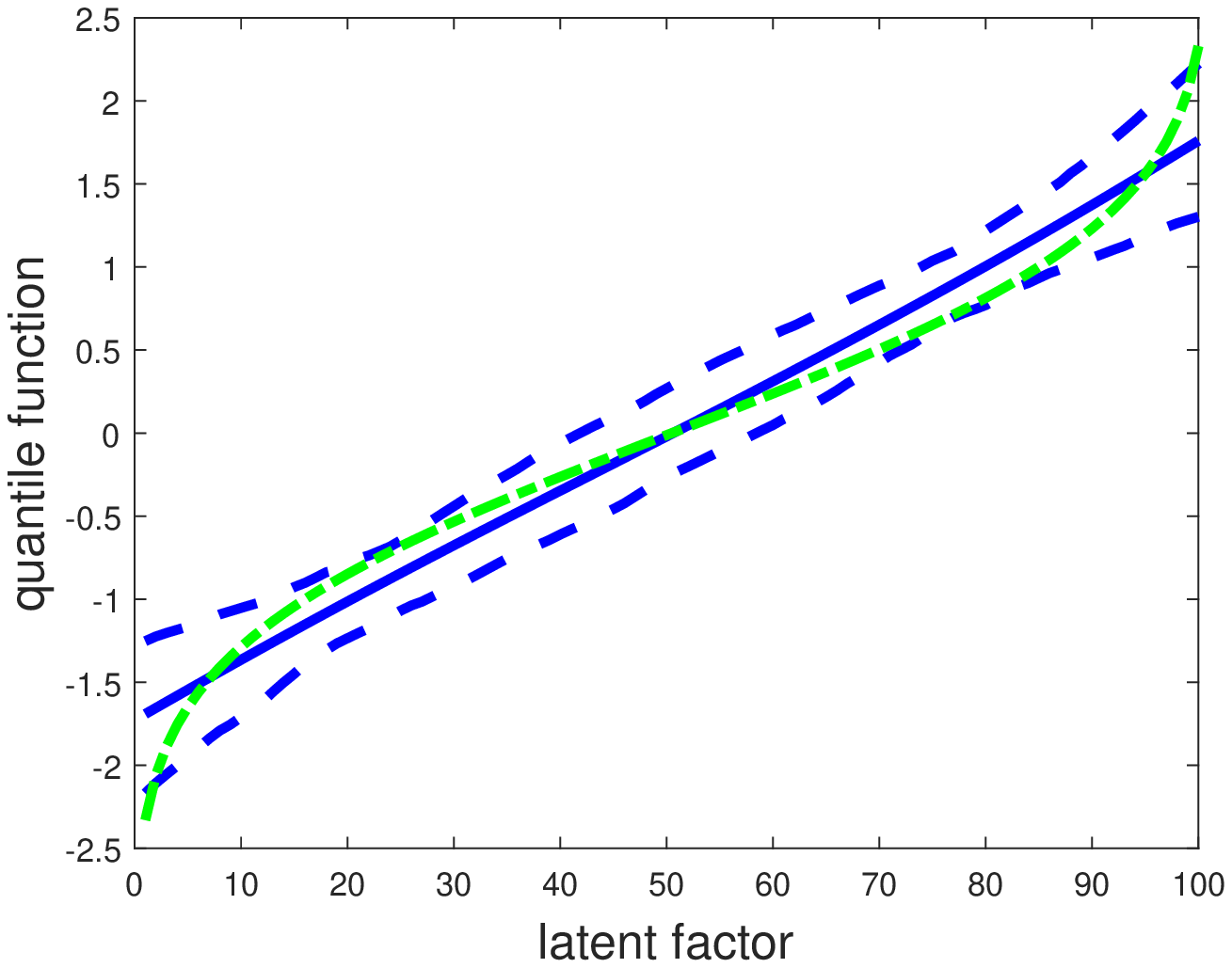} & %
	 			\includegraphics[width=40mm, height=30mm]{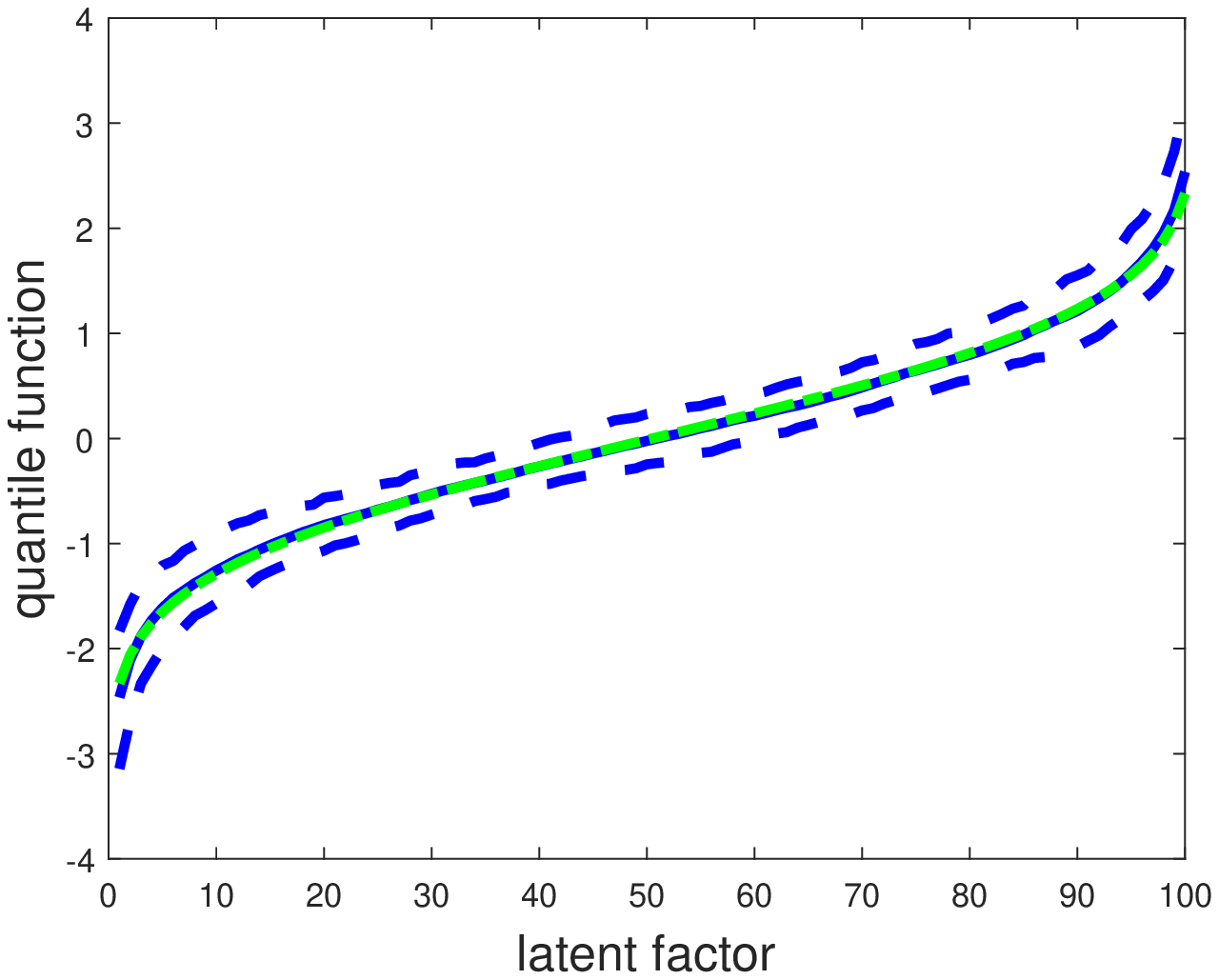} & 
	 			\includegraphics[width=40mm, height=30mm]{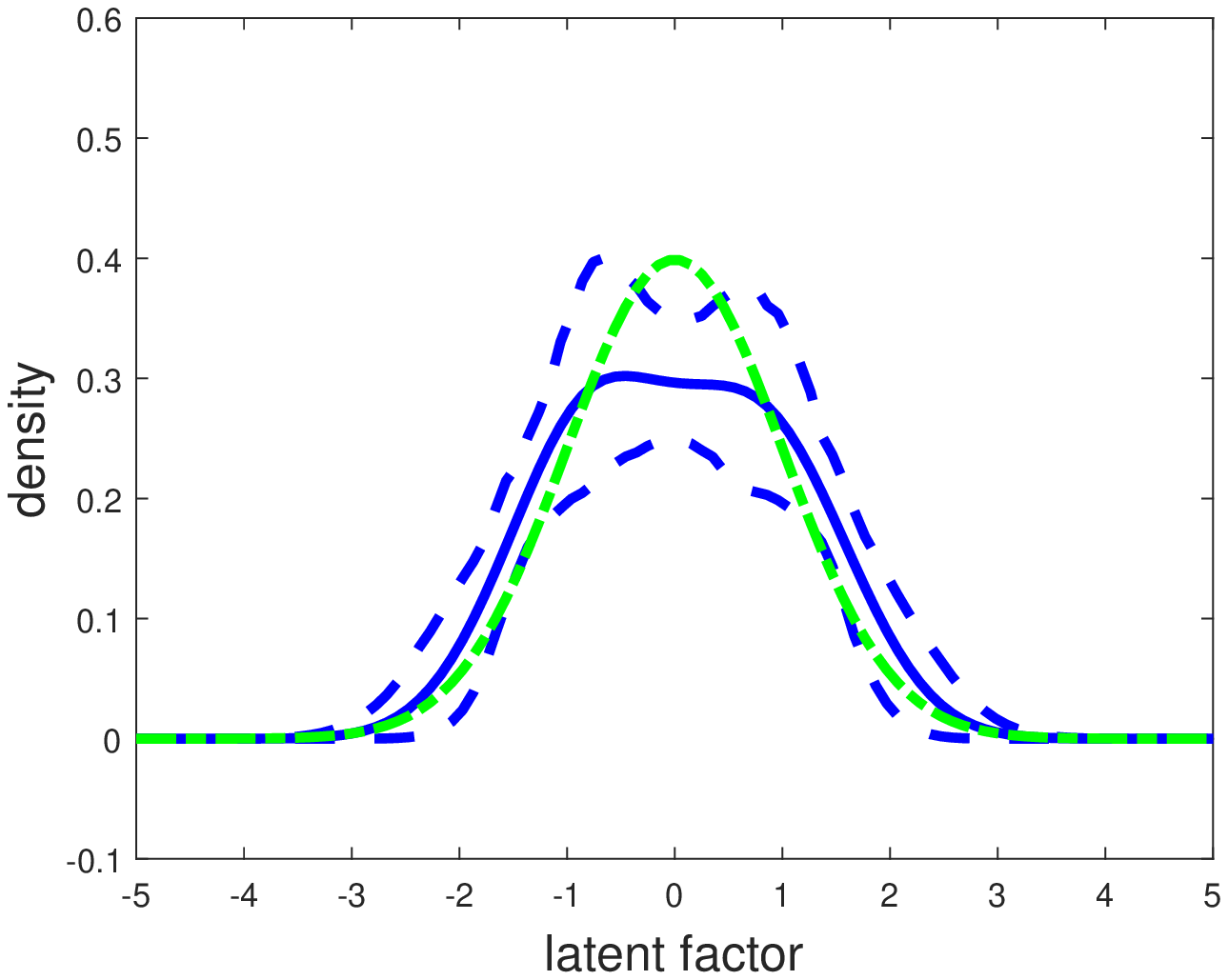} & %
	 			\includegraphics[width=40mm, height=30mm]{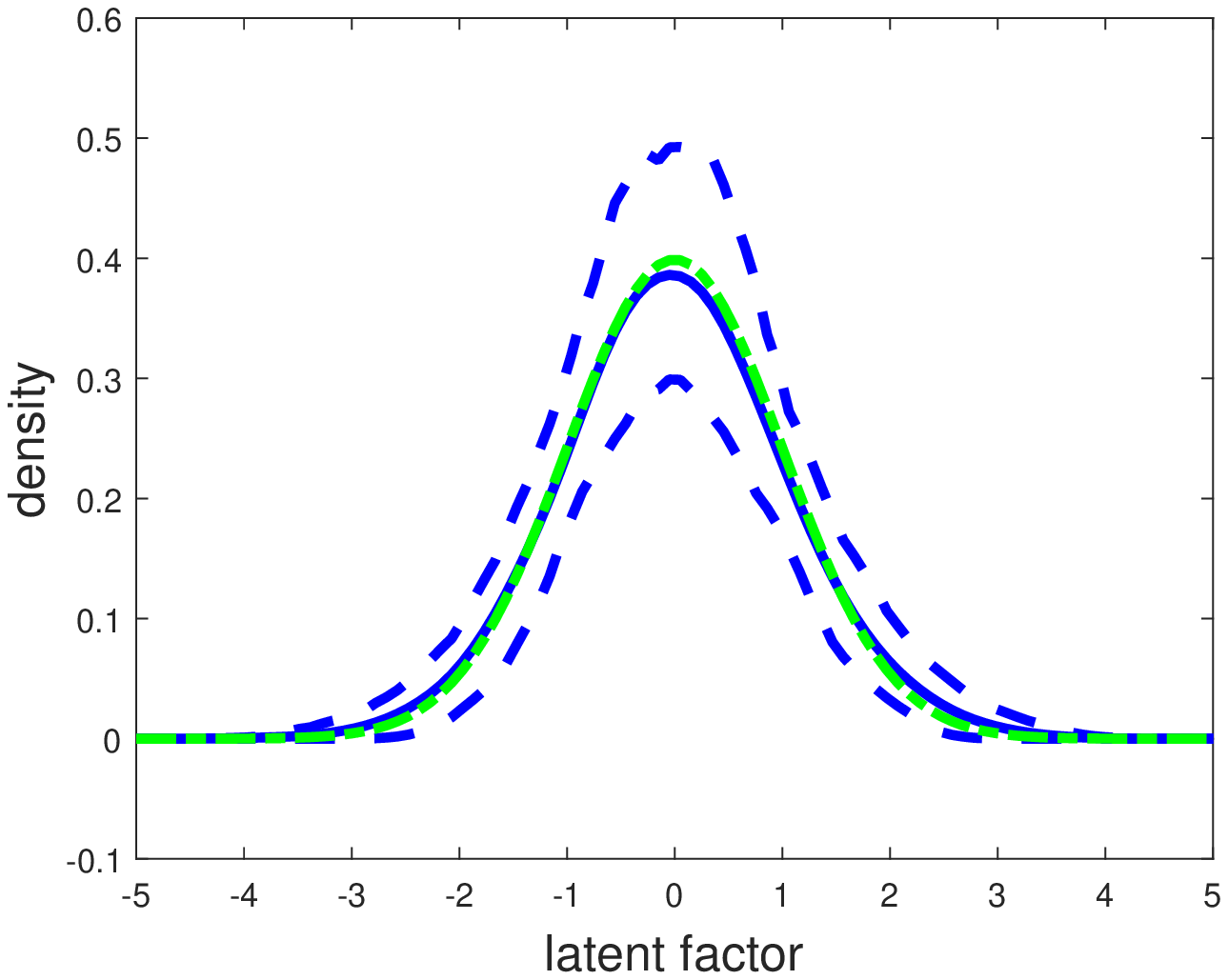}\\
	 			&  &\\
	 			\multicolumn{4}{c}{$(X_1,X_2) \sim  \limfunc{exp}{\cal{N}}(0,1)$}\\
	 			\includegraphics[width=40mm, height=30mm]{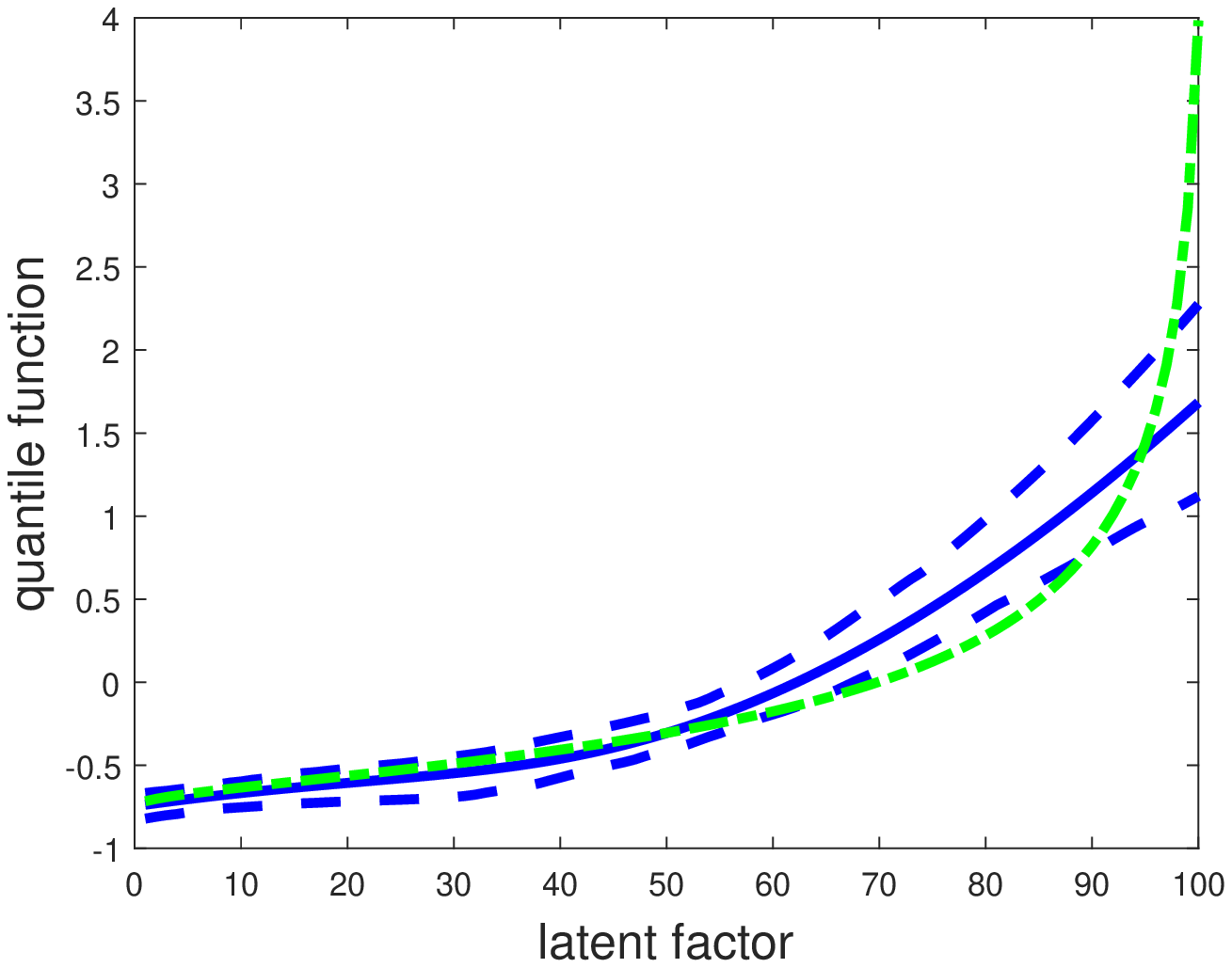} & %
	 			\includegraphics[width=40mm, height=30mm]{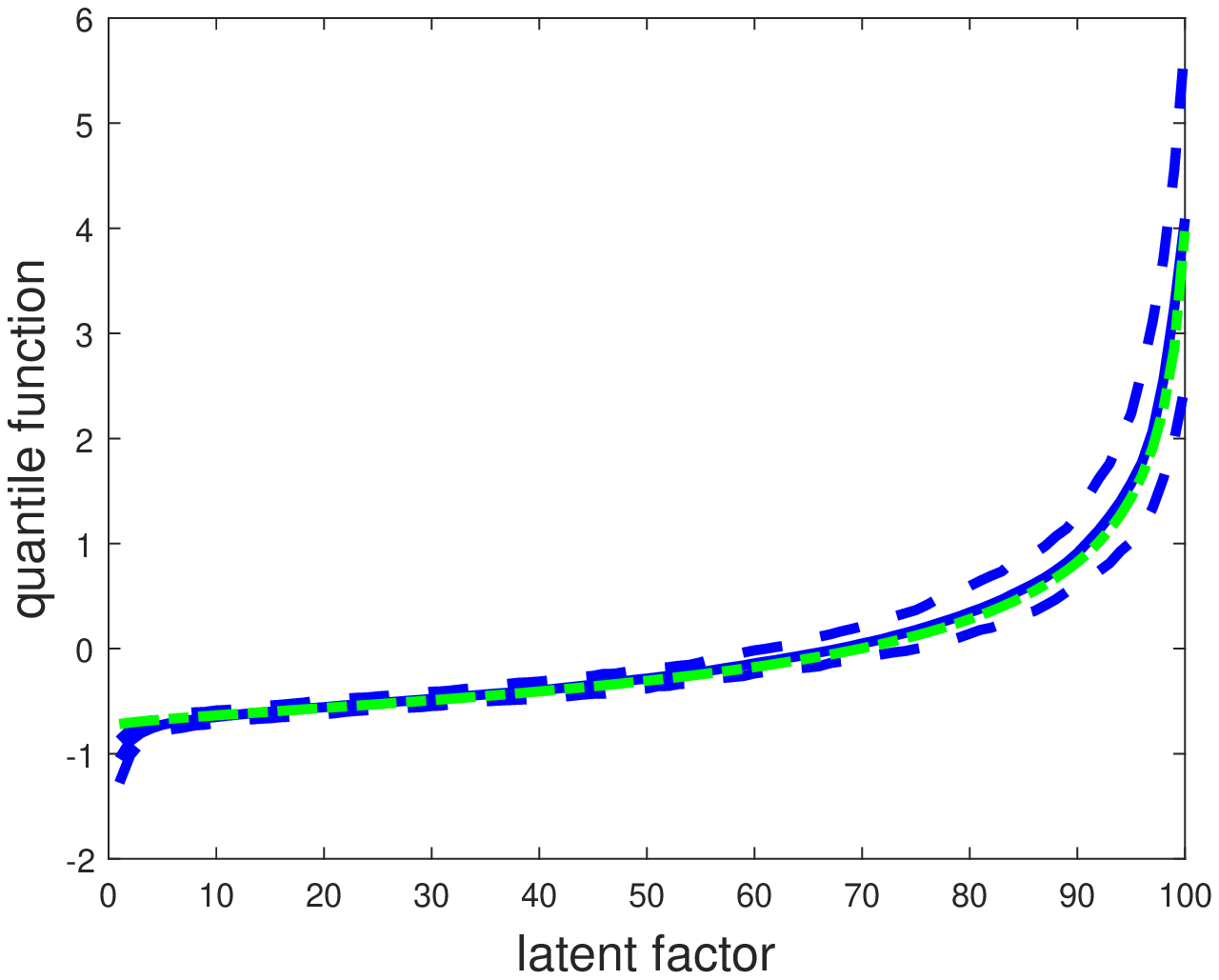} & 
	 			\includegraphics[width=40mm, height=30mm]{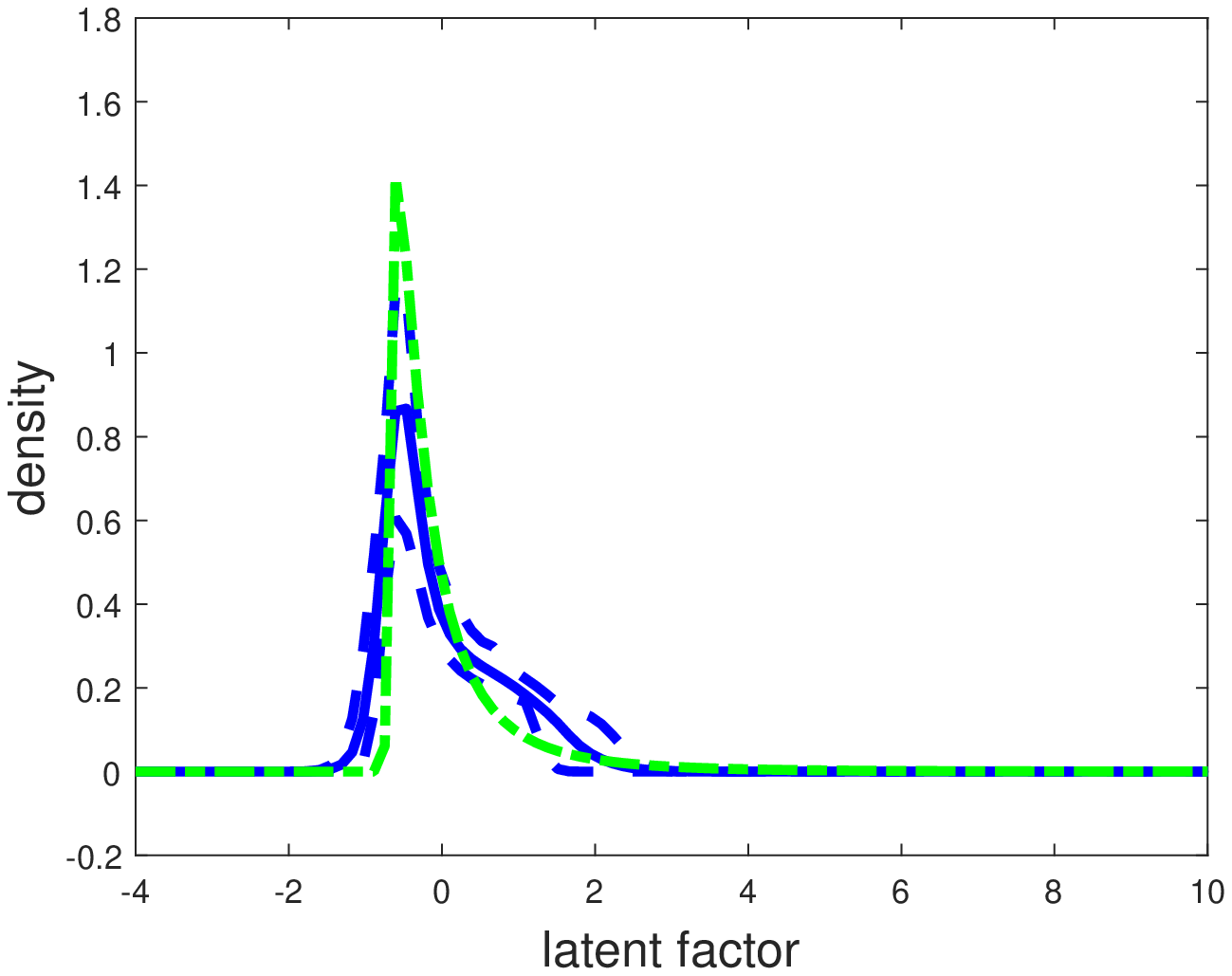} & %
	 			\includegraphics[width=40mm, height=30mm]{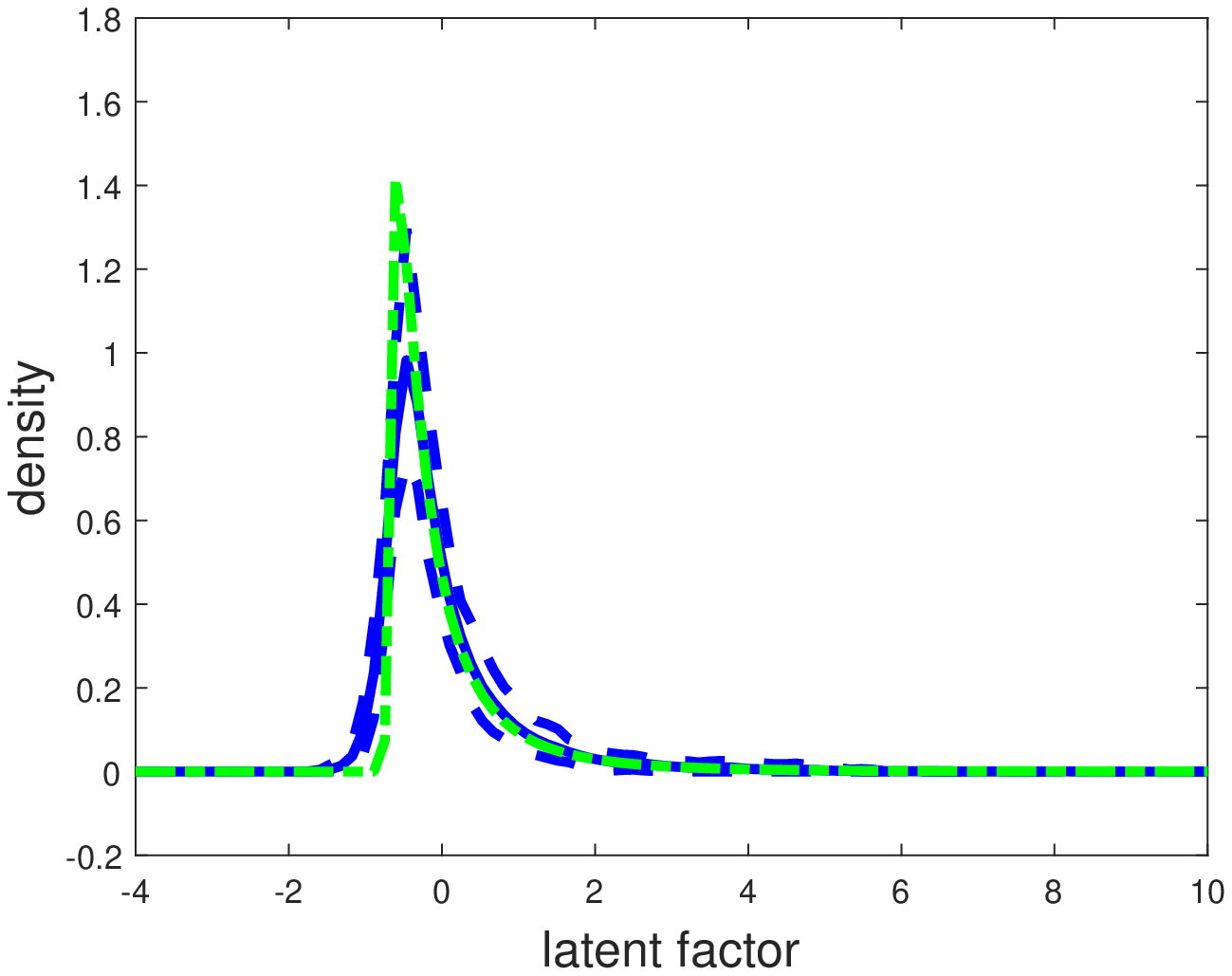}\\
	 			&  &\\
	 		\end{tabular}%
	 	\end{center}
	 	\par
	 	\textit{{\footnotesize Notes: Simulated data from the deconvolution model $Y=X_1+X_2$. The mean across simulations is in solid, 10 and 90 percent pointwise quantiles are in dashed, and the true quantile function or density of $X_1$ is in dashed-dotted. $100$ simulations. 10 averages over $\sigma$ draws.}}
	 \end{figure}

	 In the first two columns in Figure \ref{Fig_MC_Deconv} we show the estimates of the quantile functions $\widehat{X}_{i1}=\widehat{F}_{X_1}^{-1}\left(\frac{i}{N+1}\right)$, for the four specifications and both penalization parameters. The solid and dashed lines correspond to the mean, 10 and 90 percentiles across 100 simulations, respectively, while the dashed-dotted line corresponds to the true quantile function. The sample size is $N=100$. In the last two columns of Figure \ref{Fig_MC_Deconv} we show density estimates for the same specifications. The results reproduce the shape of the unknown quantile functions and densities rather well.

	 \begin{figure}[tbp]
	 	\caption{Monte Carlo results, deconvolution model, Beta(2,2), $N=100,500$}\label{Fig_MC_Deconv_N500}
	 	\begin{center}
	 		\begin{tabular}{cccc}
	 			\multicolumn{2}{c}{Quantile functions}&  \multicolumn{2}{c}{Densities}\\
	 			$N=100$ & $N=500$ & $N=100$ & $N=500$ \\
	 			\\ 
	 			\multicolumn{4}{c}{No average, strong constraint}\\
	 			\includegraphics[width=40mm, height=25mm]{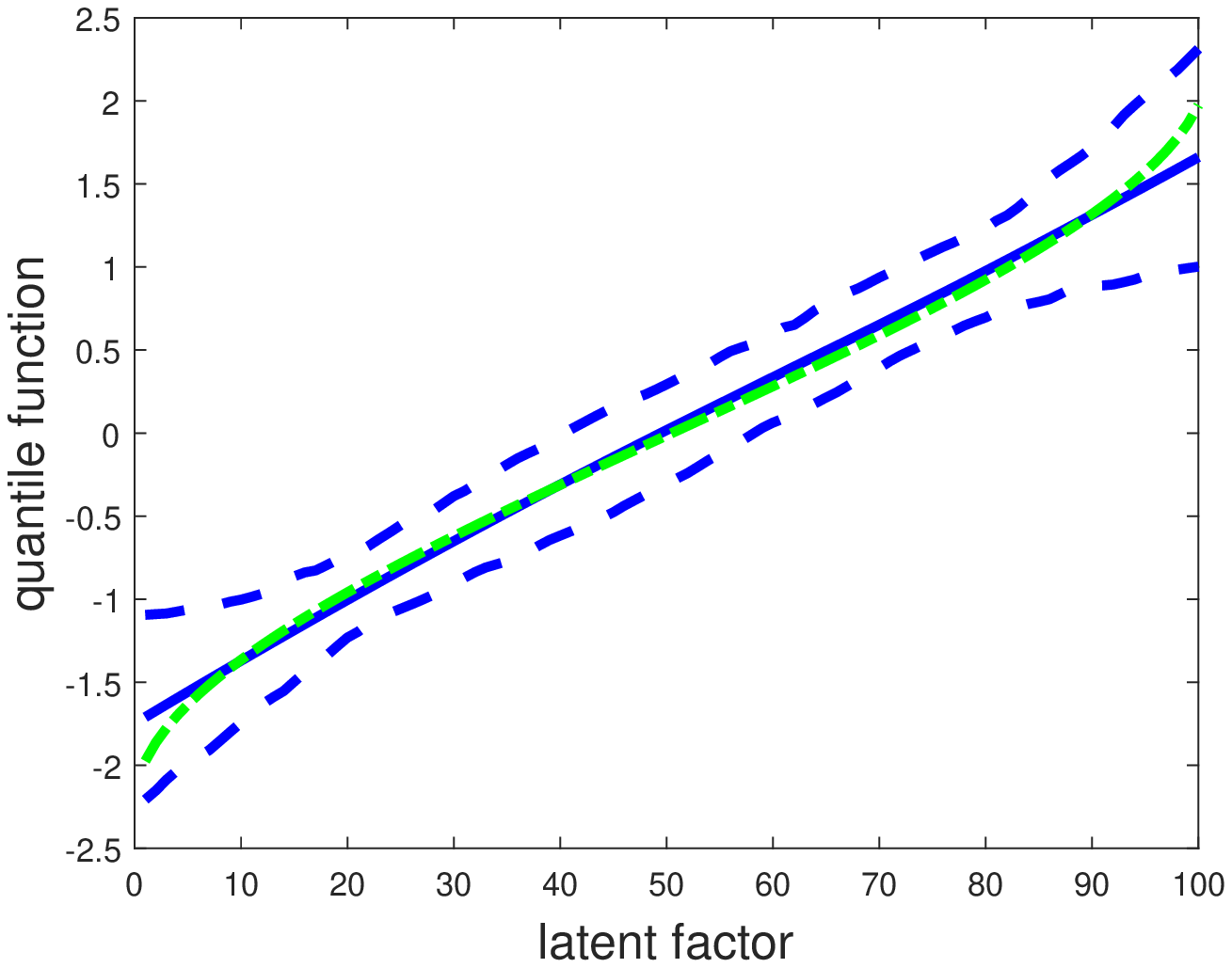} & %
	 			\includegraphics[width=40mm, height=25mm]{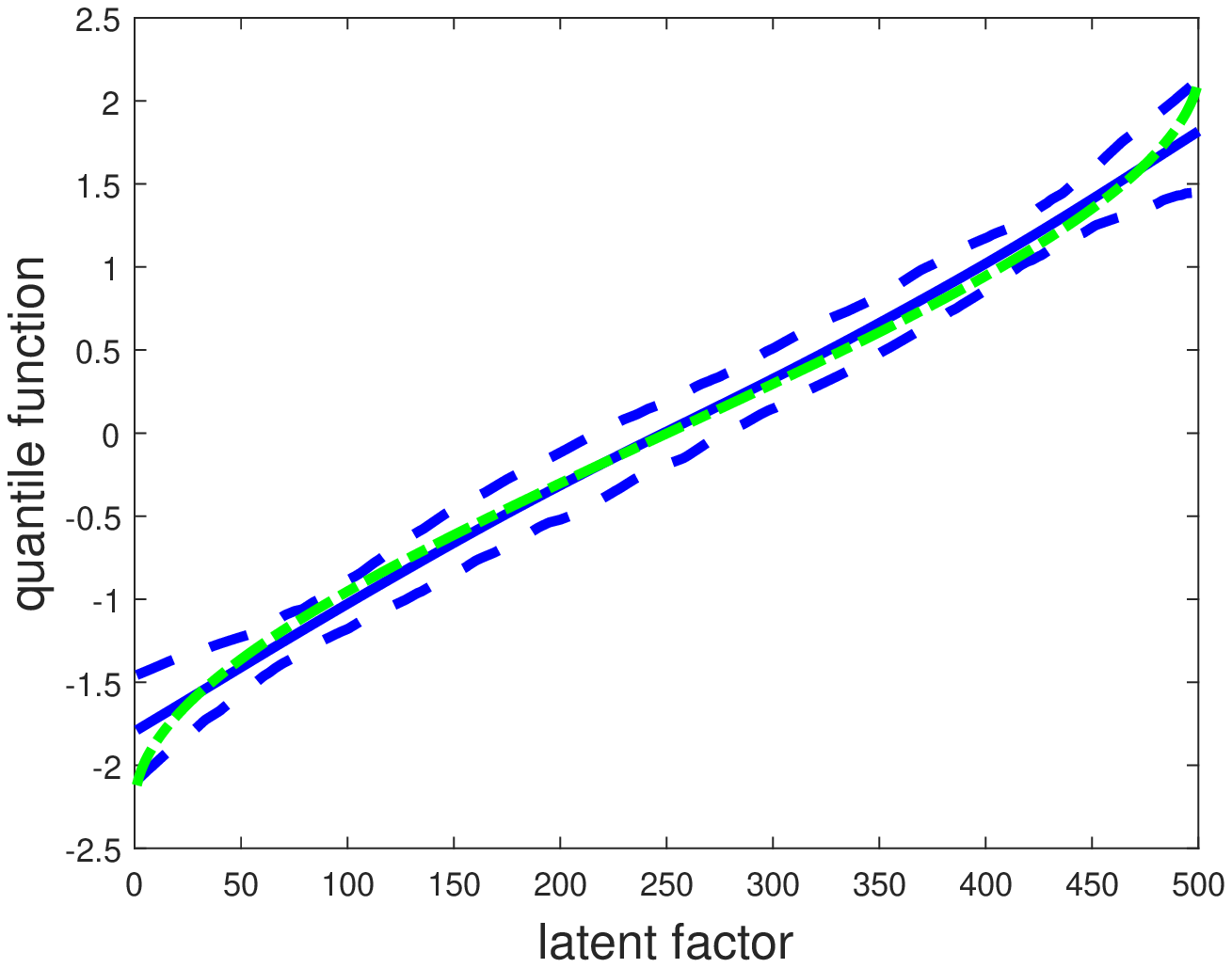} & 
	 			\includegraphics[width=40mm, height=25mm]{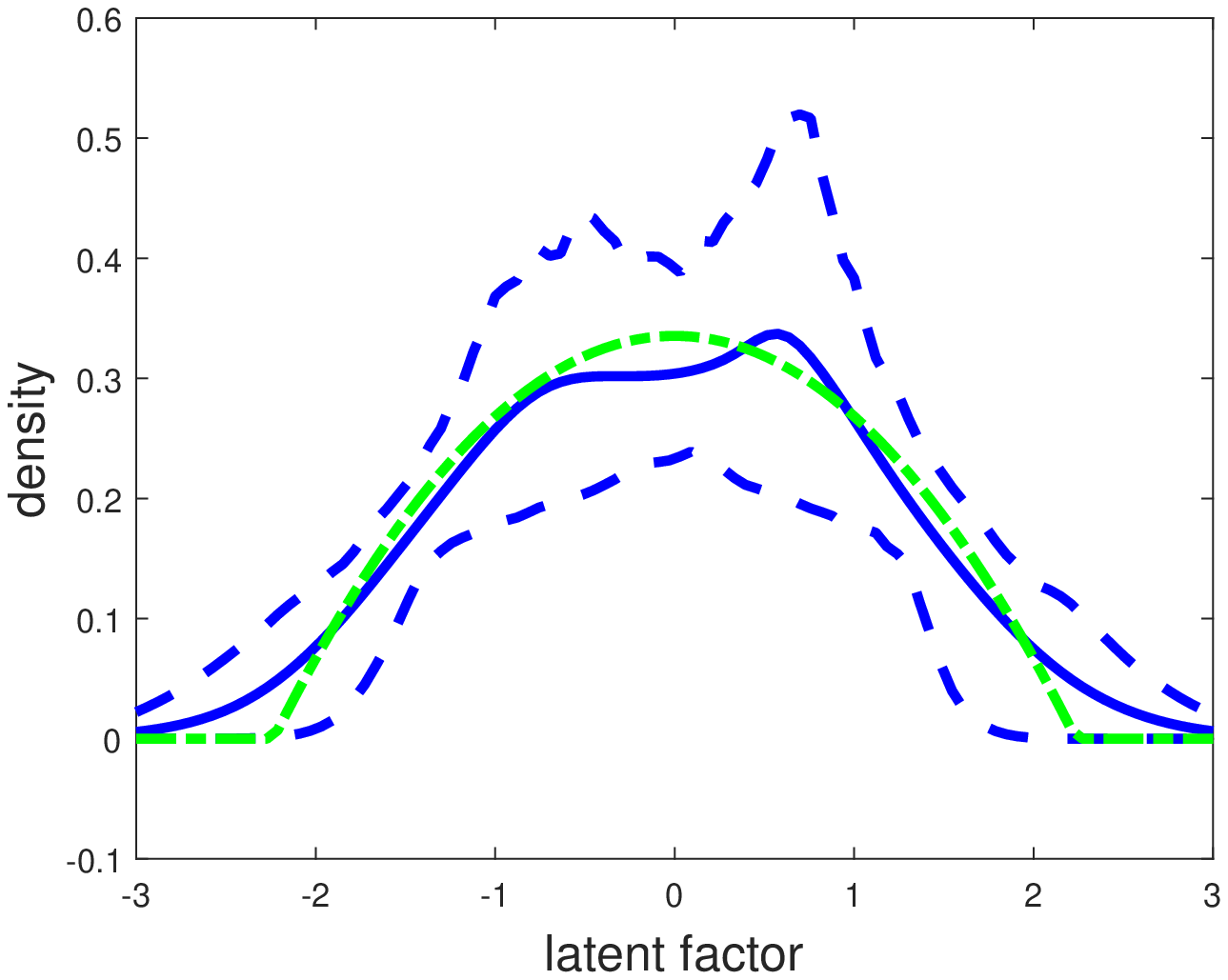} & %
	 			\includegraphics[width=40mm, height=25mm]{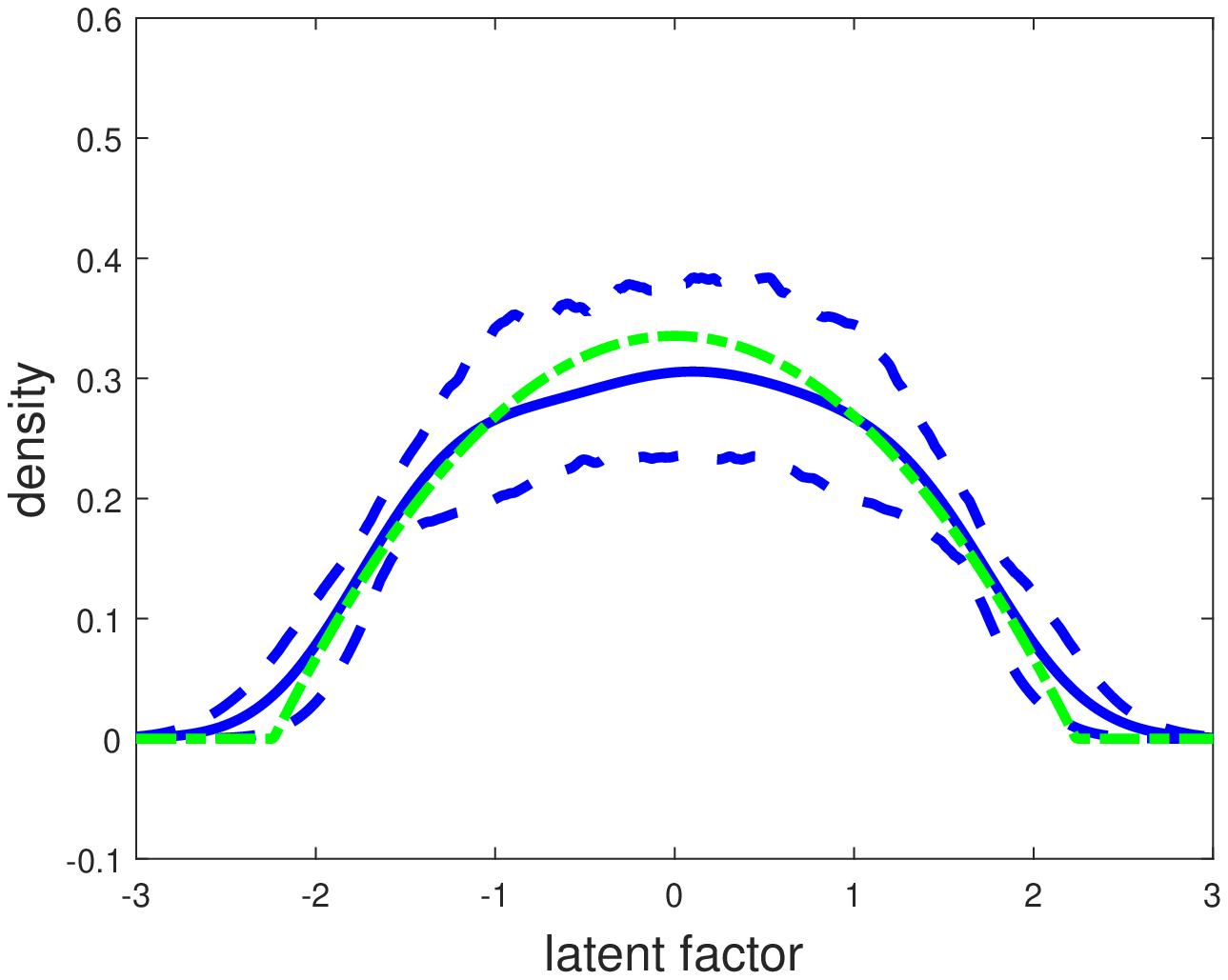}\\
	 			&  &\\
	 			\multicolumn{4}{c}{No average, weak constraint}\\
	 			\includegraphics[width=40mm, height=25mm]{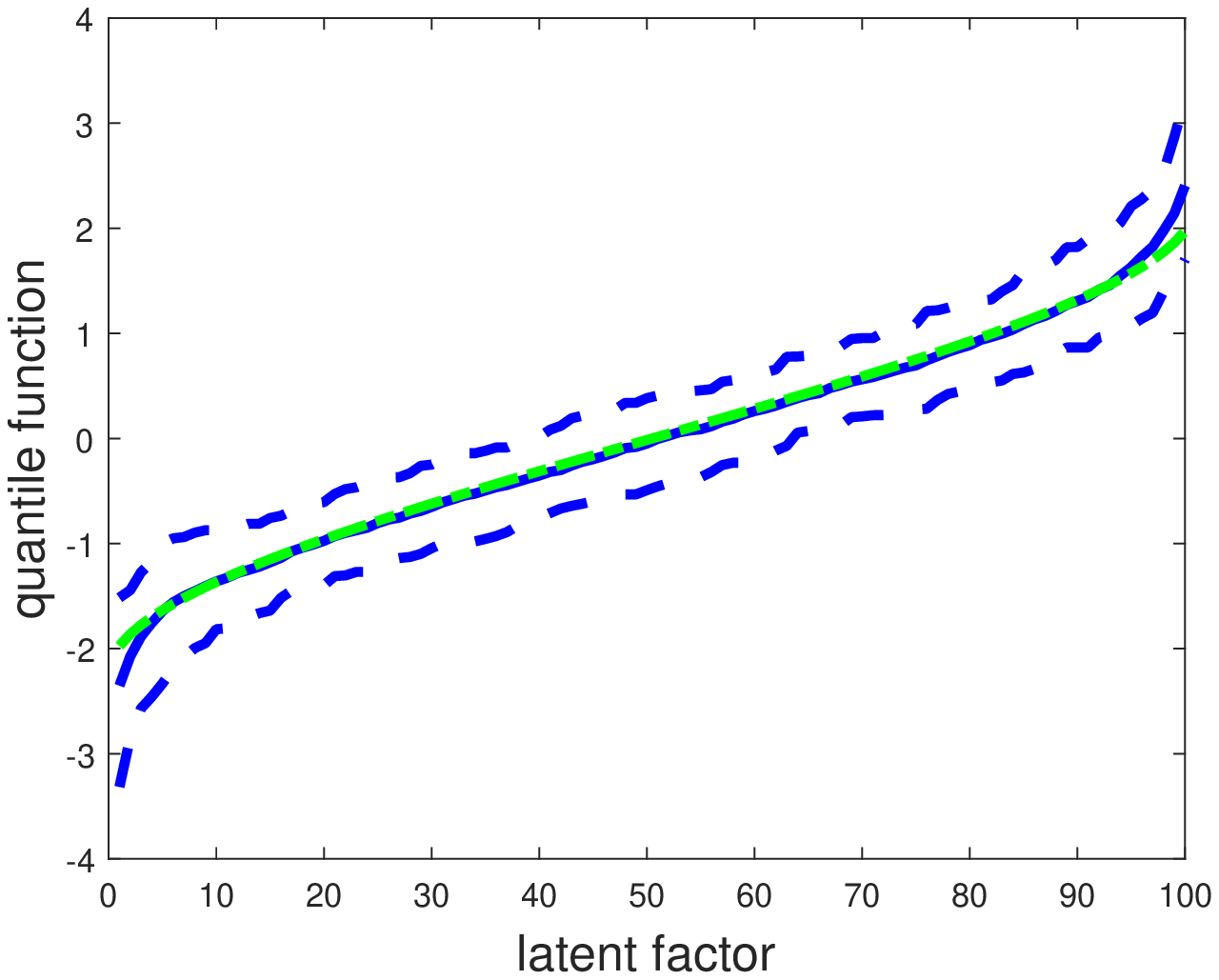} & %
	 			\includegraphics[width=40mm, height=25mm]{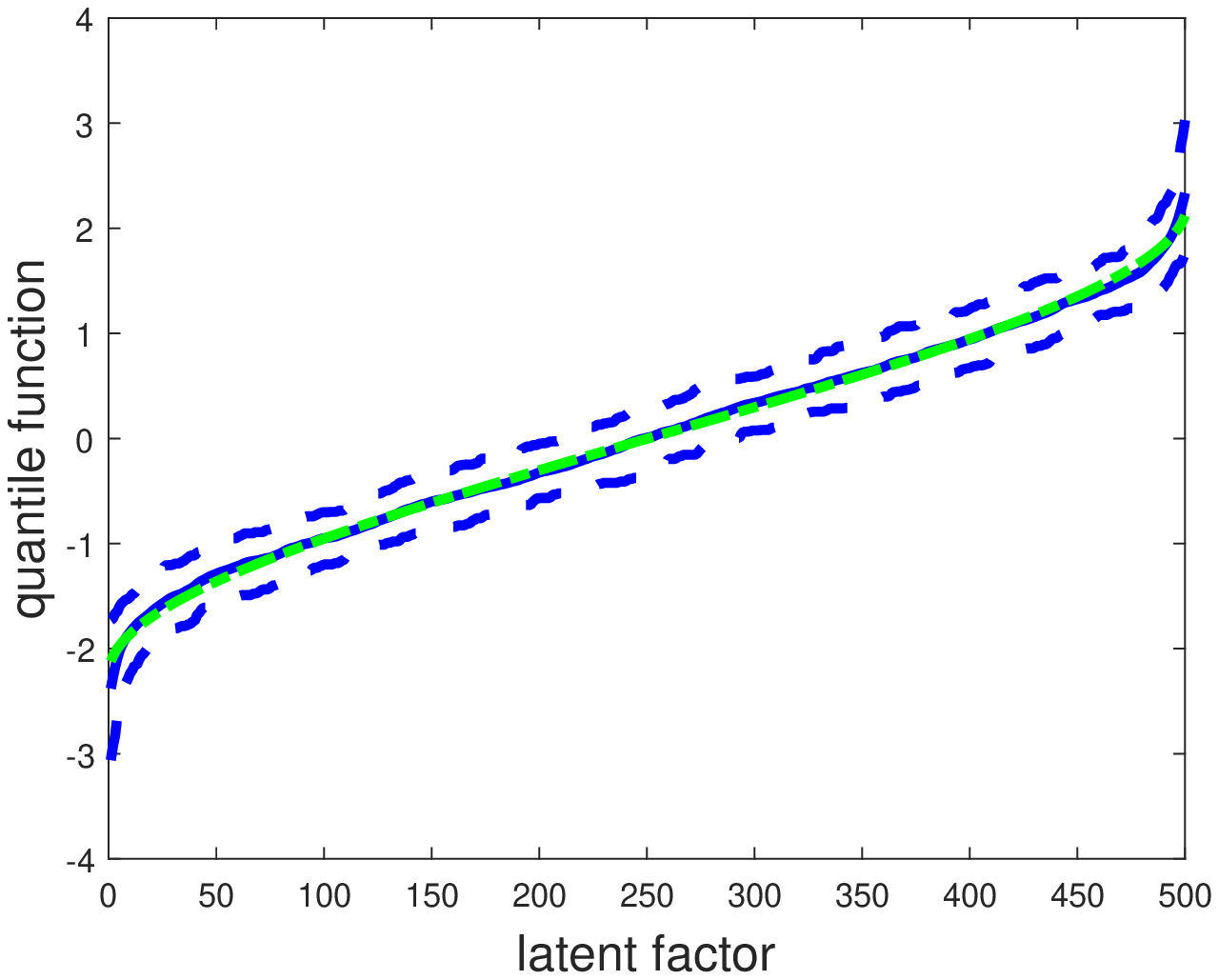} & 
	 			\includegraphics[width=40mm, height=25mm]{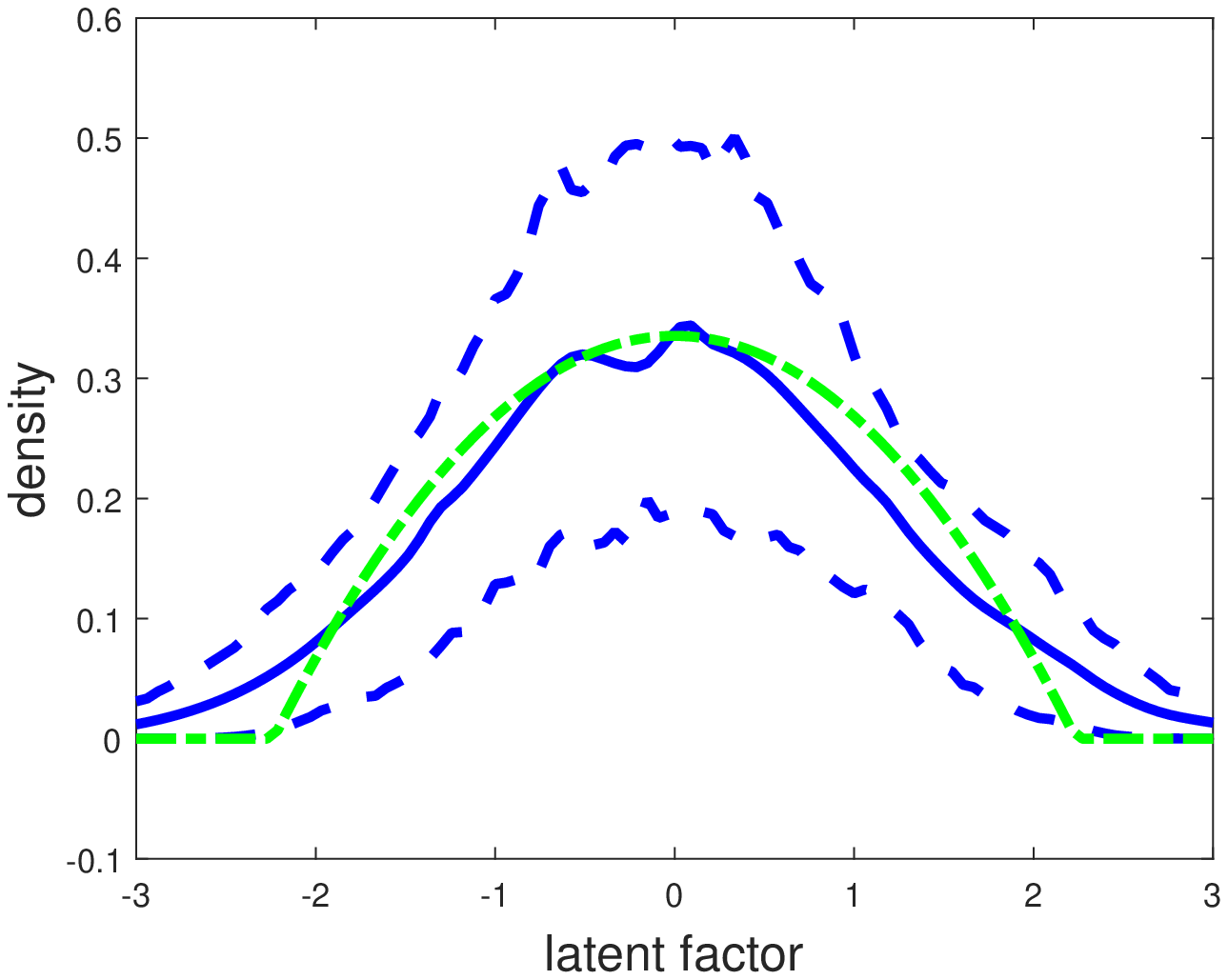} & %
	 			\includegraphics[width=40mm, height=25mm]{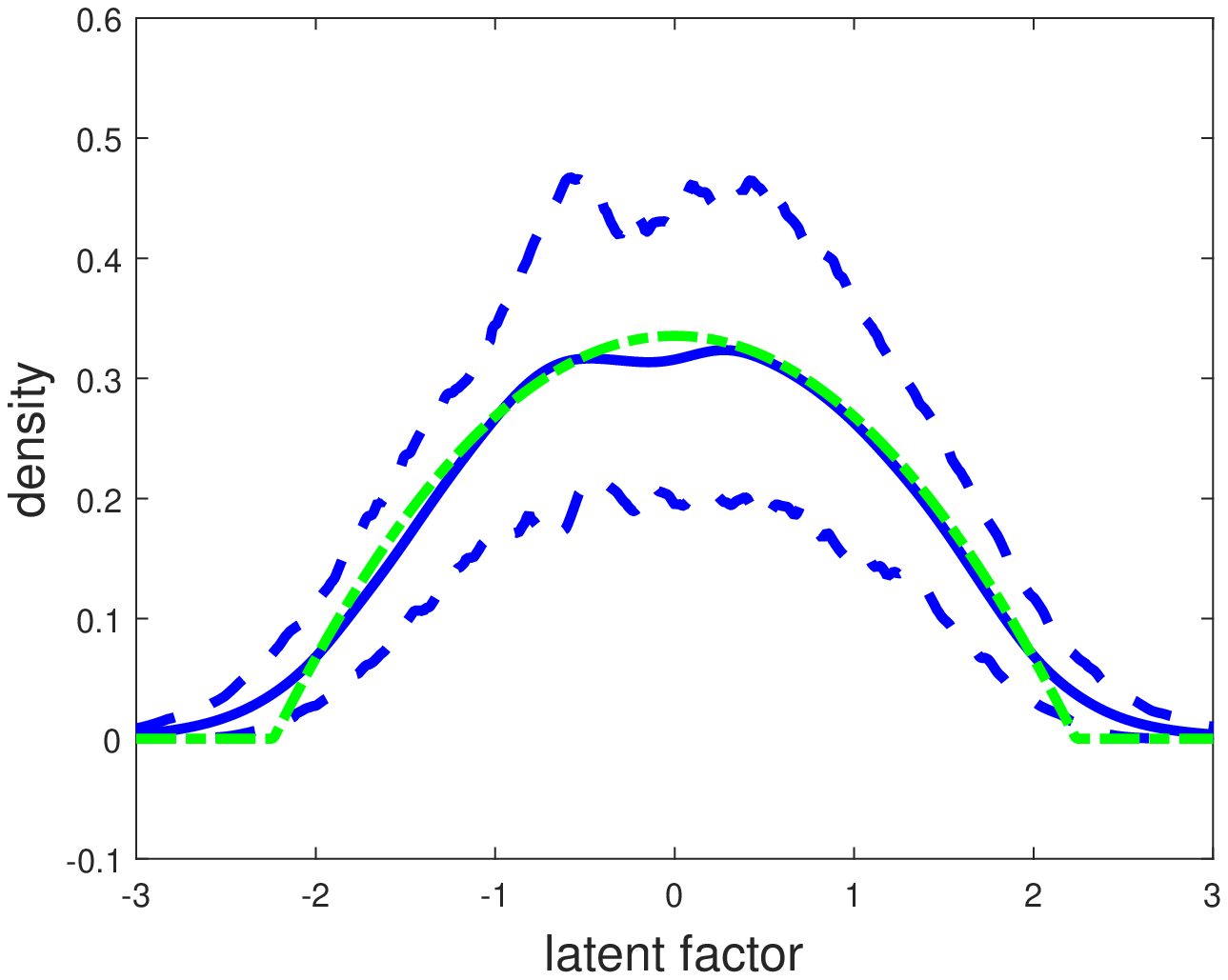}\\
	 			&  &\\
	 			\multicolumn{4}{c}{Average, strong constraint}\\
	 			\includegraphics[width=40mm, height=25mm]{Fig_beta_quant_N100.eps} & %
	 			\includegraphics[width=40mm, height=25mm]{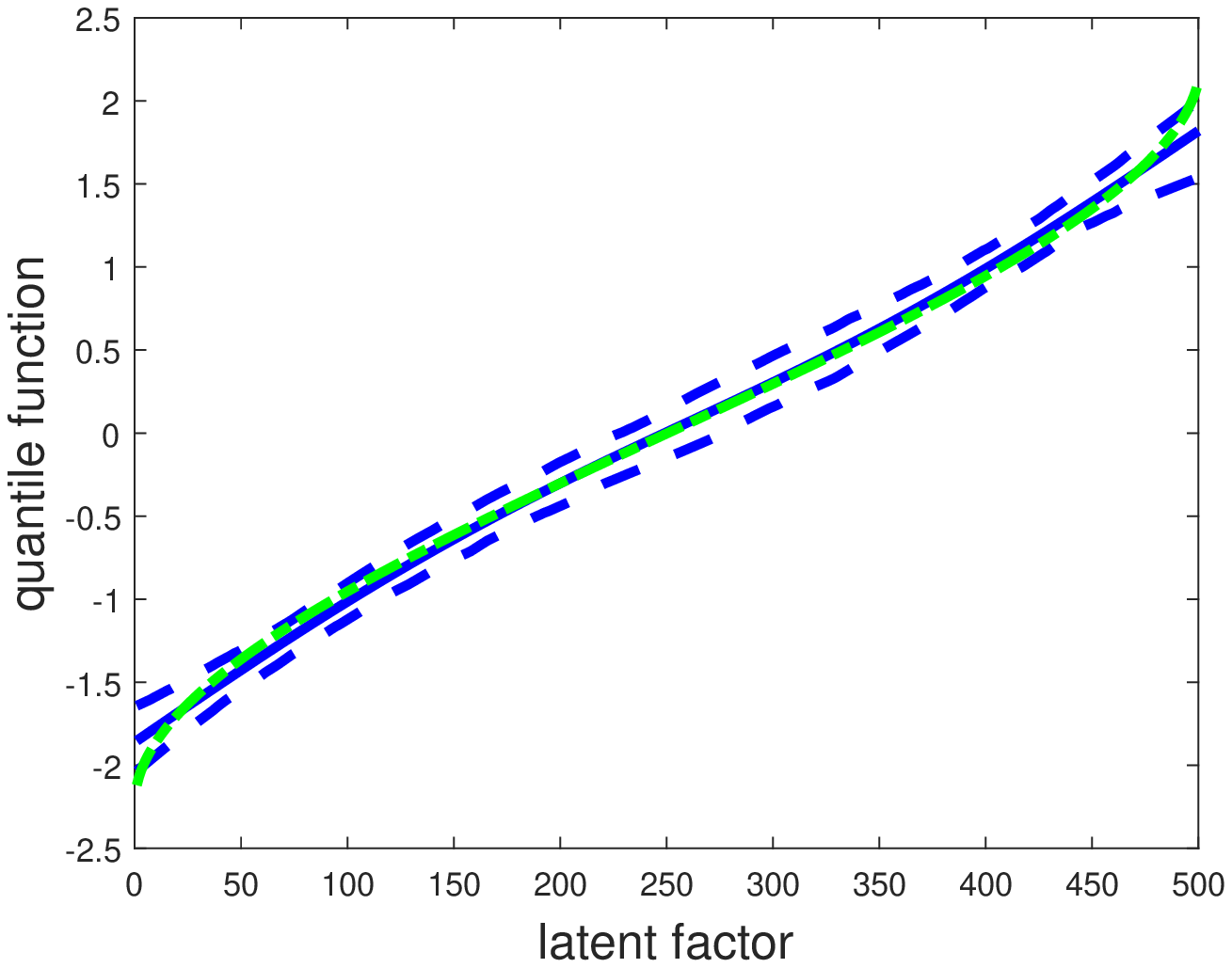} & 
	 			\includegraphics[width=40mm, height=25mm]{Fig_beta_dens_N100.eps} & %
	 			\includegraphics[width=40mm, height=25mm]{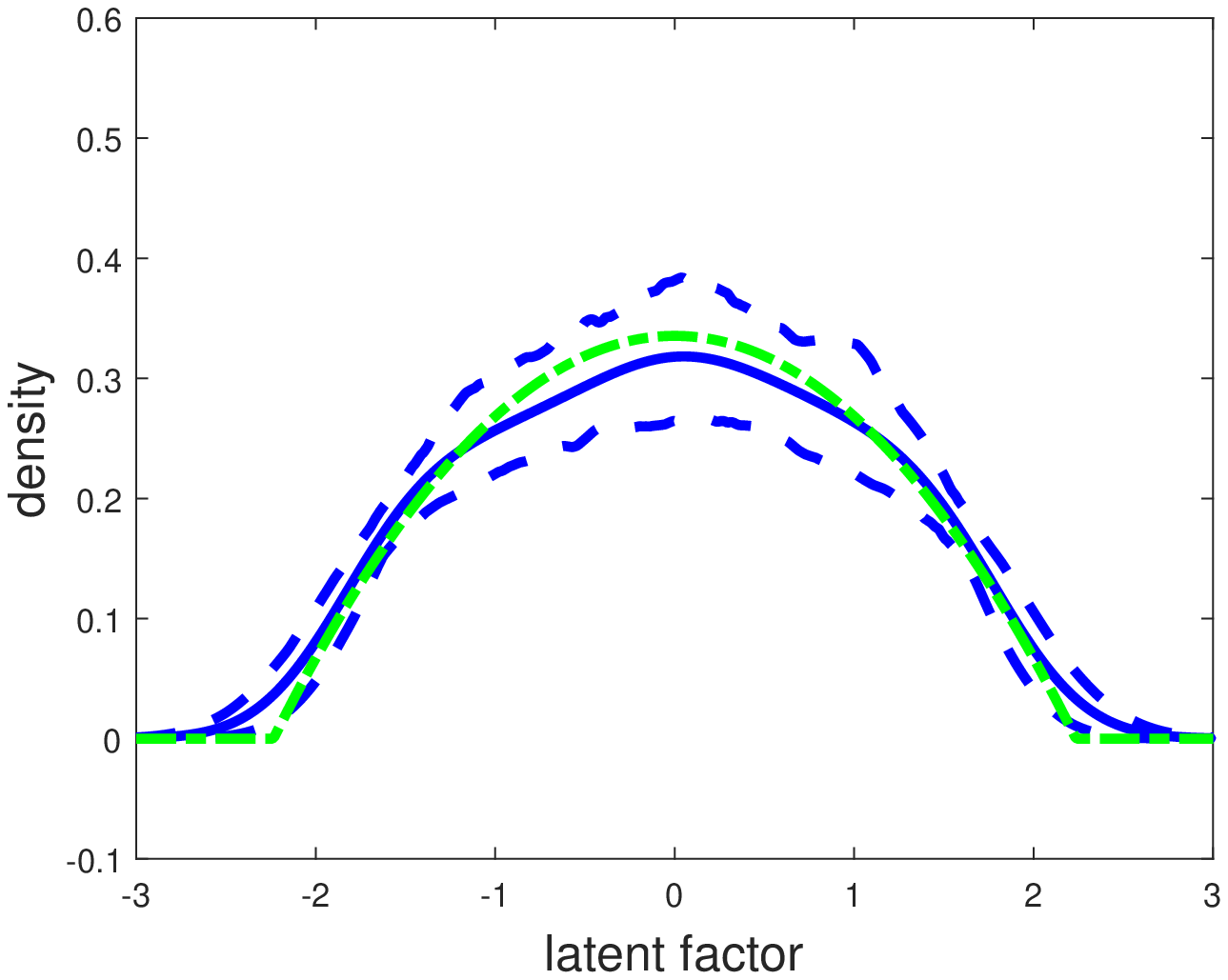}\\
	 			&  &\\
	 			\multicolumn{4}{c}{Average, weak constraint}\\
	 			\includegraphics[width=40mm, height=25mm]{Fig_beta_quant_N100_average_noconst.eps} & %
	 			\includegraphics[width=40mm, height=25mm]{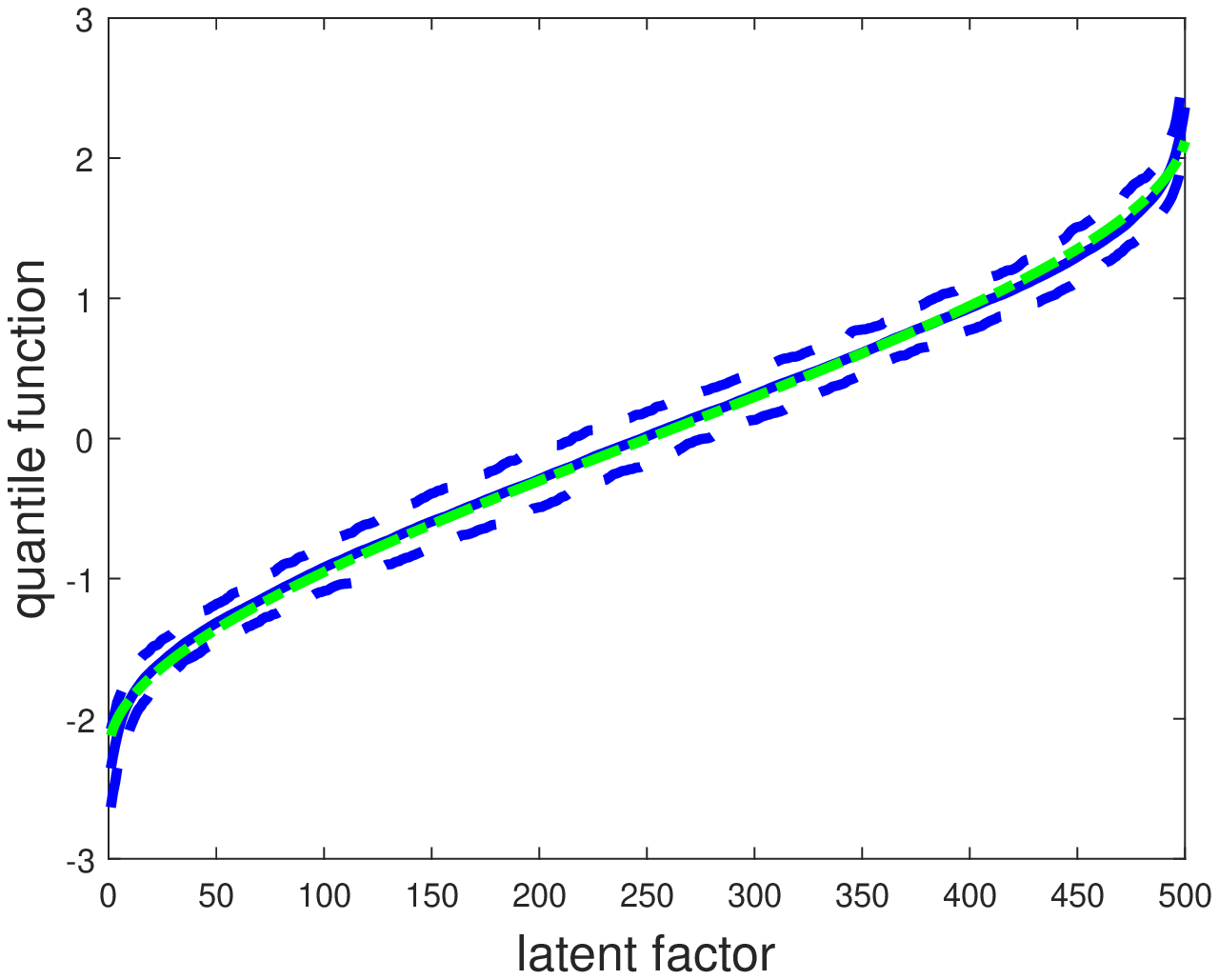} & 
	 			\includegraphics[width=40mm, height=25mm]{Fig_beta_dens_N100_average_noconst.eps} & %
	 			\includegraphics[width=40mm, height=25mm]{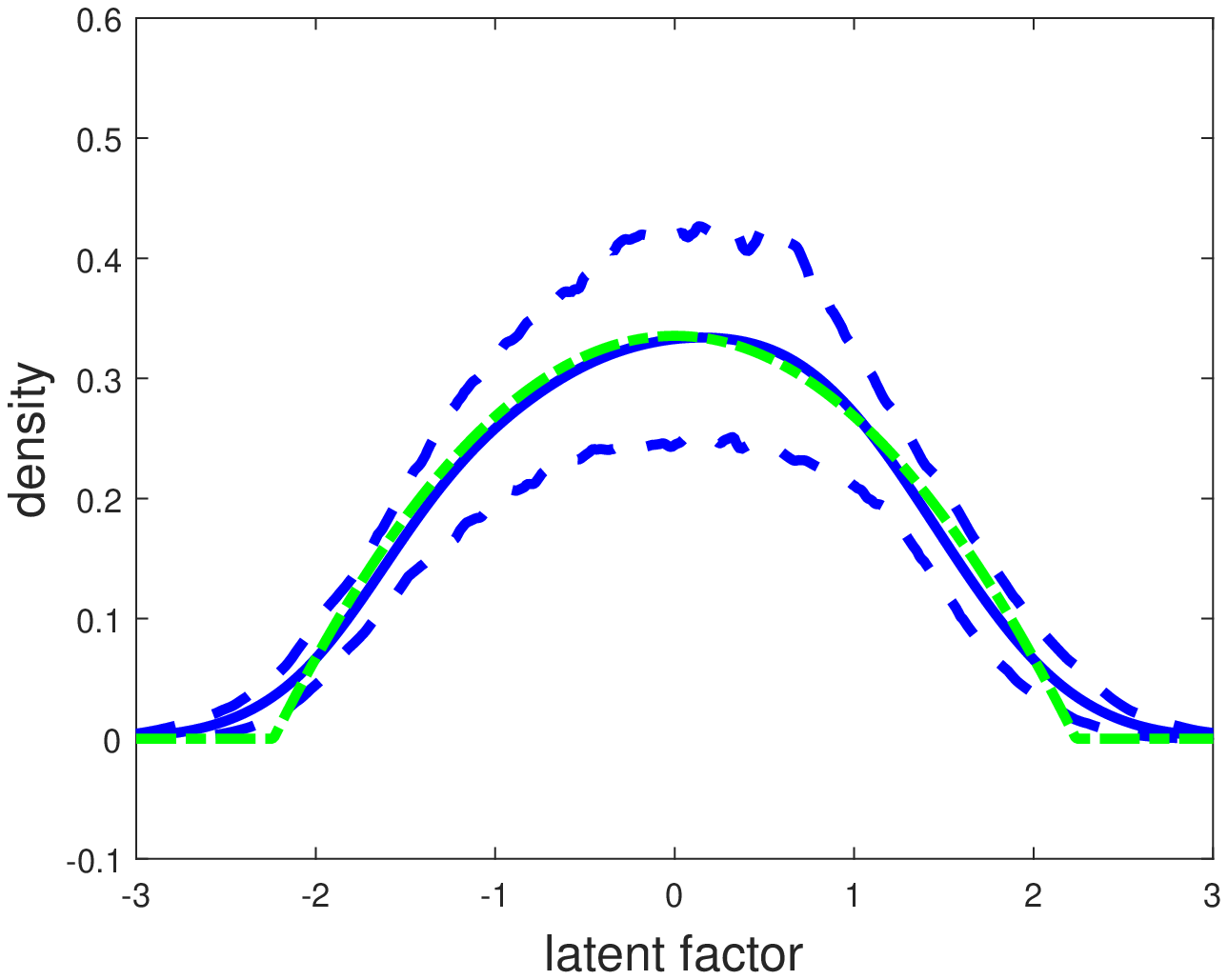}\\
	 			&  &\\
	 			\multicolumn{4}{c}{Single starting value, strong constraint}\\
	 			\includegraphics[width=40mm, height=25mm]{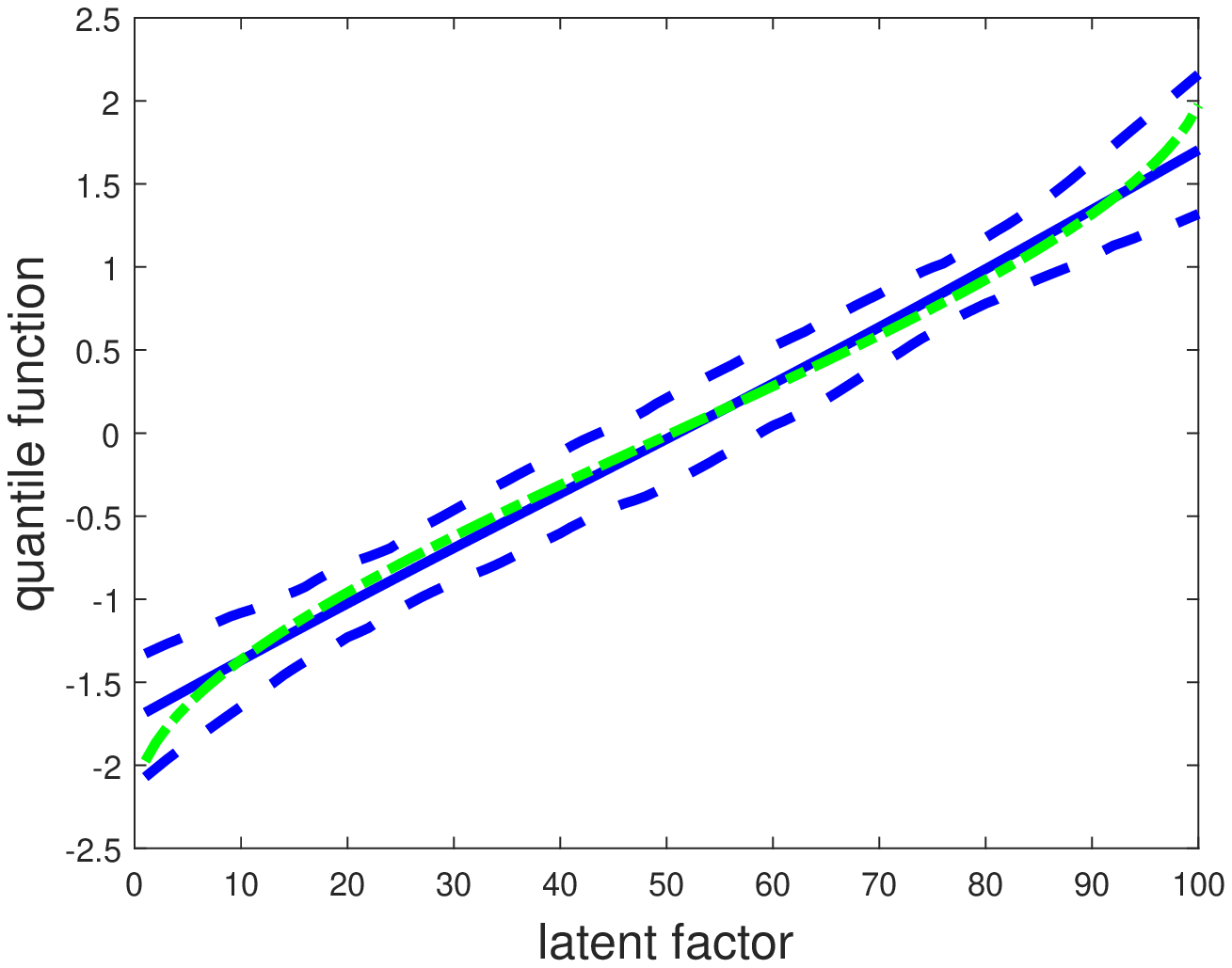} & %
	 			\includegraphics[width=40mm, height=25mm]{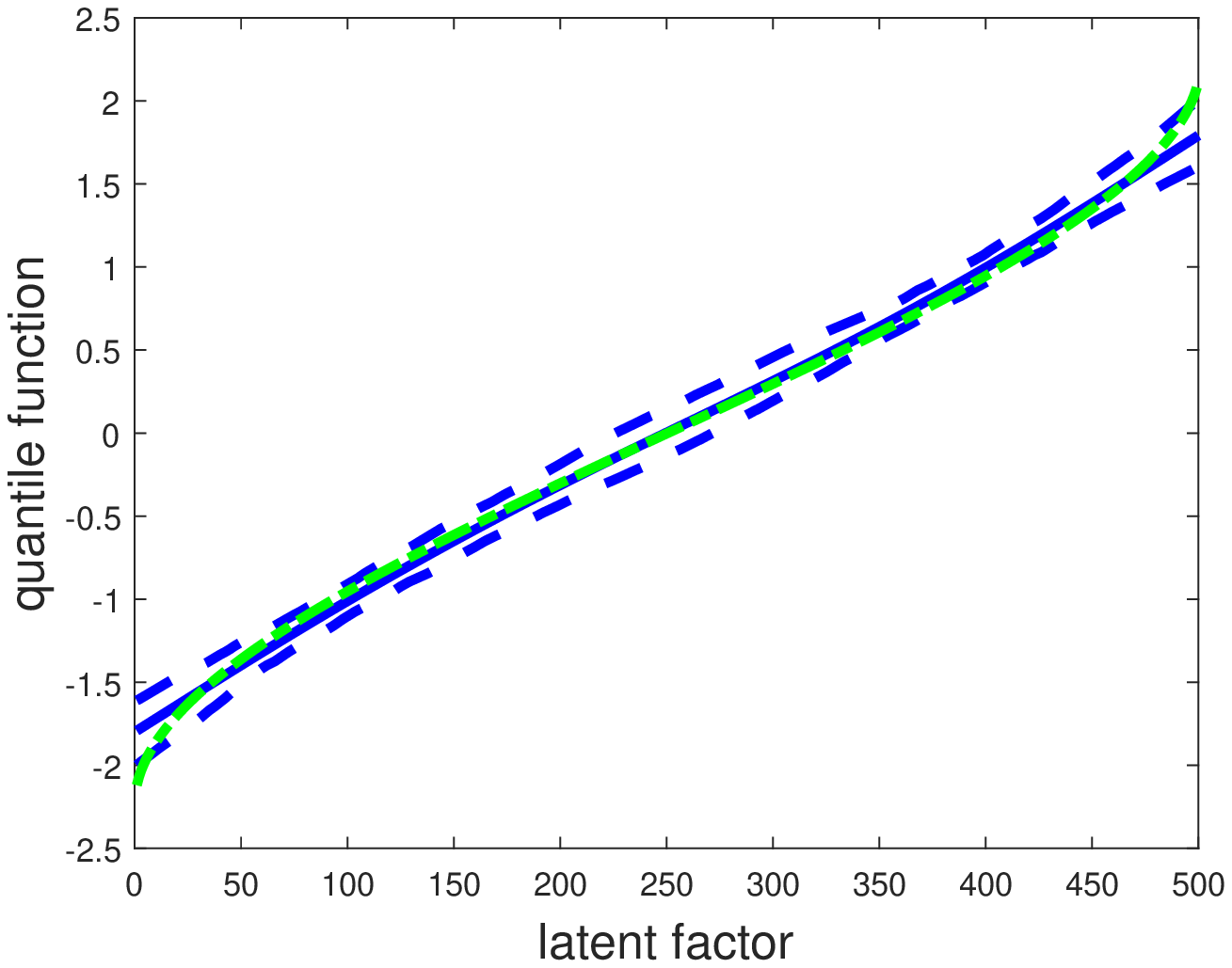} & 
	 			\includegraphics[width=40mm, height=25mm]{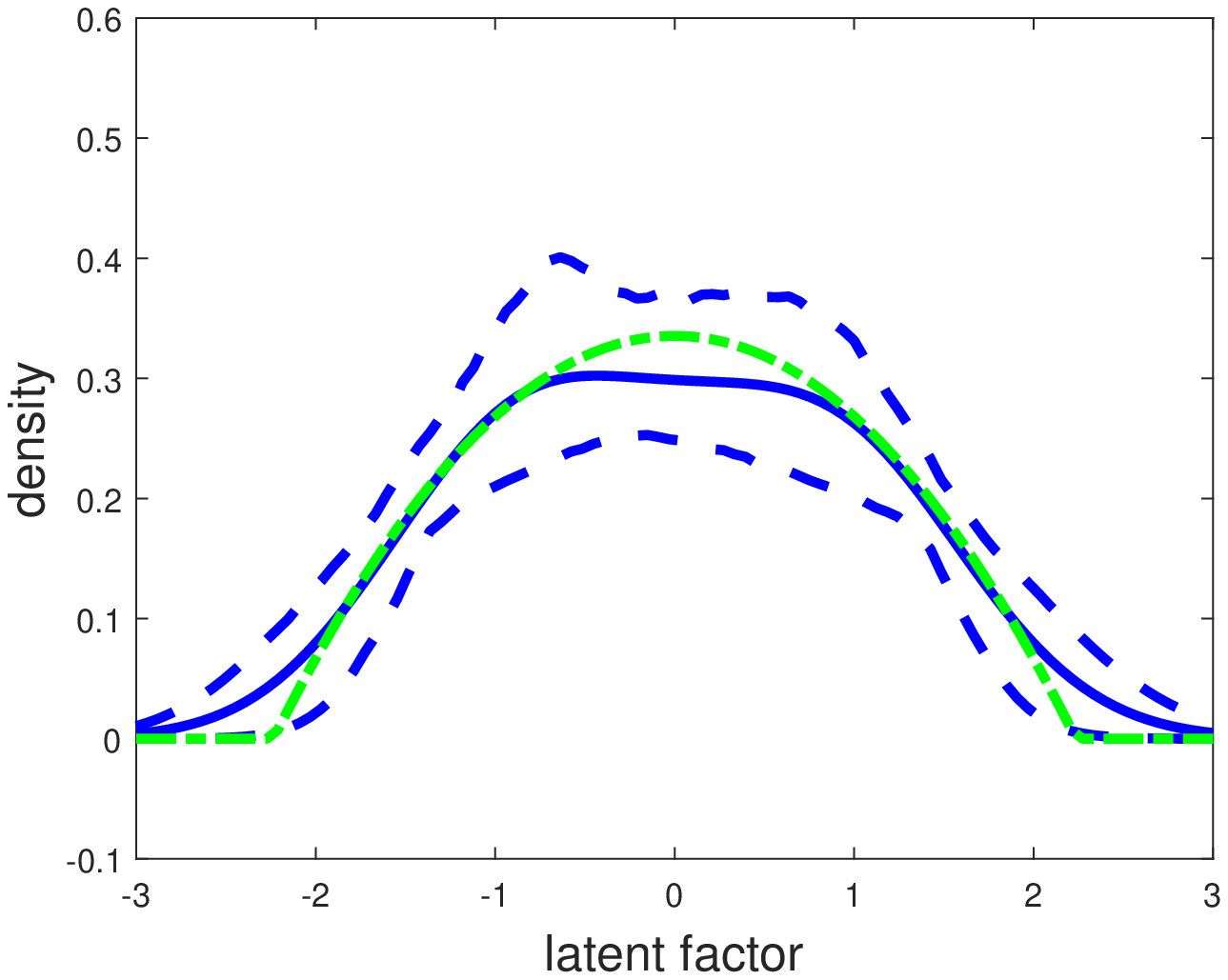} & %
	 			\includegraphics[width=40mm, height=25mm]{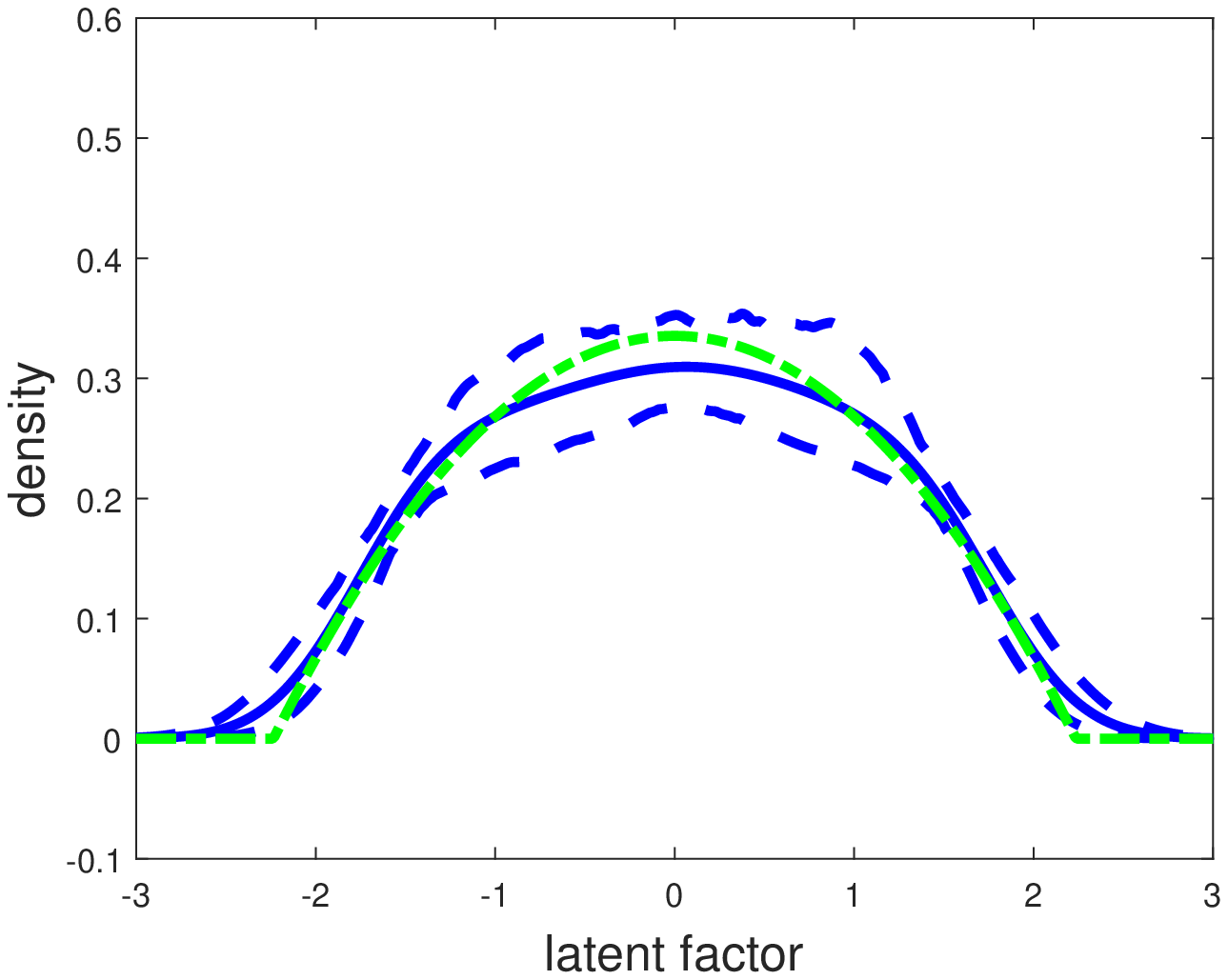}\\
	 			&  &\\
	 		\end{tabular}%
	 	\end{center}
	 	\par
	 	\textit{{\footnotesize Notes: Simulated data from the deconvolution model $Y=X_1+X_2$. The mean across simulations is in solid, 10 and 90 percent pointwise quantiles are in dashed, and the true quantile function or density of $X_1$ is in dashed-dotted. $100$ simulations.}}
	 \end{figure}

	 In Figure \ref{Fig_MC_Deconv_N500} we report additional results for the Beta$(2,2)$ specification, for $N=100$ (columns 1 and 3) and $N=500$ (columns 2 and 4). In the first two rows we report the results based on a single $\sigma$ draw per estimate (i.e., $M=1$), whereas in the next two rows we show the results for the estimator averaged over $M=10$ different $\sigma$ draws. While we see that averaging seems to slightly increase the precision of estimated quantile functions and densities, the results based on one $\sigma$ draw are comparable to the ones based on 10 draws. In the last row of Figure \ref{Fig_MC_Deconv_N500} we show results when using a single starting parameter value in our algorithm, instead of 10 values in our baseline estimates. We see that the results are very little affected, suggesting that the impact of starting values on the performance of the estimator is moderate.

	 \begin{sidewaystable}[tbp]\caption{Monte Carlo simulation, mean squared error of estimated quantiles of $X_{1}$ in the deconvolution model: 25\%, 50\%, and 75\%\label{Tab_rate_conv}}
	 	\begin{center}
	 		\begin{tabular}{c||cccccccccc|c}
	 			$N=$ &100 & 200 & 300 & 400 & 500 & 600 & 700 & 800 & 900 & 1000& Implied rate\\\hline
	 			&\multicolumn{10}{c}{Beta(2,2)}\\\hline\hline
	 			25\% perc. & 0.1019  &  0.0695 &   0.0649   & 0.0514 &   0.0436   & 0.0472   & 0.0443 &  0.0431  &  0.0387  &  0.0457 & -0.3866
	 			\\
	 			Median&0.1086  &  0.0863  &  0.0625   & 0.0624 &   0.0540  &  0.0609  &  0.0481 & 0.0494  &  0.0499   & 0.0468 &  -0.3622
	 			\\
	 			75\% perc. & 0.0946 &   0.0671 &   0.0642   & 0.0565   & 0.0466  &  0.0454  &  0.0449 &   0.0415   & 0.0392  &  0.0444 & -0.3660
	 			\\\hline\hline
	 			&\multicolumn{10}{c}{Beta(5,2)}\\\hline\hline
	 			25\% perc. & 0.1188  &  0.0916  &  0.0711  &  0.0635  &  0.0628 &   0.0625 &   0.0582  &  0.0587  &  0.0547  &  0.0572 & -0.3237 \\
	 			Median&0.0927  &  0.0757  &  0.0516  &  0.0532  &  0.0466  &  0.0456 &   0.0386  &  0.0372  &  0.0357  &  0.0387 & -0.4219\\
	 			75\% perc. & 0.0732 &   0.0503  &  0.0417 &   0.0334   & 0.0347 &   0.0319  &  0.0260 &   0.0239   & 0.0249  &  0.0246 & -0.4888
	 			\\\hline\hline
	 			&\multicolumn{10}{c}{${\cal{N}}(0,1)$}\\\hline\hline
	 			25\% perc. &  0.1146   & 0.0674    &0.0650   & 0.0559 &   0.0549  &  0.0518 &   0.0514 &   0.0396   & 0.0463  &  0.0464 &  -0.3789\\
	 			Median&0.0892  &  0.0789  &  0.0596  &  0.0584  &  0.0427  &  0.0520   & 0.0477  &  0.0444  &  0.0450  &  0.0416 &  -0.3411\\
	 			75\% perc. & 0.1036 &   0.0745  &  0.0663  &  0.0605 &   0.0599 &   0.0461  &  0.0516  &  0.0459 &   0.0430 &   0.0451 &  -0.3704	\\\hline\hline
	 			&\multicolumn{10}{c}{$\exp[{\cal{N}}(0,1)]$}\\\hline\hline
	 			25\% perc. &  0.0114   & 0.0086   & 0.0052 &   0.0052 &   0.0050   & 0.0048  &  0.0049  &  0.0050   & 0.0042   & 0.0047 &  -0.3925\\
	 			Median&0.0265  &  0.0195    &0.0133    &0.0108  &  0.0085   & 0.0118  &  0.0070  &  0.0070  &  0.0071  &  0.0084 &   -0.5885\\
	 			75\% perc. &  0.0761 &   0.0586 &    0.0384  &  0.0333 &   0.0289   & 0.0243 &   0.0227   & 0.0186   & 0.0196 &   0.0192 & -0.6555	\\\hline\hline
	 		\end{tabular}
	 	\end{center}
	 	{\textit{\footnotesize Notes: Mean squared error across $500$ simulations from the deconvolution model $Y=X_1+X_2$. No average, single starting value, weak constraint. The implied rate in the last column is the regression coefficient of the log-mean squared error on the log-sample size.}}
	 \end{sidewaystable}
	 
	 
	 In Table \ref{Tab_rate_conv} we attempt to quantify the rate of convergence of our quantile function estimator in a simulation experiment. We report the mean squared error at various quantiles (25\%, median, and 75\%) for the four distributional specifications. We focus on the weak constraint case, and rely on a single $\sigma$ draw and single starting parameter value in each replication. We report the results of 500 simulations. In the last column of Table \ref{Tab_rate_conv} we report a numerical rate of convergence based on these results, which we compute by regressing the log-mean squared error on the log-sample size. The results suggest the rate ranges between $N^{-\frac{3}{10}}$ and $N^{-\frac{7}{10}}$. From Theorem 3.7 in Hall and Lahiri (2008), when characteristic functions of $X_1$ and $X_2$ are converging at polynomial rates of order $b$ and $a$, respectively, the optimal rate of convergence for quantile estimation is $N^{-\frac{2b}{2a+2b-1}}$. As an example, in the case of the Beta(2,2) and Beta(5,2) distributions, characteristic functions converge at the quadratic rate, so the corresponding optimal rate is $N^{-\frac{4}{7}}$.

	 \begin{figure}[tbp]
	 	\caption{Monte Carlo simulation, mean squared error of estimated quantiles of $X_{1}$ as a function of the penalization parameter}\label{Fig_MC_Deconv_regu}
	 	\begin{center}
	 		\begin{tabular}{cccc}
	 			\multicolumn{4}{c}{$N=100$}\\
	 			Beta$(2,2)$ &  Beta$(5,2)$ & ${\cal{N}}(0,1)$ &  $\exp[{\cal{N}}(0,1)]$\\
	 			\includegraphics[width=40mm, height=30mm]{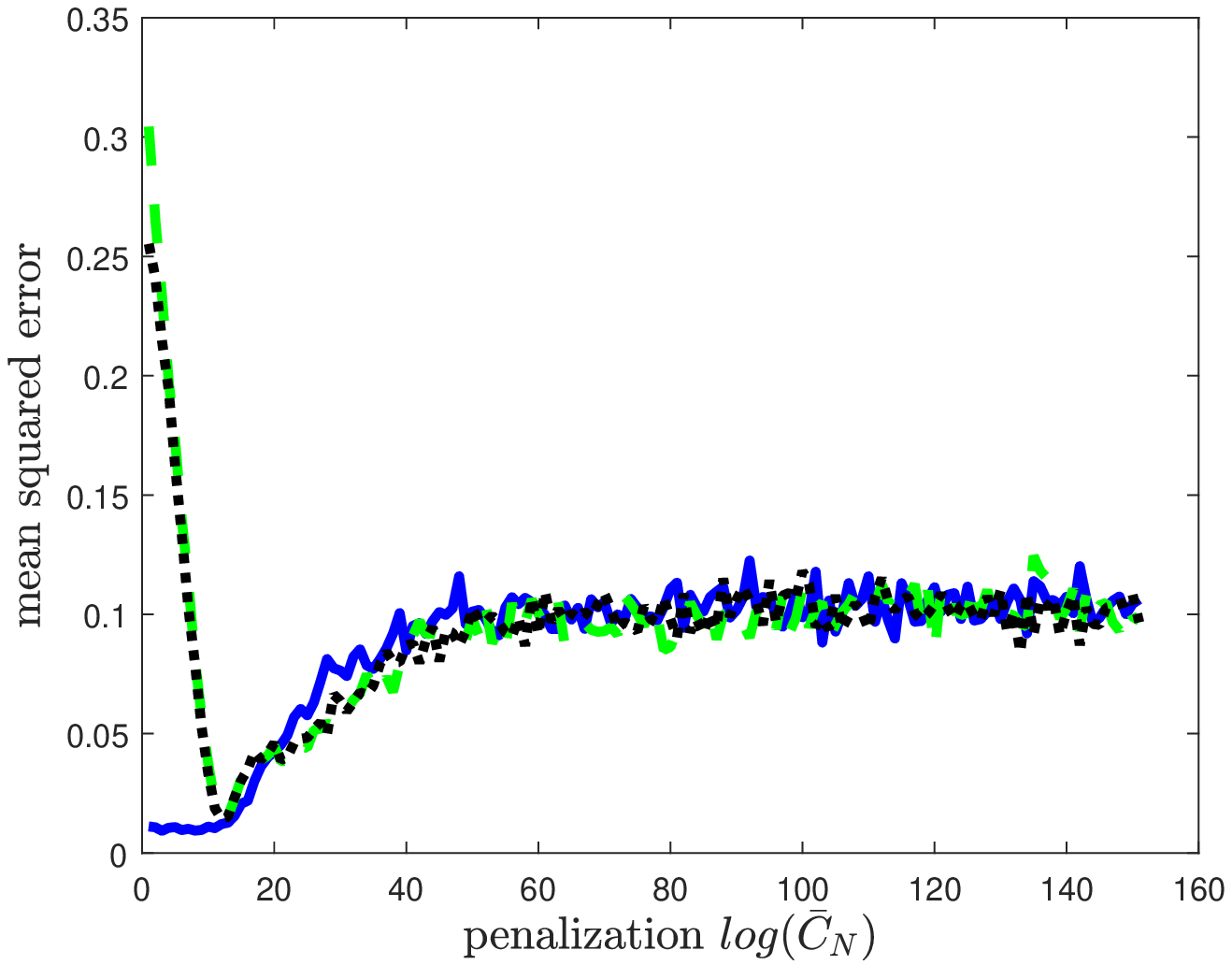} & 	\includegraphics[width=40mm, height=30mm]{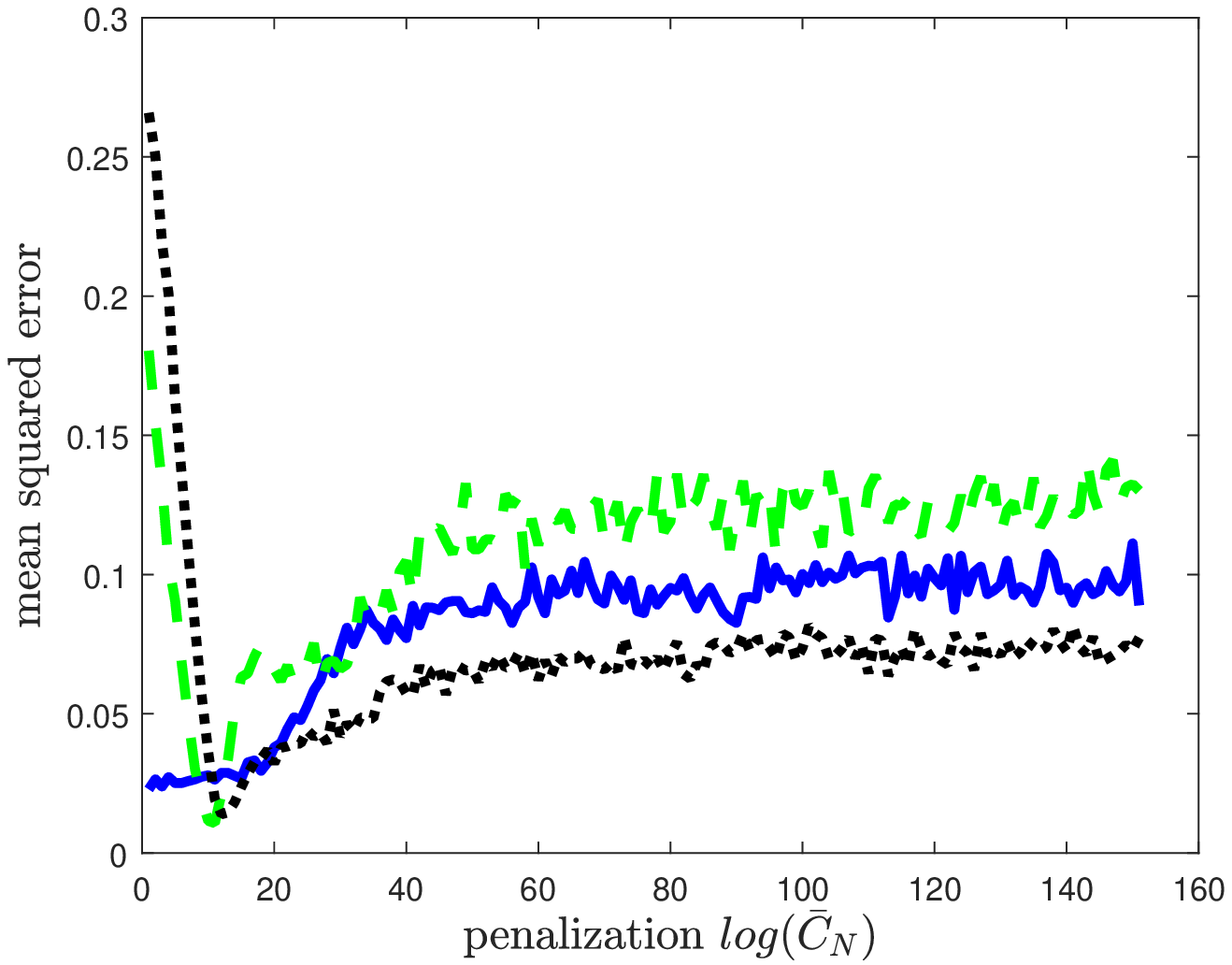}&
	 			\includegraphics[width=40mm, height=30mm]{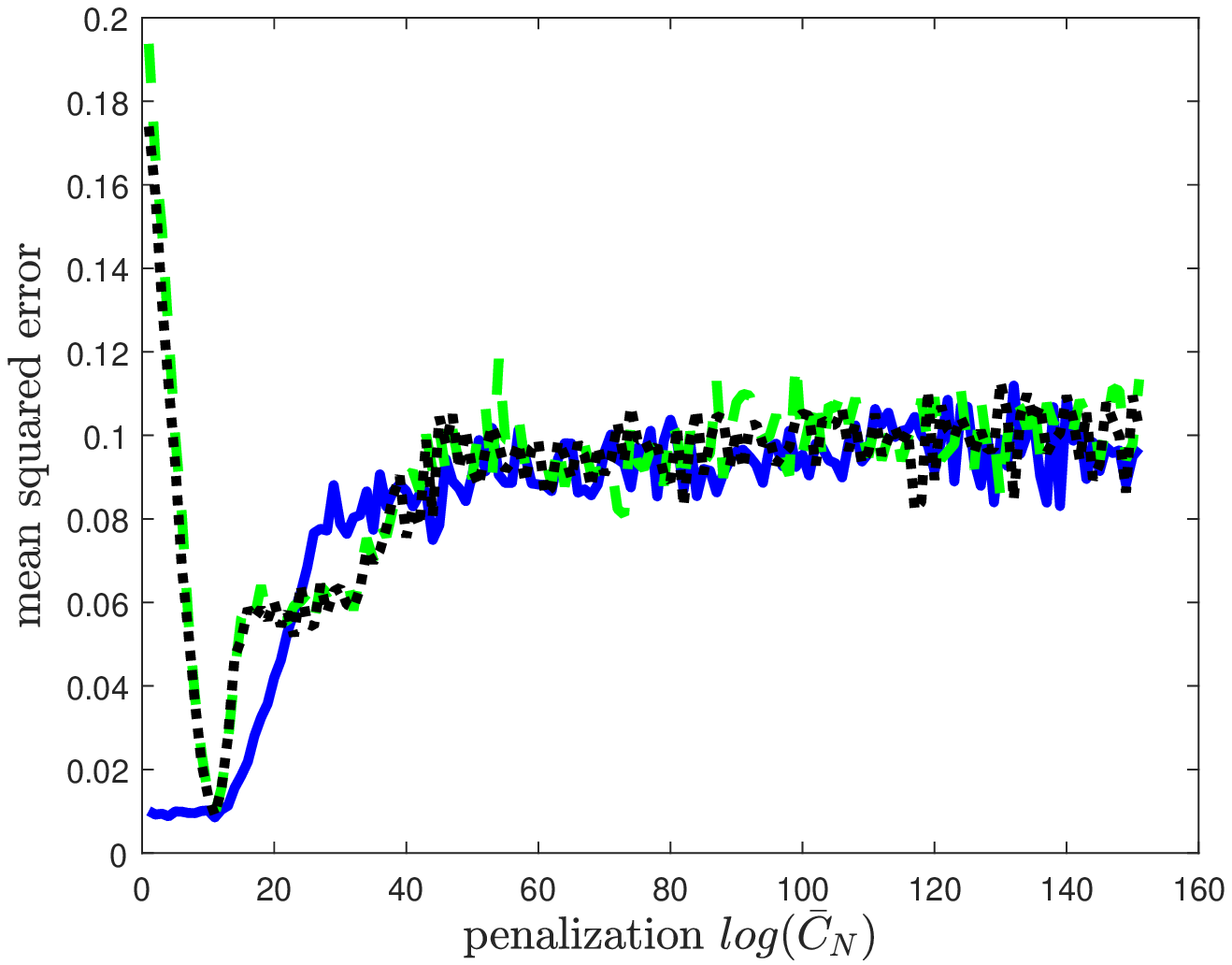} 	&	\includegraphics[width=40mm, height=30mm]{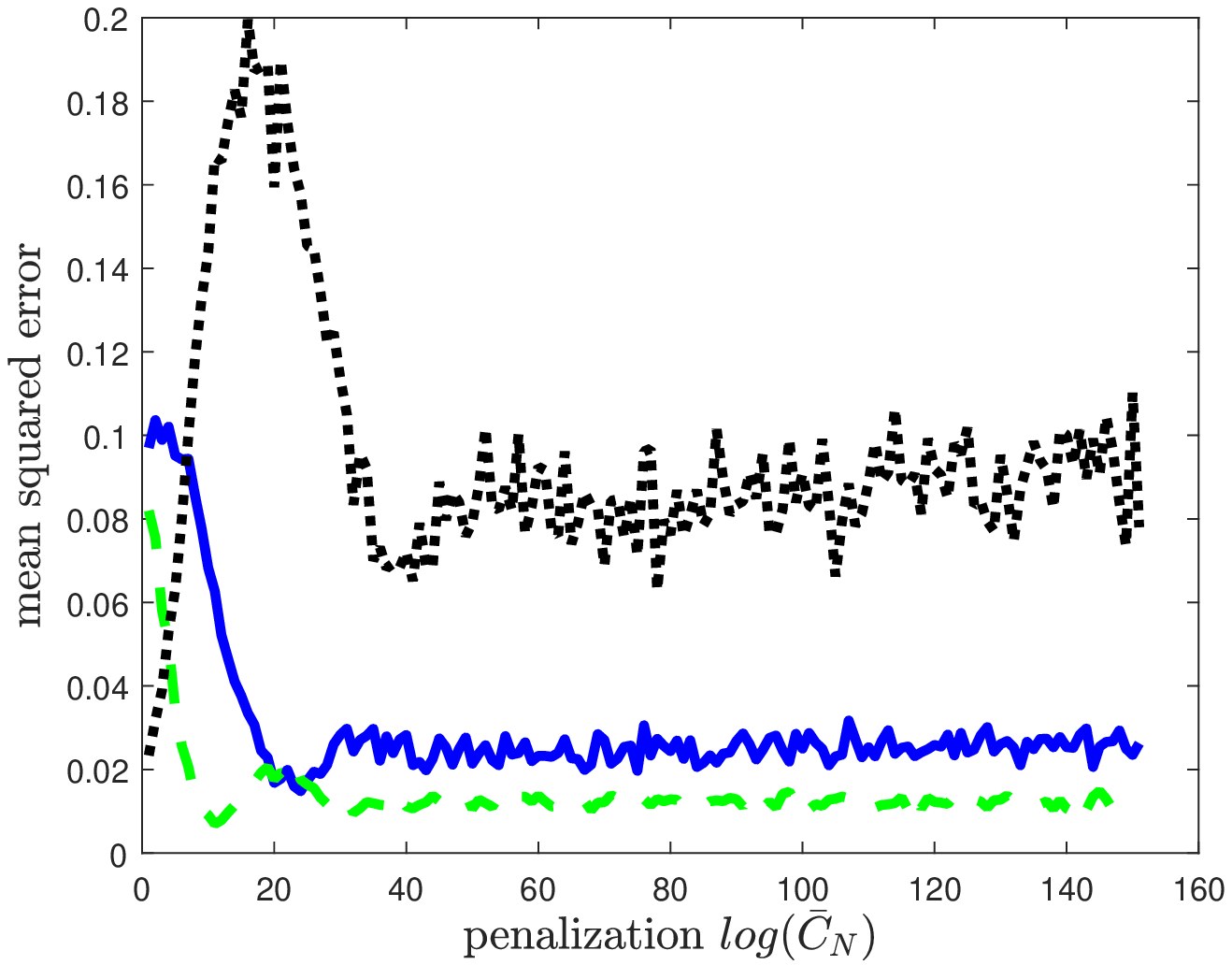}  \\
	 			\multicolumn{4}{c}{$N=500$}\\
	 			Beta$(2,2)$ &  Beta$(5,2)$ & ${\cal{N}}(0,1)$ &  $\exp[{\cal{N}}(0,1)]$\\
	 			\includegraphics[width=40mm, height=30mm]{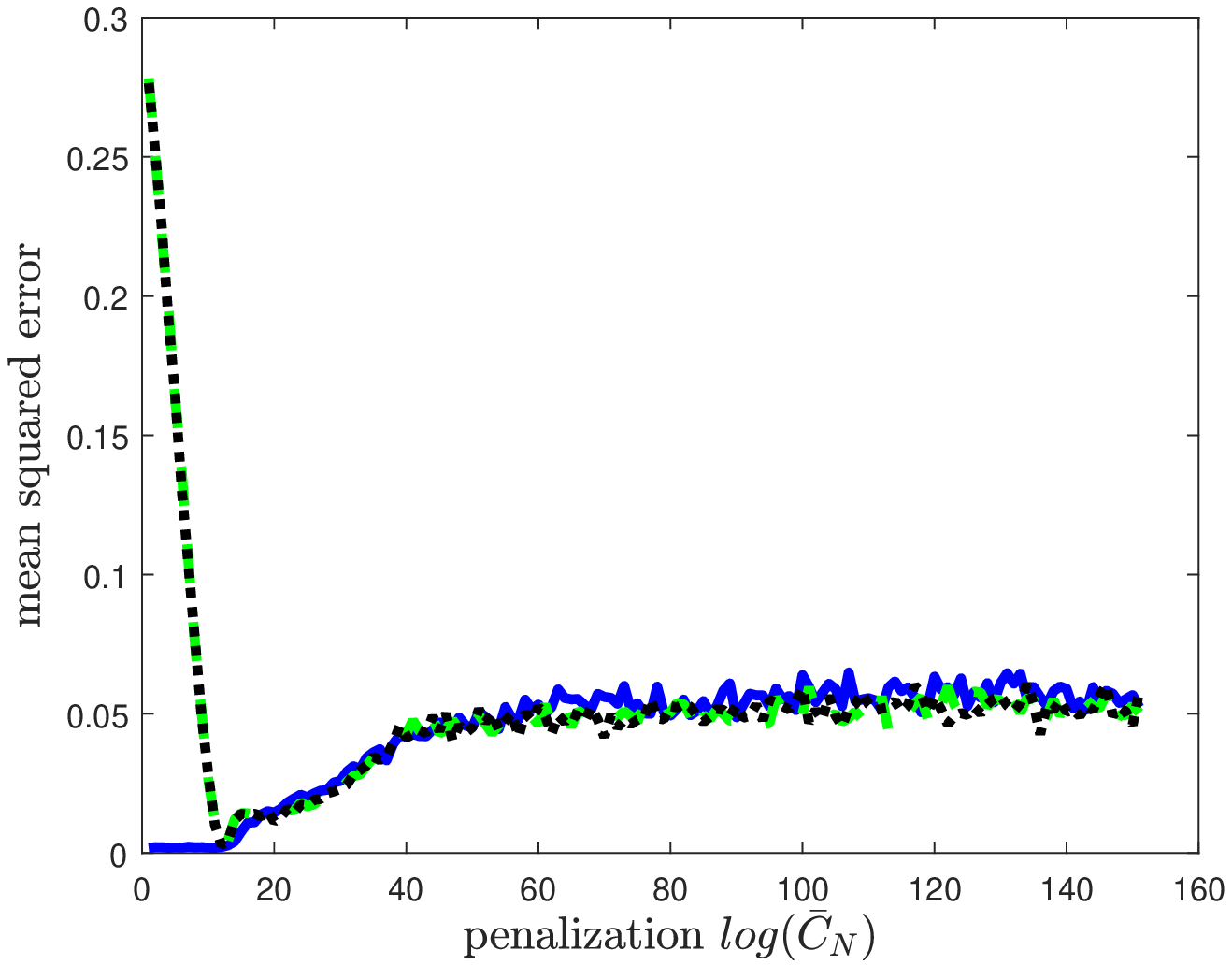} & 	\includegraphics[width=40mm, height=30mm]{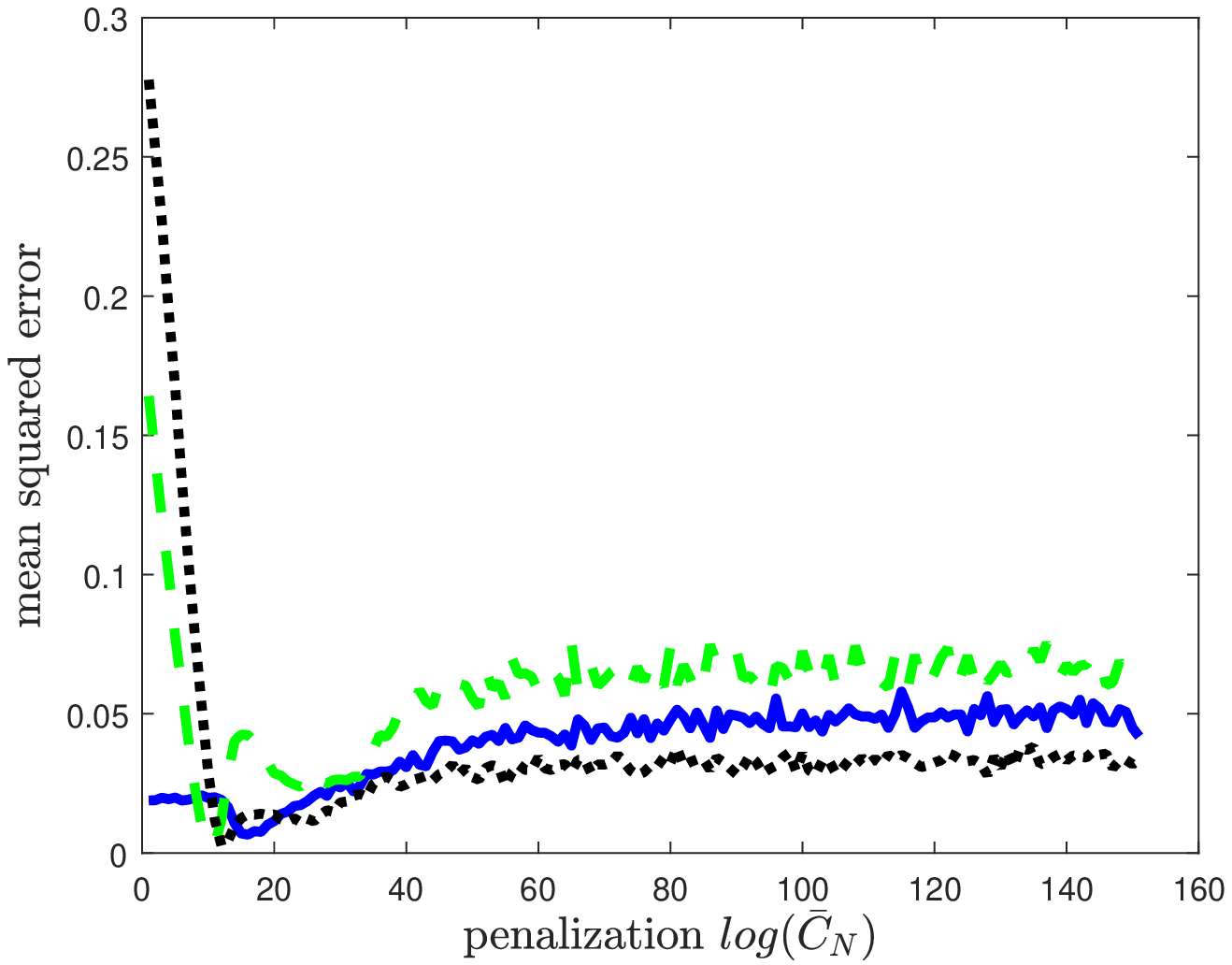}&
	 			\includegraphics[width=40mm, height=30mm]{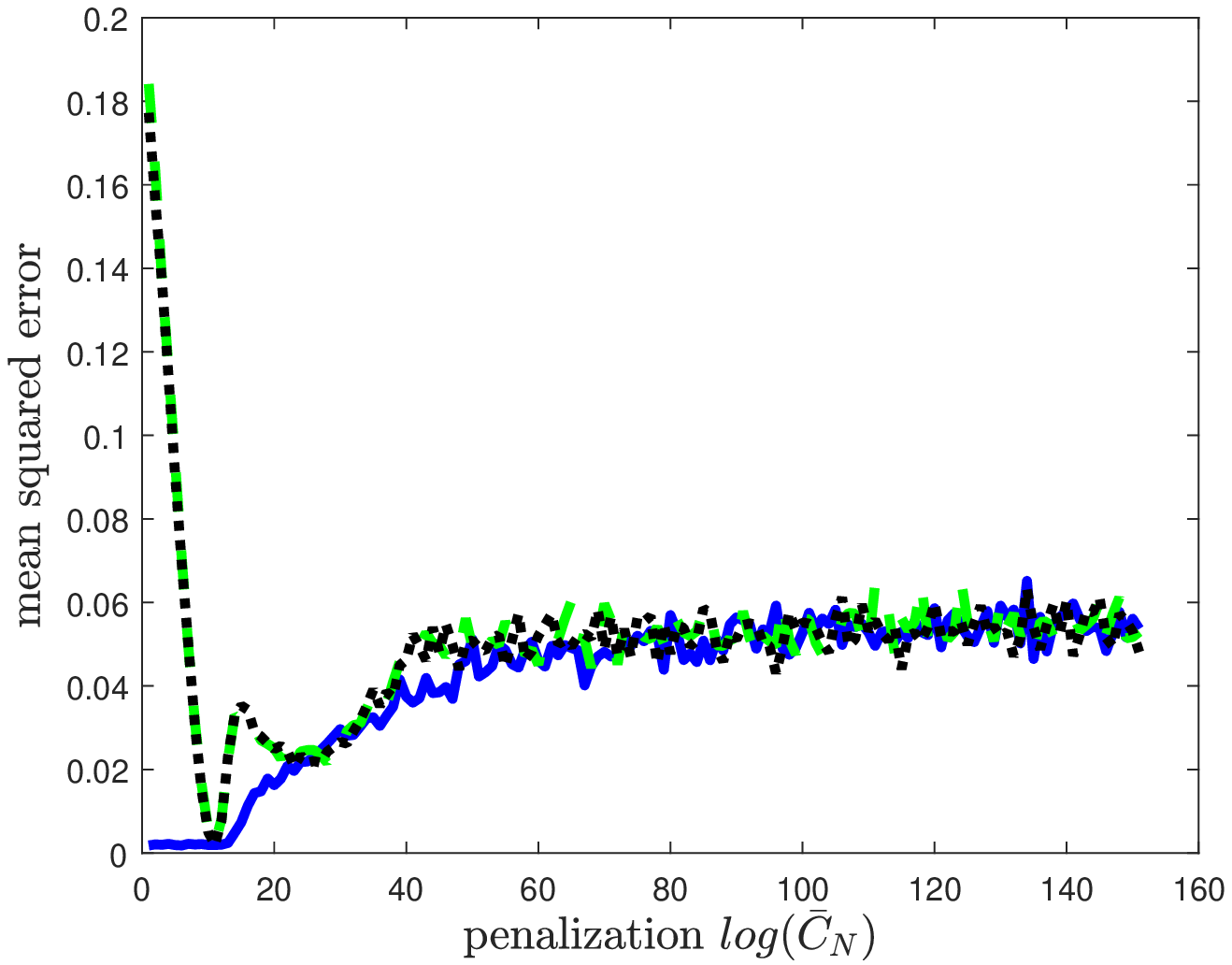} 	&	\includegraphics[width=40mm, height=30mm]{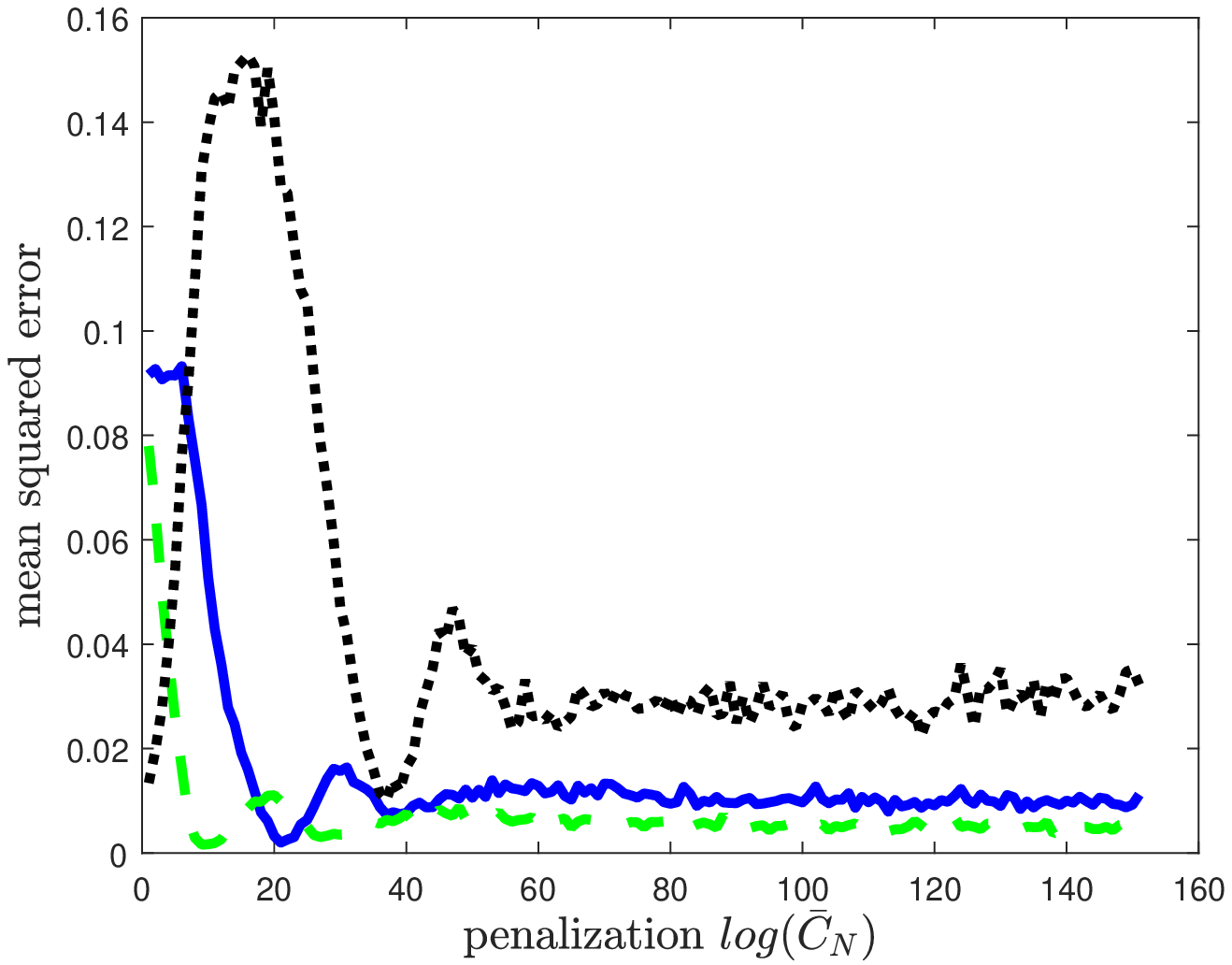}  %
	 			\\ 			   			 
	 		\end{tabular}%
	 	\end{center}
	 	\par
	 	\textit{{\footnotesize Notes: Simulated data from the deconvolution model $Y=X_1+X_2$. Log of penalization $\overline{C}_N$ (x-axis) against mean squared error (y-axis). $\underline{C}_N$ is set to $\overline{C}_N^{-1}$. Solid corresponds to the median, dashed to the 25\% quantile, dotted to the 75\% quantile. No average, single starting value, weak constraint. $N=100$ (top panel) and $N=500$ (bottom panel), $500$ simulations.}}
	 \end{figure}
	 
	 Next, we assess the impact of the penalization parameters $\overline{C}_N$ and $\underline{C}_N$ on the mean squared error of quantile estimates, at the median and 25\% and 75\% percentiles. In Figure \ref{Fig_MC_Deconv_regu} we show the results for the four specifications, when varying the logarithm of $\overline{C}_N$ between $0$ and $150$ and setting $\underline{C}_N=\overline{C}_N^{-1}$, for two sample sizes: $N=100$ (top panel) and $N=500$ (bottom panel). Two features emerge. First, setting $\overline{C}_N$ to a very large number, which essentially fully relaxes the constraints, still results in a well-behaved estimator. This is in contrast with popular regularization methods for ill-posed inverse problems such as Tikhonov regularization or spectral cut-off (e.g., Carrasco \textit{et al.}, 2007), for which decreasing the amount of penalization typically causes large increases in variance. The high sensitivity of characteristic-function based estimators to the choice of regularization parameters is also well documented. We interpret this feature of our estimator as reflecting the fact that the matching-based procedure induces an implicit regularization, even in the absence of additional constraints on parameters. Second, the results show that fully removing the penalization may not be optimal in terms of mean squared error. This raises the question of the optimal choice of the penalization parameters.

	 \begin{figure}[tbp]
	 	\caption{Monte Carlo results, deconvolution model, Efron-Koenker-Gu specification, $N=1000$ }\label{Fig_MC_Deconv_Efron}
	 	\begin{center}
	 		\begin{tabular}{cc}
	 			\\
	 			Average ($10\times$) & No average\\
	 			\includegraphics[width=60mm, height=40mm]{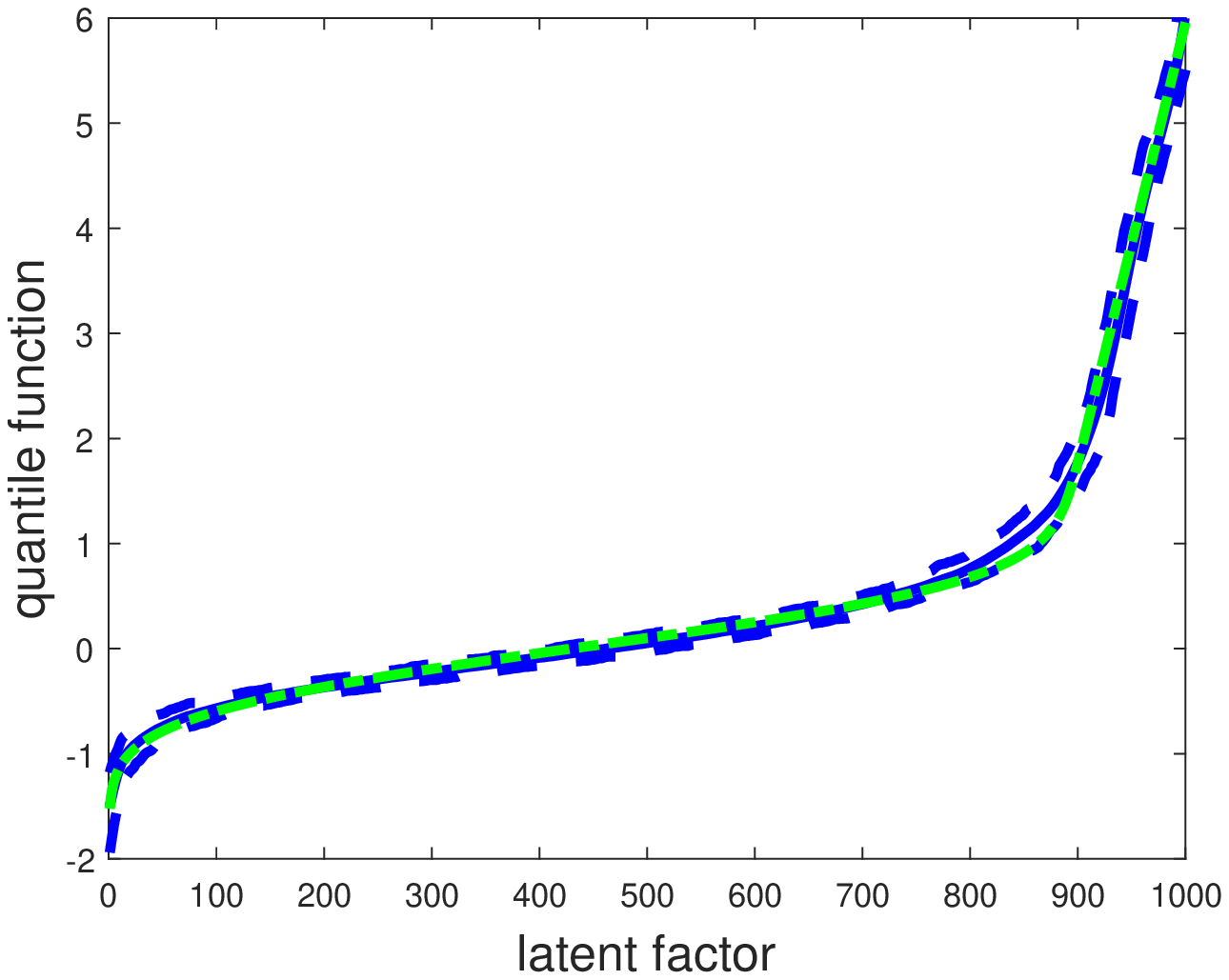} & %
	 			\includegraphics[width=60mm, height=40mm]{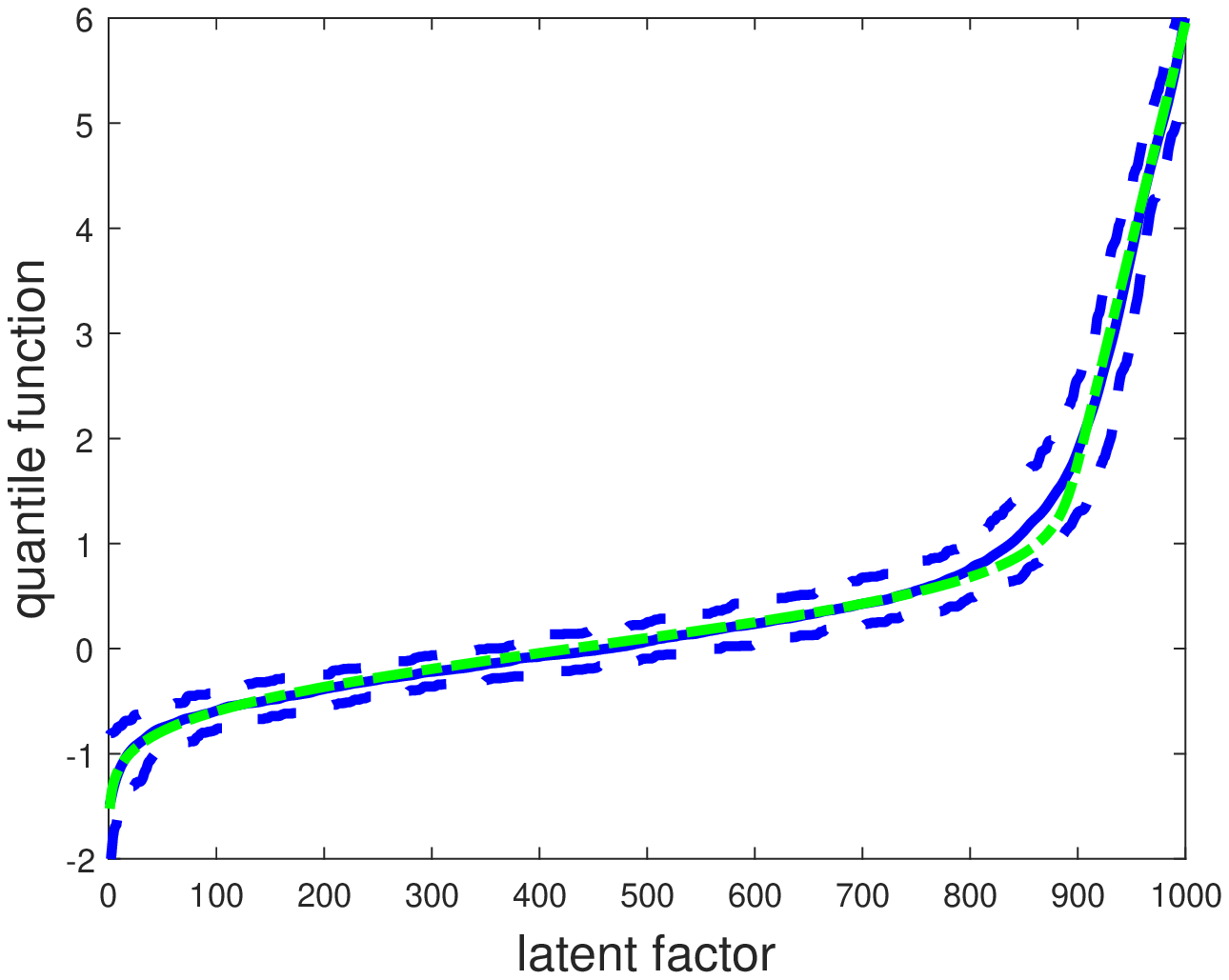}\\
	 			&\\
	 		\end{tabular}%
	 	\end{center}
	 	\par
	 	\textit{{\footnotesize Notes: Simulated data from the specification of the deconvolution model $Y=X_1+X_2$ used in Koenker and Gu (2019), which is a slight variation on a DGP used in Efron (2016). The mean across simulations is in solid, 10 and 90 percent pointwise quantiles are in dashed, and the true quantile function of $X_1$ is in dashed-dotted. Weak constraint. $100$ simulations.}}
	 \end{figure}

	 \begin{figure}[tbp]
	 	\caption{Monte Carlo results, deconvolution model, Beta(2,2), Mallows' (2007) algorithm }\label{Fig_MC_Deconv_Mallows}
	 	\begin{center}
	 		\begin{tabular}{cccc}
	 			\multicolumn{2}{c}{Quantile functions}&  \multicolumn{2}{c}{Densities}\\
	 			$N=100$ & $N=500$ & $N=100$ & $N=500$ \\
	 			\includegraphics[width=35mm, height=30mm]{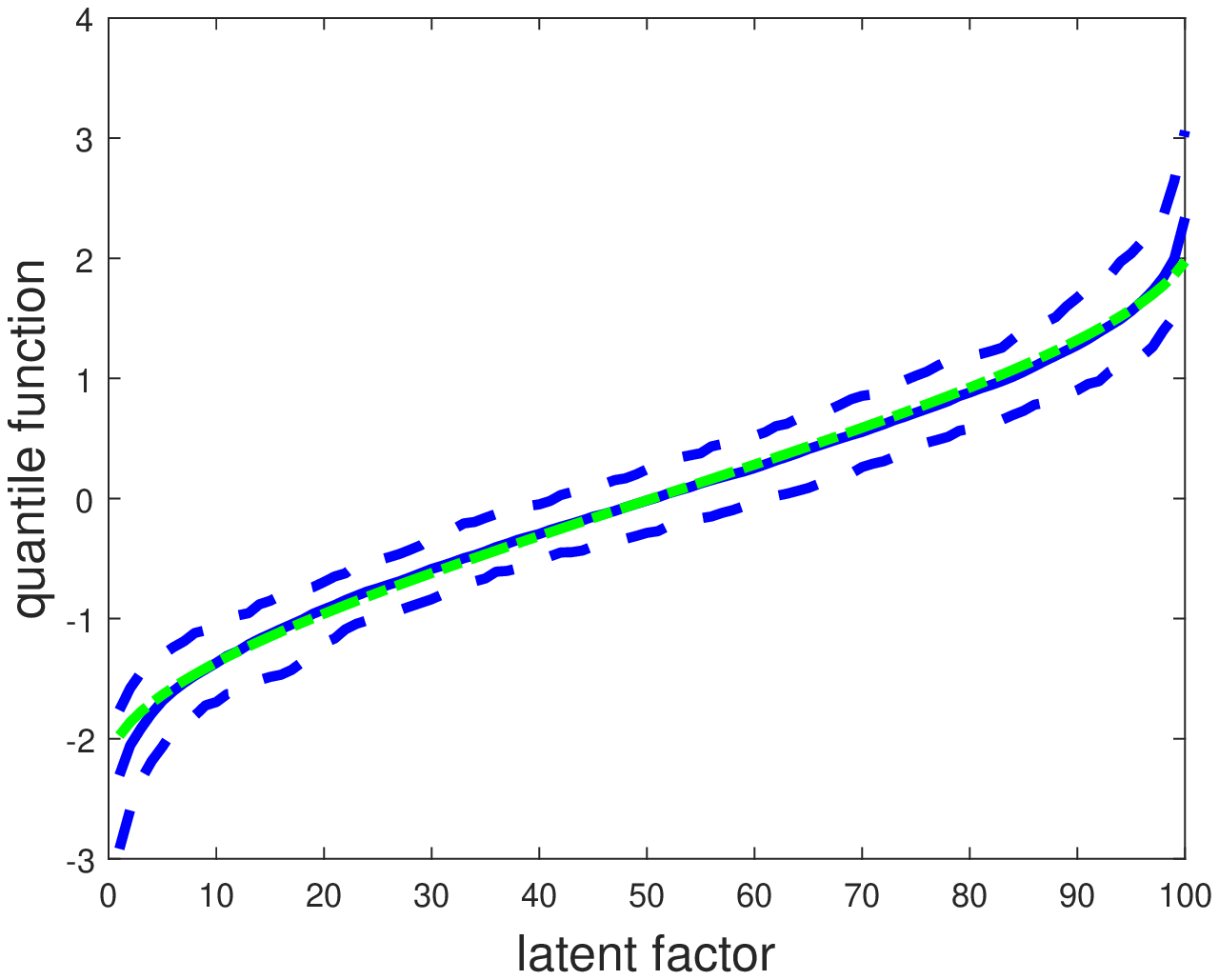} & %
	 			\includegraphics[width=35mm, height=30mm]{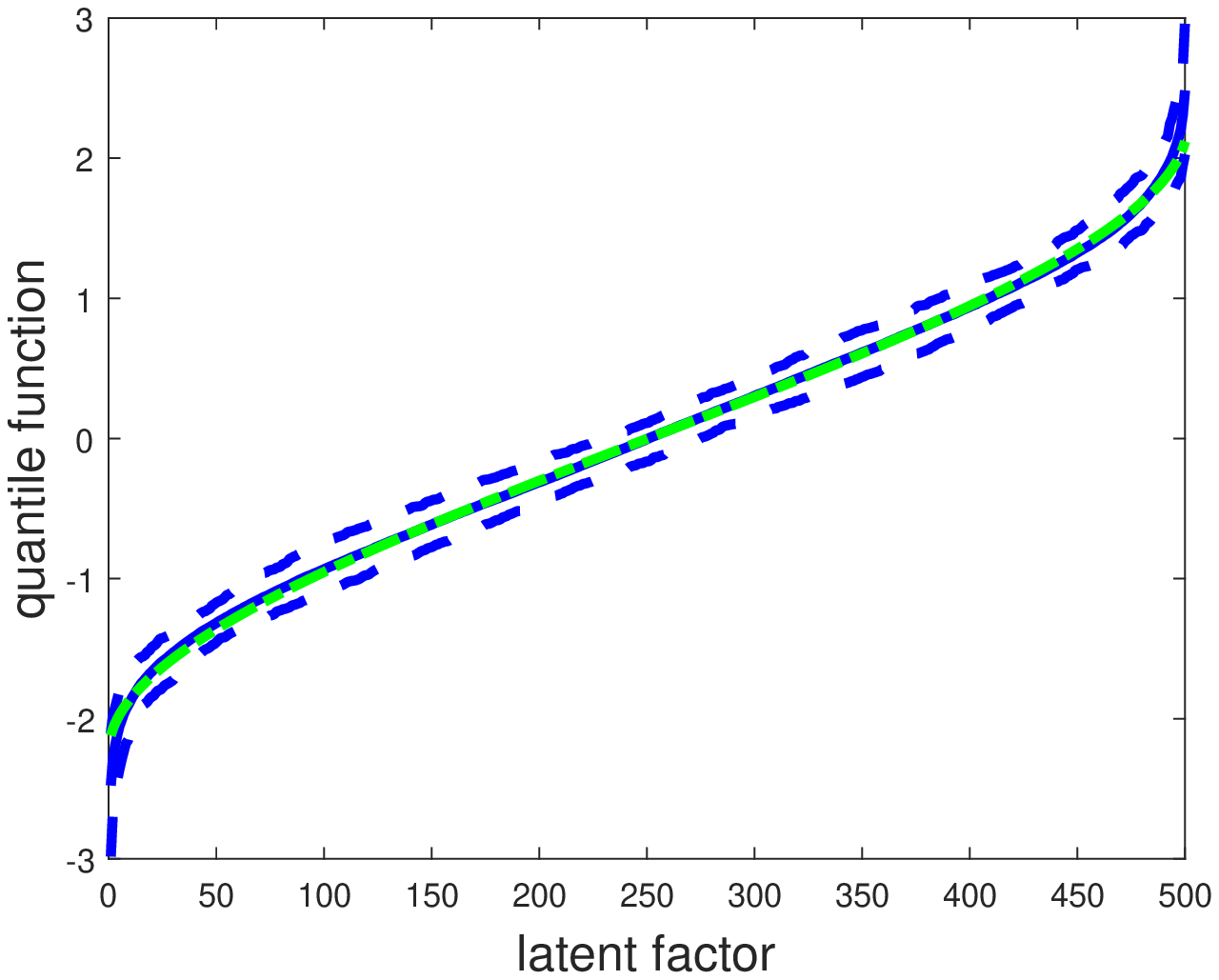} & 
	 			\includegraphics[width=35mm, height=30mm]{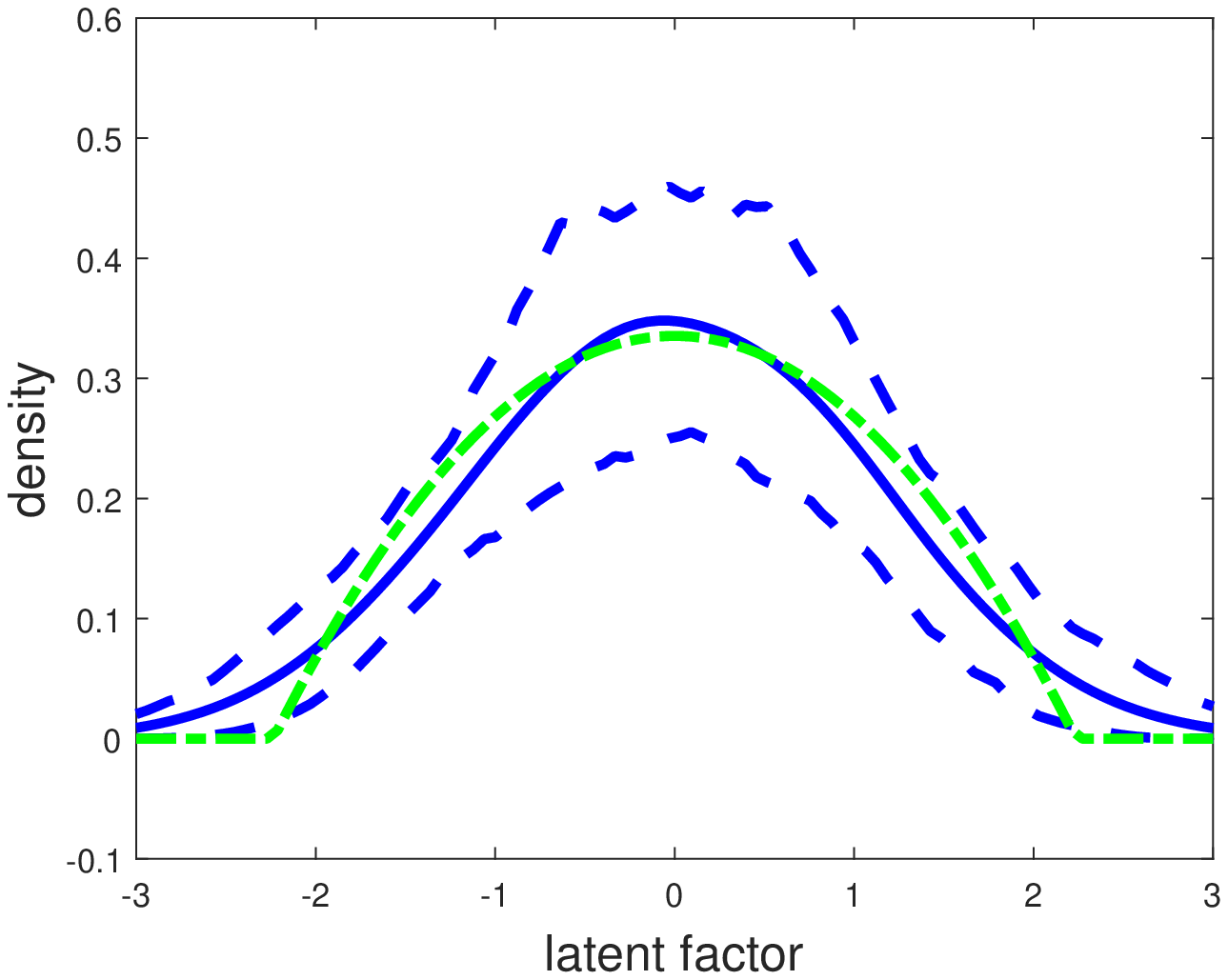} & %
	 			\includegraphics[width=35mm, height=30mm]{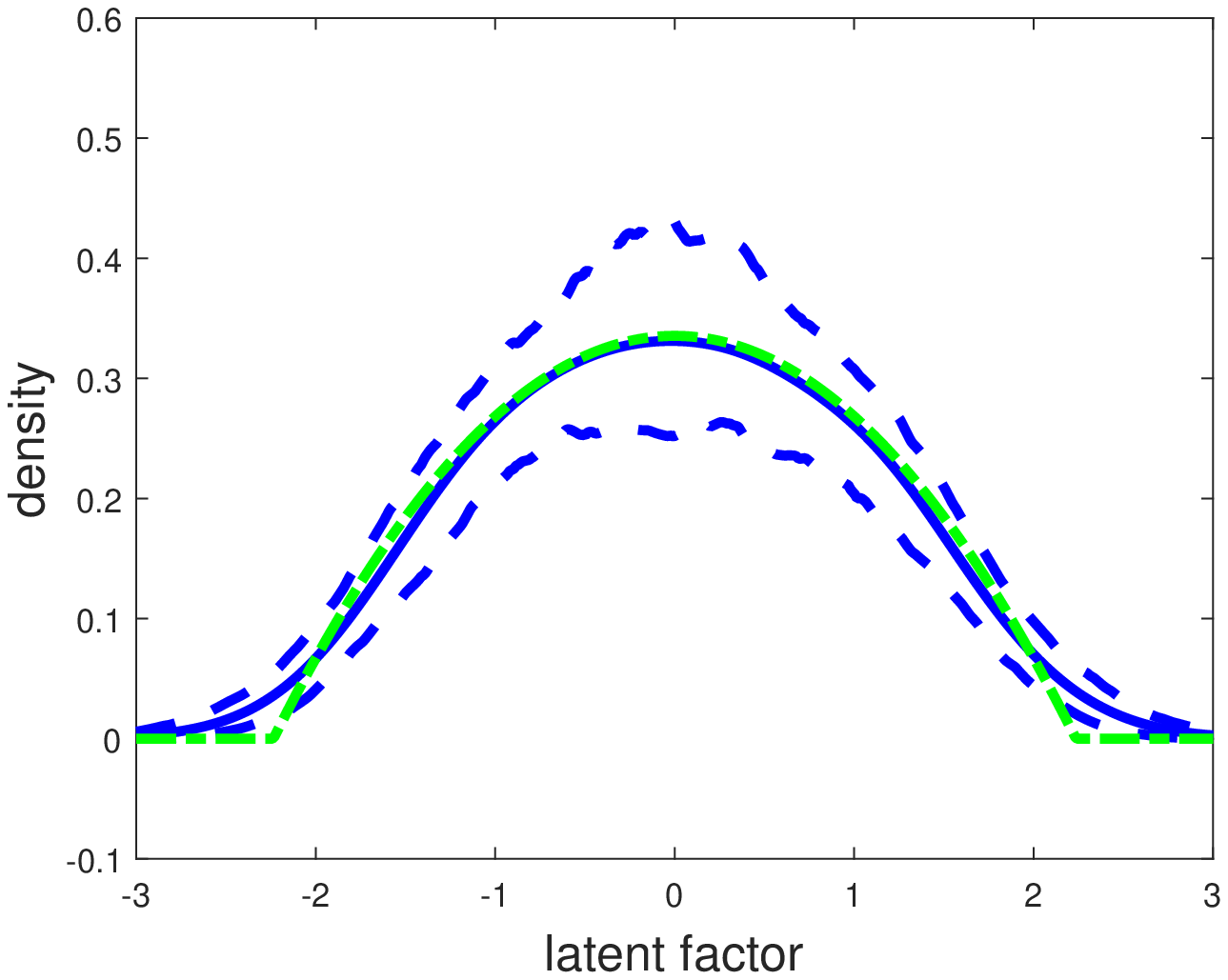}\\
	 			&  &\\
	 		\end{tabular}%
	 	\end{center}
	 	\par
	 	\textit{{\footnotesize Notes: Simulated data from the deconvolution model. The mean across simulations is in solid, 10 and 90 percent pointwise quantiles are in dashed, and the true density is in dashed-dotted. Mallows' (2007) algorithm. $100$ simulations.}}
	 \end{figure}

	 We next consider a data generating process (DGP) which has been previously used to assess the finite-sample behavior of several estimators in the nonparametric deconvolution model. This DGP was used in Koenker and Gu (2019), and it is a slight variation of a DGP introduced by Efron (2016). Let $Y=X_1+X_2$, where $X_2$ is distributed as a standard normal, and $X_1$ is distributed as a mixture of two distributions: a normal $\left(0,\frac{1}{2}\right)$ with probability $\frac{6}{7}$, and a uniform on the $[0,6]$ interval with probability $\frac{1}{7}$. Koenker and Gu report that the Stefanski and Carroll (1990) characteristic-function based estimator performs quite poorly on this DGP, distribution functions estimated on a sample of 1000 observations showing wide oscillations. In Figure \ref{Fig_MC_Deconv_Efron} we apply our estimator to this DGP, and report the results of 100 simulations. On the left graph we show quantile function estimates averaged 10 times, whereas on the right the results correspond to a single $\sigma$ draw per estimation. We see that nonparametric estimates are very close to the true quantile function. This performance stands in sharp contrast with that of characteristic-function based estimates, and is similar to the performance of the parametric estimator analyzed in Efron (2016).

	 Lastly, in Figure \ref{Fig_MC_Deconv_Mallows} we report simulation results for Mallows' (2007) stochastic estimator, in the case of the Beta$(2,2)$ specification. As we pointed out in Section \ref{Comput_sec}, this algorithm is closely related to ours, with the key difference that new random permutations are re-drawn in every step. We draw 100 such permutations, and keep the results corresponding to the last 50. The results are similar to the ones obtained using our estimator under the weak constraint, as can be seen by comparing Figures \ref{Fig_MC_Deconv_N500} and \ref{Fig_MC_Deconv_Mallows}.

	\section{Extensions\label{App_Additional}}
	
	In this section of the appendix we show simulation results for a nonparametric finite mixture model, and an empirical application of heteroskedastic deconvolution to the estimation of neighborhood effects in Chetty and Hendren (2018). 
	
	 \subsection{Simulations in a nonparametric finite mixture model}

\begin{figure}[tbp]
	 	\caption{Monte Carlo results, finite mixture model with two components}
	 	\label{Fig_FM_sim}
	 	\begin{center}
	 		\begin{tabular}{ccc}
	 			\multicolumn{3}{c}{Normal design}\\ 
	 			\includegraphics[width=50mm, height=40mm]{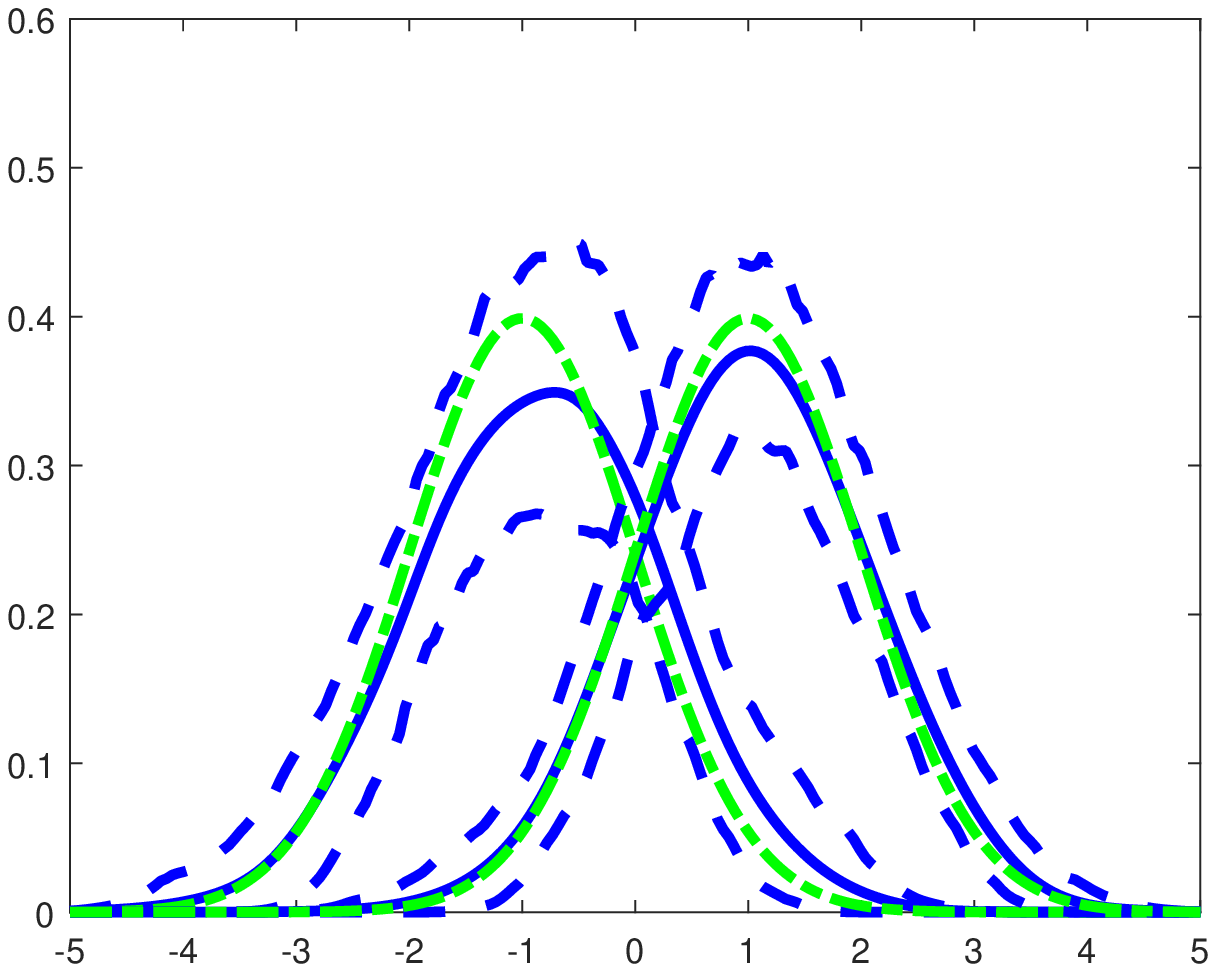} & 	\includegraphics[width=50mm, height=40mm]{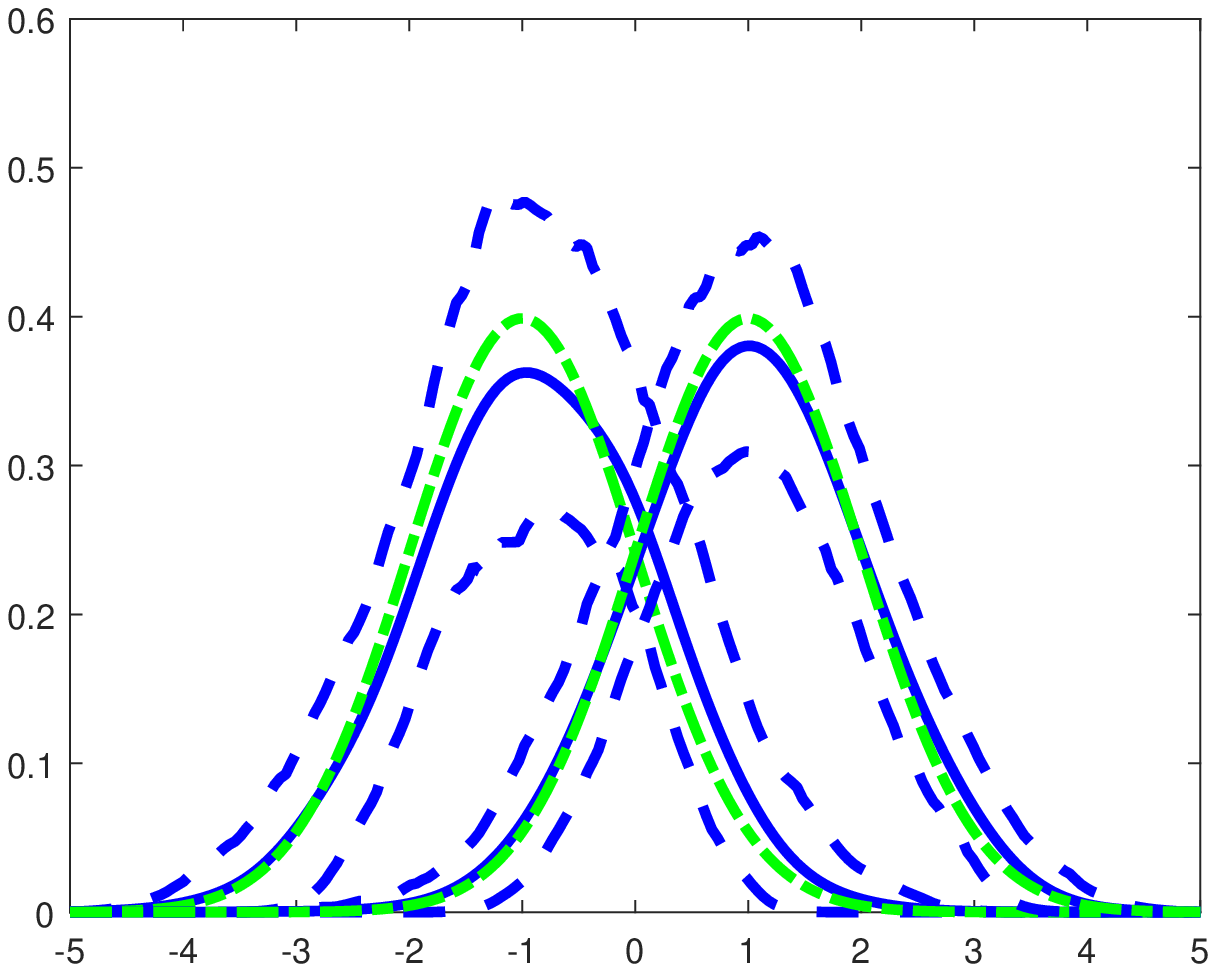} & 	\includegraphics[width=50mm, height=40mm]{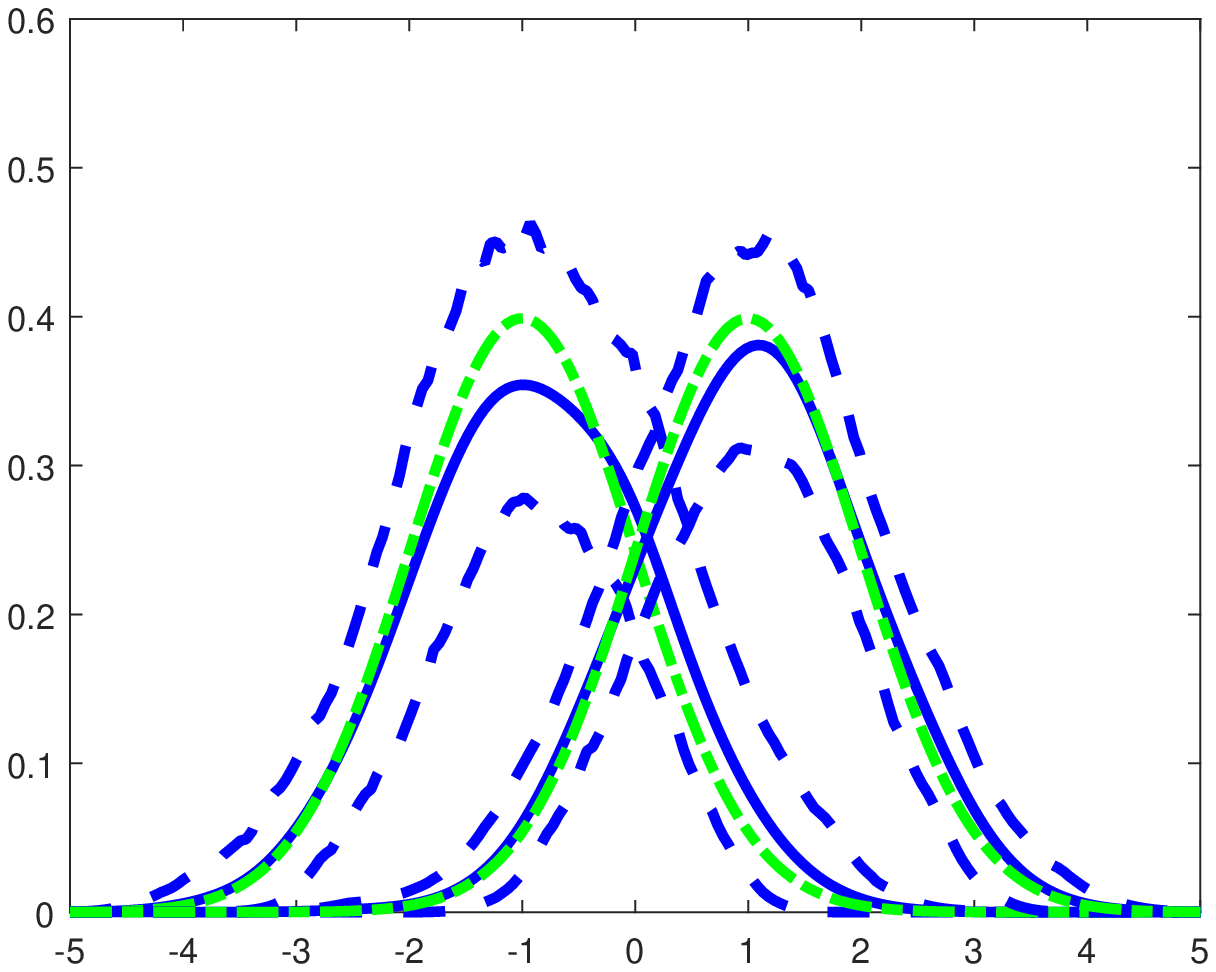}\\
	 			& &\\
	 			\multicolumn{3}{c}{Log-normal design}\\
	 			\includegraphics[width=50mm, height=40mm]{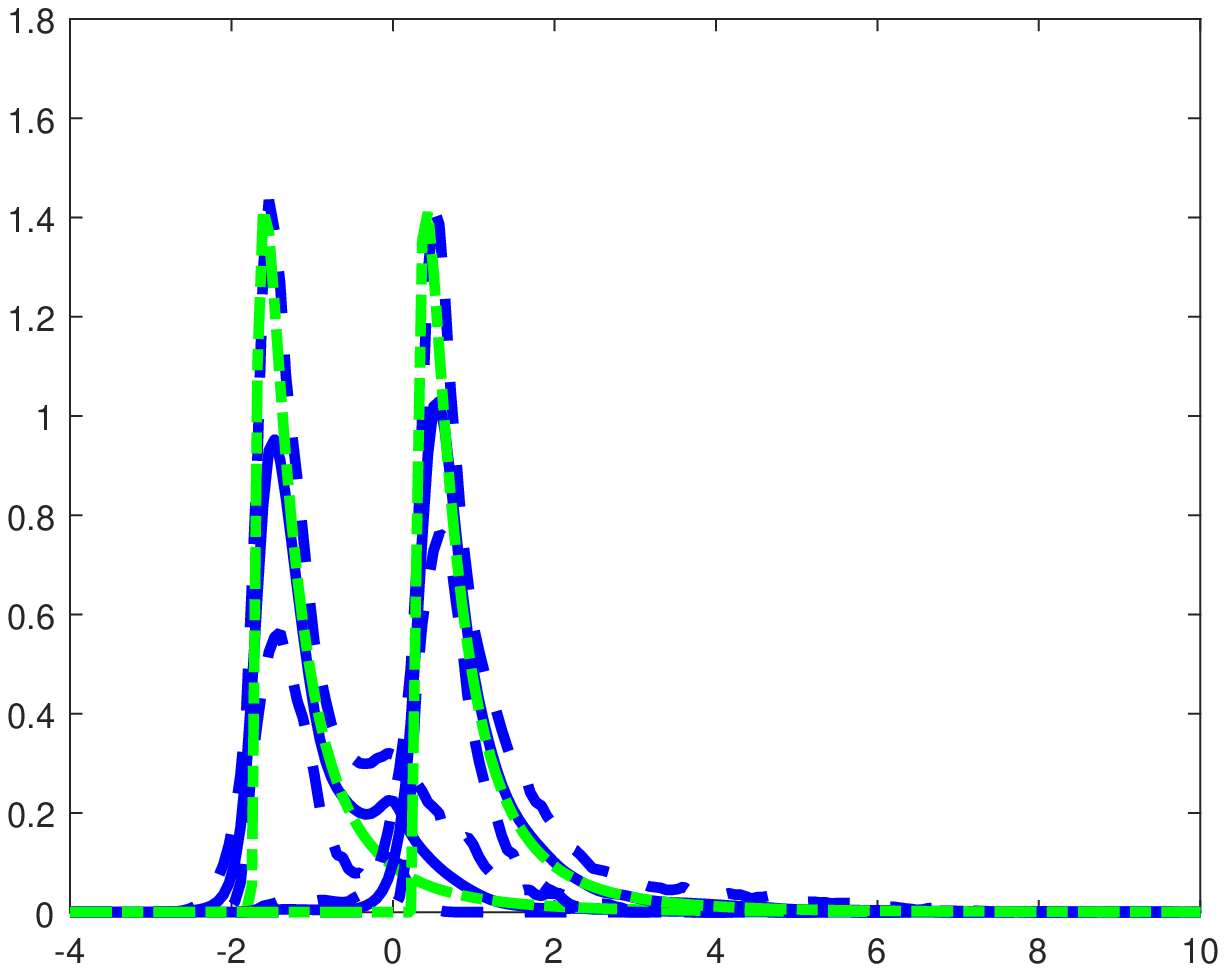} &\includegraphics[width=50mm, height=40mm]{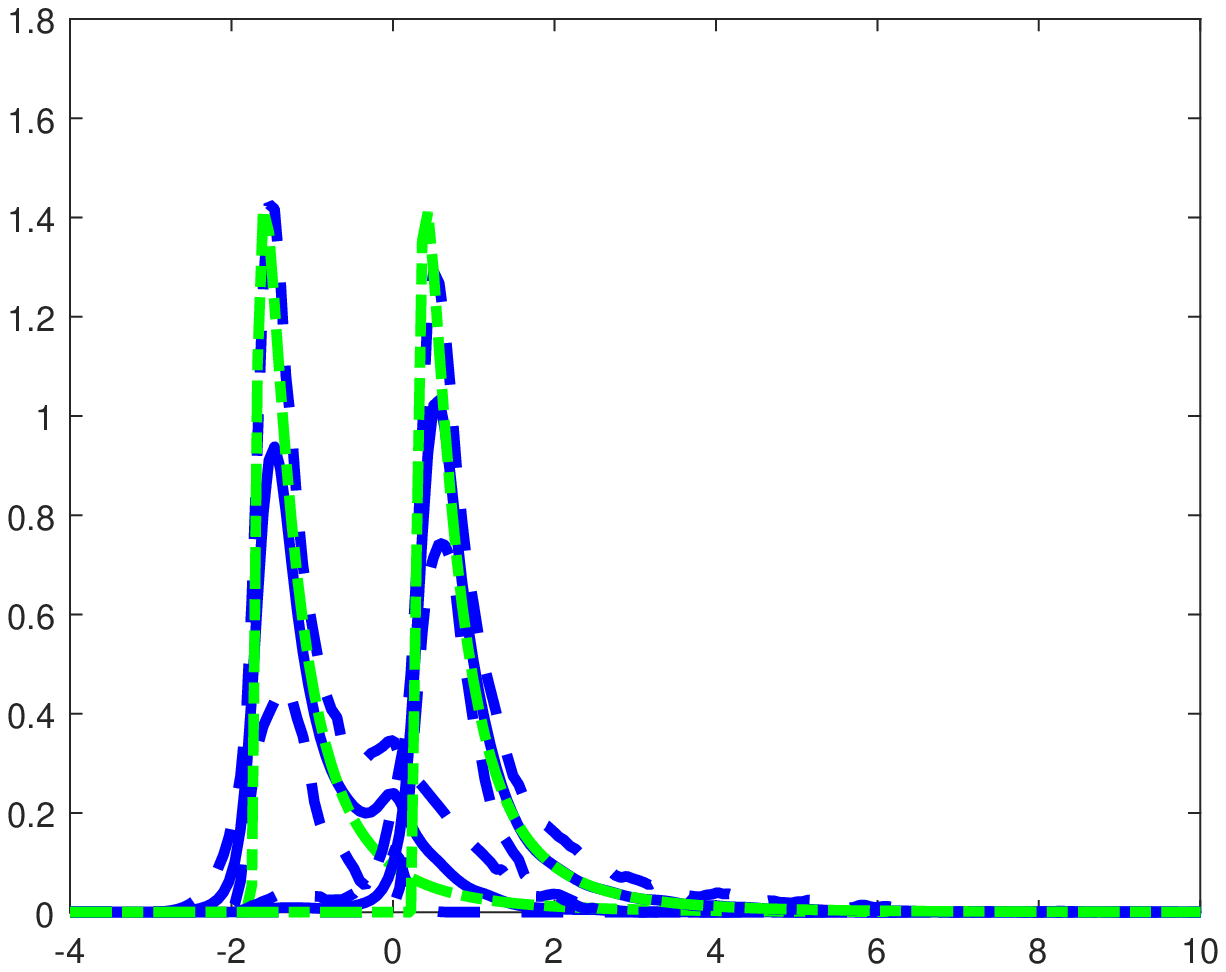}&\includegraphics[width=50mm, height=40mm]{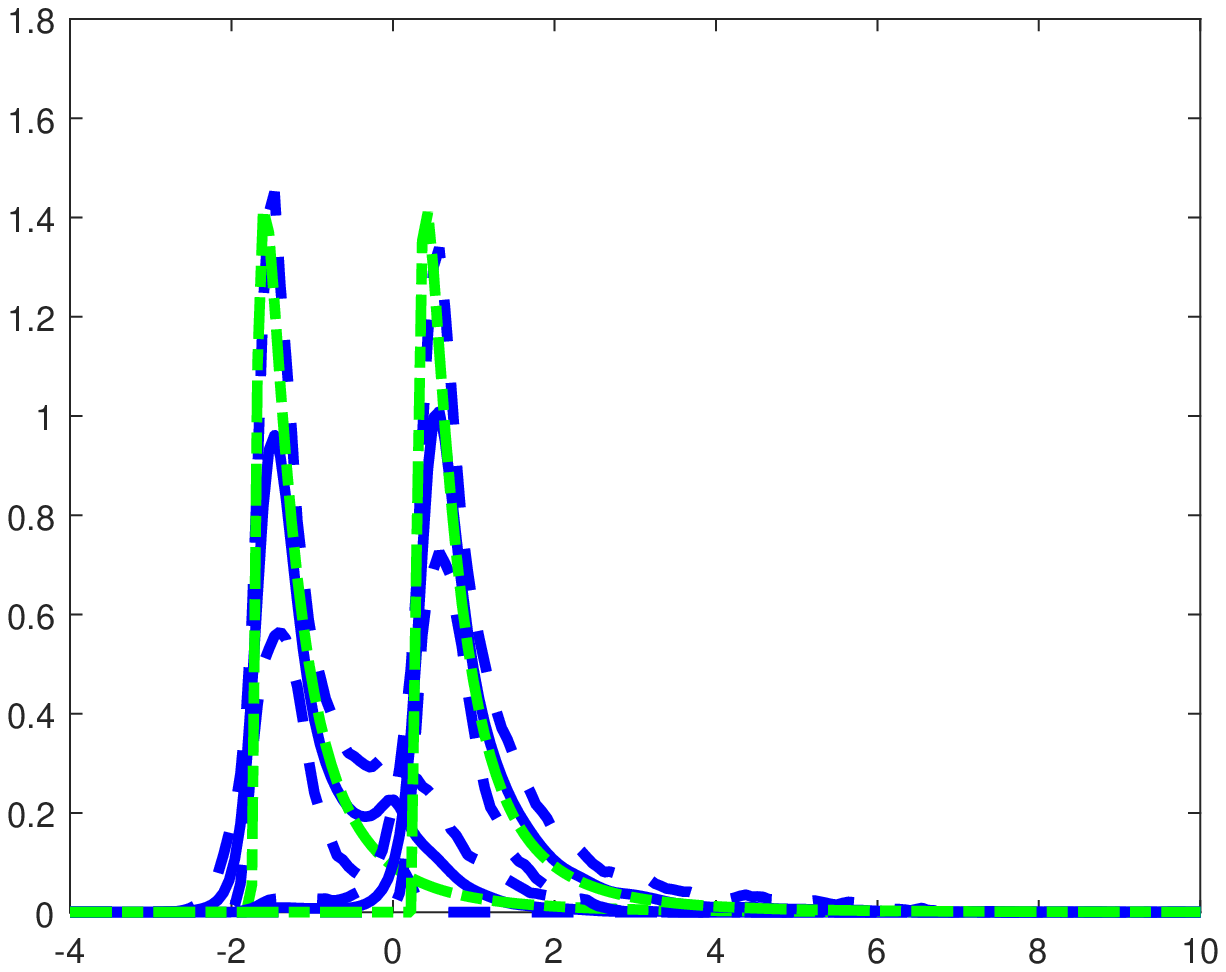}
	 		\end{tabular}%
	 	\end{center}
	 	\par
	 	\textit{{\footnotesize Notes: Simulated data from a finite mixture model with $G=2$ components. The mean across simulations is in solid, 10 and 90 percent pointwise quantiles are in dashed, and the true density is in dashed-dotted. The two components have means $-1$ and $1$ and unitary variances. Gaussian (top panel) and log-Gaussian (bottom panel) components. $N=100$, $T=3$, $100$ simulations. $R=10$ simulations per observation.}}
	 \end{figure}

		 In Figure \ref{Fig_FM_sim} we report the results of 100 simulations, for two DGPs, both of which are finite mixture models with $G=2$ components with independent measurements. We consider a normal DGP and a log-normal DGP. To fix the labeling across simulations, we order the components by increasing means. We use a version of (\ref{est_FM_model_simu}) with multiple draws $\sigma_{gt}(i,r)$ for all $i$, with $R=10$ simulations by observation. We use $3$ starting values in every inner loop, and perform an outer loop for $10$ equidistant values of the first group's probability. The results in Figure \ref{Fig_FM_sim} are encouraging, and suggest that matching estimators can perform well in nonparametric finite mixture models too.

	\subsection{Neighborhood effects in the US}

	Here we estimate the density of neighborhood effects across US commuting zones, using data made available by Chetty and Hendren (2018). For every commuting zone, Chetty and Hendren report an estimate $Y_i$ of the causal income effect of $i$, alongside an estimate $\widetilde{S}_i$ of its standard error. We compute a heteroskedastic Gaussian deconvolution estimator of the density of the latent neighborhood effects. As a by-product, we obtain an estimate of the joint density of neighborhood effects and their standard errors. To implement the calculation we set $\lambda=10$, trim the top 1\% percentile of $\widetilde{S}_i$, and weigh all results by population weights. To accommodate the presence of weights in a simple way, we draw subsamples of 500 observations from the weighted empirical distribution of $(Y_i,\widetilde{S}_i)$. We then average the results across $M=10$ subsamples.

	\begin{figure}[tbp]
		\caption{Density of neighborhood effects\label{Fig_Neighb}}
		\begin{center}
			\begin{tabular}{cc}
				Marginal density & Joint density \\
				\includegraphics[width=80mm, height=60mm]{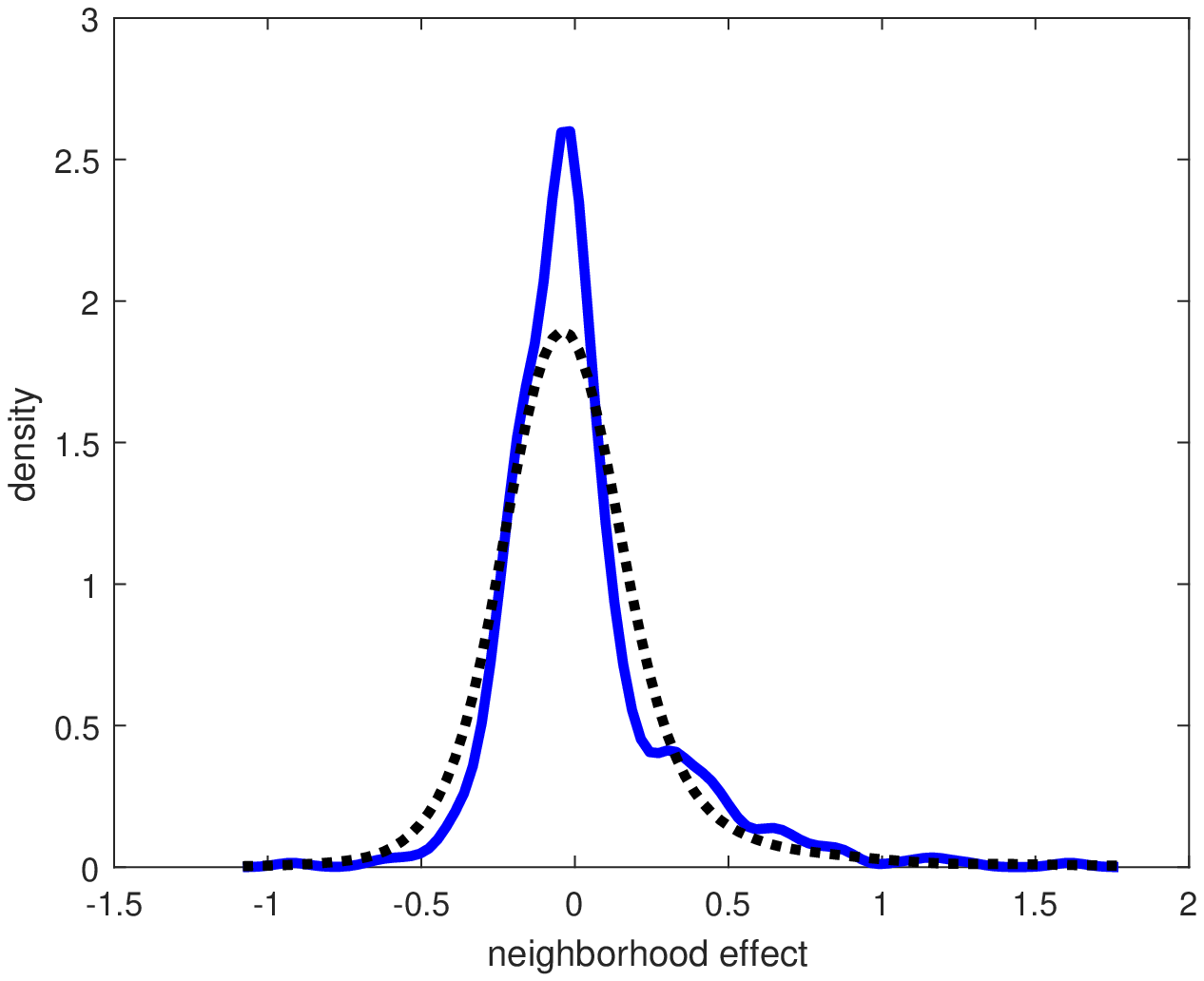} & %
				\includegraphics[width=80mm, height=60mm]{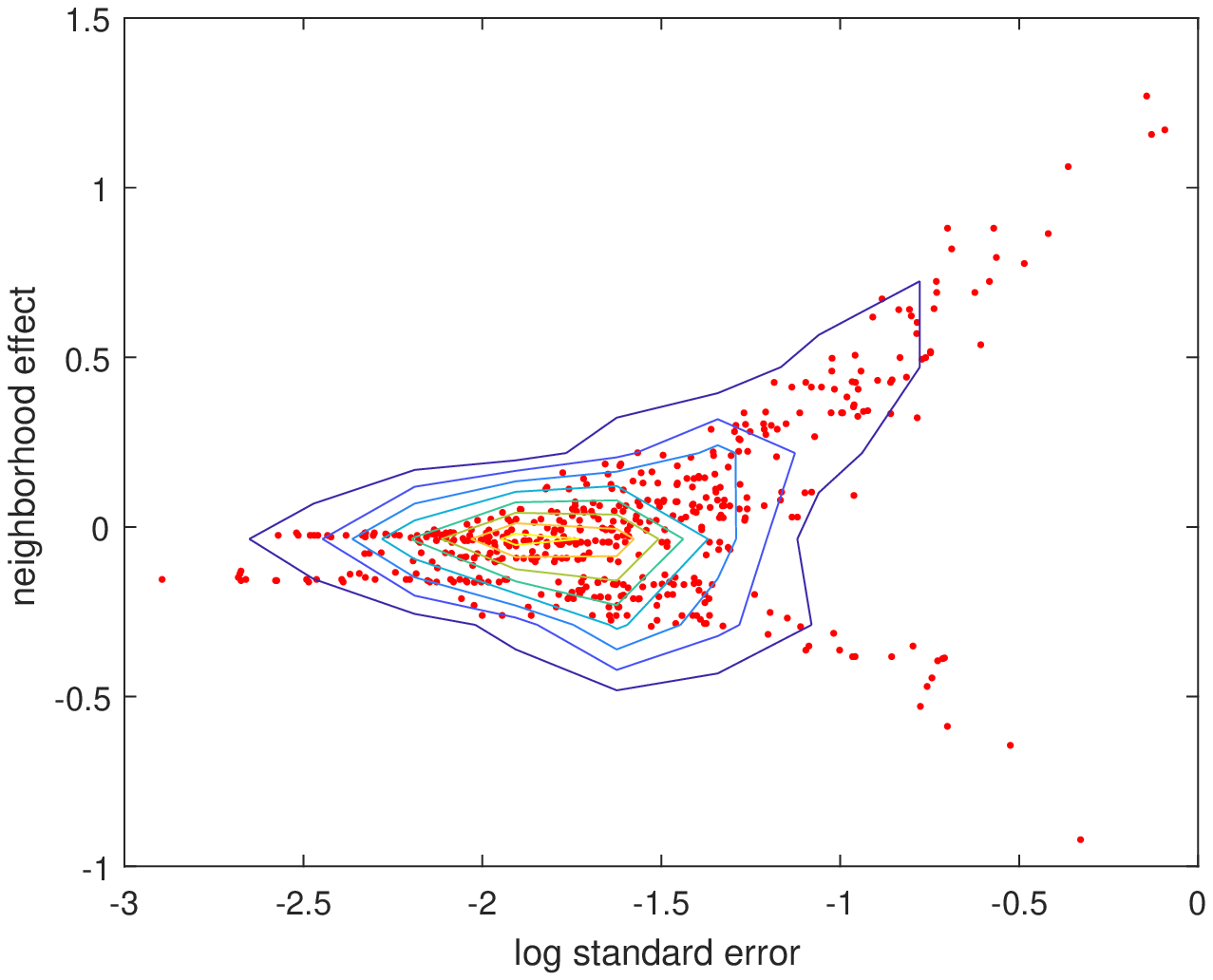} \end{tabular}%
		\end{center}
		\par
		\textit{{\footnotesize Notes: In the left graph we show the density of commuting zone effects $X_{i1}$ in model (\ref{eq_CH}) in solid, and the density of neighborhood fixed-effects $Y_i$ in dashed. In the right graph we show contour plots of the joint density of $(X_{i1},S_i)$, where $S_i$ is the standard deviation of $Y_i$. Calculations are based on statistics available on the Equality of Opportunity website.}}
	\end{figure}

		We show the results in Figure \ref{Fig_Neighb}. We see that neighborhood effects are not normally distributed. They show right skewness, and excess kurtosis. Estimates of Bowley-Kelley skewness and Crow-Siddiqi kurtosis of $X_{i1}$ are 0.33 and 4.75, respectively. This evidence of non-normality confirms results obtained by Bonhomme and Weidner (2019) using posterior estimators. The joint density of neighborhood effects and standard errors suggests that less populated commuting zones with less precise estimates tend to have higher income premia. The rank correlation between neighborhood effects and standard errors is 0.39. The joint density also shows a high degree of non-Gaussianity.

	\end{document}